\documentclass[a4paper,twoside, nofootinbib,10pt]{book}
\usepackage{amsmath,amssymb,amsthm}
\usepackage{graphicx}
\usepackage[francais]{babel}
\usepackage{psfrag}
\usepackage{epsfig,epic,eepic}

\allowdisplaybreaks

\def\ni{\noindent}

\newcommand{\Ref}[1]{(\ref{#1})}
\def\tr{\textrm}

\def\beq{\begin{equation}}
\def\eeq{\end{equation}}
\def\beqs{\begin{eqnarray}}
\def\eeqs{\end{eqnarray}}

\def\lrarr{\leftrightarrow}
\def\arr{\rightarrow}
\def\darr{\Longleftrightarrow}
\def\dmu{\tr{d}\mu}
\def\tl{\tilde}
\def\wtl {\widetilde}
\def\G{{\cal G}}
\def\Gtl{\wtl{\G}}
\def\H{{\cal H}}
\def\Htl{\wtl{\H}}
\def\gt{G^{(T)}}
\def\gu{G^{(U)}}
\def\bA{\bar{A}}
\def\om{\omega}
\def\C{{\mathbb C}}
\def\R{{\mathbb R}}
\def\N{{\mathbb N}}
\def\Z{{\mathbb Z}}
\def\su{{\rm SU}(2)}
\def\slc{{\rm SL}(2,\C)}
\def\slr{{\rm SL}(2,\R)}

\def\A{{\cal{A}}}
\def\G{{\cal{G}}}
\def\I{{\cal I}}

\def \la{\langle}
\def \ra{\rangle}
\def\s#1{|#1\rangle}
\def\cs#1{\langle#1|}
\def\sc#1#2{\langle#1|#2\rangle}
\def \f{\frac}
\def \w{\wedge}
\def \imm{\gamma}
\def \dd{\partial}
\def \hE{\hat{E}}
\def \sh{\sinh}
\def \ch{\cosh}
\def \what{\widehat}
\newcommand{\matt}[2]{\left(\begin{array}{#1} #2 \end{array}\right)}
\def \eps{\epsilon}

\newcommand{\Cm}[3]{C^{j_{#1}j_{#2}j_{#3}}_{m_{#1}m_{#2}m_{#3}}}
\newcommand{\Cg}[4]{C^{j_{#1}j_{#2}j_{#3}}_{#4_{#1}#4_{#2}#4_{#3}}}
\newcommand{\Cq}[6]{C^{J_{#1}J_{#2}J_{#3}J_{#4}#6}_{#5_{#1}#5_{#2}#5_{#3}#5_{#4}}}

\def\lg{{\it Loop Gravity} }
\def\lqg{{\it Loop Quantum Gravity} }
\def\rg{relativit{\'e} g{\'e}n{\'e}rale }

\def \cA{{\cal A}}
\def \bA{{\bf A}}
\def \harr{\hookrightarrow}

\newcommand{\mat}[4]{
\left( \begin{array}{cc} #1 &#2\\ #3 &#4\end{array} \right)
}

\def\PBOX[#1]#2{\mbox{\setlength{\unitlength}{#1 pt} #2}}
\def\CBOX[#1]#2{\begin{array}{c} \PBOX[#1]{#2} \end{array}}
\def\MPIC[#1,#2]#3{{\begin{picture}(#1,#2) {#3} \end{picture}}}
\def\FIGThetaNet#1#2#3{{\MPIC[40,40]{{
        \put(14,32){$#1$}
        \put( -5,15){\line(1,0){40}} \put(14,17){$#2$}
        \put(16,15){\oval(40,30)}   \put(14, 2){$#3$}
        \put( -4,15){\circle*{3}}    \put(36,15){\circle*{3}}
}}}}
\def\looptwo#1#2{{\MPIC[40,40]{{
        \put(14,32){$#1$}
        \put(16,15){\oval(40,30)}   \put(14, 2){$#2$}
        \put( -4,15){\circle*{3}}    \put(36,15){\circle*{3}}
}}}}
\def\loopone#1{{\MPIC[40,40]{{
        \put(14,32){$#1$}
        \put(16,15){\oval(40,30)}
        \put( -4,15){\circle*{3}}
}}}}

\newcommand{\doubleY}[5] 
{
 { \raise6pt\hbox{$#1$}\atop 
\lower6pt\hbox{$#2$}}
  { \hbox to20pt{$\hfill #5\hfill $}\overwithdelims>< {} }
 { \raise6pt\hbox{$#3$}\atop 
\lower6pt\hbox{$#4$}}
 }

\newtheorem{theo}{Th\'eor\`eme}
\newtheorem{lemma}{Lemma}
\newtheorem{prop}{Proposition}
\newtheorem{defi}{D\'efinition}

\begin{document}





\thispagestyle{empty}

\begin{center}

\huge{\bf TH\`ESE}\\

\vskip 1truecm

\large{pr\'esent\'ee par}\\

\vskip 0.5truecm

\Large{\bf Etera R. LIVINE}\\

\vskip 0.4truecm

{\large (Juin 2003)}

\vskip 0.6truecm

\large{pour obtenir le grade de}\\

\vskip 0.5truecm

\Large{Docteur de l'Universit\'e de la M\'editerrann\'ee}\\

\vskip 0.5truecm

\large{\underline{Sp\'ecialit\'e:}}\\

\vskip 0.5truecm

\large{Physique des Particules, Physique Math\'ematique et Mod\'elisation}\\

\vskip 0.5truecm {\bf \'Ecole Doctorale:} Physique et Science de
la Mati\`ere

\vskip 1.5truecm

{\bf \LARGE Boucles et Mousses de Spin en Gravit\'e Quantique:}\\

\vspace{3mm}

{\bf \large une Approche Covariante  
\`a la Quantification Non-Perturbative}\\

\vspace{1mm}

{\bf \large de la Relativit\'e G\'en\'erale}

\vskip 2truecm
\large{Soutenue le 24 Juin 2003, devant le Jury
compos\'e de}

\end{center}

\begin{center}
{\large
\begin{tabular}{l}
K. Gawedzki (Pr\'esident du Jury) \\
J. Barrett (Rapporteur) \\
L. Freidel \\
P. Roche (Rapporteur)\\
C. Rovelli (Directeur de th\`ese) \\
R. Williams
\end{tabular}}
\end{center}
\vspace{1cm}


\begin{center}
CENTRE DE PHYSIQUE TH\'EORIQUE
\\ CNRS-UPR 7061
\end{center}
\newpage

\thispagestyle{empty} ${}$
\newpage

{\bf\LARGE Remerciements}

\vspace{2cm}

Je tiens \`a remercier toutes les personnes qui m'ont entour\'e durant
ces trois ann\'ees de th\`ese et plus g\'en\'eralement
tous les membres du CPT pour m'avoir accueilli.

Tout d'abord, j'aimerais remercier Carlo Rovelli, mon directeur de th\`ese,
pour ses conseils, ses encouragements et son constant
int\'er\^et pour mon travail,
et Lee Smolin pour son encadrement durant ma premi\`ere ann\'ee de th\`ese
et pour m'avoir fait d\'ecouvrir le monde de la recherche.

J'ai eu le plaisir de collaborer avec Laurent Freidel, qui
m'a beaucoup appris et aid\'e au cours de ma th\`ese.
Je lui suis tr\`es reconnaissant pour sa patience, sa motivation
et son enthousiasme.
Je remercie \'egalement chaleureusement mon ami Daniele Oriti,
pour nos nombreuses sessions
de travail, qui m'ont beaucoup apport\'e, ponctu\'ees
d'aussi nombreuses parties de ping-pong, toujours aussi relaxantes.

Je tiens \`a remercier tout particuli\`erement
Florian Girelli, qui a eu la patience de (re)lire
toute ma th\`ese, pour ses conseils et les nombreuses discussions.
Je dois beaucoup aux discussions que j'ai eu avec Karim Noui
et Philippe Roche, et avec mes collaborateurs
Serge\"i Alexandrov, Robert Oeckl et David Louapre. Et
je tiens \`a remercier mon co-bureau Rapha\"el Zentner pour
la bonne ambiance de bureau "productive".

J'aimerais \'egalement remercier mes amis Simone, toujours pr\^et
\`a faire une partie de Worms, Lo\"ic et S\'everine, pour leur 
hospitalit\'e \`a Lyon,
Mike, qui m'a si souvent h\'eberg\'e \`a Cambridge, et Anahita,
pour son continuel soutien.

\medskip

Je tiens \`a remercier mes parents, Calixte et Jenny, pour m'avoir toujours
soutenu et aid\'e, et mes (toujours petites) soeurs Daphn\'e et Charlotte,
toujours aussi dr\^oles.

Finalement, je remercie de tout coeur Anne-Laure pour avoir
\'et\'e l\`a, pour son soutien et son \'ecoute, pour sa joie
et son enthousiasme, qui m'ont permis de m'enfuir de ma th\`ese
chaque fois que je me suis retrouv\'e submerg\'e par mon travail,
pour sa patience, qui lui a permis de me supporter...

\newpage

\thispagestyle{empty} ${}$
\newpage

{\bf\LARGE R\'esum\'e}

\vspace{2cm}

{\bf \Large Boucles et Mousses de Spin en Gravit\'e Quantique:}

\vspace{2mm}

{\large Une approche covariante \`a la quantification
non-perturbative de la relativit\'e g\'en\'erale}

\vspace{1cm}

La {\it gravit\'e \`a boucles} est une quantification canonique
de la gravit\'e 4d. Les \'etats quantiques de g\'eom\'etrie 3d
sont les {\it r\'eseaux de spin} $SU(2)$ et ont une interpr\'etation
naturelle en termes de vari\'et\'e triangul\'ee.
Une {\it mousse de spin} est une histoire de tels r\'eseaux:
c'est un formalisme d'int\'egrale de chemin et permet
d'\'etudier la dynamique des mod\`eles. Le {\it mod\`ele de
Barrett-Crane} est le mod\`ele le plus prometteur. Il est
issu des th\'eories topologiques et de quantification
g\'eom\'etrique, et utilise les repr\'esentations du
groupe de Lorentz $SL(2,C)$.

Je d\'eveloppe les r\'eseaux de spin pour groupes non-compacts
tel le groupe de Lorentz. Ceci s'applique \`a la gravit\'e
quantique 3d et je d\'erive un spectre des longueurs,
continu pour distance spatiale et discret pour intervalle
temporel. En 4d, j'explore la g\'eom\'etrie quantique du mod\`ele,
d\'ecris le r\^ole de la causalit\'e et d\'erive la th\'eorie
canonique correspondante utilisant des r\'eseaux de spin
$SL(2,C)$ dont je discute l'interpr\'etation en tant que g\'eom\'etrie 3d.

\vspace{1cm}

mots cl\'es: \\
relativit\'e g\'en\'erale, gravit\'e quantique,
gravit\'e \`a boucles, quantification canonique,
r\'eseaux de spin, mousses de spin

\vspace{1.5cm}

{\bf \large Loop Gravity and Spin Foam:}

\vspace{2mm}

{\large Covariant methods for the non-perturbative
quantization of general relativity}

\vspace{1cm}

{\it Loop gravity} is a canonical quantization of 4d gravity.
Quantum states of 3d geometry are the $SU(2)$ {\it spin networks}
and have a natural interpretation as triangulated manifolds.
{\it Spin Foams} are histories of spin networks: it is a path
integral formulation, allowing to study the dynamics of
the theory. The {\it Barrett-Crane model} is the most promising model.
It originates in topological field theory, uses geometrical
quantization and is based on representations of the Lorentz
group $SL(2,C)$.

I develop spin networks for non-compact groups such as the
Lorentz group. I apply these techniques to 3d quantum gravity
and derive a length spectrum, which turns out to be continuous
for spacelike distances and discrete for timelike intervals.
In 4d, I explore the details of the quantum geometry of the
model, analyse the issue of causality and derive the corresponding
canonical theory based on $SL(2,C)$ spin networks whose
interpretation as 3d geometry I discuss.

\vspace{1cm}

key words:\\
general relativity, quantum gravity,
loop gravity, canonical quantization, spin networks, spin foams

\newpage
\thispagestyle{empty} ${}$

\tableofcontents
\newpage


\newpage

{\renewcommand{\thechapter}{}\renewcommand{\chaptername}{}
\addtocounter{chapter}{-1}
\chapter{Introduction}\markboth{\sl INTRODUCTION}{\sl INTRODUCTION}}

\vspace{-1cm}

\hspace{7.5cm}
\begin{minipage}{10cm}
{\small 
\noindent
{\it Douter de tout ou tout croire, \\
ce sont deux solutions \'egalement commodes, \\
qui l'une et l'autre nous
dispensent de r\'efl\'echir.}
\vspace{1mm}

Poincar\'e, {\bf la Science et l'Hypoth\`ese}

\vspace{3mm}

\noindent
{\it Quiconque s'est jamais avis\'e de sp\'eculer sur ces 4 questions:
\begin{itemize}
\item Qu'y a-t-il au-dessus?
\item Qu'y a-t-il en-dessous?
\item Qu'y avait-il avant le monde?
\item Qu'y aura-t-il apr\`es?
\end{itemize}
Il aurait mieux valu pour lui qu'il ne fut jamais n\'e.}
\vspace{1mm}

{\bf Talmud de Babylone}}
\end{minipage}

\vspace{5mm}

Depuis plusieurs ann\'ees, nous avons deux th\'eories physiques
v\'erifi\'ees exp\'erimentalement: la m\'ecanique quantique et
la relativit\'e g\'en\'erale.
La m\'ecanique quantique, \`a travers la th\'eorie quantique des champs,
est notre unique cadre pour d\'ecrire la physique au niveau
microscopique, la physique des particules et de leurs interactions.
Elle  nous a forc\'e \`a revoir notre conception de la mesure
et d'affiner les notions d'observateurs et de ph\'enom\`enes observ\'es,
qui sont trait\'es dans la th\'eorie
de mani\`eres singuli\`erement diff\'erentes.
De l'autre c\^ot\'e, la relativit\'e g\'en\'erale nous permet
de faire de la physique \`a grande \'echelle. Elle d\'ecrit l'univers
dynamique: l'espace-temps lui-m\^eme n'est plus un cadre fixe dans lequel
la mati\`ere et l'\'energie \'evoluent, mais un cadre qui \'evolue
avec son contenu.
Alors, la seule mani\`ere
de d\'ecrire des \'ev\`enements est de mani\`ere relative.

Malheureusement, ces deux th\'eories nous
semblent physiquement incompl\`etes:
la relativit\'e g\'en\'erale \`a cause de ses singularit\'es,
la th\'eorie quantique \`a cause de ses probl\`emes d'interpr\'etations
(l'observateur observ\'e) et parce qu'elle suppose encore un espace-temps
fixe et immuable.
De plus, ces deux th\'eories semblent math\'ematiquement
incompatibles. Prendre en compte la constante de Planck $\hbar$
et la constante de Newton $G$ dans une m\^eme th\'eorie
relativiste n'est pas facile. Mais nous sommes convaincu qu'il devrait
exister une  th\'eorie physique se r\'eduisant \`a l'une
ou l'autre dans des r\'egimes particuliers et r\'esolvant les probl\`emes
des deux. On l'appelle la {\it gravit\'e quantique}, une th\'eorie
d\'ecrivant l'interaction gravitationnelle \`a toute \'echelle d'\'energie.
Enfin, \`a presque toute \'echelle d'\'energie... En effet, l'effet
de gravit\'e quantique de base, illustrant le probl\`eme
d'un mariage forc\'e entre les deux th\'eories, est que
d\`es qu'on regarde quelque chose de trop pr\`es
essayant d'atteindre une r\'esolution inf\'erieure
\`a la longueur de Planck $l_P\equiv\sqrt{\hbar G/c^3}\sim 1,6.10^{-35}m$,
l'objet se cache derri\`ere un horizon gravitationnel.
Cela d\'efinit une \'energie limite $E_P = \sqrt{\hbar c^5 /G}$
\`a partir de laquelle les lois physiques que nous connaissons ne peuvent
plus s'appliquer.

\medskip

Il n'y a pas une unique mani\`ere de chercher la th\'eorie de la gravit\'e
quantique mais plusieurs chemins que nous explorons depuis
plus d'un demi-si\`ecle.
En l'absence de donn\'ees exp\'erimentales ``utiles'' (c'est-\`a-dire
reconnues comme ph\'enom\`enes de gravit\'e quantique), notre principal
guide est notre sens physique et notre sens de l'esth\'etisme,
l'\'el\'egance d'une th\'eorie: notre propre conception de la r\'ealit\'e et
notre perception du monde physique. 
Ainsi travailler sur le sujet de la gravit\'e quantique n\'ecessite
une certaine recherche philosophique pour comprendre son propre point
de vue sur la structure du monde. Ensuite, cela implique \'egalement
un travail math\'ematique pour d\'evelopper et apprendre \`a utiliser
les nouveaux outils math\'ematiques n\'ecessaires pour appr\'ehender
les nouveaux concepts physiques.

Il en d\'ecoule une multitude de tentatives de th\'eories. La th\'eorie
la plus \'etudi\'ee aujourd'hui est la th\'eorie des cordes ou
son extension en une possible {\it th\'eorie M}, qui explore
la possibilit\'e de remplacer nos particules ponctuelles
par des objets \'etendus.
Elle emprunte un chemin
d\'etourn\'e et ambitieux tentant d'unifier toute notre vision de la physique
en une seule th\'eorie globale d\'ecrivant toutes les interactions.

N\'eanmoins, le plus simple pour commencer
est d'essayer d'appliquer
directement les principes de quantification que nous
connaissons \`a la relativit\'e g\'en\'erale. Si cela fonctionne alors
et permet d'obtenir une th\'eorie coh\'erente math\'ematiquement, c'est
gagn\'e et nous avons une th\'eorie de gravit\'e quantique v\'erifiable
(ou plut\^ot falsifiable) exp\'erimentalement. Sinon, cela nous
permet d'explorer notre compr\'ehension de ces th\'eories physique
et de leur principes sous-jacents, pour essayer de trouver lesquels peuvent
survivre au niveau de la th\'eorie unificatrice et lesquels
nous devons abandonner: cette m\'ethode, conservatrice, permet de tester
notre vision du monde physique petit \`a petit.

La  gravit\'e quantique \`a boucles, ou {\it Loop Quantum Gravity}, est une
telle tentative de quantifier directement la relativit\'e g\'en\'erale.
Elle se place dans un contexte canonique o\`u on regarde une hypersurface
-notre espace- \'evoluant dans le temps et cr\'eant ainsi notre espace-temps.
Elle d\'efinit des \'etats quantiques -les {\it r\'eseaux de spin}-
de l'hypersurface, o\`u
la g\'eom\'etrie elle-m\^eme est quantifi\'ee et vient par quanta
discrets d'aire et de volume. Ils s'interpr\'etent comme une hypersurface
discr\'etis\'ee avec un nombre fini de points et l'\'etat quantique
d\'ecrit le r\'eseau de relations entre ces points.  Puis ce r\'eseau
va \'evoluer g\'en\'erant le temps. Cette th\'eorie est le point de d\'epart
de ma th\`ese.

Ma motivation est de comprendre la dynamique des
\'etats quantiques dans le contexte de la {\it Loop Quantum Gravity},
c'est-\`a-dire la structure de l'espace-temps cr\'e\'e par ces
r\'eseaux de spin \'evoluant dans le temps.
Ainsi l'objet central de mon travail est l'\'etude des {\it mousses de spin},
qui sont des histoires de r\'eseaux de spin.
Un mod\`ele de mousse de spin est l'attribution d'une amplitude
\`a chaque histoire. D'un point de vue canonique, cela revient
\`a choisir une dynamique particuli\`ere. D'un point de vue
covariant, cela revient \`a impl\'ementer une int\'egrale
de chemin pour la gravit\'e.
Le point \'el\'egant est qu'il est possible de d\'efinir ces mousses
de spin d'une mani\`ere enti\`erement alg\'ebrique, d\'efinissant
une g\'eom\'etrie quantique \`a partir d'alg\`ebre et de combinatoire.

Le mod\`ele de mousse de spin le plus \'etudi\'e est le
mod\`ele de Barrett-Crane. Son c\^ot\'e combinatoire est
que les histoires consid\'er\'ees sont duales \`a des d\'ecompositions
simpliciales d'un espace-temps quadri-dimensionnel.
Son c\^ot\'e alg\'ebrique est que toutes les donn\'ees m\'etriques et
g\'eom\'etriques sont d\'efinies par la th\'eorie des repr\'esentations
du groupe de Lorentz $SO(3,1)$.
Replac\'e dans ce contexte, le but de ma th\`ese est de
comprendre les fondations de ce mod\`ele.
Tout d'abord comprendre comment il est
peut \^etre interpr\'et\'e dans un cadre canonique (des \'etats
de g\'eom\'etrie d'une hypersurface \'evoluant dans le temps)
car, apr\`es tout, notre vision (exp\'erimentale)
de l'espace physique autour de nous est encore donn\'ee par
des tranches \`a temps constant \'evoluant avec nous dans le temps.
Puis comprendre s'il fournit un mod\`ele l\'egitime de gravit\'e quantique,
c'est-\`a-dire
\`a quel point il peut \^etre consid\'er\'e comme d\'efinissant
une int\'egrale de chemin pour la relativit\'e g\'en\'erale. 

\bigskip

Dans une premi\`ere partie, je d\'ecrirai la th\'eorie de la gravit\'e
quantique \`a boucles, ou {\it Loop Quantum Gravity}, qui
fournit une quantification non-perturbative de la relativit\'e g\'en\'erale.
Cela s'effectue en deux \'etapes. Tout d'abord, il s'agit de proc\'eder
\`a l'analyse canonique de la th\'eorie classique pour comprendre
ses sym\'etries et son Hamiltonien d\'efinissant sa dynamique.
Puis, on quantifie la th\'eorie et je d\'ecrirai
l'espace d'Hilbert des \'etats quantiques
de la th\'eorie et d\'efinirai la base des r\'eseaux de spin.
Mon but est  de donner une version covariante de cette th\'eorie
d\'ecrivant l'espace-temps quantique.

\medskip

Tout d'abord, le formalisme de la {\it Loop Quantum Gravity} est
fond\'e sur le groupe de sym\'etrie $SO(3)$ des rotations
de l'espace tri-dimensionnel, alors que l'on s'attend \`a ce que le groupe
de sym\'etrie locale d'une th\'eorie d'espace-temps soit
le groupe de Lorentz, tout comme dans le mod\`ele de Barrett-Crane.
Travailler avec le groupe de Lorentz qui est non-compact se r\'ev\`ele
plus dur que travailler avec la sym\'etrie compacte $SO(3)$.
Ainsi, dans une seconde partie, je m'int\'eresserai \`a comment
utiliser un groupe de sym\'etrie non-compact et \'ecrire des observables
(invariantes sous le groupe de sym\'etrie), tel que
les r\'eseaux de spin, dans ce cadre.

\medskip

Puis, dans une troisi\`eme partie, je d\'evelopperai un formalisme canonique
invariant sous le groupe de Lorentz $SO(3,1)$ et non plus seulement
sous le groupe $SO(3)$. Cela dans le but d'obtenir une th\'eorie
canonique fond\'ee sur le groupe $SO(3,1)$ pouvant fournir un cadre
canonique au mod\`ele de mousse de spin de Barrett-Crane
(qui est invariant sous le groupe de Lorentz).
Dans un premier temps, j'\'etudierai la quantification \`a boucles
de la th\'eorie de la gravit\'e en $2+1$ dimensions, invariante sous
$SO(2,1)$. Il appara\^\i t alors une g\'eom\'etrie quantique o\`u l'espace
reste continu mais o\`u le temps est quantifi\'e.
Apr\`es ce ``toy-model'', je d\'ecrirai une quantification canonique
 de la relativit\'e g\'en\'erale invariante sous le groupe de Lorentz.
Dans ce cadre covariant, on pourra formuler la {\it Loop Quantum
Gravity} d'une mani\`ere invariante sous $SO(3,1)$, mais \'egalement
d\'evelopper une th\'eorie quantique qui a les m\^emes \'etats quantiques
cin\'ematiques que le mod\`ele de Barrett-Crane, ce qui cr\'ee un pont
entre formalisme canonique et formalisme covariant en mousse de spin.

\medskip

Dans une quatri\`eme et derni\`ere partie, j'\'etudierai les mod\`eles
de mousse de spin eux-m\^emes. Dans un premier temps, je d\'ecrirai
les mod\`eles en basse dimension (deux et trois dimensions). Ce qui permet
d'une part de comprendre le contexte et la logique de ces mod\`eles, ce
que nous pouvons attendre d'eux,
d'autre part de se familiariser avec la g\'eom\'etrie de l'espace-temps
d\'efinie par les mousses de spin. Dans ce cadre, je m'efforcerai de
d\'ecrire les structures non-perturbatives de la th\'eorie.
Dans un second temps, je m'attaquerai au mod\`ele de Barrett-Crane lui-m\^eme.
Je montrerai comment l'obtenir en tant que discr\'etisation et quantification
(g\'eom\'etrique) de la relativit\'e g\'en\'erale et comment le reformuler
en tant qu'int\'egrale de chemin de la gravit\'e.
Ceci dans le but de convaincre qu'il constitue un mod\`ele l\'egitime
de gravit\'e quantique.

\newpage

\part{La \lg en bref: Quantifier la Gravit{\'e} non-perturbativement}

La gravit{\'e} quantique {\`a} boucles, ou
\lqg, est
une quantification canonique de la relativit{\'e} g{\'e}n{\'e}rale.
C'est une quantification directe, par opposition {\`a} des approches
telles que la th{\'e}orie des cordes qui modifient la th{\'e}orie de base
ou encore d'autres approches {\`a} la gravit{\'e} quantique qui
construisent une gravit{\'e} quantique plus ou moins ind{\'e}pendemment
de la relativit{\'e} g{\'e}n{\'e}rale classique puis esp{\`e}rent la
retrouver dans une limite semi-classique. C'est {\'e}galement une
quantification non-perturbative, par opposition aux
d{\'e}veloppements perturbatifs des int{\'e}grales de chemin {\`a} la Feynman.
Le lecteur peut trouver un court historique de la recherche de la
gravit{\'e} quantique par Carlo Rovelli dans \cite{carlo:hist1,carlo:hist2}
et {\'e}galement une description par Lee Smolin des approches
poursuivies aujourd'hui dans \cite{lee:book}.

La \lg s'inscrit dans la continuit{\'e} du programme ADM
(d'apr{\`e}s Arnowit, Deser et Misner),
qui tentait de quantifier canoniquement la
relativit{\'e} g{\'e}n{\'e}rale se fondant sur la variable m{\'e}trique.
Le formalisme canonique repose sur une d{\'e}composition $3+1$ de
l'espace-temps distinguant l'espace -une hypersurface de genre espace-
et le temps: il s'agit de d{\'e}crire l'{\'e}volution dans le temps de
l'hypersurface canonique et de sa g{\'e}om{\'e}trie.
Le programme, initi{\'e} par Dirac et Bergman, avait pour but de
construire une th{\'e}orie quantique avec un espace d'Hilbert portant
une r{\'e}pr{\'e}sentation d'op{\'e}rateurs correspondant {\`a} la m{\'e}trique.
Les {\'e}quations de la th{\'e}orie furent {\'e}crites par DeWitt et Wheeler.
Malheureusement la th{\'e}orie {\'e}tait mal d{\'e}finie.
Ces probl{\`e}mes de d{\'e}finition
ne furent r{\'e}solus  que dans le cadre de la \lg par l'utilisation de variables
alternatives {\`a} la m{\'e}trique: la g{\'e}om{\'e}trie est alors d{\'e}crit
par les variables dites d'Ashtekar, une connexion et une triade.
Dans ces variables, la th{\'e}orie prend la forme d'une th{\'e}orie de jauge,
similaire aux th{\'e}ories de Yang-Mills, et la quantification canonique
est bien d{\'e}finie. Aujourd'hui, le travail semble presque
achev{\'e} du point de vue math{\'e}matique, bien qu'il reste encore
beaucoup de travail pour en extraire le sens physique et des pr{\'e}dictions
physiques r{\'e}alistes pr{\'e}cises.

Les r{\'e}sultats majeurs de la th{\'e}orie
sont la d{\'e}rivation d'un spectre discret des op{\'e}rateurs
g{\'e}om{\'e}triques, comme l'aire et le volume \cite{carlo&lee},
et une explication de l'entropie des trous noirs \cite{carlo:bh1,carlo:bh2}.  
Les {\'e}tats de g{\'e}om{\'e}trie de l'hypersurface canonique sont les r{\'e}seaux
de spins, ou commun{\'e}ment appell{\'e}s {\it spin networks}. Ces {\'e}tats forment
un espace d'Hilbert bien d{\'e}finis et les op{\'e}rateurs d'aire et de volume
permettent de leur donner une interpr{\'e}tation g{\'e}om{\'e}trique d'une
vari{\'e}t{\'e} discr{\`e}te (en morceaux), que l'on peut d{\'e}crire comme
une g{\'e}om{\'e}trie polym{\'e}rique \cite{ash:qg}. Ce sont des graphes (ensemble
de points li{\'e}s par les liens),
dessin{\'e}s sur l'hypersurface canonique, et habill{\'e}s par des 
``spins''
vivant sur les liens.
 Ces spins sont des nombres demi-entiers correspondant chacun
{\`a} une r{\'e}pr{\'e}sentation du
groupe $SU(2)$ des rotations en 3d. Ils contiennent l'information
m{\'e}trique et d{\'e}finissent l'aire d'une surface intersectant un lien donn{\'e}.
C'est ainsi que la g{\'e}om{\'e}trie semble {\^e}tre d{\'e}finie {\`a}
l'{\'e}chelle de Planck.

\medskip

Le lecteur pourra trouver des revues et cours sur la \lg, plus ou moins
math{\'e}matiquement d{\'e}taill{\'e}es dans
\cite{carlo:lqg,carlo:cours,thiemann:long,thiemann:short,lee:lqg}.
En particulier, \cite{thiemann:short} donne une pr{\'e}sentation
compl{\`e}te de la \lg sans s'encombrer de trop de d{\'e}tails math{\'e}matiques. 
La \lqg donne souvent l'impression d'une th{\'e}orie unique et bien
d{\'e}finie. N{\'e}anmoins, comme dans toute th{\'e}orie de champs, on est face
{\`a} des probl{\`e}mes de r{\'e}gularisation.
Mais plus important, il existe deux formulations distinctes,
une r{\'e}elle et une complexe,
que je vais expliquer dans cette partie, et qui ont chacunes leur force et
leurs petits probl{\`e}mes.
Et je concluerai par r{\'e}sumer les questions
pos{\'e}es par la \lqg, dont la principale est la dynamique de la th{\'e}orie,
 et les nouvelles approches qui tentent d'y r{\'e}pondre
 dont j'ai explor{\'e} les d{\'e}tails pendant ma th{\`e}se.
Ces nouvelles approches tendent {\`a} d{\'e}finir une \lg covariante, dont
les spectres des op{\'e}rateurs g{\'e}om{\'e}triques,
c'est-{\`a}-dire le r{\'e}sultat majeur de la th{\'e}orie,
seront modifi{\'e}s. Ainsi, au niveau cin\'ematique, on obtiendra
un spectre des aires (spatiales) continu
au lieu du spectre discret de la \lqg. De plus, un op\'erateur volume
n'aura plus vraiment de sens et, \`a la place, les \'etats quantiques
de g\'eom\'etrie 3d seront d\'efinis \`a travers la normale \`a l'hypersurface
(c'est-\`a-dire le plongement de l'espace dans l'espace-temps).

\medskip
Dans cette partie,
je passe en revue le
formalisme existant de la \lqg.
Je commencerais par d{\'e}crire l'action exacte que
nous nous proposons de quantifier. Puis j'expliquerai le formalisme r{\'e}el
de la \lqg, dit de Barbero, en d{\'e}crivant l'alg{\`e}bre des contraintes
{\`a} travers l'analyse de la structure canonique de l'action.
Cette alg{\`e}bre est au coeur de la th{\'e}orie puisque la \rg est une
th{\'e}orie compl{\`e}tement contrainte et enti{\`e}rement d{\'e}finie par
les contraintes. La quantification nous m{\`e}nera naturellement
{\`a} l'introduction des r{\'e}seaux de spins et
{\`a} la construction de l'op{\'e}rateur aire.
Je montrerai que le spectre de cet op{\'e}rateur est discret et
d{\'e}pend d'un param{\`e}tre,
dit d'Immirzi, qui semble {\^e}tre une nouvelle constante qu'il s'agira
a priori de d{\'e}terminer exp{\'e}rimentalement (car il ne semble pas {\^e}tre
n{\'e}cessaire de le contraindre pour assurer la coh{\'e}rence de la th{\'e}orie).

Je passerai alors {\`a} la th{\'e}orie complexe, son alg{\`e}bre des contraintes
(qui sera la m{\^e}me que celle de la th{\'e}orie r{\'e}elle) et {\`a}
ses particularit{\'e}s, comme une contrainte Hamiltonienne simplifi{\'e}e
et aux conditions de r{\'e}alit{\'e} {\`a} imposer pour retrouver une
m{\'e}trique/g{\'e}om{\'e}trie r{\'e}elle {\`a} partir des variables complexes.
Dans ce formalisme, le param{\`e}tre d'Immirzi n'intervient pas mais
la quantification se retrouve compliqu{\'e}e par ces conditions
de r{\'e}alit{\'e}.

Tout ce formalisme {\'e}tant fond{\'e} sur la description d'une hypersurface
(canonique), je terminerai sur des questions
li{\'e}s au temps, {\`a} la dynamique
et {\`a} l'espace-temps, une autre curiosit{\'e} {\'e}tant le param{\`e}tre
dit d'Immirzi dont la pertinence et la signification physique
nous {\'e}chappent encore. Ces interrogations justifient la recherche
d'un formalisme covariant, un formalisme d\'ecrivant directement d'espace-temps, ce
qui est le th{\`e}me de ma th{\`e}se.

\chapter{L'action de Palatini et sa structure canonique}

L'action que nous nous proposons d'analyser et de quantifier
est une action de Palatini g{\'e}n{\'e}ralis{\'e}e. Les variables sont
une $1-$forme, ou connexion, $\om=\om^{IJ}_\mu J_{IJ}dx^\mu$
{\`a} valeur dans l'alg{\`e}bre
de Lorentz $so(3,1)$ et un champ de t{\'e}trades $e^I=e^I_\mu dx^\mu$, tous deux
d{\'e}finis sur notre espace-temps donn{\'e} par une vari{\'e}t{\'e} ${\cal M}$
pseudo-Riemannienne de dimension 4 et de signature $(-+++)$.
$I,J$ sont des indices internes, vivant dans l'espace de Minkowski
 tangent {\`a} l'espace-temps, et les $J_{IJ}$ sont les g{\'e}n{\'e}rateurs
de l'alg\`ebre de Lorentz.
$\mu$ est un indice d'espace-temps.
Cette action s'{\'e}crit:

\beq
S[\om,e]=\f{1}{2}\int_{\cal M}
\epsilon_{IJKL}e^I\w e^J \w F^{KL}(\om)
-\f{1}{\imm}\int_{\cal M} e^I\w e^J \w F_{IJ}(\om).
\label{palatini}
\eeq
$F(\om)=d\om +\om\w\om$
est la courbure de la connexion $\om$. Les indices $I,J$ sont
descendus {\`a} l'aide de la m{\'e}trique plate $\eta_{IJ}$ sur l'espace de
Minkowski.

L'action  dite de Palatini consiste seulement en le premier terme
de l'action ci-dessus. Ses {\'e}quations du mouvement reproduisent les
{\'e}quations d'Einstein d{\'e}crivant la \rg quand la t{\'e}trade est
non-d{\'e}g{\'e}n{\'e}r{\'e}e. La m{\'e}trique (pseudo-Riemannienne) s'exprime alors
$g_{\alpha\beta}=e^I_\alpha e^J_\beta \eta_{IJ}$, c'est-{\`a}-dire que la
t{\'e}trade diagonalise la m{\'e}trique. Le second terme est sans effet sur
les {\'e}quations du mouvement, c'est-{\`a}-dire qu'il est sans importance
classiquement. La constante de couplage $\gamma$
correspond au param{\`e}tre dit d'Immirzi dans le cadre de la \lqg.

\medskip

On aimerait quantifier canoniquement cette action. Il s'agit dans un premier
temps de proc{\'e}der {\`a} un d{\'e}coupage $3+1$ de ${\cal M}$ en identifiant
une direction temps et une hypersurface dite canonique. Puis le but est de
d{\'e}crire l'{\'e}volution de cette hypersurface avec le temps. Pour cela,
on {\'e}crit l'action en mettant en valeur les termes avec des
d{\'e}riv{\'e}es temporelles. Ces termes d{\'e}finissent les variables conjugu{\'e}es
dans la th{\'e}orie. Les termes restants d{\'e}finissent le Hamiltonien de la th{\'e}orie
et il consistera dans notre cas d'une somme de contraintes
(avec leur multiplicateur de Lagrange respectif).
Choisissant l'hypersurface d{\'e}finie par les coordonn{\'e}es $\alpha=1,2,3$
et le temps donn{\'e} par la coordonn{\'e}e $\alpha=0$, il est facile de voir que
les termes avec d{\'e}riv{\'e}e temporelle sont dus {\`a} la courbure et que la
connexion $\om^{IJ}_a$ est conjugu{\'e}e {\`a}
$\pi_{IJ}^a=\epsilon^{abc}\epsilon_{IJKL}e^K_be^L_c$.
La difficult{\'e} rencontr\'ee lors de
l'analyse canonique et de sa quantification r{\'e}side dans ce que les variables
$\pi_{IJ}^a$ ne sont pas ind{\'e}pendantes. En effet, elles satisfont
des contraintes dites de seconde classe
(dans le jargon du programme de Dirac\cite{dirac}):

\beq
\forall a,b,\,\epsilon^{IJKL}\pi_{IJ}^a \pi_{KL}^b =0.
\label{lqg:2nde}
\eeq
Ce sont ces contraintes qui nous compliquent la t{\^a}che.

Heureusement, il est possible  de contourner ce probl{\`e}me en se pla\c cant
dans une jauge particuli{\`e}re nomm{\'e}e ``jauge temps''
({\it time gauge} en anglais). C'est ce qui permet d'obtenir une alg{\`e}bre
de contraintes (de premi{\`e}re classe) relativement simple
et dont l'interpr{\'e}tation physique est claire. Cela aboutira au formalisme
r{\'e}el de la \lqg. La question est alors: a-t-on bris{\'e}
irr{\'e}versiblement une sym{\'e}trie lors de la quantification?
D'un autre c{\^o}t{\'e}, le formalisme complexe
permet d'{\'e}viter les complications du formalisme r{\'e}el.
L'alg{\`e}bre des contraintes reste
simple et, de plus, l'expression de ces contraintes est m{\^e}me plus simple!
Malheureusement, les contraintes de seconde classe ressurgissent plus tard
sous la forme de contraintes de r{\'e}alit{\'e} {\`a} imposer {\`a} la th{\'e}orie
pour obtenir une m{\'e}trique r{\'e}elle.

Malgr{\'e} ces difficult{\'e}s, au final, nous nous retrouvons avec une
th{\'e}orie de champs (relativement) simple, invariante sous diff{\'e}omorphismes,
dont les seuls degr{\'e}s de libert{\'e} sont une connexion $SU(2)$ et
son champ conjugu{\'e}. L'alg{\`e}bre des contraintes est de premi{\`e}re
classe. Les degr{\'e}s de libert{\'e} canoniques sont au nombre de deux
par point de l'espace et, apr{\`e}s lin{\'e}arisation, on retrouve
directement les lois de propagation d'un champ de masse nulle et de
spin 2, ce qui correspond bien {\`a} la relativit{\'e} g{\'e}n{\'e}rale.

\section{Le formalisme r{\'e}el de Barbero}
\label{loopgravity}

Avant de proc{\'e}der {\`a} l'analyse canonique, on commence par la
d{\'e}composition $3+1$ de l'espace-temps. Il s'agit de distinguer la
direction temporelle des trois dimensions usuelles d'espace:
on se donne une fonction temps $t$ ({\`a} valeur dans $\R$)
sur notre espace-temps de topologie
$\Sigma \times \R$. Les hypersurfaces que nous allons consid{\'e}rer
comme l'espace seront les hypersurfaces {\`a} $t$ constant.
Dans ce cadre, nous pouvons d{\'e}composer la t{\'e}trade:
\beq
e_{0I}=Nn_I +N^ae_{aI},
\eeq
o{\`u} $n_I$ est le vecteur normalis{\'e} gradient de la fonction $t$ et
par cons{\'e}quent orthogonal aux hypersurfaces {\`a} temps constant.
Plus pr{\'e}cis{\'e}ment, nous avons $n^Ie_{aI}=0$ et $n^In_I=-1$.
$N$ et $N^a$ sont d{\'e}nomm{\'e}s respectivement le {\it lapse} et le
{\it shift}.

Pour proc{\'e}der {\`a} l'analyse canonique, nous suivons l'approche de
Holst \cite{holst}. Tout d'abord, nous nous pla\c cons dans une jauge 
particuli{\`e}re appel{\'e}e usuellement {\it time gauge} et nous choisissons:
\beq
n_I=(1,0,0,0).
\eeq
Cela revient {\`a} annuler les composantes temporelles (internes)
$e_{a0}=0$: les vecteurs $e_{\mu i}$ ($i=1,2,3$)
forment une base de l'espace tangent {\`a} une hypersurface ({\`a} temps
constant) donn{\'e}e et la t{\'e}trade se r{\'e}duit d'une mani{\`e}re
effective {\`a} une triade. Cette fixation de jauge n'impose bien s\^ur
aucune restriction sur la m{\'e}trique spatiale.

Il est alors possible de proc{\'e}der {\`a} l'analyse canonique de
l'action de Palatini g{\'e}n{\'e}ralis{\'e}e \Ref{palatini}. Dans ce cadre,
nous introduisons la triade:
\beq
E^{ck}=\f{1}{2}\epsilon^{abc}\epsilon^{ijk}e_{ai}e_{bj}
\eeq
et des nouvelles variables de connexion:
\beq
^\pm A_{ck}=\om_{ck0}+\f{1}{2\imm}\epsilon_{kij}\om_c^{ij}.
\eeq
Avec ces nouvelles notations, les variables $\om_{t}^{IJ}$ et $^+A$ sont
des multiplicateurs de Lagrange imposant des contraintes (pour une
expression explicite, voir \cite{holst}). Ces contraintes impliquent
que
\beq
^+A_{ai}=^-A_{ai}-\f{2}{\imm}\Gamma_{ai}.
\label{A+A-}
\eeq
$\Gamma$ est une fonction de la triade $E$, c'est la connexion de
Levi-Civita (spatiale)
\beq
\Gamma_{ai}=-\f{1}{2}\epsilon_{ijk}e_b^j\nabla_ae^{bk},
\eeq
et v{\'e}rifie ${\cal D}_a(\Gamma)E_i^b=0$ (o{\`u} la d{\'e}riv{\'e}e
covariante ${\cal D}$ est donn{\'e}e par $\Gamma$).
La relation \Ref{A+A-} peut {\^e}tre trait{\'e}e comme une contrainte de
seconde classe et permet de se d{\'e}barrasser de la variable $^+A$.
Finalement, introduisant la connexion $A_{ai}=\imm {}^-A_{ai}$,
l'action se r{\'e}-{\'e}crit:
\beq
S=\int -\f{1}{\imm}\dot{A}_{ai}E^{ai}+\Lambda_{i}\nabla_aE ^{ai}+
N^aE ^{bi}F_{abi}+NC,
\label{action:holst}
\eeq
o{\`u} $F$ est la courbure de la connexion $A$ et le terme $C$ est
sp{\'e}cifi{\'e} plus bas dans l'{\'e}quation \Ref{lqg:contraintes}.
Les variables conjugu{\'e}es sont alors $A_{ai}$ et $E^{ai}$ et leur
crochet de Poisson est:
\beq
\{A_{ai}(x),E^{bj}(y)\}=\imm\delta_a ^b\delta_i ^j\delta ^{(3)}(x,y).
\label{lqg:bracket}
\eeq
La connexion peut {\^e}tre exprim{\'e}e en tant que
\beq
A_{ai}=\Gamma_{ai}(E)+\imm K_{ai}
\eeq
o{\`u} $K_{ai}=e^\alpha_i\nabla_ae_{\alpha 0}$ est la courbure
extrins{\`e}que. C'est la {\it connexion d'Ashtekar-Barbero}.

La th{\'e}orie est un syst{\`e}me enti{\`e}rement contraint. En tout,
elle est constitu{\'e}e de 7 contraintes (de premi{\`e}re classe):
\beq
\begin{array}{ccl}
G_i&=&\nabla_aE_i^a=\partial_aE_i ^a +\epsilon_{ijk}A_a ^jE ^{ak} \\
V_a&=&E ^{bi}F_{abi} \\
C&=&\epsilon^{ijk}E_i ^aE_j ^b \left(F_{abk}(A)-\left(1+\f{1}{\imm^2}\right)R_{abk}(\Gamma)\right)
\label{lqg:contraintes}
\end{array}
\eeq
o{\`u} $R$ est la courbure de la connexion $\Gamma(E)$. Ces contraintes
sont impos{\'e}es par les multiplicateurs de Lagrange $N$,$N^a$ et
$$
\Lambda_i=\f{1}{\imm}\left(\f{1}{\imm}\om_{ti0}+\f{1}{2}\epsilon_{ijk}A^{ij}_t\right).
$$
$G_i$ g{\'e}n{\`e}re des transformations de jauge $SU(2)$ agissant sur la
triade et la connexion. Par similitude avec la th{\'e}orie de
l'{\'e}lectromagn{\'e}tisme, on l'appelle {\it Loi de Gauss}.
$V_a$ est une contrainte vectorielle qui g{\'e}n{\`e}re les diff{\'e}omorphismes
spatiaux (diff{\'e}omorphismes de $\Sigma$). $C$ est une contrainte
scalaire, la contrainte
Hamiltonienne, qui dicte/contraint l'{\'e}volution de la g{\'e}om{\'e}trie de notre
hypersurface canonique. L'expression de cette derni{\`e}re contrainte
est assez compliqu{\'e}e {\`a} cause du terme $R(\Gamma)$, qui est
non-polynomial en les variables $E$ et $A$. N{\'e}anmoins, on peut
remarquer qu'il dispara{\^\i}t pour le choix $\imm=\pm i$; mais nous nous
occupons du formalisme r{\'e}el $\imm\in\R$ dans la pr{\'e}sente section,
le cas $\imm=\pm i$ sera pr{\'e}sent{\'e} dans la section suivante et
la structure de la th{\'e}orie sera "beaucoup" plus simple.

Plus pr{\'e}cis{\'e}ment, le Hamiltonien de la th{\'e}orie s'{\'e}crit
\beq
\begin{array}{ccl}
H=-\int_\Sigma d^3x\,{\cal H}&=&-\int_\Sigma d^3x\,\left(\Lambda^iG_i+N^aV_a+NC\right) \\
&=&-\int_\Sigma d^3x\,\left((\Lambda_i-N^aA^i_a)G_i+N^a\Delta_a+NC\right) \\
&=&-(G_{[\Lambda-N^aA_a]}+\Delta_{[\vec{N}]}+C_{[N]})
\end{array}
\eeq
o{\`u} $\Delta_a=V_a+A^i_aG_i$. Alors l'alg{\`e}bre des contraintes se
lit:
\beq
\begin{array}{ccl}
\{G_{\Lambda_1},G_{\Lambda_2}\}&=&-G_{[\Lambda_1,\Lambda_2]} \\
\{G_{\Lambda},\Delta_{\vec{N}}\}&=& G_{{\mathcal L}_{\vec{N}}\Lambda}\\
\{\Delta_{\vec{N}_1},\Delta_{\vec{N}_2}\}&=& -\Delta_{[\vec{N}_1,\vec{N}_2]}\\
\{G_{\Lambda},C_N\}&=& 0\\
\{\Delta_{\vec{N}},C_N\}&=& -C_{{\mathcal L}_{\vec{N}}N}\\
\{C_{N_1},C_{N_2}\}&=&\Delta_{\vec{K}}-G_{K^aA_a} 
\end{array}
\eeq
avec $K^a(N_1,N_2)=E^{ia}E^b_i(N_1\dd_bN_2-N_2\dd_bN_1)$,
$[\Lambda_1,\Lambda_2]^i=\epsilon^i_{jk}\Lambda_1^j\Lambda_2^k$ et
$[\vec{N}_1,\vec{N}_2]^a=N_1^b\dd_bN_2^a-N_2^b\dd_bN_1^a$.
Maintenant, la transformation g{\'e}n{\'e}r{\'e}e par
\beq
J_\xi=\int_\Sigma d^3x\,\xi^0{\cal H}-\xi^a\Delta_a
\eeq
est un 4-diff{\'e}omorphisme infinit{\'e}simal param{\'e}tr{\'e} par le champs
de vecteur $\xi^\mu$ \cite{reis:constraints}. En particulier, quand
$\xi^a=0$, $J_\xi$ g{\'e}n{\`e}re une reparam{\'e}trisation du temps $t\arr
t-\xi^0$, et quand $\xi^0=0$, $J_\xi$ induit un diff{\'e}omorphisme
spatial par $\vec{\xi}$. Nous remarquons que nous avons besoin d'un
m{\'e}lange des transformations de jauge $G_i$ et des contraintes
$V_a,C$ pour g{\'e}n{\'e}rer les diff{\'e}omorphismes.

\medskip

Si on compte le nombre de variables en chaque point de l'espace, nous
avons $9$ paires de variables conjugu{\'e}es soumises {\`a} 7 contraintes
(de premi{\`e}re classe), d'o{\`u} les 2 degr{\'e}s de libert{\'e} de la
gravit{\'e} (sans mati{\`e}re). De plus, on peut d{\'e}montrer que
si nous nous pla\c cons dans le secteur invariant de jauge $SU(2)$,
c'est-{\`a}-dire si nous r{\'e}solvons la contrainte $G_i=0$, alors
l'espace de phase se r{\'e}duit {\`a} celui de l'approche ADM (approche
canonique m{\'e}trique) \cite{thiemann:long, thiemann:short}. 

\medskip

Apr{\`e}s tout ce travail, nous pouvons nous demander o{\`u}
sont pass{\'e}es les contraintes de seconde classe \Ref{lqg:2nde} et
si la fixation de jauge utilis{\'e}e est bien l{\'e}gitime.
Une r\'eponse est fournie par Barros e Sa dans \cite{barros}.

Tout d'abord, on peut consid{\'e}rer que les contraintes de seconde
classe se traduisent par la contrainte \Ref{A+A-} reliant $^+A$ et
$^-A$ \footnotemark. Ensuite, l'analyse de Barros repose sur la
r{\'e}solution explicite des contraintes de seconde classe
\Ref{lqg:2nde}:
\beq
\begin{array}{ccl}
\pi ^{a0i}&=&E ^{ai}\\
\pi ^{aij}&=&E ^{a[i}\chi ^{j]}
\end{array}
\eeq
o{\`u}
$$
\chi^i=-\f{e^{ti}}{e^{tt}}
$$
d{\'e}finit la direction temps (dans l'espace de Minkowski interne)
en tant que $(1,\chi^i)$.
La {\it time gauge} se traduit alors simplement par $\chi^i=0$.
\footnotetext{Cette intuition est due {\`a} la ressemblance que cette
  relation a avec les conditions de r{\'e}alit{\'e} du formalisme self-dual
  et des relations issues des contraintes de seconde classe dans le
  cadre covariant qui sera pr{\'e}sent{\'e} en partie III.}
Dans ce cadre, l'action d{\'e}finit deux couples de variables
conjugu{\'e}es: une connexion avec la triade, et le champ $\chi$ avec
un autre champ $\zeta$ (en conservant les notations
de \cite{barros}).
Ces variables sont soumis {\`a} 10 contraintes de premi{\`e}re classe. On
retrouve les 7 contraintes $G_i$, $V_a$, $C$, dont l'expression en
fonction de $A,E,\chi,\zeta$ est substantiellement plus compliqu{\'e}e
que dans l'analyse de Holst. Plus, 3 nouvelles contraintes $G^{(boost)}_i$
qui g{\'e}n{\`e}rent grosso modo les boosts dans l'espace tangent/interne.
Elles permettent de tourner le vecteur $(1,\chi^i)$ jusqu'{\`a}
$(1,0,0,0)$ et d'atteindre la fixation de jauge $\chi^i=0$. Dans cette
jauge, on oublie ces nouvelles contraintes, on est r{\'e}duit au couple
de variables $(A,E)$ et on peut v{\'e}rifier que
l'expression des contraintes $G_i$, $V_a$, $C$ se r{\'e}duit bien en fin
de compte {\`a} \Ref{lqg:contraintes}.
Pour les d{\'e}tails des
calculs, le lecteur peut se r{\'e}f{\'e}rer {\`a} \cite{barros}.

Finalement, la question de la l{\'e}gitimit{\'e}  de la fixation de jauge
devient "la sym\'etrie de Lorentz classique est-elle bris{\'e}e lors de
la quantification de la th{\'e}orie canonique d{\'e}crie {\`a} l'aide de
cette fixation de jauge?" Je m'efforcerai de r{\'e}pondre \`a cette question
dans la partie III en proc\'edant \`a  une analyse
canonique covariante
et \`a sa quantification invariante de Lorentz d{\'e}crite dans \cite{3+1}.

\section{Le formalisme complexe d'Ashtekar}
\label{lqg:complexe}

Le formalisme complexe self-dual d'Ashtekar correspond {\`a} un choix
$\imm=\pm i$. Ce choix particulier permet de simplifier
(consid{\'e}rablement) la th{\'e}orie. La structure canonique de la
th{\'e}orie est alors tr{\`e}s semblable {\`a} celle de la th{\'e}orie
Euclidienne, c'est-{\`a}-dire que nous retrouvons les 7 contraintes
$G_i,V_a,C$ avec une expression simple pour $C$ qui ne contient pas le
terme non-polynomial $R(\Gamma)$. La diff{\'e}rence principale est que
le crochet de Poisson \Ref{lqg:bracket} se retrouve avec un facteur
$i$.
Le lecteur peut trouver une
introduction simple du formalisme et des contraintes dans
\cite{lee:lqg} et une pr{\'e}sentation relativement compl{\`e}te dans
\cite{carlo:report}, dont nous suivrons le raisonnement. 

\medskip

L'action de d{\'e}part s'{\'e}crit
\beq
S[\om,e]=\int_{\cal M}
\epsilon_{IJKL}e^I \w e^J \w \left(
 \f{1}{2} F^{KL}(\om)
-\f{i}{4}\epsilon^{KL}_{MN}  F^{MN}(\om)
\right).
\label{action:selfdual1}
\eeq
Pour simplifier les notations, nous introduisons l'op{\'e}rateur de
Hodge (sur l'espace de Minkowski interne):
$$
\star T^{IJ} = \f{1}{2}\epsilon^{IJ}_{KL}T^{KL}
$$
o{\`u} les indices sont descendus {\`a} l'aide de la m{\'e}trique plate.
Cet op{\'e}rateur v{\'e}rifie $\star^2=-1$.
Il permet de d{\'e}composer des tenseurs de rang 2 en partie self-duale
et anti-self-duale:
$$
T^{IJ}=\f{1-i\star}{2}T^{IJ}+\f{1+i\star}{2}T^{IJ}=
{}^+T^{IJ}+{}^-T^{IJ}
$$
avec $(-i\star){}^+T={}^+T$ et $(-i\star){}^-T=-{}^-T$.
$^+T$ et $^-T$ sont respectivement la partie self-dual et
anti-self-dual du tenseur $T$.
Nous introduisons alors la connexion self-duale
\beq
^4A ^{IJ}_\mu=\f{1}{2}(1-i\star)\om^{IJ}_\mu
=\f{1}{2}\om^{IJ}_\mu-\f{i}{4}\epsilon^{IJ}_{KL}\om^{KL}_\mu.
\eeq
Elle v{\'e}rifie $(-i\star)^4A^{IJ}_\mu={}^4A^{IJ}_\mu$.
Il est int\'eressant de remarquer que la courbure de cette connexion
est la partie self-dual de la courbure de $\om$:
\beq
F^{IJ}_{\mu\nu}(^4A)=
\f{1}{2}(1-i\star)F^{IJ}_{\mu\nu}(\om)
=\f{1}{2}F^{IJ}_{\mu\nu}(\om)-
\f{i}{4}\epsilon^{IJ}_{KL}F^{KL}_{\mu\nu}(\om).
\eeq
L'action \Ref{action:selfdual1} peut alors se r{\'e}-{\'e}crire
\beq
S(e,\om)=\f{1}{2}\int
\epsilon_{IJKL}e^I \w e^J \w  F^{KL}(^4A(\om))=
\f{-i}{2}\int e^I \w e^J \w F_{IJ}(^4A(\om))
\label{action:selfdual2}
\eeq
et peut {\^e}tre consid{\'e}r{\'e}e comme une action $S(e,{}^4A)$ de la t{\'e}trade
et d'une connexion $^4A$ complexe et self-duale.
En d{\'e}veloppant tous les indices, on trouve:
\beq
S(e,^4A)=-i\int d^4x \epsilon^{abc}
(e_{ai}e_{bj}F^{ij}_{tc}(^4A)+e_{tI}e_{aj}F^{Ij}_{bc}(^4A)).
\eeq
Nous pouvons alors proc{\'e}der {\`a} l'analyse canonique. Nous commen\c
cons par effectuer le d{\'e}coupage $3+1$ et la fixation de jauge
$n^\mu=e^\mu_0$ comme pr{\'e}c{\'e}demment. Nous introduisons une
connexion spatiale et une triade (de densit{\'e} +1)\footnotemark:
\beq  
\left\{
\begin{array}{ccl}
A ^i_a&=&2i {}^4A ^{0i}_a=\epsilon^i_{jk}{}^4A ^{jk}_a \\
E ^a_i&=& \f{1}{2}\epsilon^{abc}\epsilon_{ijk}e_b^je_c^k
\end{array}
\right.
\eeq
\footnotetext{C'est l'inverse de la triade $e^i_a$ au
  d{\'e}terminant $E$ de $e^i_a$ pr{\`e}s.}
Alors l'action de la relativit{\'e} g{\'e}n{\'e}rale se lit:
\beq
S=\int dt\int_\Sigma d^3x
\left(
-iE ^c_kA_c^k-A_t^iG_i-N^aV_a-(NE^{-1})C
\right)
\eeq
o{\`u} $E$ est le d{\'e}terminant de la triade $e^i_a$.
Le crochet de Poisson est alors
\beq
\{A_a^i(x),E^b_j(y)\}=i\delta_a ^b\delta_j ^i\delta ^{(3)}(x,y),
\label{lqg:bracket2}
\eeq
et les contraintes impos{\'e}es par les multiplicateurs de Lagrange
$A_t^i,N^a,N$ sont:
\beq
\left\{
\begin{array}{ccl}
G_i&=&i\nabla_cE_i^c=i(\partial_cE_i^c+\epsilon_{ij}^kA^j_cE_i^c)\\
V_a&=& iE_i^bF^i_{ba}(A)\\
C&=&-\f{1}{2}\epsilon^{ij}_kE_i^aE_j^bF^k_{ab}(A)
\end{array}
\right.
\eeq
Le syst{\`e}me est bien de premi{\`e}re classe, les $G_i$ g{\'e}n{\`e}rent
les transformations de jauge $SU(2)$; $V_a$ et $C$ les
diff{\'e}omorphismes de l'espace-temps ${\cal M}$. On remarque que la
structure de la th{\'e}orie est identique ({\`a} des d{\'e}tails pr{\`e}s) au
formalisme r{\'e}el.  N{\'e}anmoins l'expression de la contrainte
Hamiltonienne est polynomial en $A$ et $E$, ce qui la rend plus simple
{\`a} g{\'e}rer. A part cette diff{\'e}rence, toute la structure de l'alg{\`e}bre
des contraintes est la m{\^e}me que dans le formalisme r{\'e}el
pr{\'e}sent{\'e} dans la section pr{\'e}c{\'e}dente.

\medskip

Le prix {\`a} payer est que nous avons des "contraintes"
suppl{\'e}mentaires {\`a} prendre en compte. En effet, nous voulons que la
m{\'e}trique spatiale (de $\Sigma$) reste r{\'e}elle. Puisque
$q^{ab}=E_i^aE^{bi}$ \footnotemark, cela revient {\`a} supposer 
que la triade reste r{\'e}elle:
\beq
\textrm{Im}\,E_i^a=0.
\eeq
\footnotetext{En fait, $\tl{\tl{q}}^{ab}=E_i^aE^{bi}$ est une
  densit{\'e} +2, et c'est la m{\'e}trique $q$ {\`a} un facteur $E^2$
  pr{\`e}s.  En particulier $det[\tl{\tl{q}}^{ab}]=q^2$.}
De plus, pour que cette condition soit v{\'e}rifi{\'e}e lors de
l'{\'e}volution dans le temps, il faut {\'e}galement imposer que
$\{E_i^aE^{bi},H\}$ soit r{\'e}el. Apr{\`e}s calcul
(\cite{carlo:report}, page 13), cela se traduit par
\beq
\textrm{Re}\,\epsilon^{ijk}E_i^{(a}\nabla_c[E_j^{b)}E_k^{c}]=0.
\label{Ereality}
\eeq
Ceci avec la constrainte de Gauss $G_i$ implique que
\beq
\textrm{Re}\,A_a^i=\f{1}{2}\epsilon^i_{jk}\Gamma^{jk}_a(e),
\label{Areality}
\eeq
o{\`u} $\Gamma(e)$ est la connexion de spin induite par la t{\'e}trade $e$
comme dans la section pr{\'e}c{\'e}dente. Ceci correspond {\'e}galement {\`a}
ce que la connexion $A$ s'exprime en fonction de $\Gamma(E)$ et de la
courbure extrins{\`e}que $K$:
\beq
A_{ai}=\Gamma_{ai}(E)+i K_{ai}.
\eeq
Ces conditions de r{\'e}alit{\'e}  ne compliquent pas
 la th{\'e}orie classique. Elles interviendront par contre dans
la th{\'e}orie quantique pour contraindre le produit scalaire de la
th{\'e}orie comme nous le verrons dans le prochain chapitre. 

De plus, nous pouvons remarquer que ces contraintes de r{\'e}alit{\'e}
correspondent aux contraintes de seconde classe. En effet, dans un
formalisme canonique sans {\it time gauge} \cite{sergei1,3+1}, la
contrainte \Ref{Ereality} s'obtient par le crochet de Poisson de la
contrainte Hamiltonienne avec les contraintes \Ref{lqg:2nde} de seconde classe
$\pi.\star\pi=0$. De plus, dans ce m{\^e}me contexte, on
en d{\'e}rive {\'e}galement une contrainte sur la connexion (de Lorentz)
similaire {\`a} \Ref{Areality} \cite{3+1}.

\section{Le param{\`e}tre d'Immirzi $\gamma$}

La pr{\'e}sentation que j'ai faite du formalisme de la \lg n'est
pas la d{\'e}rivation originelle de la th{\'e}orie. Chronologiquement, le
formalisme self-dual a {\'e}t{\'e} d{\'e}riv{\'e} en premier par Ashtekar. Puis, le
formalisme r{\'e}el a {\'e}t{\'e} {\'e}crit par Barbero (d'o{\`u} le nom de
connexion d'Ashtekar-Barbero) {\`a} travers une transformation
canonique.
En effet, les variables de la \lg self-dual sont la connexion $A$
self-duale et la triade r{\'e}elle. En utilisant la relation
$A^{(ash)}=\Gamma(E)+iK$, il est possible d'effectuer une
transformation canonique de $(E,A^{(ash)})$ vers les variables $(E,K)$. Le
g{\'e}n{\'e}rateur de cette transformation est $i\int
\Gamma_a^iE^a_i$. Puis, en utilisant une transformation g{\'e}n{\'e}r{\'e}e
par $\int\Gamma_a^iE^a_i$ cette fois, il est possible d'obtenir les
variables $(E,A^{(bar)}=\Gamma+\imm K)$ \cite{barbero}. Cette astuce
permet d'obtenir une th{\'e}orie r{\'e}elle, donc sans contraintes
de r{\'e}alit{\'e} compliquant la quantification de la th{\'e}orie, au prix
de compliquer le Hamiltonien (la contrainte Hamiltonienne plus
pr{\'e}cis{\'e}ment).

Puis Holst proposa sa d{\'e}rivation de la th{\'e}orie r{\'e}elle {\`a} partir
de l'action de Palatini g{\'e}n{\'e}ralis{\'e}e \Ref{palatini}.
Ce n'est que plus tard qu'Immirzi argumenta que ce param{\`e}tre,
inoffensif dans la th{\'e}orie classique, devient d{\'e}terminant dans la
th{\'e}orie quantique: il modifie le spectre des op{\'e}rateurs
g{\'e}om{\'e}triques et induit l'existence d'une famille de quantification
in{\'e}quivalente \cite{immirzi}. Il faudrait alors le fixer d'une
mani{\`e}re ou d'une autre. Depuis, on s'interroge constamment sur
l'origine th{\'e}orique \cite{carlo&tom} et surtout son r{\^o}le
physique. Peut-{\^e}tre qu'il n'existe qu'une seule th{\'e}orie
coh{\'e}rente. Mais cela ne semble pas vraisemblable et il a {\'e}t{\'e}
propos{\'e} que le param{\`e}tre d'Immirzi $\imm$ soit fix{\'e} en fonction de
la relation entre l'aire et l'entropie d'un trou noir (voir par
exemple \cite{bh,olaf}).

D'un autre c{\^o}t{\'e}, il peut {\^e}tre jug{\'e} qu'un tel param{\`e}tre n'a
pas de sens physique. Dans ce cas, la critique est dirig{\'e}e contre le
formalisme r{\'e}el, {\'e}levant des doutes quant {\`a} l'interpr{\'e}tation
des variables triade/connexion comme des variables d'espace-temps
\cite{samuel} et {\`a} la covariance de la th{\'e}orie (plus
pr{\'e}cis{\'e}ment l'impl{\'e}mentation de l'invariance par le groupe
Lorentz et des diff{\'e}omorphismes d'espace-temps, par constraste avec
les diff{\'e}omorphismes d'espace dont l'action est bien d{\'e}finie en
\lqg) \cite{3+1,sergei2,sergei3}. Dans ce cadre, il a {\'e}t{\'e} prouv{\'e}
que l'int{\'e}grale de chemin ne devrait pas d{\'e}pendre de $\imm$
\cite{sergei1} et qu'il est possible de construire une th{\'e}orie
totalement covariante \cite{3+1,sergei2} o{\`u} le param{\`e}tre d'Immirzi
n'intervient pas dans le spectre des op{\'e}rateurs g{\'e}om{\'e}triques.
Cette formulation covariante sera explicit{\'e}e dans la partie III.

\chapter{Quantifier avec des Boucles}

L'analyse canonique de l'action de Palatini (avec la fixation de jauge telle
que la direction temps soit normale {\`a} l'hypersurface canonique) nous
fournit un formalisme de g{\'e}om{\'e}trodynamique, fond{\'e} non sur la
m{\'e}trique mais sur une connexion $SU(2)$ (et sa triade
conjugu{\'e}e). Ce formalisme est tr{\`e}s similaire {\`a} une th{\'e}orie de
jauge de type de Yang-Mills et peut {\^e}tre vu (du point de vue
math{\'e}matique) comme une g{\'e}n{\'e}ralisation de
la th{\'e}orie quantique du champs {\'e}lectromagn\'etique (QED)
{\'e}changeant le groupe de jauge $U(1)$ contre le groupe non-ab{\'e}lien
$SU(2)$: la connexion $A$ est l'analogue du potentiel
{\'e}lectromagn{\'e}tique et la triade $E$ est l'analogue du champ {\'e}lectrique.

Apr{\`e}s avoir proc{\'e}d{\'e} {\`a} l'analyse de la structure canonique, il
s'agit maintenant de trouver des observables de la th{\'e}orie, puis d'en
construire une repr{\'e}sentation sur laquelle pourront agir des
op{\'e}rateurs correspondant aux variables classiques connexion et triade.
Une observable est une quantit{\'e} qui commute avec toutes les
contraintes, ici $G_i$, $V_a$ et $H$. Nous allons proc{\'e}der par
{\'e}tape et commencer par l'invariance de jauge $G_i$ avant de nous
pr{\'e}occuper de l'invariance sous diff{\'e}omorphismes (spatiaux) et
d'impl{\'e}menter la contrainte Hamiltonienne.

\medskip

Une observable naturelle dans les th{\'e}ories de jauge est la boucle de
Wilson. Elle consiste en la trace de l'holonomie de la connexion
(de jauge) autour d'une boucle; soit dans notre cas, en notant $e$ une boucle sur
notre hypersurface $\Sigma$:
\beq
W_e(A)=\textrm{Tr}\,U_e(A)=
\textrm{Tr}\,e^{\int_e A}=
\textrm{Tr}\,e^{\int_0^1 dt A^i_a(e(t))\tau_i\dot{e}^a(t)},
\eeq
o{\`u} $t\in[0,1]\arr e(t)$ est une param{\'e}trisation de la boucle $e$
et $\tau_i$ sont les g{\'e}n{\'e}rateurs de $SU(2)$ (les  matrices
de Pauli).
Une transformation de jauge $SU(2)$ est d{\'e}finie par une fonction
$g:x\in\Sigma \arr g(x)\in SU(2)$. Son action sur
la connexion est  $^gA=g^{-1}Ag+g^{-1}\partial g$
et l'holonomie le long d'une courbe $\gamma$ se transforme en
$^gU_\gamma(A)=g(\gamma(0))^{-1}U_\gamma(A)g(\gamma(1))$. Par
cons{\'e}quent, notant $x_0=e(0)=e(1)$ le point de d{\'e}part de la
boucle, $U_e$ se transforme en $g(x_0)^{-1}U_eg(x_0)$ et sa trace est
invariante sous l'action du groupe de jauge.
Il est possible de g{\'e}n{\'e}raliser la boucle de Wilson {\`a} un graphe
quelconque (au lieu d'une boucle). Cela nous fournit des {\it
  fonctions cylindriques} en la connexion $A$, qui ne d\'ependent de la
connexion qu'{\`a} travers un nombre fini de variables. Plus
pr{\'e}cis{\'e}ment, soit un graphe orient{\'e} et ferm{\'e} $\Gamma$ avec $E$
liens et $V$ vertex. Nous construisons les $E$ holonomies de la connexion
$A$ le long des liens de $\Gamma$ puis nous consid{\'e}rons des
fonctions sur $SU(2)^E$:
\beq
\varphi_{\Gamma(A),\phi}=\phi(U_{e_1},\dots,U_{e_E})\in\C.
\eeq
\begin{figure}[t]
\begin{center}
\psfrag{a}{$U_1$}
\psfrag{b}{$U_2$}
\psfrag{c}{$U_3$}
\psfrag{d}{$U_4$}
\psfrag{e}{$U_5$}
\psfrag{f}{$U_6$}
\psfrag{v1}{$g_A$}
\psfrag{v2}{$g_B$}
\psfrag{v3}{$g_C$}
\includegraphics[width=6.5cm]{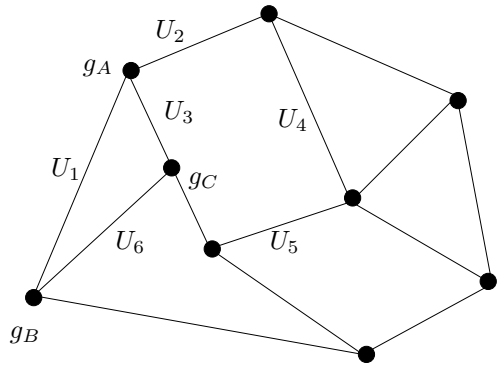}
\end{center}
\caption{Support du graphe d'une fonction cylindrique d\'ependant
des holonomies le long des liens $1,2,..$ 
et invariantes de jauge aux vertex $A,B,..$.}
\end{figure}
L'action d'une transformation de jauge agit sur chaque holonomie $U_e$
par une rotation {\`a} la source $s(e)$ du lien $e$ et {\`a} sa cible
$c(e)$:
\beq
U_e\arr g_{s(e)}^{-1}U_eg_{c(e)},
\eeq
donc agit sur la fonctionnelle $\varphi_{\Gamma,\phi}$ au niveau des vertex du
graphe $\Gamma$ {\`a} travers $V$ {\'e}l{\'e}ments $g_1,\dots g_V$ de
$SU(2)$. Par cons{\'e}quent, demander que $\varphi_{\Gamma,\phi}$ soit
invariante sous transformation de jauge revient {\`a} imposer que la
fonction $\phi$ d{\'e}finie sur $SU(2)^{E}$ soit invariante sous
l'action de $SU(2)^{V}$:
\beq
\forall g_v\in SU(2),\,
\phi(U_{e_1},\dots,U_{e_E})=
\phi(g_{s(e_1)}^{-1}U_{e_1}g_{c(e_1)},\dots,g_{s(e_E)}^{-1}U_{e_E}g_{c(e_E)}).
\label{invjauge}
\eeq

\medskip

Les fonctions cylindriques que nous avons construites sont invariantes
sous l'action de $SU(2)$ i.e sous l'action de la loi de Gauss
$G_i$. Par contre, elles ne sont pas invariantes sous l'action de
$V_a$ et de $C$. Puisque $V_a$ induit les diff{\'e}omorphismes sur
$\Sigma$, il est intuitif que son action sur une fonctionnelle
$\varphi_{\Gamma,\phi}$ {\`a} support sur un graphe $\Gamma$ va {\^e}tre de
d{\'e}placer le graphe $\Gamma$. Par contre, l'action de $H$ va {\^e}tre
plus subtile {\`a} impl{\'e}menter.

Il est {\'e}galement possible de construire des "observables" \footnotemark
invariantes sous $SU(2)$ {\`a} partir de la triade. La triade se
transforme de mani{\`e}re homog{\`e}ne sous le groupe de jauge $E(x)\arr
g(x)^{-1}E(x)g(x)$. Se choisissant une surface ${\cal S}\subset\Sigma$
et une fonction $f:{\cal S}\arr su(2)$, on peut d{\'e}finir un flux de
la densit{\'e} $E^a_i$ {\`a} travers la surface  orient{\'e}e ${\cal S}$ par
\beq
E({\cal S},f)=\f{1}{2}\int_{{\cal S}} E_if^i=
\f{1}{2}\int_{{\cal S}} E^a_if^i\epsilon_{aa_1a_2}dx^{a_1}\w dx^{a_2}
\eeq
o{\`u} $x^1,x^2,x^3$ est un syst{\`e}me de coordonn{\'e}es sur $\Sigma$.
Le champs $E$ {\'e}tant conjugu{\'e} {\`a} la connexion, ces flux
$E({\cal S},f)$ agiront comme des d{\'e}rivations sur les fonctionnelles
$\varphi_\Gamma(A)$ de la connexion. L'alg{\`e}bre form{\'e}e par les
$E({\cal S},f)$
et les $\varphi_\Gamma$ peut {\^e}tre consid{\'e}r{\'e}e comme l'alg{\`e}bre
fondamentale en \lg (voir par exemple \cite{holo-flux}).

\footnotetext{Le mot "observables" est ici un abus de language puisque
les fonctionnelles consid{\'e}r{\'e}es ne
sont pas invariantes sous toutes les contraintes: ce ne sont que des observables
partielles.}

N{\'e}anmoins, je ne regarderai pas en d{\'e}tails cette alg{\`e}bre-l{\`a} et je
concentrerai mon attention sur d'autres "observables" fonctions de
la triade, qui permettent de construire l'op{\'e}rateur d'aire d'une surface.
Choisissons une surface ${\cal S}$ de coordonn{\'e}es $\sigma ^1$ et
$\sigma ^2$ et commen\c cons par construire la quantit{\'e}
\beq
E_i({\cal S})=\int_{{\cal S}}d\sigma^1d\sigma^2 \epsilon_{abc}
\f{\dd x^a(\vec{\sigma})}{\dd\sigma^1}\f{\dd
  x^b(\vec{\sigma})}{\dd\sigma^2}
E^c_i(x(\vec{\sigma})).
\eeq
Alors la quantit{\'e} $\sqrt{E_i({\cal S})E_i({\cal S})}$ est invariante
sous $SU(2)$. Justement, cela permet de construire l'aire d'une surface
${\cal S}$ en prenant la limite d'une fine partition de ${\cal S}$ en
petites surface ${\cal S}^{(n)}_k$ \cite{carlo:area}:
\beq
{\cal A}({\cal S})=\lim_{n\arr\infty}\sum_k\sqrt{E_i({\cal S}^{(n)}_k)E_i({\cal S}^{(n)}_k)}.
\label{aire:def}
\eeq
En effet, par d{\'e}finition de l'int{\'e}grale de Riemann, le terme de droite
se lit
$$
\int_{{\cal S}}d\sigma^1d\sigma^2
\sqrt{
\epsilon_{abc}
\f{\dd x^a(\vec{\sigma})}{\dd\sigma^1}\f{\dd
  x^b(\vec{\sigma})}{\dd\sigma^2}
E^c_i(x(\vec{\sigma}))
\epsilon_{def}
\f{\dd x^d(\vec{\sigma})}{\dd\sigma^1}\f{\dd
  x^e(\vec{\sigma})}{\dd\sigma^2}
E^f_i(x(\vec{\sigma}))
}.
$$
En choisissant $x^3=0$ sur la surface ${\cal S}$ et donc
$\sigma^1=x^1$, $\sigma^2=x^2$, cela se r{\'e}duit {\`a}
$$
\int_{{\cal S}}d\sigma^1d\sigma^2
\sqrt{E^{3i}(x)E^{3i}(x)}=
\int_{{\cal S}}d\sigma^1d\sigma^2
\sqrt{det(g)g^{33}}=
\int_{{\cal S}}d\sigma^1d\sigma^2
\sqrt{det^2(g)},
$$
o{\`u} nous avons utilis{\'e} le fait que $E^a_i$ est la densit{\'e} inverse de la
triade $e^i_a$ (comme vu dans le chapitre pr{\'e}c{\'e}dent) et par cons{\'e}quent
$E^a_iE^b_i=det(g)g^{ab}$ avec $g^{ab}$ l'inverse de la m{\'e}trique.

\medskip

Maintenant, je vais proc{\'e}der {\`a} la quantification 
canonique de l'action de Palatini formul\'ee dans les variables de connexion $SU(2)$ 
et de triade. La \lqg  est fond{\'e}e sur le choix des fonctions cylindriques 
de la connexion $A$ comme {\it fonctions d'onde}.
L'espace des fonctions cylindriques $L^2$ est alors l'espace de Hilbert 
des {\'e}tats quantiques de la g{\'e}om{\'e}trie de l'hypersurface canonique. Je 
montrerai comment construire la base des r{\'e}seaux de spin de cet espace 
$L^2$, puis j'expliquerai comment impl{\'e}menter un op{\'e}rateur d'aire, 
diagonal sur les r{\'e}seaux de spins, et un op{\'e}rateur de volume.
Cela d{\'e}finira la \lqg, une quantification non-perturbative de la
relativit{\'e} g{\'e}n{\'e}rale cens{\'e}e nous d{\'e}crire l'espace-temps {\`a} l'{\'e}chelle de Planck.

\section{Des r{\'e}seaux de spins}

Dans cette section, je vais expliquer comment les fonctions
cylindriques nous fournissent un espace d'Hilbert pour la th{\'e}orie
quantique, qui sera l'espace des {\'e}tats quantiques de notre
hypersurface canonique. Je commencerai par regarder l'espace des
fonctions cylindriques {\`a} support sur un graphe $\Gamma$ donn{\'e}. Je
d{\'e}finirai l'espace d'Hilbert des fonctions $L^2$ et je pr{\'e}senterai sa
base donn{\'e}e par les r{\'e}seaux de spins. Puis, dans une seconde {\'e}tape, je
sommerai sur tous les graphes et consid{\'e}rerai l'ensemble de toutes les
fonctions cylindriques. Cet espace pourra {\^e}tre vu comme l'espace des
fonctions $L^2$ sur un espace de connexions g{\'e}n{\'e}ralis{\'e}es {\`a} l'aide de
la mesure d'Ashtekar-Lewandowski. Cet espace aura comme base
l'ensemble de tous les r{\'e}seaux de spins ({\`a} support sur n'importe quel graphe).
Il fournit une repr\'esentation (des fonctionnelles cylindriques) de la 
connexion $A$ (l'op{\'e}rateur multiplication) 
et des quantit{\'e}s g{\'e}om{\'e}triques fonction de la triade (comme les 
flux, l'aire et le volume) en tant qu'op{\'e}rateurs hermitiens.

\subsection{La Base des R{\'e}seaux de Spin}

Pla\c cons nous donc sur un graphe fix{\'e} $\Gamma$
plong\'e dans $\Sigma$ et consid{\'e}rons
l'ensemble des fonctions cylindriques {\`a} support sur $\Gamma$.
Nous aimerions construire un espace d'Hilbert \`a partir de ces fonctionnelles,
plus pr{\'e}cis{\'e}ment un espace $L^2$. Pour cela, il nous faut choisir
une mesure. Une mesure naturelle est la mesure de Haar
$d\mu_\Gamma=\prod_{e}dg_e$ 
sur $SU(2)^{E}$. Ainsi la mesure d'une fonctionnelle
$\varphi_{\Gamma,\phi}=\phi(U_{e_1},\dots,U_{e_E})$ sera:
\beq
\mu_\Gamma(\varphi_{\Gamma,\phi})=\int \prod_{i=1}^Edg_i \,
\phi(g_1,\dots,g_E).
\label{lqg:mesure}
\eeq
Cette mesure ne faisant pas mention explicite du graphe $\Gamma$,
elle permettra une impl\'ementation simple de l'action des
diff\'eomorphismes comme nous le verrons plus loin.
Egalement,
cette mesure sera l'unique mesure pour laquelle  les op{\'e}rateurs
construits {\`a} partir de la triade seront hermitiens.

Ayant choisi la mesure, on peut donc construire l'espace d'Hilbert
$\H_\Gamma=L^2(d\mu_\Gamma)$ sur laquelle le produit scalaire
s'{\'e}crit:
\beq
\sc{\varphi_\phi}{\varphi_\psi}=\sc{\phi}{\psi}=
\int \prod_{i=1}^Edg_i \,
\overline{\phi}(g_1,\dots,g_E)\psi(g_1,\dots,g_E).
\label{lqg:prodscal}
\eeq

\medskip

A partir de l{\`a}, il serait int{\'e}ressant d'exhiber une base de cet
espace d'Hilbert pour avoir une id{\'e}e de sa structure. C'est l'espace
des fonctions $L^2$ sur $SU(2)^{E}$ qui sont invariantes sous
l'action de $SU(2)^{V}$. On pourra donc commencer par regarder
l'espace $L^2(SU(2)^{E})$, puis imposer l'invariance sous
$\su^{V}$. Cette approche est valide car $SU(2)$ est
compact: si le groupe de jauge {\'e}tait de volume infini (comme le
groupe de Lorentz), alors l'espace $L^2(G^{E})$ ne
contiendrait aucune fonction invariante sous l'action de $G^{V}$.

$L^2(SU(2)^{E})$ s'obtient en utilisant la formule de
d{\'e}composition de Peter-Weyl. C'est une sorte de transform{\'e}e de
Fourier qui exprime une fonction $f\in L^2(\su)$ comme une somme sur
les repr{\'e}sentations de $\su$:
\beq
\forall g\in \su,\,
f(g)=\sum_{j,m,n}f^j_{mn}D^{(j)}_{mn}(g),
\eeq
o{\`u} $j\in\N/2$ labelle les repr{\'e}sentations irr{\'e}ductibles (de
dimension finie) de $\su$. Appelons $V^j$ l'espace associ{\'e} {\`a} la
repr{\'e}sentation de spin $j$.
$m,n$ labellent une base de $V^j$ (la base usuelle $\s{m}$) et varient
sur $-j,-j+1,\dots,j-1,j$. $D^{(j)}(g)$ 
est la matrice repr{\'e}sentant l'{\'e}l{\'e}ment de groupe $g$ dans la
repr{\'e}sentation de spin $j$. Les $f^j_{mn}$ sont l'{\'e}quivalents des
composantes de Fourier de la fonction $f$.

A l'aide de cette formule, on peut d{\'e}composer une fonction
$f(g_1,\dots,g_E)\in L^2(\su^E)$ sur les repr{\'e}sentations de
$\su$. On va associer {\`a} chaque lien $e$ du graphe une
repr{\'e}sentation $j_e$ et des vecteurs $m_e,n_e$. Ces deux vecteurs
sont en fait associ{\'e}s aux deux vertex source $v=s(e)$ et cible
$v=c(e)$ du lien $e$. Puis une base de
$L^2(\su^E)$ est fournie par les vecteurs:
\beq
f^{\{j_e,m_e,n_e\}}(g_1,\dots,g_E)=\prod_e D^{(j_e)}_{m_en_e}(g_e)=
\prod_e \cs{j_em_e}g_e\s{j_en_e}.
\eeq

\begin{figure}[t]
\begin{center}
\psfrag{a}{$j_1$}
\psfrag{b}{$j_2$}
\psfrag{c}{$j_3$}
\psfrag{d}{$j_4$}
\includegraphics[width=7cm]{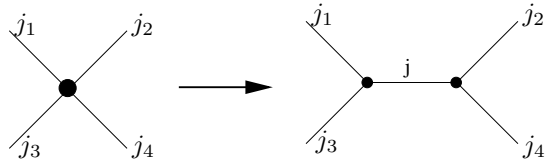}
\end{center}
\caption{D\'ecomposition d'un entrelaceur 4-valent
en vertex trivalents: on ``d\'eplie'' le vertex 4-valent
en un lien ``virtuel''.}
\label{4val}
\end{figure}

\begin{figure}[t]
\begin{center}
\psfrag{a}{$j_1$}
\psfrag{b}{$j_2$}
\psfrag{c}{$j_3$}
\psfrag{d}{$j_4$}
\psfrag{e}{$j_5$}
\psfrag{f}{$j_6$}
\psfrag{v1}{$I_{v_1}$}
\psfrag{v2}{$I_{v_2}$}
\psfrag{v3}{$I_{v_3}$}
\includegraphics[width=7cm]{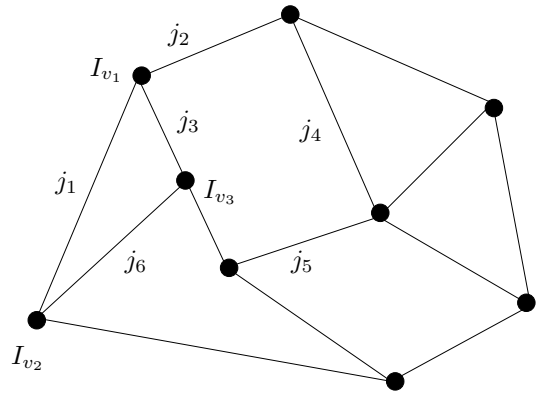}
\end{center}
\caption{R\'eseau de spin labell\'e par les repr\'esentations
$j_1,j_2,..$ sur les liens et par les entrelaceurs $I_v$ aux vertex.}
\label{spinnetwork}
\end{figure}

Il s'agit maintenant d'identifier le secteur invariant de jauge i.e
satisfaisant la relation \Ref{invjauge}. Pour cela, on veut imposer
une invariance sous $\su$ {\`a} chaque vertex. Cela est r{\'e}alis{\'e} en
choisissant pour chaque vertex $v$ un tenseur $\prod_{e\, \textrm{in}}
V^{j_e}\arr\prod_{e\,\textrm{out}} V^{j_e}$, ou d'une mani{\`e}re {\'e}quivalente
$\prod_{e\, \textrm{in}} V^{j_e}\otimes\prod_{e\,\textrm{out}}
\overline{V^{j_e}}\arr\C$ ,
invariant sous $\su$. Les liens $e\, \textrm{in}$ sont les liens incidents sur
$v$ et les liens $e\,\textrm{out}$ sont les liens partant de $v$.
De tels objets sont appell{\'e}s entrelaceurs.
\def \twin{entrelaceur}
Si nous avons un vertex trivalent, l'entrelaceur  entre 3
repr{\'e}sentations de $\su$  est unique  {\`a} une normalisation pr{\`e}s:
il est donn{\'e} par les coefficients de Clebsh-Gordan. Une fois donn{\'e}s
des entrelaceurs $I_v$ pour tous les vertex $v$, on peut construire
une fonction invariante de jauge en contractant les {\'e}l{\'e}ments de
matrices $D^{(j_e)}(g_e)$ au niveau des vertex {\`a} l'aide des $I_v$:
\beq
\begin{array}{ccl}
f^{j_e,I_v}(g_1,\dots,g_E)&=&
\prod_e D^{(j_e)}(g_e) \otimes
\prod_v (I_v)^{j_{e\, \textrm{in}},j_{e\,\textrm{out}}}\\
&=& \prod_e \cs{j_em_e}g_e\s{j_en_e}
\prod_v
\cs{j_{e\,\textrm{in}}n_e}
(I_v)^{j_{e\,\textrm{in}},j_{e\,\textrm{out}}}
\s{j_{e\,\textrm{out}} m_e}
\end{array}.
\eeq
Finalement, comme nous voulons une base de l'espace $L^2$ des
fonctions invariantes, il ne reste plus qu'{\`a} choisir une base
d'entrelaceurs pour chaque vertex.
Pour trouver une base d'entrelaceurs pour un vertex n-valent, il
suffit de d{\'e}plier le vertex en un graphe (avec n pattes
ext{\'e}rieures) ne contenant que des vertex (virtuels) trivalents:
les entrelaceurs des vertex trivalents sont donn{\'e}s par les
Clebsh-Gordan et on labelle les liens virtuels avec des spins
internes aux vertex (voir fig.\ref{4val}
pour l'exemple de l'entrelaceur 4-valent).
On obtient alors une base de $\H_\Gamma$ par
fonctions labell{\'e}es par des spins $j_e$ sur chaque lien et des
entrelaceurs $I_v$ {\`a} chaque vertex: c'est la base des r{\'e}seaux de
spin ou {\it spin networks} (voir fig.\ref{spinnetwork}).

\subsection{Connexion G{\'e}n{\'e}ralis{\'e}e et
Mesure d'Ashtekar-Lewandowski}
\label{ashlewMeasure}

Il s'agit maintenant de consid{\'e}rer l'ensemble des fonctions
cylindriques, c'est-{\`a}-dire en quelque sorte de sommer sur l'ensemble
des graphes $\Gamma$, pour obtenir un espace d'Hilbert repr{\'e}sentant
tous les {\'e}tats quantiques de l'hypersurface (canonique).
Dans ce cadre, l'approche d'Ashtekar et Lewandowski est d'utiliser des
techniques de limites projectives pour d{\'e}finir un espace de
connexion g{\'e}n{\'e}ralis{\'e}e sur lequel on peut introduire une mesure,
la mesure dite d'Ashtekar-Lewandowski \cite{ash-lew}.
L'espace $L^2$ en r{\'e}sultant sera
une somme, ou limite projective, des espaces d'Hilbert $\H_\Gamma$.
Le lecteur pourra trouver une revue d{\'e}taill{\'e}e de ce formalisme
dans \cite{thiemann:long}.

\medskip

Tout d'abord on d{\'e}finit l'espace des connexions discr{\`e}tes
invariantes de jauge {\`a} support sur le graphe $\Gamma$ (plong{\'e} dans
l'hypersurface $\Sigma$):
\beq
\begin{array}{ccl}
A_{\Gamma}&=&\su^{E}/\su^{V}\\& =&
\left\{ \left[
(g_{e_1},\dots,g_{e_E})\right] _{\su^{V}} \right\} =\left\{
\{ (k^{-1}_{s(e_i)}g_{e_i}k_{t(e_i)},i=1\dots E),k_v \in \su \}
\right\}.
\end{array}
\eeq
On introduit un ordre partiel $\prec$ sur l'ensemble des graphes: 
$\Gamma_1 \prec \Gamma_2$ ssi $\Gamma_1$ peut {\^e}tre obtenu {\`a} partir
de $\Gamma_2$ en enlevant des liens et des vertex bivalents. On
peut alors d{\'e}finir des projections $p_{\Gamma_2
  \Gamma_1}:A_{\Gamma_2}\arr A_{\Gamma_1}$ pour $\Gamma_1\prec
\Gamma_2$ en enlevant les liens et vertex bivalents suppl{\'e}mentaires:
\beq
\left\{
\begin{array}{cccc}
\tr{enlever le lien $e_i$} &
(g_1,\dots,g_i,\dots,g_E) & \arr
& (g_1,\dots,g_{i-1},g_{i+1},\dots,g_E) \\
\tr{vertex bivalent entre $e_1$ et $e_2$} &
(g_1,g_2,\dots,g_E)
& \arr
& (g_1g_2^\epsilon,\dots,g_E)
\end{array}
\right.
\eeq
o{\'u} $\epsilon=\pm 1$ depend de l'orientation relative de $e_1$ et
$e_2$.
Illustrons ces projections par l'exemple de la r{\'e}duction du graphe $\Theta$ {\`a}
une simple boucle:
\beq
\begin{array}{ccc}
\CBOX[1]{\FIGThetaNet{1}{2}{3}} & \longrightarrow &
\CBOX[1]{\looptwo{1}{3}} \\
(g_1,g_2,g_3)\sim (h^{-1}g_1k,h^{-1}g_2k,h^{-1}g_3k)
& \longrightarrow & (g_1,g_3) \sim (h^{-1}g_1k,h^{-1}g_3k) \\
\\
\CBOX[1]{\looptwo{1}{3}} & \longrightarrow &
\CBOX[1]{\loopone{1}} \\
(g_1,g_3) \sim (h^{-1}g_1k,h^{-1}g_3k) &
\longrightarrow &
G_1=g_1g_3^{-1}\sim h^{-1}G_1h
\end{array}
\eeq
On d{\'e}finit ensuite la limite projective  $\bar{A}$
comme l'ensemble des familles d'{\'e}l{\'e}ments de $A_{\Gamma}$
coh{\'e}rentes avec les projections:
\beq
\bA=\left\{
(a_{\Gamma})_{\Gamma \mathrm{ graphe}}
\in \times_{\Gamma} A_{\Gamma} \, / \,
\forall \, \Gamma_{1,2} \, , \,
\Gamma_1\prec \Gamma_2 \Rightarrow
p_{\Gamma_2 \Gamma_1}a_{\Gamma_2}=a_{\Gamma_1}
\right\}.
\eeq
Dans le cas du groupe compact $SU(2)$,
les espaces $A_{\Gamma}$ sont topologiques, compacts et Haussdorf, et les
projections sont continues, par cons{\'e}quent $\bA$, muni de la
topologie de Tychonov (topologie produit) est compact et Haussdorf.
On construit donc les fonctions continues sur $\bA$.
On commence par d{\'e}finir les espaces
\beq
C^0(A_{\Gamma})=\left\{
f \in {\cal F}(A_{\Gamma},{\mathbf C}), f\tr{ continues}
\right\}.
\eeq
Les projections $p$ induisent des injections entre les espaces de fonctions
$C^0(A_{\Gamma_1})$ et $C^0(A_{\Gamma_2})$ d{\`e}s que $\Gamma_1 \prec \Gamma_2$:
\beqs
i_{\Gamma_1\Gamma_2}:  C^0(A_{\Gamma_1}) & \arr & C^0(A_{\Gamma_2})\\
\phi(\{g_e\}_{e\in\Gamma_1})
& \rightarrow & \tl\phi(\{g_e\}_{e\in\Gamma_2})=
\phi(p_{\Gamma_2\Gamma_1}\{g_e\}_{e\in\Gamma_2}).
\eeqs
On d{\'e}finit alors la relation d'{\'e}quivalence suivante:
\beq
\begin{array}{ccc}
f_{\Gamma_1}\in C^0(A_{\Gamma_1}) \sim
f_{\Gamma_2}\in C^0(A_{\Gamma_2})
& \Leftrightarrow &
\exists \, \Gamma_3 \succ \Gamma_1 , \Gamma_2, \,
i_{\Gamma_1\Gamma_3}f_{\Gamma_1}=
i_{\Gamma_2\Gamma_3}f_{\Gamma_2} \\
& \Leftrightarrow &
\forall \, \Gamma_3 \succ \Gamma_1 , \Gamma_2, \,
i_{\Gamma_1\Gamma_3}f_{\Gamma_1}=
i_{\Gamma_2\Gamma_3}f_{\Gamma_2}
\end{array}
\eeq
Ceci nous permet de d{\'e}finir l'espace des {\it fonctions cylindriques}:
\beq
Cyl(\bA)=\bigcup_{\Gamma}C^0(A_{\Gamma}) {\Big /} \sim.
\eeq
Nous avons divis{\'e} par la relation d'{\'e}quivalence pour enlever les
redondances dues {\`a} l'existence des injections.
Sur $Cyl(\bA)$, on d\'efinit la norme 
\begin{equation}
\| [ f_{\Gamma} ]_\sim \| = \sup_{x_\Gamma \in A_\Gamma} |f_\Gamma (x_\Gamma)|.
\end{equation}
Alors l'espace compl{\'e}t{\'e} est
une $C^*$ alg{\`e}bre ab{\'e}lienne,
{\`a} laquelle nous pouvons appliquer le th{\'e}or{\`e}me de Gelfand-Naimark.
Ainsi, c'est l'alg{\`e}bre des fonctions continues
sur un certain espace compact et Haussdorf
d{\'e}nomm{\'e} le spectre de Gelfand de la $C^*$ alg{\`e}bre.
Dans \cite{ash-lew},  Ashtekar et Lewandowski ont d{\'e}montr{\'e}
que son spectre de Gelfand est tout simplement  $\bA$
i.e que nous avons l'isomorphisme suivant:
\beq
Cyl(\bA)\approx C^0(\bA).
\eeq
$Cyl(\bA)$ est notre espace de {\it connexion g{\'e}n{\'e}ralis{\'e}e}. Nous
allons construire une mesure sur cet espace.
Pour cela, si on choisit des mesures $\dmu^{(\Gamma)}$
sur les espaces des connexions discr{\`e}tes $A_{\Gamma}$
de telle sorte qu'elles soient compatibles avec les injections:
\beq
\forall \, \Gamma_1 \prec \Gamma_2 \, ,
\, i_{\Gamma_1\Gamma_2}\dmu^{(\Gamma_2)}=\dmu^{(\Gamma_1)}.
\eeq
Il est alors possible de d{\'e}finir une mesure $\overline{\tr{d}\mu}$
sur $\bA$ en consid{\'e}rant la limite projective.
Dans notre cas, les mesures $\dmu^{(\Gamma)}$ sont les mesure de Haar
et elles satisfont trivialement la condition de compatibilit{\'e}. Leur
limite projective est la mesure d'Ashtekar-Lewandowski et l'espace
d'Hilbert final est donc:
$$
\H_{\mathrm{cyl}}=L^2(\bar{A},\overline{\tr{d}\mu}).
$$

\bigskip

Un mani{\`e}re alternative et {\'e}l{\'e}gante de construire cet espace
d'Hilbert est d'utiliser la construction GNS (d'apr{\`e}s
Gelfand-Naimark-Segal) \cite{ash-lew2,gns}.

On consid{\`e}re l'alg{\`e}bre ${\cal A}$ de toutes les fonctions
cylindriques $f_\Gamma$ ({\`a} support sur tous les graphes $\Gamma$)
munie de la multiplication entre fonctions.
On d{\'e}finit la norme sup sur cet espace
comme pr{\'e}c{\'e}demment
\beq
\| f_\Gamma \| =\sup_{A_\Gamma}|f_\Gamma|.
\eeq
Puis, on compl{\`e}te ${\cal A}$ en une $C^*$ alg{\`e}bre $\bar{{\cal A}}$.
Sur $\bar{{\cal A}}$, on definit un {\it {\'e}tat} $\om$ -une forme
lin{\'e}aire positive normalis\'ee-
qui est ici simplement l'int{\'e}gration:
\beq
\om(f_\Gamma)=\int_{A_\Gamma} \dmu^{(\Gamma)} f_\Gamma
=\int_{SU(2)^E}\tr{d}g_1 \dots\tr{d}g_E f_\Gamma(g_1,\dots,g_E).
\eeq
$\om$ induit un produit scalaire d{\'e}g{\'e}n{\'e}r{\'e}
$\langle f_{\Gamma_1}|f_{\Gamma_2} \rangle=\om(f^*_{\Gamma_1}f_{\Gamma_2})$
en se souvenant que le produit de deux fonctions cylindriques
est une fonction cylindrique {\`a} support sur un graphe plus grand contenant
{\`a} la fois $\Gamma_1$ et $\Gamma_2$. On d{\'e}finit l'id{\'e}al de Gelfand:
\beq
{\cal I}=\{ a\in \bar{{\cal A}} | \om(a^*a)=0\}.
\eeq
On obtient alors un vrai produit scalaire Hermitien (d{\'e}fini positif)
sur l'espace ${\cal H}_{\mathrm{gns}}=\bar{{\cal A}}/{\cal I}$.
Puis on obtient un espace d'Hilbert en compl{\'e}tant
cet espace en $\overline{{\cal H}_{\mathrm{gns}}}$.
Il est facile de se convaincre que la relation d'{\'e}quivalence
$\sim$ est la m{\^e}me que celle d{\'e}finie par ${\cal I}$ et donc que
les deux m{\'e}thodes m{\`e}nent au m{\^e}me r{\'e}sultat:
$\overline{{\cal H}_{\mathrm{gns}}}=\H_{\mathrm{cyl}}$.

\bigskip

Faisons cette construction explicitement. La base des r{\'e}seaux de spin
nous fournit une d{\'e}composition de $\H_\Gamma$ avec laquelle il est
ais{\'e} d'impl{\'e}menter la relation d'{\'e}quivalence
$\sim$. Si un lien $e$ de $\Gamma$ est labell{\'e} par la
repr{\'e}sentation triviale $j=0$, alors la fonctionnelle r{\'e}seau de
spin correspondante ne d{\'e}pend en fait pas de l'{\'e}l{\'e}ment de groupe
$g_e$ vivant sur ce lien: elle sera \'egale au r\'eseau de spin
d\'efini sur $\Gamma'=\Gamma\setminus\{e\}$ avec les m\^emes labels et
entrelaceurs. Par cons\'equent, nous pouvons d\'ecomposer
$\H_\Gamma$ en la somme des espaces d'Hilbert 
$\tl{\H}_{\Gamma'}$, $\Gamma'\subset\Gamma$, des r\'eseaux de spin
\`a support sur $\Gamma'$ {\bf ne contenant aucune repr\'esentation
triviale $j=0$}.
De plus, si nous consid\'erons un graphe arbitraire
$\Gamma_1$ (qui n'est pas une simple boucle avec un unique vertex) et
un graphe $\Gamma_2$ obtenu en enlevant un vertex bivalent de $\Gamma_1$,
les espaces $\tl{\H}_{\Gamma_1}$ et $\tl{\H}_{\Gamma_2}$ sont
isomorphes par la restriction de l'injection
$i_{\Gamma_2\Gamma_1}$ \`a $\tl{\H}_{\Gamma_2}$.
Cela veut dire que nous d\'ecomposons l'espace $\H_\Gamma$
en tant que  somme directe des espaces
$\H_\Gamma'$ avec $\Gamma'\subset \Gamma $ ne contenant pas de vertex
bivalents (plus les boucles simples avec un unique vertex).
Pour r\'esumer, on d\'efinit l'ensemble $\G$ de tous les
graphes et l'ensemble $\Gtl$ des graphes ne contenant pas le vertex
bivalent. On a alors:
\beq
\H_\Gamma=\bigoplus_{\Gamma' \in \Gtl,\,
\Gamma'\subset\Gamma}(i_{\Gamma'\Gamma}) \Htl_{\Gamma'}.
\eeq
A l'aide de ceci, on obtient que
$\H_{\mathrm{gns}}=
\bigoplus_{\Gamma \in {\Gtl}}\Htl_{\Gamma}$,
qui impl\'emente en pratique la somme non-directe
$+_{\Gamma \in {\Gtl}}\H_{\Gamma}\equiv
\bigoplus_{\Gamma \in {\G}}\H_{\Gamma}/\sim$.

\bigskip
Nous avons donc construit notre espace d'Hilbert des fonctions
d'onde invariantes sous l'action du groupe de jauge $SU(2)$. Il reste
\`a impl\'ementer l'action des diff\'eomorphismes spatiaux et de la
contrainte Hamiltonienne. Reste \'egalement \`a fournir un sens physique,
une interpr\'etation g\'eom\'etrique aux \'etats construits. Cela sera
obtenu \`a travers l'impl\'ementation des op\'erateurs aires et volumes
agissant sur la base des r\'eseaux de spin, comme explicit\'e dans la
prochaine section.

\section{Quantifier la relativit\'e g\'en\'erale}

Je vais d\'ecrire la quantification de la th\'eorie r\'eelle, en d\'ecrivant
principalement le cadre cin\'ematique de la th\'eorie.
La quantification
de la formulation complexe self-duale est identique \`a l'exception de
 l'impl\'ementation des conditions de r\'ealit\'e,
que je discuterai \`a part.

\subsection{L'espace d'Hilbert de la \lqg}

Nous avons construit un espace d'Hilbert d'\'etats invariants
de jauge $SU(2)$, c'est-\`a-dire qu'ils r\'esolvent la
contrainte $G_i=0$. On note cet espace $\H_{inv}$.
Maintenant, nous voudrions nous occuper des contraintes $V_a$ et $C$.
Tout d'abord, nous avons vu que les diff\'eomorphismes spatiaux
sont g\'en\'er\'es par les contraintes $\Delta_a$ qui sont
des combinaisons des $G_i$ et $V_a$. Par cons\'equent, sur
un espace invariant de jauge, les diff\'eomorphismes spatiaux
sont induits par $V_a$, c'est-\`a-dire que pour r\'esoudre $V_a=0$, il faut
se placer sur un espace d'\'etats invariants sous diff\'eomorphismes.
En d'autres termes, il s'agit de quotienter $\H_{inv}$ par
l'action des diff\'eomorphismes. 

Quelle est donc l'action des diff\'eomorphismes (spatiaux) sur
les fonctionnelles de r\'eseaux de spins? Elle est simple,
elle d\'eplace le graphe support de la fonctionnelle cylindrique.
Plus pr\'ecis\'ement, consid\'erons un diff\'eomorphisme $\phi$ sur $\Sigma$.
Puisque son action sur une holonomie se lit
$\phi\cdot U_e(A)\arr U_e(\phi^{-1}A)=U_{\phi\cdot e}(A)$
son action sur une fonctionnelle cylindrique $\varphi_{\Gamma,f}(A)$ est:
\beq
\phi\cdot\varphi_{\Gamma,f}(A)\arr
\varphi_{\Gamma,f}(\phi^{-1}A)=\varphi_{\phi\cdot\Gamma,f}(A).
\eeq
On consid\`ere donc l'espace des classes d'\'equivalence des
fonctions cylindriques sous l'action des diff\'eomorphismes. 
La mesure \Ref{lqg:mesure} et le produit scalaire
\Ref{lqg:prodscal} passe sans probl\`eme au quotient.
La diff\'erence est que pr\'ec\'edemment deux fonctions cylindriques
\`a support sur des graphes distincts \'etaient orthogonaux par
d\'efinition et que maintenant ils sont comparables si les deux graphes
sont \'equivalents sous diff\'eomorphismes. Ainsi on construit
l'espace d'Hilbert $\H_{diff}$. Une base de cet espace est donn\'ee
formellement par les r\'eseaux de spin labell\'es par les spins
et par les classes d'\'equivalence de graphes (sous diff\'eomorphismes)\footnotemark.
Ces classes d'\'equivalence d\'ependent explicitement de la topologie de $\Sigma$
et rendent compte de la mani\`ere dont les graphes s'enroulent autour de $\Sigma$.
Notons que, d\`es qu'on autorise des vertex de valence sup\'erieure ou \'egale \`a 5, 
la base de $\H_{diff}$ est labell\'ee par un param\`etre continu et $\H_{diff}$
devient non-s\'eparable.
\footnotetext{Ce ne sont pas les r\'eseaux de spins abstraits ou
{\it abstract spin networks}
(comme on utilise dans les mod\`eles de mousse de spin par exemple)
qui sont d\'efinis enti\`erement par leur structure combinatoriale
car les classes d'\'equivalence sous diff\'eomorphismes d\'ependent
encore de la topologie de l'hypersurface $\Sigma$.} Le lecteur pourra
trouver des d\'etails techniques dans
\cite{carlo:cours,thiemann:long,thiemann:short,grot}.

Maintenant, pour compl\`etement d\'efinir la th\'eorie, il faudrait
impl\'ementer l'action de la contrainte $C$ (induisant les
diff\'eomorphismes dans la direction temps). Puis on projeterait
sur l'espace des \'etats invariants par cette action (ou les classes
d'\'equivalence sous cette action si elles ont un sens physique), ce
qui nous donnerait l'espace $\H_{phys}$ des \'etats physiques
satisfaisant toutes les contraintes classiques. Malheureusement, il n'y
a pas encore de concensus sur l'action de $C$ sur les $\H_{diff}$
(ou $\H_{inv}$). N\'eanmoins, tout le monde s'accorde plus ou moins
sur le fait qu'elle agit sur les vertex des r\'eseaux de spin,
modifiant les spins mais aussi le graphe support en cr\'eant de
nouveaux vertex. Et il existe
plusieurs propositions d'actions possibles
\cite{thiemann:qsd,borissov,lee:hamil,gaul} se distinguant par
les modifications exactes induites sur le graphe support (action dite
ultra-locale agissant uniquement au niveau des vertex comme dans
\cite{thiemann:qsd} ou action visant des corr\'elations \`a longues distances
comme dans \cite{lee:hamil}) et les amplitudes (d\'ependantes des spins)
associ\'ees \`a ces modifications. Le but \'etant de produire une bonne
limite semi-classique avec des corr\'elations \`a longue distance
(similaire \`a la propagation des ondes gravitationelles). 
Des simulations num\'eriques dans des mod\`eles de dynamique simplifi\'ee
ont \'et\'e effectu\'ees
mais rien de d\'efinitif
n'a encore \'et\'e prouv\'e. Ainsi un flou r\`egne encore sur la
dynamique des r\'eseaux de spin. D'o\`u l'introduction  des mod\`eles de
mousse de spin, qui sont des mod\`eles d'espace-temps pouvant \^etre
repr\'esent\'es comme des histoires de r\'eseaux de spin, pour
rem\'edier \`a ce probl\`eme et d\'efinir une ``bonne'' dynamique pouvant
induire une ``bonne'' limite semi-classique.

Malgr\'e ce probl\`eme au niveau de la dynamique, une caract\'eristique
attrayante de la \lqg est l'interpr\'etation g\'eom\'etrique des \'etats
de r\'eseaux de spin de $\H_{diff}$. Elle est obtenue \`a travers
les op\'erateurs g\'eom\'etriques d'aire et de volume repr\'esent\'es sur
$\H_{inv}$.

\subsection{La g\'eom\'etrie des r\'eseaux de spin}

Tout d'abord, en utilisant l'expression de l'aire d'une surface \Ref{aire:def}
en fonction de la triade, il est possible d'en donner une repr\'esentation
op\'erationelle sur l'espace $\H_{inv}$. Ceci est obtenu en
rempla\c cant la quantit\'e
classique $E^a_i$ par la d\'eriv\'ee par rapport \`a la connexion
vu le crochet de
Poisson \Ref{lqg:bracket}:
\beq
\{A_a^i(x),E^b_j(y)\}=\imm\delta_a ^b\delta^i_j\delta ^{(3)}(x,y)
\quad \Rightarrow\quad
E^a_i(x)\arr \hE^a_i(x)=-i\imm \f{\dd}{\dd A_a^i(x)}.
\label{hatE}
\eeq
Par cons\'equent, pour \'etudier l'action sur les fonctions cylindriques
d'un op\'erateur form\'e \`a partir de $\hE^a_i$,
nous avons besoin de l'expression de la d\'eriv\'ee d'une holonomie
\cite{holovariation}:
\beq
\f{\dd}{\dd A_a^i(x)} U_e(A)=
\f{\dd}{\dd A_a^i(x)} e^{\int_0^1 ds \,\dot{e}^b(s) A_b^j \tau_j}=
\int ds\, \dot{e}^a(s)\delta^{(3)}(e(s),x)
U_{e_1(s)}(A)\tau_iU_{e_2(s)}(A)
\eeq
o\`u la courbe $e$ est coup\'ee \`a la coordonn\'ee en $e_1(s)$ et $e_2(s)$.
En d'autres termes, l'action de $\hE^a_i(x)$ revient \`a ins\'erer
le g\'en\'erateur $\tau_i$ de $\su$ au milieu de l'holonomie.

Maintenant, suivant la logique de \cite{carlo:area}, on en d\'eduit l'action
de l'op\'erateur $\hE_i({\cal S})$ sur un r\'eseau de spin. Pour simplifier
les expressions, nous supposons la surface ${\cal S}$ n'intersecte le graphe
support $\Gamma$ du r\'eseau de spin qu'une seule fois, en un point qui
n'est pas un vertex. On note $e$ le lien intersect\'e
et $j$ la repr\'esentation l'indexant.
Il s'agit alors de d\'eriver (la matrice de repr\'esentation de)
l'holonomie $R^j[U_e]$. Plus pr\'ecis\'ement:
\beqs
\hE_i({\cal S}).R^j(U_e)&=&
-i\imm\int_{{\cal S}}
d\sigma^1d\sigma^2 \int_e ds\, \epsilon_{abc}
\f{\dd x^a(\vec{\sigma})}{\dd\sigma^1}\f{\dd
  x^b(\vec{\sigma})}{\dd\sigma^2}
\f{\dd e^a(s)}{\dd s}
\delta^{(3)}(e(s),x) \nonumber \\
&&\times
R^j[U_{e_1(s)}(A)]R^j[\tau_i]R^j[U_{e_2(s)}(A)].
\eeqs
Maintenant, en utilisant la r\'egularisation \'evidente
$\lim_{y\arr x}(\hE_i({\cal S})(x).\hE_i({\cal S})(y))$,
on peut regarder l'action de
l'op\'erateur $\hE_i({\cal S})\hE_i({\cal S})$:
\beqs
\hE_i({\cal S})\hE_i({\cal S}). R^j(U_e)&=&
-i\imm^2\int_{{\cal S}}
d\sigma^1d\sigma^2 \int_e ds\, \epsilon_{abc}
\f{\dd x^a(\vec{\sigma})}{\dd\sigma^1}\f{\dd
  x^b(\vec{\sigma})}{\dd\sigma^2}
\f{\dd e^a(s)}{\dd s}
\delta^{(3)}(e(s),x) \nonumber \\
&&\times
R^j[U_{e_1(s)}(A)]R^j[\tau_i]R^j[\tau_i]R^j[U_{e_2(s)}(A)].
\eeqs
$R^j[\tau_i]R^j[\tau_i]$ est le Casimir de $\su$ dans la
repr\'esentation $j$ et c'est juste le nombre $j(j+1)$. \\
Alors $R^j[U_{e_1(s)}(A)]R^j[\tau_i\tau_i]R^j[U_{e_2(s)}(A)]=R^j(U_e)$
ne d\'epend plus de $s$ et sort de l'int\'egrale. Il appara\^it alors
que
$$
\epsilon_{abc}
\f{\dd x^a(\vec{\sigma})}{\dd\sigma^1}\f{\dd
  x^b(\vec{\sigma})}{\dd\sigma^2}
\f{\dd e^a(s)}{\dd s}
$$
est le Jacobien du changement de coordonn\'ees $\sigma_1,\sigma_2,s$
\`a $x_1,x_2,x_3$. Donc l'int\'egrale restante compte simplement
le nombre (orient\'e) d'intersections entre ${\cal S}$ et le lien
$e$, que nous avons choisi d'\^etre 1.
Par cons\'equent l'action
de $\hE_i({\cal S})\hE_i({\cal S})$ est diagonale sur l'holonomie
$R^j(U_e)$:
\beq
\hE_i({\cal S})\hE_i({\cal S}).R^j(U_e)=
\imm^2 j(j+1)R^j(U_e).
\eeq
Nous g\'en\'eralisons tout de suite cette relation
\`a la fonctionnelle r\'eseau de spin:
\beq
\hE_i({\cal S})\hE_i({\cal S}).
\phi^{(j_1,\dots,j_e,\dots,j_E)}_\Gamma(U_1,\dots,U_e,\dots,U_E)=
\imm^2j_e(j_e+1)\phi^{(j_1,\dots,j_e,\dots,j_E)}_\Gamma.
\label{EE}
\eeq
On peut alors d\'efinir l'op\'erateur $\sqrt{\hE_i({\cal S})\hE_i({\cal S})}$
dont l'action sera \'egalement diagonale avec comme valeur propre
$\sqrt{j_e(j_e+1)}$.

\begin{figure}[t]
\begin{center}
\psfrag{e1}{$e_1$}
\psfrag{e2}{$e_2$}
\psfrag{S}{${\cal S}$}
\psfrag{s1}{$\sigma_1$}
\psfrag{s2}{$\sigma_2$}
\psfrag{s}{$s$}
\psfrag{A}{${\cal A}_{{\cal S}}\equiv\gamma\sqrt{j_e(j_e+1)}$}
\includegraphics[width=7cm]{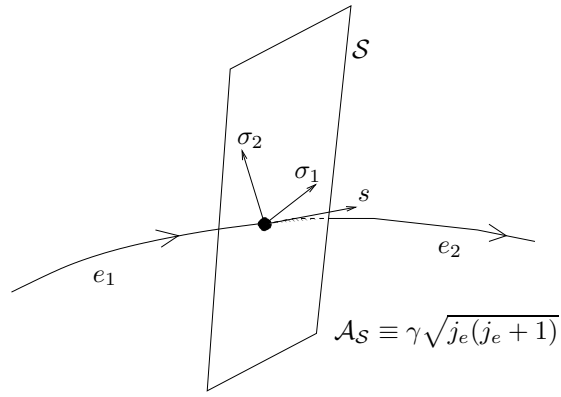}
\end{center}
\caption{Aire d'une surface ${\cal S}$ intersectant le lien $e$.}
\end{figure}

Maintenant, on utilise la formule \Ref{aire:def} de
l'aire d'une surface ${\cal S}$
pour \'elever ${\cal A}_{{\cal S}}$ au rang
d'op\'erateur $\hat{{\cal A}}_{{\cal S}}$.
Calculant l'action de l'op\'erateur aire $\hat{{\cal A}}$
sur une fonctionnelle de r\'eseau de spin \`a support sur un graphe $\Gamma$,
on peut choisir une partition telle que la surface ${\cal S}$
soit d\'ecoup\'ee en petits morceaux ${\cal S}_k$ qui
n'intersectent le graphe $\Gamma$ qu'en
un seul point chacun. Pour le moment, nous ne consid\'erons que le cas
o\`u le Jacobien $(\sigma_1,\sigma_2,s)\arr (x_1,x_2,x_3)$ est non nul \`a
toutes les intersections. Alors $\hat{{\cal A}}_{{\cal S}}$ est diagonal
sur les $\phi_\Gamma$ et:
\beq
\hat{{\cal A}}_{{\cal S}}.
\phi^{(j_1,\dots,j_E)}_\Gamma=
\imm\sum_{e |e\cap{\cal S}\ne\emptyset}
\sqrt{j_e(j_e+1)}\,\phi^{(j_1,\dots,j_E)}_\Gamma.
\eeq
Dans le cas d'une surface ${\cal S}$ intersectant
le graphe $\Gamma$ en un vertex,
la valeur propre n'est
plus $\sqrt{j_e(j_e+1)}$ et
il faut distinguer les liens transversaux {\it up} (le Jacobien est non nul
et positif), les liens transversaux {\it down} (le Jacobien est n\'egatif) et
les liens tangents (le Jacobien est nul donc l'op\'erateur $E_iE_i$
construit ci-dessus est nul). On regroupe tous les spins
des liens {\it up} en un lien (virtuel) de spin $j^u$, de m\^eme avec
les liens {\it down} qui sont regroup\'es en $j^d$ et les liens tangents
en $j^t$. Alors $\hat{{\cal A}}_{{\cal S}}$ est toujours diagonale et la valeur
propre est \cite{carlo:area}:
\beq
\hat{{\cal A}}_{{\cal S}}.\phi_\Gamma=
\imm\f{1}{2}\sqrt{2j^u(j^u+1)+2j^d(j^d+1)-j^t(j^t+1)}\,\phi_\Gamma.
\eeq

\medskip

Les fonctionnelles r\'eseaux de spin sont donc la base diagonalisant
les op\'erateurs d'aire (des surfaces de $\Sigma$).
Cela fournit une interpr\'etation g\'eom\'etrique
aux \'etats r\'eseaux de spin en inversant le raisonnement.
En effet, un r\'eseau de spin donn\'e d\'efinit un \'etat quantique
de la g\'eom\'etrie de l'hypersurface $\Sigma$. Cette g\'eom\'etrie
est reconstruite par le fait que l'aire d'une surface intersectant
le graphe support est donn\'ee en fonction des spins de l'\'etat. Toute autre
surface n'existe pas!

Notons que le spectre de l'aire contient un facteur $\imm$. Par cons\'equent,
ce param\`etre innofensif au niveau classique intervient d'une mani\`ere
cruciale au niveau quantique modifiant le spectre des op\'erateurs. Par
cons\'equent, nous obtenons toute une famille de quantification
{\bf in\'equivalente}\footnotemark
labell\'ee par le param\`etre d'Immirzi $\imm$.

\footnotetext{En fait, elles ne sont pas \'equivalentes dans le context cin\'ematique
o\`u on ignore la contrainte Hamiltonienne. Par contre, si on prend en
compte la contrainte Hamiltonienne $C$ (ce que l'on ne sait pas faire explicitement),
les op\'erateurs aires/volumes ne repr\'esentent pas des observables et on ne peut
pas conclure a priori quant \`a l'\'equivalence ou la non-\'equivalence
de ces quantifications.}

Il existe une autre ambigu\"\i t\'e lors de la quantification des op\'erateurs
triades et aires, li\'ee \`a la r\'egularisation du produit d'op\'erateur 
triade $\hE^i_a(x)\hE^i_a(x)$ au m\^eme point dans l'expression \Ref{EE}.
Il est possible d'utiliser une autre r\'egularisation, se fondant sur
une quantification sym\'etrique des op\'erateurs $\hE$. Alors l'action de
$\hE^i_a(x)\hE^i_a(x)$ ne va pas se traduire par un facteur Casimir de $SU(2)$
$C=j(j+1)$, mais par un facteur $C+1/4$ \cite{alekseev}.
Prenant la racine de cette nouvelle
expression pour obtenir les valeurs propres de l'op\'erateur aire, on trouve
que $\sqrt{j(j+1)}$ est remplac\'e par l'expression plus simple:
\beq
{\cal A}_{{\cal S}}=\gamma\sqrt{j(j+1)}
\quad\arr\quad
{\cal A}_{{\cal S}}=\gamma\left(j+\f{1}{2}\right)
\eeq
Ceci est coh\'erent avec la vision d'un spectre de l'aire \'egalement
espac\'e et avec certains r\'esultats obtenus dans le contexte des mod\`eles
de mousse de spin (int\'egrale de chemin de la \lg) \cite{laurent:sf}.

\medskip

Il est possible de pousser cette interpr\'etation g\'eom\'etrique plus
loin en associant un volume \`a chaque vertex du graphe. Ceci est obtenu
par la construction d'un op\'erateur volume $\hat{\cal V}$.
Pour cela, on consid\`ere la formule classique du volume d'une
r\'egion ${\cal R}$:
\beq
\hat{\cal V}_{{\cal R}}=
\int_{{\cal R}}d^3x\,\sqrt{det^{(3)} g}
=
\int_{{\cal R}}d^3x\,\sqrt{\f{1}{3!}
\left|\epsilon_{abc}\epsilon_{ijk}E^{ai}E^{bj}E^{ck}\right|}
\eeq
puis on utilise la correspondance \Ref{hatE} rempla\c cant $E$ par
l'op\'erateur d\'erivation au niveau quantique.
Cela fournit un op\'erateur bien d\'efini \cite{thiemann:qsd}. Il n'agit
qu'au niveau des vertex du graphe: on associe un volume \`a chaque vertex du
graphe, d\'ependant des repr\'esentations vivant sur les liens partant
de ce vertex, et le volume d'une r\'egion ${\cal R}$ est la somme
des contributions des vertex contenus dans cette r\'egion.
Le volume associ\'e \`a un vertex bivalent ou trivalent est nul.
Le volume associ\'e \`a un vertex n-valent avec $n\ge4$
est non trivial et d\'epend
de l'entrelaceur vivant au vertex en question. Cela est intuitif
car cr\'eant des surfaces duales aux liens adjacents au vertex
(dont les aires sont donn\'ees par les spins vivant sur ces liens), les cas
bivalent et trivalent sont d\'eg\'en\'er\'es. Le premier cas non-trivial
est le vertex 4-valent auquel on peut associer un t\'etra\`edre dual.
Dans ce cas, le volume du t\'etra\`edre n'est pas enti\`erement d\'etermin\'e
par les aires de ces 4 faces. Il faut des donn\'ees en plus, qui sont ici
fournit par l'entrelaceur. Malheureusement, dans la base canonique
des entrelaceurs 4-valents (donn\'ee par la d\'ecomposition du 4-valent
vertex en deux vertex trivalents reli\'es par un lien virtuel interne dont
le spin labelle la base), cet op\'erateur volume n'est pas diagonal.
De plus, on ne conna\^it pas le spectre complet de cet op\'erateur.
Les premi\`eres valeurs propres sont donn\'ees dans \cite{carlo:vol}.
Le lecteur int\'eress\'e peut aussi d'autres r\'esultats analytiques et num\'eriques
dans \cite{vol:thesis}.
\begin{figure}[t]
\begin{center}
\psfrag{a}{$j_1$}
\psfrag{b}{$j_2$}
\psfrag{c}{$j_3$}
\psfrag{d}{$j_4$}
\psfrag{e}{$j_5$}
\psfrag{f}{$j_6$}
\psfrag{j}{$j$}
\psfrag{aire}{${\cal A}\equiv l_P^2\gamma\sqrt{j(j+1)}$}
\psfrag{vol}{${\cal V}(j,j_1,j_2)$}
\includegraphics[width=8.5cm]{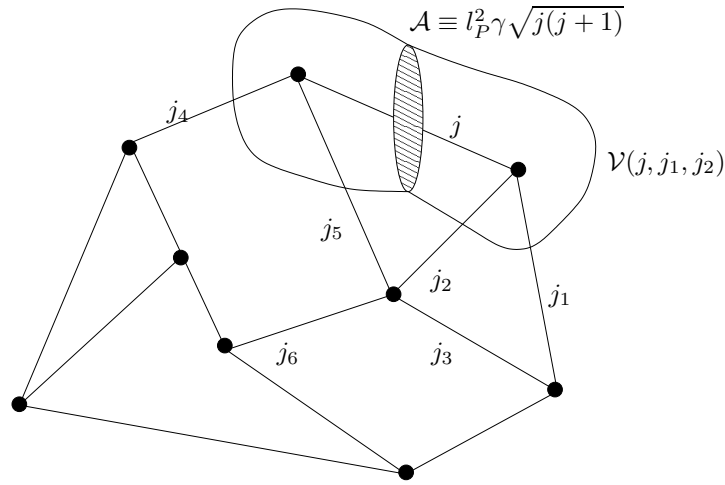}
\end{center}
\caption{Interpr\'etation g\'eom\'etrique d'un r\'eseau de spin:
les vertex repr\'esentent des morceaux d'espace et les liens
d\'ecrivent leurs surfaces fronti\`eres,
le volume et l'aire \'etant donn\'es par les spins
port\'es par les liens.}
\end{figure}

\medskip
Au final, nous avons plus ou moins d\'ecrit la g\'eom\'etrie d'un \'etat
quantique de l'hypersurface associ\'e \`a un r\'eseau de spin.
En passant au dual, cela donne l'image d'un espace par bloc.
Chaque vertex et son entrelaceur d\'efinissent un morceau de 3-volume.
Chaque lien et son spin associ\'e d\'efinissent la surface entre les
deux 3-volumes associ\'es \`a ces points source et cible. Ils d\'ecrivent
la relation entre deux vertex.
Puis ce r\'eseau de relations va \'evoluer dans le temps et se modifier. Cette
\'evolution est donn\'ee par la contrainte Hamiltonienne.

\medskip
Pour terminer, un petit mot sur
le param\`etre d'Immirzi $\imm$ et l'ambigu\"\i t\'e dans la quantification
qui en r\'esulte. Ce param\`etre peut \^etre vu comme soit d\'ecoulant
d'un terme dans
l'action initial ne modifiant en rien les \'equations du mouvement,
soit issu d'une
transformation canonique. Dans les deux cas, il n'intervient pas classiquement.
N\'eanmoins, il appara\^it en facteur du crochet de Poisson de la th\'eorie
et intervient donc au niveau de la th\'eorie quantique, modifiant le spectre
de l'aire et du volume par un facteur d'\'echelle $\imm$ au niveau
cin\'ematique. Il d\'efinit donc toute une famille de quantification
unitairement in\'equivalente.
Mais cela ne se limite pas au cadre cin\'ematique puisqu'il intervient
\'egalement dans la dynamique de la th\'eorie
(dans la contrainte Hamiltonienne).
Reste par cons\'equent \`a comprendre sa signification physique\dots

\subsection{A propos de la th{\'e}orie complexe}

La proc\'edure de quantification de la formulation complexe est
exactement identique \`a celle de la formulation r\'eelle.
De nouveau, on peut consid\'erer
les fonctionnelles cylindriques invariantes
de jauge de la connexion (cette fois-ci complexe et self-duale),
les flux et l'aire. Les fonctionnelles cylindriques d\'efinissent
les fonctions d'onde de la th\'eorie quantique et une base de
l'espace d'Hilbert est fournie par les r\'eseaux de spin.
Puis la contrainte Hamiltonienne d\'efinit la dynamique de ces
r\'eseaux de spin. Compar\'ee \`a la th\'eorie r\'eelle, son expression
est plus simple. Lors de la quantification, elle va \^etre plus simple
\`a impl\'ementer \cite{thiemann:qsd} bien qu'un certain flou subsiste quant \`a
d\'eterminer la ``bonne'' impl\'ementation.

La complication r\'eside alors dans le fait que les variables que
nous consid\'erons sont {\it a priori} complexes. Et que demander
que la m\'etrique reste r\'eelle impose les conditions de
r\'ealit\'e, \Ref{Ereality} et \Ref{Areality},
discut\'ees dans la section \ref{lqg:complexe}. Lors de la quantification,
une quantit\'e classiquement r\'eelle est contrainte \`a devenir un
op\'erateur auto-adjoint. Par cons\'equent, on aimerait que
\beq
(\hE^a_i)^\dagger = \hE^a_i,
\label{opEreality}
\eeq
\beq
(\hat{A}_a^i)^\dagger+\hat{A}_a^i=2\Gamma_a^i(\hE).
\label{opAreality}
\eeq
La premi\`ere condition impose directement que la m\'etrique soit r\'eelle.
La seconde correspond \`a \Ref{Ereality} et \Ref{Areality}, et donc
revient \`a imposer que la r\'ealit\'e de la m\'etrique est conserv\'ee lors
de l'\'evolution dans le temps.

Ces relations d'hermicit\'e contraignent la mesure et le produit scalaire
de la th\'eorie comme d'habitude en m\'ecanique quantique.
Dans la repr\'esentation donn\'ee par les fonctionnelles cylindriques
de la connexion, la condition \Ref{opEreality} sur $\hE$ correspond
\`a l'hermicit\'e de la d\'eriv\'ee par rapport \`a la connexion.
La condition \Ref{opAreality} sur $\hat{A}$ est plus compliqu\'ee \`a prendre
en compte. Pour l'impl\'ementer,
on peut se restreindre aux fonctionnelles holomorphes et il est possible
de construire explicitement le produit scalaire de la th\'eorie. Reste que
sa structure n'est pas intuitive i.e. on ne conna\^it pas explicitement
le produit scalaire de deux fonctionnelles r\'eseaux de spin. Par contre,
il est facile de voir \`a quoi cela correspond dans une repr\'esentation o\`u
on utiliserait des fonctionnelles de la triade (au lieu de la connexion).
La condition \Ref{opAreality} serait alors une constrainte sur les
op\'erateurs d\'eriv\'ees $\hat{{\cal A}}\sim\dd/\dd E$.

On peut remarquer que dans ce formalisme complexe, bien qu'ayant fix\'e
la jauge la t\'etrade, nous avons conserv\'e une connexion $A$ de jauge 
$SU(2)^\C\sim SL(2,\C)$ et donc il semble que la th\'eorie est un formalisme
covariant invariant sous Lorentz. Cependant, bien que les contraintes $G_i$
impose invariance sous le groupe complexifi\'e $SU(2)^\C$,
cette sym\'etrie est bris\'ee par les conditions de r\'ealit\'e
\Ref{opEreality} et \Ref{opAreality} et r\'eduit au groupe r\'eel et
compact $SU(2)$. Cela refl\`ete la fixation de jauge initiale.
En fait, l'approche complexe revient \`a complexifier
la th\'eorie, qui est alors plus simple \`a manier,
puis de prendre une section r\'eelle de la th\'eorie quantique complexe pour
obtenir la ``vraie'' th\'eorie quantique r\'eelle. La section r\'eelle
est d\'etermin\'ee par le choix de jauge (dans le cadre usuel,
la {\it time gauge}), et serait chang\'ee/tourn\'ee si on choisissait une
autre fixation de jauge.

Le lecteur peut trouver une discussion d\'etaill\'ee
de la quantification de la formulation complexe dans \cite{kodama}, et une
pr\'esentation simple et claire du produit scalaire et une mani\`ere
de l'exploiter pour avancer dans l'\'etude de la \lqg self-duale
dans \cite{soo}.

\chapter*{Conclusion: et la Dynamique?}

La \lqg proc\`ede explicitement
\`a la quantification canonique de la relativit\'e
dans des variables connexion $SU(2)$ et triade conjugu\'ee.
C'est une quantification math\'ematiquement bien d\'efinie et
non-pertubative de la gravit\'e, qui pr\'edit \`a quoi ressemblerait
l'espace(-temps) \`a l'\'echelle de Planck. Les \'etats quantiques
de l'espace (l'hypersurface canonique) sont donn\'es par les r\'eseaux
de spin, qui d\'efinissant une sorte de g\'eom\'etrie discr\`ete.
Il est possible d'impl\'ementer des op\'erateurs aire et volume et
leur spectres est discret, confirmant notre intuition que l'espace vient
par quanta \`a l'\'echelle de Planck!

\medskip

N\'eanmoins, la dynamique de la th\'eorie n'est pas explicitement connue
bien qu'il en existe plusieurs propositions. De plus, peu de solutions
exactes de toutes les contraintes sont connues. A part les boucles
sans intersection (les graphes triviaux), la
seule solution connue explicitement
est l'\'etat de Kodama \cite{kodama} dans la cas d'une constante cosmologique
$\Lambda>0$ (que nous n'avons pas discut\'e ici) et il semble
possible de d\'evelopper une th\'eorie des perturbations autour de
cet \'etat quantique \cite{soo,lee:lqg}.

Ce probl\`eme de la dynamique est li\'e au probl\`eme plus profond
de comprendre ce qu'est le {\bf temps} en gravit\'e quantique. Sur ce point,
la \lqg, en tant que formalisme canonique reposant sur un d\'ecoupage $3+1$
de l'espace-temps, ne nous \'eclaire pas beaucoup. Pour y rem\'edier, il a
\'et\'e introduit la notion de {\bf mousses de spin} ou {\it spin foams}
\cite{carlo&reis,baez:sf}
qui correspondent \`a des histoires de r\'eseaux de spin: ce sont des
mod\`eles d'espace-temps.
Plus pr\'ecis\'ement, un mod\`ele particulier correspond \`a assigner
une amplitude \`a chaque histoire possible. A l'aide
de ces mod\`eles, nous esp\'erons comprendre un peu mieux la notion de temps
et la dynamique des r\'eseaux de spin.
Je consacrerai la partie IV de cette th\`ese
\`a d\'ecrire ces objets, qui forment le sujet principal de ma th\`ese.
Je me suis particuli\`erement attach\'e \`a comprendre
la g\'eom\'etrie quantique qu'ils d\'efinisssent
et comment les utiliser en gravit\'e quantique non-perturbative.

Les mousses de spin sont des structures d'espace-temps
et le groupe de sym\'etrie usuellement utilis\'e dans leur construction
est le groupe de Lorentz. De l\`a naissent plusieurs difficult\'es.
Premi\`erement, il faut adapter le cadre des r\'eseaux\
de spin au cas d'un groupe
non-compact, ce qui ne semble pas \'evident \`a premi\`ere vue.
Pendant ma th\`ese, en collaboration avec Laurent Freidel,
je me suis attach\'e \`a dev\'elopper une telle th\'eorie
\cite{spinnet}, que
je d\'ecrirai dans la partie suivante.
Mais \'egalement, ces mousses de spin, bien qu'issues du formalisme canonique
de la \lqg, en semblent finalement assez \'eloign\'e. En effet, la \lg
repose sur la {\it time gauge}. Cette fixation de jauge
fixe la direction temps orthogonalement \`a l'hypersurface et r\'eduit
la t\'etrade \`a une triade: elle r\'eduit la sym\'etrie de Lorentz \`a une
sym\'etrie effective $SU(2)$ en gelant le plongement de l'hypersurface 
dans l'espace-temps. Ceci est \`a l'oppos\'e des mousses de spin,
dont la structure covariante repose sur la sym\'etrie de Lorentz et
qui d\'ecrit une hypersurface \`a travers son plongement dans l'espace-temps.
Reli\'e les mousses de spin au formalisme canonique est par cons\'equent
une t\^ache non-triviale. J'ai donc explor\'e au courant de ma th\`ese
des formalismes canoniques, en 3d et 4d,
ne reposant pas sur une fixation de jauge (et donc
conservant le groupe de Lorentz comme groupe de jauge), pour tenter de
d\'evelopper un formalisme canonique correspondant aux mousses de spin
et de comprendre leur lien avec la \lqg \cite{3+1,psn}.

\medskip
D'autres points sensibles en \lqg concernent le param\`etre d'Immirzi et
la construction d'une limite semi-classique de la th\'eorie.
En ce qui concerne $\imm$, il s'agit de comprendre son sens physique.
Ceci est encore une question ouverte. Dans le cadre d'un formalisme canonique
covariant et des mousses de spin, j'argumenterai (dans les parties III et IV)
qu'il n'intervient pas et donc qu'il ne semble pas avoir de sens physique.
En ce qui concerne la limite semi-classique, je ne me suis pas
attaqu\'e \`a la question durant ma th\`ese. Il s'agit de v\'erifier que l'on
retrouve bien la relativit\'e g\'en\'erale (et la m\'ecanique quantique)
dans une limite de basse \'energie et de pr\'evoir des effets physiques
mesurables de gravit\'e quantique. La question de retrouver la relativit\'e
g\'en\'erale est ardue et encore ouverte.

Li\'ee au probl\`eme de la limite semi-classique, la question de la
renormalisation de la \lqg est tout aussi fondamentale.
En effet, nous d\'ecrivons l'espace(-temps) \`a l'\'echelle de Planck. Et
pour obtenir des pr\'edictions physiques (v\'erifiables), il nous faut une
proc\'edure ({\it coarse graining}) pour passer de cette \'echelle de Planck
\`a des \'echelles plus raisonnables et ainsi tirer de la th\'eorie des 
valeurs d'observables pertinentes.


\part{Construire des r{\'e}seaux de spin}

Les r{\'e}seaux de spin, construits {\`a} partir du groupe $SU(2)$,
sont fondamentaux dans la construction de la \lqg comme expliqu\'e dans 
la partie pr{\'e}c{\'e}dente.
Ils consistent en des graphs dont les liens sont labell{\'e}s par des
repr{\'e}sentations de $SU(2)$ (et dont les noeuds sont labell{\'e}s par
des tenseurs invariants sous $SU(2)$). Ils forment une base des observables
de la \lg.
Plus g{\'e}n{\'e}ralement,
ce sont des observables des th{\'e}ories
de jauge, g{\'e}n{\'e}ralisation naturelle des boucles de Wilson.
Le cas des groupes compacts est suffisant pour l'{\'e}tude des
th{\'e}ories de Yang-Mills, mais l'extension aux groupes non-compacts
est n{\'e}cessaire quand on s'int{\'e}resse aux th{\'e}ories de la gravit{\'e} pour
lesquelles le groupe symm{\'e}trie locale est le groupe de Lorentz, qui est
non-compact.
La \lqg a motiv{\'e} l'{\'e}tude du cas compact ({\`a} travers des
g{\'e}n{\'e}ralisations du cas $SU(2)$), dont la th{\'e}orie est maintenant solide
du point de vue math{\'e}matique. Le but de cette partie est d'introduire
les notions n{\'e}cessaires {\`a} la d{\'e}finition rigoureuse des r{\'e}seaux
de spin pour des groupes non-compacts et d'explorer dans quelle mesure ils
forment une base des observables (invariantes de jauge). Au final,
la d{\'e}finition sera compatible avec les d{\'e}finitions d{\'e}j{\`a} connues
dans le cas compact et
nous obtiendront des graphs avec leurs liens labell{\'e}s par des
repr{\'e}sentations unitaires
(donc de dimension infinie dans le cas des groupes non-compacts)
du groupe consid{\'e}r{\'e}.
Le mat{\'e}riel pr{\'e}sent{\'e} dans cette partie
est issu d'une collaboration avec Laurent Freidel et 
a {\'e}t{\'e} pr{\'e}sent{\'e} dans \cite{spinnet}.
Les r{\'e}sultats obtenus pourront {\^e}tre appliqu{\'e}s {\`a}
la quantification canonique en $2+1$ dimensions, au formalisme
covariant en $3+1$ dimensions, aux mod{\`e}les de mousse de spins
({\it spin foams}) Lorentziens, et sans doute {\`a} d'autres th{\'e}ories.
En particulier, ils interviendront dans la partie suivante de cette th{\`e}se
dans le cadre de la \lqg en $2+1$ dimensions et de la quantification canonique
propos{\'e}e de la gravit{\'e} en $3+1$ dimensions.

\medskip

Rappellons tout d'abord le cadre du travail. On se place sur une
vari{\'e}t{\'e} $\Sigma$ et un $G$-fibr{\'e} principal
au-dessus de $\Sigma$, o{\`u} $G$ est un groupe
semi-simple. On note ${\cal A}$ l'espace des $G$-connexions
et ${\cal G}$ le groupe de jauge dont l'action sur $A$ est d{\'e}finie par
$A^k=k^{-1}Ak+k^{-1}dk$. Les th{\'e}ories auxquelles nous nous
int{\'e}ressons sont du type Yang-Mills dans le sens que les variables
conjugu{\'e}es dans l'espace des phases sont une $G$-connexion $A$ et
un champs de vecteur densit{\'e} $E$ {\`a} valeur dans $ad(P)$. L'espace
des phases correspondant est le fibr{\'e} cotangent $T^*({\cal A})$ {\`a}
l'espace des connexions. Il s'agit ensuite d'imposer l'invariance de jauge
(et {\'e}ventuellement l'invariance sous diff{\'e}omorphismes) sur cette espace.
Les op{\'e}rateurs sont repr{\'e}sent{\'e}s comme agissant sur des fonctions d'onde
ne d{\'e}pendant que de la connexion et l'espace d'Hilbert correspondant est
formellement $L^2({\cal A}/{\cal G})$, reste {\`a} d{\'e}finir une
bonne mesure invariante de jauge sur ${\cal A}/{\cal G}$. Dans cette partie,
on s'int{\'e}resse {\`a} la structure de cet espace.

En fait, nous allons nous concentrer sur des fonctionelles
invariantes de jauge particuli{\`e}res, les fonctionelles cylindriques,
et nous t{\^a}cherons de  munir l'espace de ces fonctionelles
d'une structure d'espace d'Hilbert. Ces fonctionelles cylindriques
ne d{\'e}pendent de la connexion qu'{\`a} travers un nombre fini de variables.
En effet, comme dans le cas $SU(2)$ de la \lqg, elles sont d{\'e}finies sur
des graphes. Plus pr{\'e}cis{\'e}ment, soit un graphe $\Gamma$ orient{\'e} et
${\cal C}^\infty$, compos{\'e} de $V$ vertex et de $E$ liens orient{\'e}s,
on peut d{\'e}finir l'application associant {\`a} une connexion l'ensemble 
de ses holonomies le long des liens:
\beq
\begin{array}{cccc}
\Gamma : & {\cal A} &\arr & G^{E} \\
& A &\arr& (g_{e_1},\dots,g_{e_E}).
\end{array}
\eeq
L'espace des fonctions cylindriques associ{\'e} {\`a} $\Gamma$ est le pullback de
 ${\cal C}^\infty(G^{E})$ d{\'e}fini par:
\beq
\Gamma^{*}\phi(A)=\phi(g_e(A)).
\eeq
L'action du groupe de jauge sur $\A$ se traduit par une action
de $G$ aux vertex du graphe $\Gamma$. En effet, notant $s(e)$ et
$c(e)$ les points {\bf s}ource et {\bf c}ible du lien $e$, l'action est
donn{\'e}e par:

\beq g_e(A^k)=
k_{s(e)}^{-1}g_e(A)k_{c(e)}.
\eeq
L'espace des {\it connexions discr{\`e}tes} (d{\'e}finies sur le graphe)
est not{\'e} $A_\Gamma=G^{E} /G^{V}$ et les fonctionelles
cylindriques sont des fonctions sur cet espace. Le but est de construire
une mesure $d\mu_\Gamma$ sur $A_\Gamma$, qui nous fournirait l'espace
d'Hilbert $\H_\Gamma=L^2(\Gamma,d\mu_\Gamma)$. Cet espace d'Hilbert
se doit de porter une repr\'esentation de l'alg{\`e}bre des op{\'e}rateurs 
de la th{\'e}orie (de Yang-Mills) {\'e}tudi{\'e}e restreinte aux graphes $\Gamma$.
Cet alg{\`e}bre  est obtenue par quantification de la structure cotangente
$T^*(G^{E}/G^{V})$. Elle est  g{\'e}n{\'e}r{\'e}e par les multiplications
par des fonctions invariantes de jauge sur $G^{E}$ et les op{\'e}rateurs
de d{\'e}rivation (invariants de jauge {\'e}galement). Ce que l'on attend
de la mesure $d\mu_\Gamma$ est qu'elle fournisse une
quantification
des quantit{\'e}s classiques r{\'e}elles en des op{\'e}rateurs Hermitiens.

Dans le cas compact, cette mesure est unique {\`a} un facteur pr{\`e}s. C'est,
comme utilis{\'e}e dans la partie pr{\'e}c{\'e}dente, le produit des mesures de Haar
sur chaque {\'e}l{\'e}ment de groupe (correspondant {\`a} un lien):
\beq
d\mu_\Gamma=\prod_{e\in E_\Gamma}dg_e.
\eeq
Cette mesure peut alors {\^e}tre {\'e}tendue d'une mani{\`e}re coh{\'e}rente
{\`a} l'espace de toutes les fonctionelles cylindriques (somme sur les graphes)
et d{\'e}finit une mesure -la mesure d'Ashtekar-Lewandowski,
d{\'e}finie dans la section pr{\'e}c{\'e}dente- sur l'espace des connexions
g{\'e}n{\'e}ralis{\'e}es modulo l'invariance de jauge.

Dans le cas non-compact, il ne suffit plus de prendre la mesure de Haar.
En effet, puisque le groupe a un volume infini, il faut diviser par ce
volume, c'est-{\`a}-dire fixer de jauge l'action du groupe. C'est ce qui est
pr{\'e}sent{\'e} dans cette partie.

\medskip

La construction de la mesure va se faire en deux {\'e}tapes.
Tout d'abord, nous montrerons que $A_\Gamma \sim G^{h_\Gamma}/Ad(G)$
o{\`u} $Ad(G)$ d{\'e}note l'action adjointe diagonale du groupe.
$h_\Gamma$ est le genre (d{\'e}termin{\'e} par le nombre d'anses) de la surface 2d
obtenue en gonflant/{\'e}paississant le graphe $\Gamma$.
Ce sera l'objet du chapitre \ref{chap:fleur}.
Puis, nous poursuivrons la fixation de jauge en exhibant un isomorphisme
entre $G^{h_\Gamma}/Ad(G)$ et $G^{h_\Gamma -1}$. On pourra d{\'e}finir la mesure
comme le pullback de la mesure de Haar sur $G^{h_\Gamma -1}$. On v{\'e}rifiera
que cette mesure est bien ind{\'e}pendante du choix de fixation de jauge.
Cette proc{\'e}dure sera expos{\'e}e dans le chapitre \ref{chap:mesure}.
Enfin, nous pourrons d{\'e}finir les r{\'e}seaux de spin comme une base
de l'espace $L^2$ construit {\`a} l'aide de la mesure. Plus particuli{\`e}rement,
ce seront les vecteurs propres d'op{\'e}rateurs de d{\'e}rivations invariants
de jauge (op{\'e}rateurs Laplaciens) dans le chapitre \ref{chap:spinnet}.
Finalement, le chapitre \ref{chap:exemples} sera consacr{\'e}
{\`a} l'application des r{\'e}sultats aux groupes $\slc$ et $\slr$ -en vue
de l'application {\`a} la gravit{\'e} en $2+1$ dimensions
et en $3+1$ dimensions -et {\'e}galement $SU(2)$- pour v\'erifier
que l'on retrouve bien les r{\'e}sultats connus sur les r{\'e}seaux de spin
pour groupes compacts.

\chapter{Des graphes aux fleurs} \label{chap:fleur}

Consid\'erons l'espace $A_\Gamma=G^{E}/G^{V}$
des connexions discr\`etes invariantes de jauge sur le graphe $\Gamma$.
Si $\Gamma$ n'est pas connexe, on le d\'ecompose en parties connexes
$\Gamma=\cup_i \Gamma_i$ et alors $A_\Gamma$ est le produit
$\times_i A_{\Gamma_i}$. Il est donc naturel de se restreindre \`a
l'\'etude des graphes $\Gamma$ connexes.

Le graphe $\Gamma$ est compos\'e de $E$ liens orient\'es et de $V$ vertex.
Chaque lien $e$ part d'un vertex source $s(e)$  et se termine \`a un vertex
cible $c(e)$. Une fonction sur $A_\Gamma$ est une fonction sur
$G^{E}$ invariante de jauge \`a chaque vertex du graphe.
Plus pr\'ecis\'ement, pour tous \'el\'ements $k_v\in G$ , un pour
chaque vertex $v$, la fonction $\phi$ doit satisfaire:
\beq
\phi(g_{e_i})=\phi(k^{-1}_{s(e_i)}g_{e_i}k_{c(e_i)}),\,i=1\dots E.
\label{gaugeinv}
\eeq
Le but est de d\'efinir une mesure pour int\'egrer une telle fonction.
Il s'agit dans un premier temps d'identifier les ``vrais''
degr\'es de libert\'e de $\phi$. Pour cela, nous allons fixer
l'invariance de jauge \Ref{gaugeinv}.

Le moyen le plus simple pour r\'eduire l'invariance de jauge est d'\'eliminer
le maximum de variables $g_e$ en les fixant par exemple \`a l'Identit\'e
$1\in G$. Plus pr\'ecis\'ement, choisissons un {\it arbre maximal} $T$ sur
notre graphe $\Gamma$. $T$ est un sous-graphe passant par tous
les vertex de $\Gamma$ sans jamais faire de boucle. En particulier,
un tel arbre $T$ est compos\'e de $V-1$ liens.
La propri\'et\'e fondamentale de $T$ est qu'\'etant donn\'es
deux vertex  arbitraires de $\Gamma$, il existe
un chemin unique dans $T$ allant d'un vertex \`a l'autre.
Ainsi, \'etant donn\'es deux vertex $A$ et $B$,
on peut d\'efinir le produit orient\'e $h^T_{AB}$ (l'holonomie)
des \'el\'ements du groupe le long du chemin entre $A$ et $B$ dans $T$.
Maintenant, en utilisant l'invariance de jauge \Ref{gaugeinv},
il est possible de fixer tous les \'el\'ements du groupe vivant
sur les liens de $T$ \`a $1$. Pour cela, fixons-nous un vertex $A$
comme point de r\'ef\'erence pour notre proc\'edure de fixation de jauge.
Puis on utilise la formule \Ref{gaugeinv} avec 
$$
k_v=h^T_{vA}.
$$
Pour un lien $e$ quelconque, la transformation s'\'ecrit
\beq
\gt_{e}=h^T_{A s(e)} g_e  h^{T}_{c(e) A}.
\label{gtoG}
\eeq
$\gt_e$ est l'holonomie le long de la boucle partant de $A$ passant par $e$
et revenant \`a $A$.
Regardons ce qu'il advient d'un lien $e\in T$.
Il existe un unique chemin le reliant \`a $A$,
sinon il y aurait une boucle dans $T$. On distingue deux cas. Soit
le chemin connecte $A$ \`a $s(e)$, soit il connecte $A$ \`a $c(e)$.
Quitte \`a inverser les orientations, on suppose  par exemple que le chemin
relie $A$ \`a $c(e)$. Alors 
$h^T_{s(e)A}=g_eh^T_{c(e)A}$ et $h^T_{A
s(e)}=(h^T_{s(e)A})^{-1}$, de telle sorte que \Ref{gtoG} se lisent
$\gt_{e}=1$.
Par cons\'equent le choix $k_v=h^T_{vA}$ permet de fixer les \'el\'ements
du groupe $g_e$ pour $e\in T$ \`a $1$. Cela d\'efinit une fonction
$\phi_T$ d\'ependant de $h_\Gamma=E-V+1$ \'el\'ements du groupe,
correspondant aux liens qui ne sont pas dans $T$:
\beq
\phi_T( \{ \gt_e,e\notin T \} )= \phi(g_e = \gt_e \tr{ si } e\notin T
\tr{ ou } = 1 \tr{ sinon}).
\eeq
Cette fonction a une invariance de jauge r\'esiduelle tr\`es simple:
\beq \forall k\in G, \,
\phi_T(\gt_{f_i})=\phi_T(k^{-1}\gt_{f_i}k),i=1\dots h_\Gamma.
\label{flower}
\eeq
En r\'esum\'e, la fixation de jauge fournit un isomorphisme
\beq
T:G^{h_\Gamma}/Ad(G) \rightarrow A_\Gamma
\eeq
et $\phi_T$ est le pullback de $\phi$ par cet isomorphisme.

L'invariance de jauge r\'esiduelle correspond \`a celle d'un graphe
qui n'a qu'un unique vertex. On appelle un tel graphe une {\it fleur}.
En fait, nous avons contract\'e l'arbre $T$
en le point $A$, qui est le vertex unique \`a l'arriv\'ee. A priori,
la construction et donc la fonction $\phi_T$ semble
d\'ependre du choix du point $A$. Dans les faits, toute la construction est
ind\'ependante du choix de $A$. Pour montrer cela, prenons un autre
vertex $B$ et notons $h=h^T_{AB}$ le produit orient\'e des \'el\'ements
du groupe le long du chemin reliant $A$ \`a $B$. Fixant de jauge utilisant
$B$ comme point de r\'ef\'erence, on cr\'ee les variables
\beq
\tl{G}^{(T)}_{e}=h^T_{B s(e)} g_e h^T_{t(e) B}= h^{-1} h^T_{A
s(e)} g_e h^T_{t(e)A} h= h^{-1} \gt_{e} h.
\eeq
Cela d\'efinit une nouvelle fonction
$\tl{\phi}_T$ bas\'ee sur ces nouvelles variables, mais elle sera \'egale
\`a $\phi_T$ gr\^ace \`a l'invariance de jauge (r\'esiduelle)
\Ref{flower} avec $k=h$.

\medskip

Un point important pour la suite est la mani\`ere dont la fonction $\phi^T$
change lorsqu'on modifie l'arbre maximal $T$. Choisissons pour cela
un autre arbre maximal $U$. On peut suivre la m\^eme proc\'edure de fixation
de jauge pour $U$ en utilisant le m\^eme point de r\'ef\'erence $A$.
On obtient alors des variables $\gu_e$ pour chaque lien $e\notin U$
et on d\'efinit une fonction $\phi_U$ sur la fleur.
Pour relier $\phi_T$ et $\phi_U$, nous voudrions d\'ecomposer
les variables $\gu_e$ en fonction des variables $\gt_e$.
Plus g\'en\'eralement, consid\'erons une boucle orient\'ee ${\cal L}$ partant
du point $A$ et y revenant et d\'efinissons le produit
orient\'e $H$ des \'el\'ements du groupe le long de ${\cal L}$ (l'holonomie
ou transport parall\`ele
autour de ${\cal L}$). Il est possible d'exprimer $H$ en fonction
des $\gt_e$.  Une telle boucle doit contenir un lien
n'appartenant pas \`a $T$, sinon $T$ contiendrait une boucle.
Il est alors facile de se rendre compte que $H$ est le produit orient\'e
-suivant l'orientation de ${\cal L}$- des variables $\gt_e$ pour
$e$ sur ${\cal L}$ mais pas dans $T$. Maintenant, pour un lien $e\notin U$,
l'\'el\'ement du groupe $\gu_e$ est l'holonomie le long de la boucle
${\cal L}^{(U)}[e]$ suivant l'arbre $U$ et
allant de $A$ \`a $s(e)$ puis revenant de $c(e)$ \`a $A$.
Cela permet donc d'exprimer les variables $\gu_e$ en tant que produit orient\'e
de $\gt_f$. Au niveau des fonctions $\phi_T$ et $\phi_U$, la relation se lit:
\beq
\phi_T(\gt_e)=
\phi_U(\gu_e=\overrightarrow{\prod_{f\in {\cal L}[e]\setminus T}} \gt_f)
\label{TtoU}
\eeq

\medskip
Pour illustrer la proc\'edure, il est plus facile de visualiser
les relations sur des exemples. Consid\'erons par exemple
le graphe:
\begin{center}
\epsfig{figure=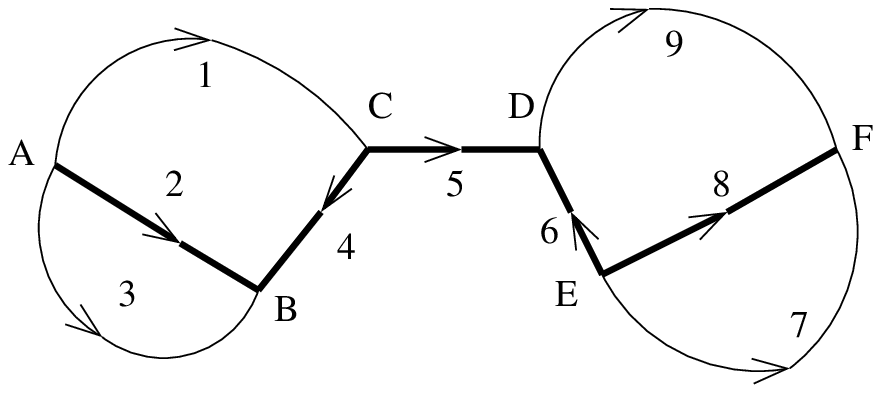,height=3.5cm}
\end{center}
o\`u l'arbre $T$ choisi pour la proc\'edure de fixation de jauge est en gras.
Une fonction invariante de jauge $\phi$ \`a support sur ce graphe satisfait
la relation suivante pour tout choix de $k_v\in G$:
\beq
\begin{array}{l}
\phi(g_1,\dots,g_9)= \\
\quad\phi(k_A^{-1}g_1k_C,k_A^{-1}g_2k_B,k_A^{-1}g_3k_B,k_C^{-1}g_4k_B,
k_C^{-1}g_5k_D,k_E^{-1}g_7k_F,k_E^{-1}g_8k_F,k_D^{-1}g_9k_F).
\end{array}
\eeq
On suit la proc\'edure de fixation de jauge en utilisant le vertex $C$
comme point de r\'ef\'erence. Cela r\'eduit le graphe \`a une fleur
\`a quatre p\'etales:
\begin{center}
\setlength{\unitlength}{0.00066667in}
\begingroup\makeatletter\ifx\SetFigFont\undefined%
\gdef\SetFigFont#1#2#3#4#5{%
  \reset@font\fontsize{#1}{#2pt}%
  \fontfamily{#3}\fontseries{#4}\fontshape{#5}%
  \selectfont}%
\fi\endgroup%
{\renewcommand{\dashlinestretch}{30}
\begin{picture}(2579,1537)(0,-10)
\path(957,750)(955,750)(951,751)
	(943,752)(931,754)(915,757)
	(896,761)(874,765)(850,771)
	(824,777)(797,784)(770,792)
	(742,802)(715,812)(687,825)
	(660,839)(632,856)(605,876)
	(578,898)(554,923)(533,949)
	(515,976)(501,1000)(490,1023)
	(482,1042)(477,1059)(473,1074)
	(470,1087)(468,1099)(467,1110)
	(467,1121)(467,1133)(467,1145)
	(468,1159)(469,1174)(471,1192)
	(474,1210)(479,1230)(486,1250)
	(496,1269)(507,1282)(519,1293)
	(531,1302)(544,1310)(556,1316)
	(567,1322)(577,1326)(587,1330)
	(596,1333)(605,1336)(613,1339)
	(621,1341)(629,1343)(637,1344)
	(646,1345)(656,1344)(668,1343)
	(680,1340)(694,1335)(710,1328)
	(727,1318)(745,1305)(765,1289)
	(784,1269)(804,1244)(823,1215)
	(840,1185)(855,1154)(869,1123)
	(880,1091)(891,1059)(901,1026)
	(909,993)(917,961)(925,928)
	(931,897)(937,867)(942,840)
	(946,815)(950,794)(953,777)
	(955,765)(956,757)(957,752)(957,750)
\path(957,750)(957,752)(958,756)
	(959,764)(961,776)(964,792)
	(968,811)(972,833)(978,858)
	(984,883)(991,910)(999,938)
	(1008,965)(1019,993)(1031,1020)
	(1046,1048)(1062,1076)(1082,1103)
	(1104,1129)(1129,1154)(1155,1175)
	(1182,1193)(1206,1207)(1228,1218)
	(1248,1226)(1265,1231)(1280,1235)
	(1293,1238)(1305,1240)(1316,1241)
	(1327,1241)(1339,1241)(1351,1241)
	(1365,1240)(1380,1239)(1398,1237)
	(1416,1234)(1436,1229)(1456,1222)
	(1475,1212)(1488,1201)(1499,1189)
	(1508,1177)(1516,1164)(1522,1152)
	(1528,1141)(1532,1130)(1536,1120)
	(1539,1111)(1542,1103)(1545,1095)
	(1547,1087)(1549,1079)(1550,1070)
	(1550,1061)(1550,1051)(1548,1040)
	(1545,1027)(1541,1013)(1533,997)
	(1524,980)(1511,962)(1495,942)
	(1475,923)(1450,903)(1421,884)
	(1392,867)(1361,852)(1329,838)
	(1297,826)(1265,816)(1232,806)
	(1200,797)(1167,790)(1135,782)
	(1104,776)(1074,770)(1046,765)
	(1022,761)(1001,757)(984,754)
	(972,752)(964,751)(959,750)(957,750)
\path(957,750)(957,748)(956,744)
	(955,736)(953,724)(950,708)
	(946,689)(942,667)(936,642)
	(930,617)(923,590)(915,562)
	(905,535)(895,507)(882,480)
	(868,452)(851,424)(831,397)
	(809,371)(784,346)(758,325)
	(731,307)(707,293)(684,282)
	(665,274)(648,269)(633,265)
	(620,262)(608,260)(597,259)
	(586,259)(574,259)(562,259)
	(548,260)(533,261)(515,263)
	(497,266)(477,271)(457,278)
	(438,288)(425,299)(414,311)
	(405,323)(397,336)(391,348)
	(385,359)(381,370)(377,380)
	(374,389)(371,397)(368,405)
	(366,413)(364,421)(363,430)
	(362,439)(363,449)(364,460)
	(367,473)(372,487)(379,503)
	(389,520)(402,538)(418,558)
	(438,577)(463,597)(492,616)
	(522,633)(553,648)(584,662)
	(616,674)(648,684)(681,694)
	(714,703)(746,710)(779,718)
	(810,724)(840,730)(867,735)
	(892,739)(913,743)(930,746)
	(942,748)(950,749)(955,750)(957,750)
\path(957,750)(959,750)(963,749)
	(971,748)(983,746)(999,743)
	(1018,739)(1040,735)(1064,729)
	(1090,723)(1117,716)(1144,708)
	(1172,698)(1199,688)(1227,675)
	(1254,661)(1282,644)(1309,624)
	(1336,602)(1360,577)(1381,551)
	(1399,524)(1413,500)(1424,477)
	(1432,458)(1437,441)(1441,426)
	(1444,413)(1446,401)(1447,390)
	(1447,379)(1447,367)(1447,355)
	(1446,341)(1445,326)(1443,308)
	(1440,290)(1435,270)(1428,250)
	(1418,231)(1407,218)(1395,207)
	(1383,198)(1370,190)(1358,184)
	(1347,178)(1336,174)(1326,170)
	(1317,167)(1309,164)(1301,161)
	(1293,159)(1285,157)(1276,156)
	(1267,155)(1257,156)(1246,157)
	(1233,160)(1219,165)(1203,172)
	(1186,182)(1168,195)(1148,211)
	(1129,231)(1109,256)(1090,285)
	(1073,315)(1058,346)(1045,377)
	(1033,409)(1022,441)(1013,474)
	(1004,507)(996,539)(989,572)
	(983,603)(977,633)(972,660)
	(968,685)(964,706)(961,723)
	(959,735)(958,743)(957,748)(957,750)
\put(784,808){\makebox(0,0)[lb]{\smash{{{\SetFigFont{10}{12.0}{\familydefault}{\mddefault}{\updefault}C}}}}}
\put(1418,1327){\makebox(0,0)[lb]{\smash{{{\SetFigFont{10}{12.0}{\familydefault}{\mddefault}{\updefault}$G^{(T)}_9$}}}}}
\put(1533,288){\makebox(0,0)[lb]{\smash{{{\SetFigFont{10}{12.0}{\familydefault}{\mddefault}{\updefault}$G^{(T)}_7$}}}}}
\put(0,1125){\makebox(0,0)[lb]{\smash{{{\SetFigFont{10}{12.0}{\familydefault}{\mddefault}{\updefault}$G^{(T)}_1$}}}}}
\put(225,0){\makebox(0,0)[lb]{\smash{{{\SetFigFont{10}{12.0}{\familydefault}{\mddefault}{\updefault}$G^{(T)}_3$}}}}}
\end{picture}
}
\end{center}
et cr\'ee les variables $\gt$:
\beqs
\gt_1=g_4g_2^{-1}g_1, \quad \gt_3=g_4g_2^{-1}g_3g_4^{-1}, \quad \nonumber \\
\gt_7=g_5g_6^{-1}g_7g_8^{-1}g_6g_5^{-1}, \quad
\gt_9=g_5g_9g_8^{-1}g_6g_5^{-1}.
\eeqs
On d\'efinit ainsi la fonction fix\'ee de jauge
\beq
\phi_T(\gt_1,\gt_3,\gt_7,\gt_9)=
\phi(\gt_1,1,\gt_3,1,1,1,
\gt_7,1, \gt_9). 
\eeq
On peut faire la m\^eme chose avec un autre arbre $U$:
\begin{center}
\epsfig{figure=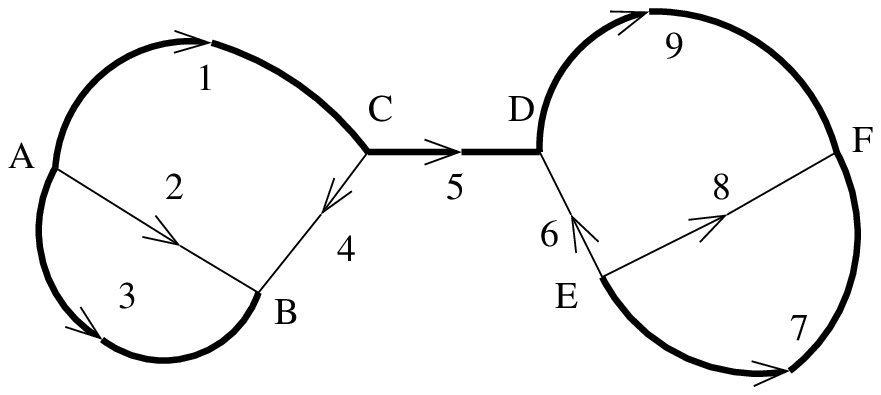,height=3.5cm}
\end{center}
Effectuant toujours la proc\'edure de fixation de jauge avec $C$ comme point
de r\'ef\'erence, on r\'eduit le graphe \`a la fleur \`a quatre p\'etales et
on d\'efinit les variables $\gu$:
\beqs
\gu_2=g_1^{-1}g_2g_3^{-1}g_1, &\quad &
\gu_4=g_4g_3^{-1} g_1, \nonumber \\
\gu_6=g_5g_9g_7^{-1}g_6g_5^{-1}, &\quad &
\gu_8=g_5g_9g_7^{-1}g_8g_9^{-1}g_5^{-1}.
\eeqs
La fonction $\phi_U$ est d\'efinit par
\beq
\phi_U(\gu_2,\gu_4,\gu_6,\gu_8)=
\phi(1,\gu_2,1,\gu_4,1,\gu_6,1,\gu_8,1). 
\eeq
Alors la d\'ecomposition des boucles $\gu$ en fonction des
variables $\gt$ s'\'ecrit:
\beqs
\gu_2=(\gt_1)^{-1}(\gt_3)^{-1}\gt_1, & \quad &
\gu_4=(\gt_3)^{-1}\gt_1, \nonumber \\
\gu_8=\gt_9(\gt_7)^{-1}(\gt_9)^{-1}, & \quad &
\gu_6=\gt_9(\gt_7)^{-1}.
\eeqs
On peut v\'erifer ces relations en passant par les variables initiales $g$.
Puis finalement, le changement de variables pour passer de $\phi_T$
\`a $\phi_U$ se lit:
\beq
\begin{array}{l}
\phi_T(\gt_1,\gt_3,\gt_7,\gt_9)=\\
\quad \phi_U((\gt_1)^{-1}(\gt_3)^{-1}\gt_1,(\gt_3)^{-1}\gt_1,
\gt_9(\gt_7)^{-1},\gt_9(\gt_7)^{-1}(\gt_9)^{-1})
\end{array}
\eeq

\medskip

Pour le moment, nous avons r\'eduit le probl\`eme de construire une mesure
sur $A_\Gamma$ au probl\`eme de construire une mesure sur les fleurs
$A_h=G^{h}/Ad(G)$, o\`u $Ad(G)$ d\'enote l'action
ajointe diagonale de $G$:
\beq
g\cdot (g_{1},\cdots,g_{h}) \rightarrow (gg_{1}g^{-1},\cdots,
gg_{h}g^{-1}).
\label{gaction}
\eeq
La mesure que nous recherchons doit \^etre sym\'etrique et invariante
sous multiplication \`a gauche et \`a droite (comme une mesure de Haar):
\beqs
d\mu(g_{\sigma_{1} },\cdots,g_{\sigma_{h}})  = d\mu(g_{1},\cdots,g_{h}) , \\
d\mu(kg_{1}h,\cdots,g_{h}) = d\mu(g_{1},\cdots,g_{h}),
\eeqs
o\`u $\sigma$ est une permutation quelconque.
De plus, la mesure $d\mu$ devra \'egalement satisfaire des conditions
de r\'ealit\'e. Plus pr\'ecis\'ement, supposons que
$P(X_1,\cdots, X_h)$ est un \'el\'ement r\'eel ($P^\dag
=P$) et $Ad(G)$-invariant de $U({\cal G}^h)$ -$U({\cal G})$ \'etant
l'alg\`ebre enveloppante universelle de $G$. $P$ peut \^etre repr\'esent\'e
en tant qu'op\'erateur diff\'erentiel sur $A_h$ en utilisant la
correspondance entre un \'el\'ement $X$ de l'alg\`ebre de Lie et
un op\'erateur d\'eriv\'ee invariant \`a gauche:
\beq
\partial_X^i \phi(g_1,\cdot,g_i,\cdots,g_h)\equiv
\phi(g_1,\cdot,g_i X,\cdots,g_h).
 \eeq
L'op\'erateur diff\'erentiel $P$ doit \^etre Hermitien par rapport
au produit scalaire induit par la mesure $d\mu$ que nous voulons construire.

Dans le cas d'un groupe compact, il y a une unique mesure satisfaisant
ces conditions. $d\mu$ est le produit des mesures de Haar (normalis\'ees),
c'est-\`a-dire la mesure de Haar (normalis\'ee) sur $G^{h}$.
Dans ce cas, la propri\'et\'e de sym\'etrie est triviale et les conditions
de r\'ealit\'e sont impl\'ement\'ees gr\^ace \`a
l'invariance \`a gauche et \`a droite de la mesure de Haar.
Plus pr\'ecis\'ement, dans le cas $h=2$,
l'int\'egrale d'une fonction $\phi$ invariante  de jauge
sur $G^{\otimes 2}/Ad(G)$ se factorise
\beq
\int_{G^{\otimes 2}}dg_{1}dg_{2}
\phi(g_{1},g_{2})=
 vol(G)\int_{A_{2}} d\mu(g_{1},g_{2}) \phi(g_{1},g_{2}),
\eeq
et, puisque le volume du groupe compact est fini, il est possible de
normaliser la mesure de Haar de telle sorte que $vol(G)=1$.

Dans le cas d'un groupe non-compact, cela n'est plus possible
puisqu'il s'agirait de diviser par le volume infini du groupe.
La construction de la mesure $d\mu$ est donc plus d\'elicate
et je vais consacrer le chapitre suivant \`a expliciter sa d\'efinition.

\chapter{Mesure pour les fonctions cylindriques} \label{chap:mesure}

Ce chapitre est consacr\'e \`a la construction d'une mesure 
sur les espaces quotients $A_\Gamma$. Dans un premier temps,
j'expliquerai comment la g\'eom\'etrie alg\'ebrique permet de d\'ecrire
les espaces quotients.
En effet, tout le probl\`eme est de construire
le ``bon'' quotient, qui aura des bonnes propri\'et\'es de r\'egularit\'e,
permettant par la suite d'induire une mesure (de Haar) sur cet espace.
La g\'eom\'etrie alg\'ebrique nous fournit une solution formelle.
Ensuite, je d\'ecrirai comment construire explicitement un ``bon'' quotient
et la mesure sur cet espace. Dans un premier temps, je me concentrerai sur
les espaces $A_h$ corespondant aux fleurs.
Puis je g\'en\'eraliserai  la construction aux espaces quelconques $A_\Gamma$
en utilisant la proc\'edure de fixation de jauge d\'ecrite dans le
chapitre pr\'ec\'edent. Enfin, je montrerai que la mesure construite sur
$A_\Gamma$ est bien ind\'ependante de l'arbre maximal choisi pour effectuer
la r\'eduction \`a la fleur, et que, par cons\'equent, nous avons bien
une unique mesure bien d\'efinie.

\section{Cadre th{\'e}orique}
\subsection{Groupes de Lie et sous-alg{\`e}bres de Cartan}

Les constructions d\'ecrites ici sont valides
pour des groupes lin\'eaires, connexes et semi-simples,
c'est-\`a-dire des sous-groupes de matrices qui sont connexes,
invariants par transconjugaison ${}^t\overline{(\,)}$ et
de centre fini. Cela contient tous les groupes compacts,
mais aussi des groupes non-compacts, parmi lesquels on distingue
les groupes complexes  ($SL(N,\C)$, $SO(N,\C)$ et
$Sp(N,\C)$) des groupes r\'eels non-compacts
(comme $SL(2,\R)$, $SO(N,1)$, $SL(N,\R)$, \dots).

Dans un premier temps, introduisons les notations n\'ecessaires
pour la suite et rappellons certains \'el\'ements de th\'eorie
des groupes.
Soit un groupe de Lie $G$ et son alg\`ebre de Lie $\G$. Une sous-alg\`ebre
de Cartan(-Lie) $\cal{H}$ est une sous-alg\`ebre ab\'elienne maximale
de $\G$ stable sous transconjugaison. Un sous-groupe de Cartan $H$
est le centralisateur d'une alg\`ebre de Cartan $\cal{H}$ i.e. le sous-groupe
des \'el\'ements de $G$ commutant avec l'ensemble des \'el\'ements de $\H$.
Pour chaque sous-groupe de Cartan, on d\'efinit le groupe de Weyl
$W(H)=N(H)/H$, o\`u $N(H)$ est le normalisateur de $H$, qui prend en compte
l'action r\'esiduelle sur $H$ de la conjugaison par $G$.
Dans le cas des groupes compacts, il y a une unique sous-alg\`ebre de Cartan
(\`a conjugaison pr\`es)
et, de plus, chaque \'el\'ement du groupe $G$ est conjugu\'e \`a un \'el\'ement
de $H$. Dans le cas des groupes non-compacts, cela n'est plus vrai. 
Tout d'abord, en g\'en\'eral, il existe un nombre fini de sous-alg\`ebres de
Cartan non-conjugu\'ees. Elles ont toutes le m\^eme rang (ex. 2 pour $SL(2,\R)$,
1 pour $SO(2N+1,1)$, $N$ pour $SL(N,\R)$). Remarquons n\'eanmoins que les
groupes complexes (ex. $SL(N,\C)$) ont tous un unique sous-groupe de Cartan.
De plus les \'el\'ements de $G$ ne sont pas tous conjugu\'es \`a un
sous-groupe de Cartan. Les \'el\'ements qui le sont, sont  appel\'es
{\it r\'eguliers} et l'ensemble de ces \'el\'ements est not\'e $G_1$.
$G_1$ consiste en les \'el\'ements $x\in G$ tels que $Ad(x)$
est diagonalisable. C'est un ensemble ouvert dans $G$ et son compl\'ementaire
a une mesure de Haar nulle. De plus,
l'action de $Ad(G)$ sur $G_1$ est r\'eguli\`ere et $G_{1}/Ad(G)$
est l'union disjointe
$\sqcup_{i} H_{i}/W(H_{i})$ des sous-groupes de Cartan
modulo leur groupe de Weyl.

Ayant choisi une sous-alg\`ebre de Cartan $\H$, la d\'ecomposition de Cartan
de ${\cal G}$ s'\'ecrit 
$\cal G= {\cal H} \oplus {\cal B(H)} $, o\`u  ${\cal B(H)}$
est la sous-alg\`ebre de Borel associ\'ee \`a $\H$.
${\cal B(H)}$ se d\'ecompose en sous-espaces propres de $ad(\H)$:
${\cal B} = \oplus_{\alpha \in \Delta(H) }{\cal B}_{\alpha} $ o\`u
$\Delta(H) $ est l'espace des racines $\Delta \subset {\cal H}^{*}$.
Les g\'en\'erateurs de ${\cal H} $ sont soient (de type) compact $H^{*}=-H$ ou
(de type) non-compact $H^{*}=H$.

Ayant choisi un sous-groupe de Cartan $H$, on d\'efinit la mesure sur $G/H$
par
\beq\label{GH}
\int_{G} f(g) dg = \int_{ G/H  }
\left[\int_{H} f(s(x)h) dh \right]  dx,
\eeq
o\`u $dg, dh$ sont les mesures de Haar invariantes sur $G$ et $H$,
et $f$ est une fonction \`a support compact sur $G$.
Remarquons que nous avons choisi une section $s:G/H\rightarrow G$ mais que la formule
de factorisation reste valable quelquesoit la section choisie.

$G_1$ se d\'ecompose en l'union des classes d'\'equivalence sous conjugaison
$G_1= \sqcup_i {G}^H_{i}$, o\`u 
${G}^H_{i}=\{ghg^{-1},h \in H_{i},g \in G \}$. Chaque classe de conjugaison
recouvre  $w(H_{i})=\#W(H_i)$ fois la composante connexe
correspondante de $G_1$. D'o\`u la formule d'int\'egration de Weyl,
qui exprime l'int\'egrale sur $G$ comme l'int\'egrale sur les classes
de conjugaison:
\beq\label{Weyl}
\int_{G} f(g) dg = \sum_{i} \f{1}{w(H_{i})}
\int_{H_{i}} \left[\int_{G/H_{i} } f(xhx^{-1}) dx \right]
|\Delta_{i}(h)|^{2} dh,
\eeq
avec
\beq
\Delta_i(e^{H})=  \prod_{\alpha \in \Delta^{+}(H_{i}) }
\sinh \f{\alpha(H)}{2},
\eeq
pour  $H \in {\cal H}_{i}^\C $, $e^{H} \in  H_{i}$.
Contrairement au cas des groupes compacts, tous les \'el\'ements du groupe
ne s'\'ecrivent pas en tant qu'exponentielle $e^X$ avec $X\in \G$ (en fait, il
faut un produit d'exponentielles). N\'eanmois, on peut \'ecrire n'importe
quel \'el\'ement en tant que $e^X$ avec $X$ dans l'alg\`ebre de
Lie complexifi\'ee $ \G^\C$.

\subsection{Un peu de G{\'e}om{\'e}trie Alg{\'e}brique}
\label{geomalg}

$A_h$ est d\'efini en tant qu'espace quotient sous l'action du groupe $G$.
En g\'en\'eral, l'espace des orbites r\'esultant
n'est jamais une ``gentille'' vari\'et\'e
Haussdorf. Plusieurs types de singularit\'e peuvent alors intervenir.
Regardons par exemple le cas de $A_2=G\times G/Ad(G)$. Si 
$(g_{1},g_{2})$ sont des \'el\'ements g\'en\'eriques (ne commutant pas)
de $G$, le groupe d'isotropie de ces points est le centre de $G$ et donc
un sous-groupe fini.
Par contre, si $g_{1}$ et $g_{2}$ commutent, leur groupe d'isotropie
est non-trivial, et c'est
l'intersection des centralisateurs de $g_{1}$ et $g_{2}$.
Si $g_1$ est r\'egulier, son centralisateur est un sous-groupe de Cartan
et la dimension du groupe d'isotropie est en
g\'en\'eral au moins le rang du groupe $G$.
Ces points particuliers peuvent agir comme des attracteurs pour
l'action de $Ad(G)$ sur $G\times G$. 
Consid\'erons l'exemple de $G=SL(2,\R)$ et le point
$(g_{1},g_{2})= (1,e^{\sigma_3})$ o\`u $\sigma_{3} = diag(+1,-1)$.
Le groupe d'isotropie de ce couple d'\'el\'ements de $G$ est
le groupe ab\'elien $\{e^{t\sigma_3},t\in\R\}$. Ce point n'est pas
Haussdorf et est un attracteur pour les orbites voisines.
Regardons cela de plus pr\`es et consid\'erons 
$g_u= (e^{u\sigma_{+}}, e^{\sigma_{3}}) $
avec $\sigma_{+}=\mat{0}{1}{0}{0}$.
Alors $\lim_{t\rightarrow \infty} e^{-t\sigma_{3}} g_u e^{t\sigma_{3}}
=(1,e^{\sigma_3})$ et l'orbite 
$Ad(G).{(e^{u\sigma_{+}},
e^{\sigma_{3}})}$ n'est pas ferm\'ee puisqu'elle contient
le point $(1,e^{\sigma_3})$.
Ainsi, deux orbites diff\'erentes associ\'ees \`a $u>0$ et $u'<0$ n'ont pas
de voisinages disjoints, ce qui signifie que l'espace
quotient n'est pas Haussdorf.
Une mani\`ere de rem\'edier \`a ce probl\`eme est d'enlever d\`es le
d\'ebut l'ensemble des couples d'\'el\'ements commutants, pour que toutes
les orbites soient ferm\'ees. Mais cela n'est toujours pas suffisant.
Ainsi, dans notre exemple,  notons $(x,y)\in\R^2$ le couple
$(e^{\sigma_{3}}e^{x\sigma_{+}},e^{\sigma_{3}} e^{y\sigma_{-}})$.
L'action de $e^{t\sigma_{3}}$ se traduit par  
$(x,y)\rightarrow (e^t x,e^{-t}y)$.
Exclure les \'el\'ements commutants se traduit par la condition
$(x,y)\ne(0,0)$. Sur ce nouvel espace, les orbites sont toutes ferm\'ees,
cependant il est facile de voir que tout voisinage de l'orbite de $(x,0)$
intersectera une voisinage de l'orbite de $(0,y)$. Et donc, l'espace quotient
n'est toujours pas Haussdorf. La solution est d'exclure \'egalement les points
 $(x,0)$ et $(0,y)$. On obtient alors un bon quotient!

Cet exemple illustre  la probl\'ematique g\'en\'erale de la
d\'efinition d'un espace quotient. En fait,
comme $G$ est un groupe alg\'ebrique (groupe de matrices) et que par
cons\'equent l'action de $Ad(G)$ est aussi alg\'ebrique, ce probl\`eme a
re\c cu beaucoup d'attention en math\'ematiques dans le cas o\`u le groupe
est complexe sous le nom de ``th\'eorie des invariants'' \cite{brion,spring}
({\it invariant theory}).

\medskip

Tout d'abord, il faut rappeler quelques notions de g\'eom\'etrie alg\'ebrique
pour pouvoir donner ensuite la d\'efinition d'un espace quotient r\'egulier ou
g\'eom\'etrique. Une vari\'et\'e alg\'ebrique affine $X$ sur $\C$ est d\'efinie
comme \'etant l'ensemble des racines d'une collection de polyn\^omes sur
$\C^N$: $X= \cap_{i}V(P_{i})$
o\`u $V(P)=\{x \in \C^N | P(x) =0 \}$.
Elle est dite
irr\'eductible si elle ne peut pas \^etre consid\'er\'ee comme
l'union de deux sous-vari\'et\'es alg\'ebriques. 
La topologie utile dans ce contexte est la topologie de Zariski dont les
ferm\'es sont les sous-vari\'et\'es alg\'ebriques de $X$, g\'en\'er\'ees par
les $X(P)=X\cap V(P)$. Les ouverts de cette topologie
sont les unions finies des ensembles $X_{P}= \{ x\in X, P(x)\neq 0\}$
ouverts pour la topologie usuelle. il est important de noter que
les ouverts dans la topologie de Zariski sont bien plus gros que
dans la topologie usuelle. Ainsi, pour une vari\'et\'e irr\'eductible,
n'importe quel ouvert
non vide de $X$ est dense dans $X$ et, plus g\'en\'eralement,
n'importe quelle intersection d'ouverts non-vides sera encore dense.
On d\'efinit l'alg\`ebre des {\it fonctions r\'eguli\`eres}, not\'ee $\C[X]$,
comme l'alg\`ebre des polyn\^omes sur $\C^N$ restreints \`a $X$. C'est
l'ensemble $\C[X]= \C[\C^N]/I(X)$ o\`u $I(X)$ est
l'id\'eal des polyn\^omes s'annulant sur $X$. La condition
d'irr\'eductibilit\'e de $X$ se traduit par
la condition que $\C[X]$ soit int\`egre.

Un th\'eor\`eme d'Hilbert stipule qu'une sous-alg\`ebre quelconque $A$ de 
$\C[\C^N]$ (ou d'une alg\`ebre commutative) admettant un nombre fini de
g\'en\'erateurs et ne contenant d'\'el\'ement nilpotent est l'alg\`ebre
des fonctions r\'eguli\`eres d'une certaine vari\'et\'e affine $X$.
Une telle alg\`ebre est appel\'ee affine.
$X$ est le spectre de $A$ et est d\'efini comme l'ensemble des homomorphismes
de $A$ dans $\C$. Ce th\'eor\`eme permet de traduire des concepts
g\'eom\'etriques dans un contexte alg\'ebrique.

Enfin, ayant d\'efini $\C[X]$,
on peut introduire le corps des fractions rationelles
$\C(X)$. Egalement, on dit qu'une application entre deux vari\'et\'es affines
$\phi:X \rightarrow
Y$  est un morphisme ssi $\phi^{*}$ envoie les fonctions r\'eguli\`eres de $Y$ sur les fonctions r\'eguli\`eres de $X$.

Nous sommes maintenant pr\^ets \`a pr\'eciser la notion de ``bon'' quotient.
Grosso modo, ce sera un espace pour lequel les orbites seront s\'epar\'ees par
les fonctions rationelles invariantes. Soit $G$ un groupe alg\'ebrique agissant
sur une vari\'et\'e affine irr\'eductible $X$. Un quotient g\'eom\'etrique
de $X$ par l'action de $G$ est une vari\'et\'e affine $Y$ munie d'un morphisme
surjectif $\pi: X\rightarrow Y$ tel que:
\makeatletter
\renewcommand{\theenumi}{\alph{enumi}}
\renewcommand{\labelenumi}{(\theenumi)}
\makeatother
\begin{enumerate}
\item  $\pi$ induit un isomorphisme entre  $\C(Y)$ and
$\C(X)^{G}$.
\item  Les fibres de $\pi$ sont les orbites de $X$ sous l'action de $G$.
\end{enumerate}
La condition $(b)$ nous dit que $Y$ est un espace quotient puisque c'est
un espace d'orbites. La condition $(a)$ nous dit que l'espace quotient est
une vari\'et\'e alg\'ebrique dont les points sont s\'epar\'es par
les fonctions rationelles.
A travers les exemples du d\'ebut de cette section, nous avons
vu qu'en g\'en\'eral, il n'existe pas de tel bon quotient.
C'est un th\'eor\`eme fondamental de {\bf Rosenlich} \cite{brion}
qui sauve la situation:
quelque soit la vari\'et\'e $X$ et l'action alg\'ebrique de $G$
sur $X$, il existe toujours un ouvert dense $X_0$ stable
sous $G$ tel que $X_0/G$ est un bon quotient!
Pour d\'emontrer ce th\'eor\`eme, on commence par restreindre $Y$
de telle sorte que $(a)$ soit vraie. Puis l'hypoth\`ese $(a)$
implique que l'orbite de $x$ est dense dans $\pi^{-1}(x)$. Mais alors
la propri\'et\'e $(b)$ n'est pas vraie en g\'en\'eral.
On se restreint ensuite \`a un sous-ensemble $X_0$ de $X$ ne contenant que des orbites de dimension maximale. Cela implique $(b)$. Alors le quotient g\'eom\'etrique $X_0/G$ existe en tant que vari\'et\'e alg\'ebrique.

\medskip

Dans le cas, d'un groupe $G$ r\'eductif, il existe un th\'eor\`eme fondamental
par Hilbert et Nagata, qui assure que si $X$ est
irr\'eductible et $G$ r\'eductif, alors 
$\C[X]^{G}$ est g\'en\'er\'e par un nombre fini d'\'el\'ements.
Comme $\C[X]^{G}$ ne contient aucun \'el\'ement nilpotent,
c'est donc une alg\`ebre affine i.e. c'est l'alg\`ebre des
fonctions r\'eguli\`eres sur son spectre, que nous notons
$X//G \equiv spec(\C[X]^{G})$. Il est muni d'un morphisme surjectif
$\pi:X\rightarrow X//G $ et appel\'e {\it quotient de $X$ par $G$}.
Ce quotient est universel dans le sens que tout morphisme $G$-invariant 
$p:X\rightarrow Y$ peut \^etre factoris\'e sur
$X//G$, c'est-\`a-dire qu'il existe un morphisme
 $q: X//G\rightarrow Y$ tel que $p= q\circ \pi$.
Il est alors possible de montrer que chaque fibre de $\pi$ contient
une {\it unique} orbite ferm\'ee.
G\'eom\'etriquement, cela veut dire que $X//G$ est l'espace des
orbites ferm\'ees de $X$ sous $G$. Cela est quelque peu dommage parce que
cela implique $X//G$ peut donner une description tr\`es impr\'ecise de
l'espace des orbites. En effet, $X//G$ n'est g\'en\'eralement pas un quotient
g\'eom\'etrique. Par exemple, pour 
$X=\C^{2}$ et $G=\C^{*}$ agissant par multiplication 
$(x,y)\rightarrow (tx,ty)$, les seuls polyn\^omes invariants sont
les polyn\^omes constants, de telle sorte que $X//G$ est r\'eduit \`a
un simple point, l'unique orbite ferm\'ee, celle de $(0,0)$.
Heureusement, la propri\'et\'e suivante est vraie quand $G$ est un
groupe lin\'eaire, connexe et {\it semi-simple}. 
Dans ce cas, l'alg\`ebre des fractions de 
$\C[X]^{G}$ (i.e. $\C(X//G)$) est \'egale \`a  $\C(X)^{G} $.
Cela affirme que $X//G$ est l'espace des orbites ferm\'ees et denses i.e.
toute fibre $\pi:X\rightarrow X//G$ contient une orbite dense et ferm\'ee.
Cela implique que, dans le cas que nous consid\'erons, il est possible
de d\'efinir un espace quotient g\'eom\'etrique comme le dual
alg\'ebrique de l'espace des polyn\^omes invariants.

\medskip

Revenant \`a notre probl\`eme, $G$ est un groupe lin\'eaire,
connexe et semi-simple agissant sur $X=G^n$ par conjugaison.
Exploitant la th\'eorie expos\'ee ci-dessus, nous savons que, quand $G$ est complexe, l'espace quotient universel $G^{n}//Ad G$ consistant
en les orbites denses s\'epar\'ees par les polyn\^omes invariants
est bien d\'efini.
De plus, par le th\'eor\`eme de Rosenlich, il est toujours possible
d'exclure de $G^n$ un ensemble ferm\'e (n\'egligeable) pour
que l'espace d'orbites soit un bon quotient g\'eom\'etrique.
N\'eanmoins, bien qu'\'el\'egante et g\'en\'erique, ces m\'ethodes
ne s'appliquent pas directement dans le cas r\'eel (sur $\R$)
et de plus ne sont pas constructives.

Il est donc n\'ecessaire de mieux comprendre (d'une mani\`ere explicite)
quel est l'ensemble ferm\'e que nous devons exclure de $G^n$ pour
obtenir une espace quotient g\'eom\'etrique (qui sera une
vari\'et\'e alg\'ebrique bien d\'efinie). De plus, nous sommes
int\'eress\'es par les propri\'et\'es vis-\`a-vis de la mesure du quotient
et nous aimerions d\'emontrer que la diff\'erence entre le quotient
g\'eom\'etrique et l'espace quotient $G^{n}//Ad G$ est de mesure nulle.

\medskip

Tout d'abord, le th\'eor\`eme de Rosenlich nous affirme
qu'il est toujours possible de construire un quotient
g\'eom\'etrique retirant un ensemble ferm\'e (pour la
topologie de Zariski) de $G^n$.

Le cas $G/Ad(G)$ est bien connu. La solution consiste \`a enlever
de $G$ les points auxquels l'action de $Ad(g)$ n'est pas r\'eguli\`ere.
L'ensemble des points r\'eguliers, not\'e $G_1$, est l'ensemble des
points $g\in G$ pour lesquels $Ad(g)$ est
diagonalisable et tel que $1$ soit une valeur propre de
multiplicit\'e \'egale au rang du groupe.
L'espace quotient $G_1/Ad(G) $ 
est alors \'egal \`a l'union des \'el\'ements r\'eguliers des sous-groupes
de Cartan modulo leur groupe de Weyl i.e. $\sqcup_{i} H_{i}/W(H_{i})$.

Dans le cas $G\times G$, la strat\'egie est identique, et il s'agit
d'exclure les points irr\'eguliers, puis on considerera l'action de $Ad(G)$
sur une sous-espace de $G\times G$.
Dans le cas d'un groupe de rang $1$, un tel sous-ensemble est
donn\'e explcitement par
\beq \label{g2def}
G_{2}\equiv \{ (g_{1},g_{2}) \in G \times G ; g_{1}\in G_{1}\,
\textrm{ ou }
\,g_{2}\in G_{1} \textrm{, et }
\textrm{det}[g_1,g_2]\ne0\},
\eeq
o\`u $[g_1,g_2]=g_1g_2-g_2g_1$ est le commutant dans l'alg\`ebre.
On a alors la proposition suivante:
\begin{prop}
\label{G2}
$G_{2}$ est un sous-ensemble dense de $G\times G$,
son compl\'ementaire est de mesure de Haar nulle et $G_{2}/Ad(G)$
est un quotient g\'eom\'etrique quand $G$ est de rang 1.
Par cons\'equent, $A_{2}=G_{2}/Ad(G)$ est une vari\'et\'e Haussdorf
de dimension $dimG$, s\'eparant les fonctions rationelles
et c'est la vari\'et\'e de base
d'un fibr\'e homog\`ene de fibre $G$ et dont l'espace total est
$G_{2}\approx A_{2}\times G $.
\end{prop}
Cette proposition sera montr\'e explicitement sur les exemples
de $\slc$, $\slr$ et $\su$ dans le chapitre \ref{chap:exemples}
o\`u nous construirons le spectre dual \`a l'espace des polyn\^omes
invariants et montrerons qu'il est isomorphe \`a $G_2/Ad(G)$. Le point central
est que la condition $\textrm{det}[g_1,g_2]\ne0$ peut \^etre impl\'ement\'ee
sous la forme d'in\'egalit\'es alg\'ebriques.

Notons que tenant compte de la d\'efinition \Ref{g2def}, le centralisateur
de n'importe quel \'el\'ement de $G_2$ est trivial. En effet, supposons
que $g$ commute avec $(g_1,g_2)$, alors, supposant que $g_1$ est r\'egulier,
$g$ peut \^etre diagonalis\'e dans la m\^eme base que $g_1$
(l'hypoth\`ese de r\'egularit\'e est essentielle ici). $g$
ne peut pas \^etre r\'egulier, car dans ce cas, car comme $g$
commute avec $g_2$, cela impliquerait que $g_2$ est
diagonal dans cette m\^eme base et donc commute aussi avec $g_1$,
ce qui est contraire aux hypoth\`eses.  Par cons\'equent,
la d\'efinition de $g$ impose que $g$ est diagonal et non-r\'egulier.
Si le rang du groupe est $1$, cela revient \`a dire que $g$ est l'identit\'e.
Dans le cas de $SL(N,\C)$, il est facile de v\'erifier que la condition
de d\'eterminant non-nul  est suffisante pour conclure que $g$ est
dans le centre de $G$.
Il est important de comprendre qu'il ne suffit pas d'exclure les points
avec un centralisateur non-trivial: pour obtenir un bon quotient, il faut
enlever plus de points, et cela est dict\'e par le fait que
l'espace quotient est le spectre de l'alg\`ebre des fonctionelles invariantes.

Dans le cas d'un groupe de rang plus \'elev\'e, on peut d\'efinir $G_2$ par
\beq \label{2G2}
G_2\equiv\{(g_{1},g_{2}) \in G \times G ; g_{1}\in G_{1}\,
\textrm{ ou}
\,g_{2}\in G_{1} \textrm{, et }
C(g_{1},g_{2}) = Z_G \},
\eeq
o\`u $C(g_{1},g_{2})$ d\'enote le centralisateur de $g_1$ et $g_2$, et $Z_G$
le centre du groupe. Nous avons vu que, dans le cas de $SL(N,\C)$, cela peut
\^etre impl\'ement\'e par une condition alg\'ebrique. On peut s'attendre
\`a ce que cela soit de m\^eme pour tout groupe, mais nous ne d\'emontrerons
pas un telle proposition.  Nous verrons dans la prochaine section que $G_2$
admet un bon quotient $A_2$ par $Ad(G)$. N\'eanmoins, contrairement
\`a la d\'efinition \Ref{g2def} (valide pour tous les groupes de rang 1,
et \'egalement $SL(N,\C)$), la d\'efinition \Ref{2G2} n'est
pas \'equivalente \`a une d\'efinition de $G_2$ en tant que dual alg\'ebrique.

\section{Mesure sur les Fleurs}

Pour construire la mesure sur les fleurs, je vais commencer par d\'ecrire le
cas \`a deux p\'etales $A_2=G_2/Ad(G)$, puis je g\'en\'eraliserai \`a un nombre
de p\'etales quelconque.

\subsection{Les fleurs \`a deux p\'etales}

\subsubsection{Cas d'un unique sous-groupe de Cartan}

Je commence par le cas o\`u le groupe $G$ contient 
un unique sous-groupe de Cartan (\`a conjugaison pr\`es).
C'est par exemple le cas des groupes complexes $SL(N,\C)$. Ce cas
sera par la suite facilement g\'en\'eralis\'e au cas
o\`u on aurait plusieurs sous-groupes de Cartan non conjugu\'es.

Consid\'erons l'application suivante de $G_1$ dans $A_2$. Etant donn\'e,
$g\in G_1$, il est possible de la conjuguer au sous-groupe
de Cartan $H\subset G$ i.e. il existe $h\in H, x \in G/H$ tel que
$g=xhx^{-1}$.
On choisit alors une  section $s:G/H \rightarrow G$ et on d\'efinit
l'application:
\begin{equation}
\begin{array}{cccc}
j_{s}: &G_{1} &\rightarrow& A_{2} \\
& g=xhx^{-1} & \rightarrow & Ad(G).(h,s(x))
\end{array}
\end{equation}
o\`u $Ad(G).(h,s(x))$ est l'orbite de $(h,s(x))$ sous l'action par
conjugaison $Ad(G)$. Ceci effectue une fixation de jauge i.e.
$j_{s}$ est surjective.
En effet, soit  $(g_1,g_2)\in A_2$.
Il est possible de conjuguer $g_1$
au sous-groupe de Cartan $H$
i-e il existe $h\in H$, $y \in G/H$ tel que
$g_1=yhy^{-1}$. Alors $Ad(G).(g_1,g_2)=Ad(G).(h,y^{-1}g_2y)$.
Ceci fixe la jauge seulement partiellement
puisque
$H$ peut encore agir sur $y$ par $y\rightarrow yk$, ce qui veut dire
que nous pouvons encore conjuguer 
$\tl{g}_2=y^{-1}g_2y$ par un \'el\'ement de sous-groupe de Cartan.
N\'eanmoins, puisque nous sommes dans $A_2$, le centralisateur
$g_1$ et $g_2$ est trivial, ce qui implique que le centralisateur de
$h\in H$ et $\tl{g}_2$ est trivial.
Puisque le centralisateur de $h \in H\cap G_1$ est $H$,
cela signifie que l'action par conjugaison de $H$
sur $\tl{g}_2$ n'a pas d'autres points fixes que les \'el\'ements
du centre de $G$.
Supposons, pour la suite, que le centre de $G$ est trivial. Alors, quitte
\`a exclure un ensemble de mesure nulle de $G$, il existe des sections
$s:G/H\arr G$ telles que nous pouvons utiliser la sym\'etrie r\'esiduelle
de l'action par conjugaison par $H$ pour imposer que 
$\tl{g}_2$ soit dans l'image de $s$.

Par exemple, dans le cas $SL(N,\C)$, nous pouvons exclure
les points $\prod_{i=1}^{N-1}a_{ii+1}=0$ o\`u l'on note
$a_{ij}$ les \'el\'ements de matrices, alors on peut choisir la section 
d\'efinie par $a_{ii+1}=1$ pour tout $i=1,\dots,N-1$.
En effet, c'est une bonne fixation de jauge \`a la fois
pour l'action par multiplication \`a gauche (ou \`a droite) par $H$ et 
pour l'action par conjugaison $Ad(H)$.
Dans le cadre plus g\'en\'eral d'un groupe semi-simple, on d\'ecompose
un \'el\'ement $g$ du groupe en ces composantes sur le Cartan et sur les racines
$$
g=e^{\sum_{\alpha>0} u_\alpha E^\alpha}he^{\sum_{\alpha<0} u_\alpha E^\alpha},
$$
o\`u l'on somme sur les racines $\alpha$ postives et n\'egatives avec $h$
un \'el\'ement du sous-groupe de Cartan et les $^\alpha$ les g\'en\'erateurs de
l'alg\`ebre de Lie. Alors une bonne fixation de jauge sera par exemple $u_\alpha=1$
pour toutes les racines simples.

Au final, la fixation de jauge revient \`a imposer que 
$(g_1,g_2)\rightarrow(h,s(x)) \in H\times G/H $,
car nous venons de d\'emontrer que tout \'el\'ement de $G_2$ peut
\^etre \'ecrit sous cette forme.
De plus, la condition que le centralisateur de $(g_1,g_2)$ soit trivial
est impl\'ement\'e si nous demandons que $g=s(x)hs(x)^{-1} \notin H$.
En effet, $s(x)hs(x)^{-1} \in H$ impliquerait que soit $s(x)\in H$, soit que
$s(x)$ est une transformation de Weyl.
La premi\`ere possibilit\'e est impossible
car $s(x)$ et $h$ ne commute pas. Ainsi $j_s$ d\'efinit
une application de $G_1\setminus H$ sur $A_2$.
La seconde possibilit\'e est li\'ee \`a un probl\`eme d'ambigu\"\i t\'e
de Gribov, qui rend la d\'efinition de $j_s$ ambig\"ue.

Cette ambigu\"\i t\'e est due au fait qu'un \'el\'ement arbitraire du groupe peut
\^etre conjugu\'e \`a diff\'erents \'el\'ements du sous-groupe de Cartan,
tous reli\'es par l'action du groupe de Weyl $W(H)$,
qui est justement l'action par conjugaison r\'esiduelle
sur le sous-groupe de Cartan $H$.
Il y a deux mani\`eres de s'affranchir de ce probl\`eme.
Tout d'abord, nous pouvons exiger un bon comportement
de la section $s$ sous l'action du groupe de Weyl:
\beq
\forall x\in G/H,\,\forall w\in W(H),\,s(xw)=w^{-1}s(x)w.
\eeq
Cela rend $j_{s}$ bien d\'efinie et c'est l'hypoth\`ese que nous utiliserons
dans la suite.
Ou bien, nous pouvons imposer que $h$ soit dans une chambre de Weyl fix\'ee.
Dans ce cas, il suffit d'enlever tous les facteurs 
$1/w(H)$ des preuves suivantes.

\medskip

En utilisant cet isomorphisme, nous pouvons tirer en arri\`ere les fonctions
sur $A_2$ (ou de mani\`ere \'equivalente les fonctions invariantes sur $G_2$)
\`a des fonctions d\'efinies sur
$G_{1}$ par $j_{s}^{*}F (xhx^{-1}) = F(h,s(x))$.

\begin{defi} \label{defmeasure}
Soit une mesure $\mu$ sur $A_{2}$ d\'efinie par
\beq
\int_{A_{2}} F(g_{1},g_{2}) d\mu(g_{1},g_{2})\equiv
\int_{G} j_{s}^{*}F (g) dg.
\eeq
\end{defi}

\begin{prop} \label{facto}
Soit une fonction $F$ $L^1$ sur $G_2$ pour la mesure de Haar.
Nous demandons \'egalement que sa version invariante de jauge soit
bien d\'efinie:
\beq
^{G}F(g_{1},g_{2})=\int_{G} F(gg_{1} g^{-1}, gg_{2} g^{-1})  dg
\eeq
Alors, nous avons:
\beq
\int_{G\times G} F(g_{1},g_{2}) dg_{1} dg_{2}
=\int_{A_{2}} {}^GF(g_{1},g_{2}) d\mu(g_{1},g_{2}),
\eeq
\end{prop}

Je vais d\'emontrer ce r\'esultat explicitement dans le cas de $SL(n,\C)$.
La d\'emonstration devrait pouvoir s'adapter \'a tous les groupes semi-simples.
Commen\c cons donc par introduire un lemme:

\begin{lemma}
Pour $g=SL(n,\C)$, le sous-groupe de Cartan est consistu\'e des matrices diagonales
$h=(\lambda_i)_{i=1..n}$ avec $\prod_i\lambda_i=1$. A condition  que
$\prod_i^{n-1} a_{i,i+1}\ne 0$, la condition $\forall i=1..n-1,\,a_{i,i+1}=1$
consistue une bonne fixation de jauge \`a la fois pour l'action de $H$
par multiplication \`a droite sur $G$ et pour l'action  adjointe $Ad(H)$. Elle d\'efinit
une section $s:G/H\rightarrow G$. On a alors une extension de la formule de factorisation
\Ref{GH} qui s'\'ecrit:
\beq
\label{HGH}
\int_{G} f(g) dg= n\times\int_{H} \left[\int_{G/H} f(h^{-1}s(x) h) dx\right] dh,
\eeq
o\`u $n$ est donc qu'un facteur multiplicatif.
\end{lemma}

Pour d\'emontrer ce r\'esultat, il s'agit d'analyser les propri\'et\'es
de la section choisie $s$ sous multiplication \`a gauche par un \'el\'ement
du Cartan. Un calcul simple prouve que pour $x\in G/H$ et $h=(\lambda_i)\in H$:
$$
s(h^{-1}x)=h^{-1}s(x)k_{(h)}
$$
avec $k_{(h)}=(\lambda_n,\lambda_1,..,\lambda_{n-1})\in H$ est une permutation circulaire
des \'el\'ements de matrice de $h$. Notons le fait important que l'\'el\'ement $k$ ne
d\'epend que de $h$ et pas de $x$.
Alors, en utilisant l'invariance de la mesure $dx$ sur $G/H$ par multiplication \`a gauche,
on obtient:
$$
\int_{H} \left[\int_{G/H} f(h^{-1}s(x) h) dx\right] dh=
\int_H \left[\int_{G/H} f(s(x)k^{-1}_{(h)}h)dx\right] dh.
$$
Il s'agit maintenant de calculer le Jacobien du changement de variables
$k^{-1}_{(h)}h\rightarrow h$. Compte tenu que la mesure $dh$ est simplement
$\prod_{i=1}^{n-1}d\alpha_i$ pour $h=(e^{\alpha_i})_i$, on trouve
$$
\int_{H} \left[\int_{G/H} f(h^{-1}s(x) h) dx\right] dh=
\f{1}{n}\int_H \left[\int_{G/H} f(s(x)h)dx\right] dh,
$$
ce qui permet de conclure.

\medskip

D\'emontrons maintenant la proposition \Ref{facto}.
Pour cela, consid\'erons une fonction $F(g_{1},g_{2}) $  $L^1$ sur $G\times G$. Alors:
\beqs
\int_{G\times G} F(g_{1},g_{2}) dg_{1}dg_{2}
&=&\f{1}{w(H)}\int_{G/H\times H\times G}
F(xhx^{-1}, g_{2}) |\Delta(h)|^2 dx dh dg_{2} \nonumber \\
&=& \f{1}{w(H)}\int_{G/H\times H\times G}
F(xhx^{-1},x g_{2}x^{-1}) |\Delta(h)|^2 dx dh dg_{2}
\eeqs
o\`u nous avons utilis\'e la formule d'int\'egration de Weyl
 (\ref{Weyl}) pour la premi\`ere \'egalit\'e, puis l'invariance
de la mesure de Haar par multiplication \`a gauche et \`a droite pour
la seconde.
Utilisant l'identit\'e (\ref{HGH}) sur l'int\'egration sur $G_{2}$,
notre int\'egrale peut \^etre  d\'evelopp\'ee:
\beqs
& & n\times\f{1}{w(H)}
\int_{G/H\times H\times H\times H\backslash G  }
F(xhx^{-1}, xkyk^{-1}x^{-1}) |\Delta(h)|^2 dx dh dk dy \nonumber\\
&=& n\times\f{1}{w(H)} \int_{ H\times H\backslash G }
\left[\int_{G/H\times H} F(xhx^{-1}, xky(xk)^{-1})  dx dk  \right]
|\Delta(h)|^2 dh dy \nonumber \\
&=& n\times\f{1}{w(H)} \int_{ H\times H\backslash G }
\left[\int_{G/H\times H} F(xkh(xk)^{-1}, xky(xk)^{-1})  dx dk  \right]
|\Delta(h)|^2 dh dy\quad,
\eeqs
o\`u nous avons utilis\'e le fait que $H$ est ab\'elien pour la
derni\`ere \'egalit\'e.
Puis, en utilisant la d\'efinition de la mesure sur
$G/H$ (\ref{GH}), nous avons finalement:
\beq
\int_{G\times G} F(g_{1},g_{2}) dg_{1}dg_{2} =
n\times\f{1}{w(H)} \int_{ H\times H\backslash G }
 {}^{G}F(h, y)
|\Delta(h)|^2 dh dy
\eeq
o\`u $^{G}F$  est la version invariante de jauge de $F$:
\beq
^{G}F(g_{1},g_{2})=\int_{G} F(gg_{1} g^{-1}, gg_{2} g^{-1})  dg.
\eeq
Quitte \`a absorber le facteur multiplicatif $n$ dans la d\'efinition de la mesure
$d\mu(g_1,g_2)$, ceci conclut la d\'emonstration de la proposition \Ref{facto}.

\begin{theo}\label{theorem}
$\mu$ est  ind\'ependante du choix de la section $s$,
sym\'etrique, invariante sous multiplication \`a droite et \`a gauche
et invariante sous passage \`a l'inverse:
\beqs
d\mu(g_{1},g_{2}) = d\mu(g_{2},g_{1}) \\
d\mu(kg_{1}h,g_{2}) = d\mu(g_{1},g_{2}) \\
d\mu(g_{1},g_{2}) = d\mu(g_{1}^{-1},g_{2})
\eeqs
\end{theo}

D\'emontrons le th\'eor\`eme ci-dessus pour $d\mu(kg_{1},g_{2})$ 
(multiplication \`a gauche)
Il sera ais\'e de d\'emontrer les autres propri\'et\'es de la m\^eme
fa\c con.
Le plus simple est d'utiliser la proc\'edure de fixation de jauge \`a la
Faddev-Popov  puis de se servir de la proposition \ref{facto}.
Consid\'erons une fonction invariante  $F$ sur $G_2$.
Maintenant choisissons une fonction
$\varphi$ sur $G_2$ tel que $^G\varphi=1$ et d\'efinissons la fonction
fix\'ee de jauge
$\tilde{F}=F\varphi$.
Dans la proc\'edure de Faddeev-Popov usuelle, nous choisirions
une fonction $\varphi$ proportionelle \`a une distribution $\delta$,
mais cela n'est pas n\'ecessaire dans notre cas.
La proposition \ref{facto} donne:
\beq
\int_{A_{2}} F(g_{1},g_{2}) d\mu(g_{1},g_{2})
=\int_{G\times G} \tilde{F}(g_{1},g_{2}) dg_{1} dg_{2},
\eeq
En utilisant la libert\'e du choix de la fonction
$\varphi$ dans la proposition \ref{facto}
et le fait que
si $\varphi(g_1,g_2)$ est une fixation de jauge alors
$\varphi_k(g_1,g_2)=\varphi(k^{-1}g_1,g_2)$ l'est aussi,
on obtient
\begin{eqnarray}
\int_{A_{2}}  F(g_1,g_2)\dmu(kg_1,g_2)
&= & \int_{G\times G}  F(k^{-1}g_1,g_2)
\varphi(k^{-1}g_1,g_2) dg_1 dg_2 \nonumber \\
&=& \int_{G\times G}  F(g_1,g_2)\varphi(g_1,g_2)dg_1 dg_2 \nonumber \\
&=& \int_{A_{2}} F(g_1,g_2)\dmu(g_1,g_2).
\end{eqnarray}
Ce qui conclut que la mesure $\dmu$ d\'efinie sur $A_2$
est bien invariante \`a gauche.

\subsubsection{Pour un nombre quelconque de sous-groupes de Cartan}

Dans le cas de plusieurs sous-groupes de Cartan, notons-les 
$H_1,H_2,\dots,H_n$.
$G_1$ se d\'ecompose en componsantes connexes
$G^{(i)}=Ad(G).H_i$,
chacune \'etant conjugu\'ee \`a un sous-groupe de Cartan $H_i$.
Pour chacune de ces composantes, on choisit une section
$s_i:G/H_i\rightarrow G$ et on d\'efinit l'application:
\begin{equation}
\begin{array}{cccc}
j_i: &G^{(i)} \subset G_1 &\rightarrow& A_{2} \\
& g=yhy^{-1} & \rightarrow & Ad(G).(h,s_i(y))
\end{array}
\end{equation}
Avec cette application, nous pouvons d\'efinir une mesure
$d\mu_i$ sur $A_2$
comme dans le cas pr\'ec\'edent \Ref{defmeasure} par
\beq
\int_{A_{2}} F(g_{1},g_{2}) d\mu_i(g_{1},g_{2})\equiv
\int_{G_i} j_i^{*}F (g) dg.
\eeq

\begin{prop}
Soit une fonction $F$ $L^1$ sur $G_{2}$. Nous avons alors 
\beq
\int_{G\times G} F(g_{1},g_{2}) dg_{1} dg_{2}
=\int_{A_{2}} {}^GF(g_{1},g_{2}) d\mu(g_{1},g_{2}),
\eeq
avec
\beq
d\mu(g_{1},g_{2})=\sum_i \f{1}{w(H_{i})}d\mu_i(g_{1},g_{2})
\eeq
et o\`u $^{G}F$ est la version de $F$ invariante de jauge:
\beq
^{G}F(g_{1},g_{2})=\int_{G} F(gg_{1} g^{-1}, gg_{2} g^{-1})  dg.
\eeq
\end{prop}

Soit une fonction $F(g_{1},g_{2})$ $L^1$ sur $G\times G$. Alors
\beqs
& &\int_{G\times G} F(g_{1},g_{2}) dg_{1}dg_{2}\\
&=&\sum_{i}\f{1}{w(H_{i})}\int_{(G/H_{i})\times H_{i}\times G}
F(xhx^{-1}, g_{2}) |\Delta(h)|^2 dx dh dg_{2}\\
&=& \sum_{i}\f{1}{w(H_{i})}\int_{(G/H_{i})\times H_{i}\times G}
F(xhx^{-1},x g_{2}x^{-1}) |\Delta(h)|^2 dx dh dg_{2},
\eeqs
o\`u nous avons utilis\'e la formule d'int\'egration de Weyl
 (\ref{Weyl}) pour la premi\`ere \'egalit\'e, puis l'invariance
de la mesure de Haar par multiplication \`a gauche et \`a droite pour
la seconde.
Utilisant l'identit\'e (\ref{HGH}) sur l'int\'egration sur $G_{2}$,
notre int\'egrale peut \^etre  d\'evelopp\'ee en
\beqs
& & \sum_{i}\f{1}{w(H_{i})}
\int_{G/H_{i}\times H_{i}\times H_{i}\times H_{i}\backslash G  }
F(xhx^{-1}, xkyk^{-1}x^{-1}) |\Delta(h)|^2 dx dh dk dy\\
&=& \sum_{i}\f{1}{w(H_{i})} \int_{ H_{i}\times H_{i}\backslash G }
\left[\int_{G/H_{i}\times H_{i}} F(xhx^{-1}, xky(xk)^{-1})  dx dk  \right]
|\Delta(h)|^2 dh dy.
\eeqs
Utilisant la d\'efinition (\ref{GH}) de la mesure sur $G/H_i$ et
le fait que $H_i$ est ab\'elien, on aboutit \`a
\beqs
\int_{G\times G} F(g_{1},g_{2}) dg_{1}dg_{2} &= &
\sum_{i}\f{1}{w(H_{i})} \int_{ H_{i}\times H_{i}\backslash G }
 {}^{G}F(h, y)
|\Delta(h)|^2 dh dy \nonumber \\
&=&\sum_{i}\f{1}{w(H_{i})} \int_{G^{(i)}}
dg \,\, j_i^* ({}^GF)(g) \nonumber \\
&=&\sum_{i}\f{1}{w(H_{i})} \int_{A_2} d\mu_i(g_1,g_2) \,
{}^GF(g_1,g_2)
\eeqs
Puis, il est direct d'utiliser cette proposition pour g\'en\'eraliser
le th\'eor\`eme \ref{theorem} au cas de plusieurs
sous-groupes de Cartan en suivant la m\^eme preuve que pr\'ec\'edemment.

\subsection{Les fleurs en g\'en\'eral}

On veut maintenant g\'en\'eraliser le cas de $A_2$ au cas $A_h$.
D'apr\'es le th\'eor\`eme de Rosenlich d\'ecrit dans la sous-section
\ref{geomalg}, il est toujours possible de choisir un ensemble dense
$G_h\subset G^h$ tel que le quotient g\'eom\'etrique 
$A_h=G_h/Ad(G)$ est bien d\'efini comme dans la proposition
\ref{G2}.
Suivant la logique de la construction dans le cas \`a deux p\'etales,
on d\'efinit $G_h$, dans le cas d'un groupe de rang 1, comme:
\begin{defi}
\beq
G_{h} \equiv \{ (g_{1},\cdots ,g_{h}) \in G^{h}|\,
  \exists
(i,j) \in [1,\cdots,h],\, (g_{i},g_{j}) \in G_{2} \}
\eeq
\end{defi}
Notons que $G_h$ est tel que le centralisateur de n'importe quel \'el\'ement
de $G_h$ est l'identit\'e.

\begin{defi} \label{defdmun}
Fixons deux p\'etales $i,j$ sur la fleur \`a  $n$ p\'etales.
Alors, nous pouvons d\'efinir une mesure sur $A_h$:
\beq
\mu^{(ij)}\left[f(g_1,g_2,\dots,g_n)\right]
=\int_{G_2^{(ij)}} d\mu(g_i,g_j)
\int\prod_{k\ne i,j}dg_k
f(g_1,g_2,\dots,g_n)
\eeq
o\`u nous avons pris la mesure fix\'e de jauge
$d\mu(g_i,g_j)$ pour les deux p\'etales fix\'ees et la mesure
de Haar sur les autres liens.
\end{defi}
Cette mesure est bien d\'efinie puisque 
$\int\prod_{k\ne i,j}dg_k f(g_1,g_2,\dots,g_n)$
est une fonction invariante sur $G_2^{(ij)}$
(et par cons\'equent une fonction sur $A_2$,
que nous pouvons int\'egrer en utilisant $d\mu(g_i,g_j)$).

\begin{prop} \label{facton}
 Soit une fonction $F$ $L^1$ sur $G^n$. Alors nous avons 
\beq
\int_{G^n} F(g_{1},\dots,g_n) dg_{1} \dots dg_n
=\int_{A_{n}} {}^GF(g_{1},\dots,g_n) d\mu(g_{1},\dots,g_n),
\eeq
o\`u  $^{G}F$ est la version invariante de jauge de $F$:
\beq
^{G}F(g_{1},\dots,g_n)=\int_{G} F(gg_1 g^{-1}, \dots,gg_n g^{-1})  dg
\eeq
\end{prop}
Cette proposition est facilement d\'emontr\'ee en utilisant la proposition
\ref{facto} et l'invariance de la mesure de Haar sous multiplication \`a gauche
et \`a droite. Elle implique le th\'eor\`eme suivant:

\begin{theo} \label{totalinv}
Les mesures $d\mu^{(ij)}$ ne d\'ependent pas du choix des liens $i,j$
et d\'efinissant donc une unique mesure 
$d\mu(g_1,g_2,\dots,g_n)$ sur $A_h$.
De plus, cette mesure est sym\'etrique sous permutations arbitraires de
$g_1,\dots,g_n$, sous multiplication \`a gauche et \`a droite et sous
passage \`a l'inverse:
\beqs
d\mu(g_{\sigma_{1} },\cdots,g_{\sigma_{n}})  = d\mu(g_{1},\cdots,g_{n}) ,
\nonumber \\
d\mu(kg_{1}h,\cdots,g_{n}) = d\mu(g_{1},\cdots,g_{n}),
\nonumber \\
d\mu(g_{1}^{-1},\cdots,g_{n})=d\mu(g_{1},\cdots,g_{n})
\eeqs
\end{theo}

\section{Mesure pour un graphe quelconque}

Pour construire la mesure sur un graphe arbitraire
$\Gamma$, on choisit un arbre maximal $T$,
on effectue la proc\'edure de fixation de jauge d\'ecrite dans le
chapitre \ref{chap:fleur} pour r\'eduire le
graphe $\Gamma$ \`a une simple fleur, puis on d\'efinit
une mesure $\dmu_T$ tel que
pour toute fonction $\phi$ invariante de jauge sur $\Gamma$, on ait:
\beq
\int \dmu_T(g_1,\dots,g_E) \phi(g_1,\dots,g_E) =
\int \wtl{\dmu}(g_1,\dots,g_F) \phi_T(g_1,\dots,g_F)
\eeq
o\`u  $\wtl{\dmu}$ est la mesure sur la fleur \`a $h_\Gamma=E-V+1$ p\'etales.

\medskip

Cette d\'efinition d\'epend {\it a priori} du choix de l'arbre $T$.
Je vais montrer que ce n'est en fait pas le cas.
Choisisson pour cela, deux arbres $T$ et $U$. Les fonctions fix\'ees
de jauge $\phi_T$ et $\phi_U$ sont reli\'ees par le changement de variables
\Ref{TtoU}:
$$
\phi_T(\gt_e)=
\phi_U(\gu_e=\overrightarrow{\prod_{f\in {\cal L}[e]\setminus T}} \gt_f)
$$
et il s'agit de d\'emontrer maintenant que:
\beq
\int \wtl{\dmu}(\gt_1,\dots,\gt_F) \phi_T(\gt_1,\dots,\gt_F)
=
\int \wtl{\dmu}(\gu_1,\dots,\gu_F) \phi_U(\gu_1,\dots,\gu_F)
\eeq
ou d'une mani\`ere \'equivalente:
\beq
\int \wtl{\dmu}(\gu_1,\dots,\gu_F) \phi_U(\gu_1,\dots,\gu_F)=
\int \wtl{\dmu}(\gt_1,\dots,\gt_F)
\phi_U(\gu_e=\overrightarrow{\prod_{f\in {\cal L}[e]\setminus T}} \gt_f).
\label{chvar}
\eeq
Nous allons prouver cette \'egalit\'e en effectuant des changements de
variables \'el\'ementaires, qui correspondent en fait \`a des changements
\'el\'ementaires d'arbres maximaux.
Nous allons dans une premier temps d\'efinir ces changements \'el\'ementaires
entre arbres et montrer qu'il est possible de passer de n'importe quel arbre
\`a un autre par une suite de tels mouvements. Puis je montrerai que la mesure
est invariante sous de tels changements de variables.

\begin{defi}
Soit un graphe $\Gamma$ et un arbre maximal $T$. Consid\'erons un vertex
$v$  tel qu'au moins un lien attach\'e \`a $v$ ne soit pas dans
l'arbre $T$. Appelons $f\notin T$ un tel lien.
Il existe un unique chemin dans $T$ liant l'autre vertex de $f$
au point $v$. Ce chemin d\'efinit un autre lien $e\in T$ attach\'e \`a $v$.
Alors un changement \'el\'ementaire d'arbre consiste \`a \'echanger
les liens $e$ et $f$ et \`a consid\'erer l'arbre 
$U=T\cup f \setminus e$.
\end{defi}

\begin{center}
\setlength{\unitlength}{0.00083333in}
\begingroup\makeatletter\ifx\SetFigFont\undefined%
\gdef\SetFigFont#1#2#3#4#5{%
  \reset@font\fontsize{#1}{#2pt}%
  \fontfamily{#3}\fontseries{#4}\fontshape{#5}%
  \selectfont}%
\fi\endgroup%
{\renewcommand{\dashlinestretch}{30}
\begin{picture}(3719,1275)(0,-10)
\thicklines
\put(675.000,457.500){\arc{615.000}{6.0619}{9.6461}}
\put(2925.000,457.500){\arc{615.000}{6.0619}{9.6461}}
\thinlines
\path(675,1050)(975,525)
\path(2925,1050)(2625,525)
\thicklines
\path(1575,600)(2250,600)
\blacken\thinlines
\path(2070.000,532.500)(2250.000,600.000)(2070.000,667.500)(2124.000,600.000)(2070.000,532.500)
\thicklines
\path(675,1050)(375,525)
\path(2925,1050)(3225,525)
\put(600,1125){\makebox(0,0)[lb]{\smash{{{\SetFigFont{10}{12.0}{\rmdefault}{\mddefault}{\updefault}$v$}}}}}
\put(2850,1125){\makebox(0,0)[lb]{\smash{{{\SetFigFont{10}{12.0}{\rmdefault}{\mddefault}{\updefault}$v$}}}}}
\put(975,675){\makebox(0,0)[lb]{\smash{{{\SetFigFont{10}{12.0}{\rmdefault}{\mddefault}{\updefault}$f\notin T$}}}}}
\put(3150,750){\makebox(0,0)[lb]{\smash{{{\SetFigFont{10}{12.0}{\rmdefault}{\mddefault}{\updefault}$f\in  U$}}}}}
\put(600,0){\makebox(0,0)[lb]{\smash{{{\SetFigFont{10}{12.0}{\rmdefault}{\mddefault}{\updefault}$\in T$}}}}}
\put(2850,0){\makebox(0,0)[lb]{\smash{{{\SetFigFont{10}{12.0}{\rmdefault}{\mddefault}{\updefault}$\in U$}}}}}
\put(2325,750){\makebox(0,0)[lb]{\smash{{{\SetFigFont{10}{12.0}{\rmdefault}{\mddefault}{\updefault}$e\notin U$}}}}}
\put(0,675){\makebox(0,0)[lb]{\smash{{{\SetFigFont{10}{12.0}{\rmdefault}{\mddefault}{\updefault}$e\in T$}}}}}
\end{picture}
}
\end{center}

L'int\'er\^et d'une telle d\'efinition repose en la proposition suivante:
\begin{prop}\label{elmoves}
Ayant choisi deux arbres maximaux $T$ et $U$ sur un graphe $\Gamma$, il existe
une suite de changements \'el\'ementaires permettant de passer de $T$ \`a $U$.
\end{prop}
Puis, le changement de variables de $G^{(U)}$ \`a $G^{(T)}$
est tr\`es simple pour un changement \'el\'ementaire et consiste
en une simple inversion ou multiplication \`a gauche, ce qui
simplifie \'enorm\'ement l'\'etude des changements d'arbres.

\medskip

Montrons tout d'abord la proposition pr\'ec\'edente. Ayant choisi
$\Gamma$ et deux arbres maximaux $T$ et $U$, on distingue 4 types de liens:
les liens appartenant \`a la fois \`a $T$ et $U$, les liens dans
$V=T\setminus U$
les liens dans $W=U\setminus T$ et les liens dans aucun des deux.
Il s'agit d'effectuer des changements \'el\'ementaires pour r\'eduire
les ensembles $V$ et $W$ \`a l'ensemble vide.
Concentrons-nous sur $V$. Tout d'abord, il peut ne pas \^etre connexe.
On peut alors travailler sur chacune des parties connexes.
Choisissons-en une et notons-la $V_1$.
$V_1$ est un arbre (puisque c'est un sous-graphe de $T$).
En particulier, $V_1$ n'est pa ferm\'e et contient des liens ouverts
(sans bouts) i.e. des liens li\'es \`a $V_1$ par un seul vertex.
Nous allons les enlever de $V_1$ en effectuant des changements
\'el\'ementaires, puis en r\'ep\'etant l'op\'eration, on pourra
vider coml\`etement $V_1$. Et finalement, en faisant de m\^eme avec toutes
les composantes connexes, on pourra absorber enti\`erement l'ensemble 
$T\setminus U$.

Choisissons donc un lien $e$ ouvert de $V_1$. Il a deux vertex:
$v$ \`a l'ext\'erieur de $V_1$ et $w$ \`a l'int\'erieur de $V_1$.
Il existe un unique chemin ${\cal P}$ dans
$U$ reliant ces deux vertex. Ce chemin ne passe pas par $e \notin U$.
Dans ${\cal P}$, il existe au moins un lien dans $U$ et non dans $T$
sinon il y aurait une boucle dans l'arbre $T$.

Supposons qu'un tel lien $f\in U\setminus T$
soit directement rattach\'e au lien $e$
(par le vertex $v$).
Alors, nous pouvons effectuer un changement \'el\'ementaire 
$e\lrarr f$ et passer \`a un arbre (maximal)
$\tl{U}=U\cup e \setminus f$ plus pr\`es de l'arbre $T$ que de l'arbre initial
$U$.

\begin{center}
\setlength{\unitlength}{0.00058333in}
\begingroup\makeatletter\ifx\SetFigFont\undefined%
\gdef\SetFigFont#1#2#3#4#5{%
  \reset@font\fontsize{#1}{#2pt}%
  \fontfamily{#3}\fontseries{#4}\fontshape{#5}%
  \selectfont}%
\fi\endgroup%
{\renewcommand{\dashlinestretch}{30}
\begin{picture}(4948,3059)(0,-10)
\put(1362.500,2451.500){\arc{617.454}{5.9102}{8.4083}}
\path(1706.842,2435.937)(1650.000,2564.000)(1632.648,2424.967)
\path(1339.852,2180.479)(1200.000,2189.000)(1315.421,2109.570)
\put(2297.635,2609.608){\arc{697.751}{0.3530}{3.2259}}
\path(2585.979,2354.432)(2625.000,2489.000)(2522.165,2393.836)
\path(2002.081,2508.928)(1950.000,2639.000)(1927.534,2500.701)
\put(3053.572,2274.714){\arc{622.906}{1.5019}{3.6052}}
\path(2935.865,1947.485)(3075.000,1964.000)(2947.281,2021.611)
\path(2778.838,2273.941)(2775.000,2414.000)(2706.052,2292.027)
\path(1125,2639)(2175,2714)
\path(2175,2714)(3150,2414)
\path(3150,1589)(2325,1064)
\path(2325,1064)(1125,1289)
\thicklines
\path(1106,1298)(1629,44)
\path(1106,1298)(479,44)
\path(1106,2655)(1106,1298)
\dashline{120.000}(3150,2414)(3150,1589)
\put(0,1889){\makebox(0,0)[lb]{\smash{{{\SetFigFont{10}{12.0}{\rmdefault}{\mddefault}{\updefault}$e\in T\setminus U$}}}}}
\put(2625,2714){\makebox(0,0)[lb]{\smash{{{\SetFigFont{10}{12.0}{\rmdefault}{\mddefault}{\updefault}$f_2$}}}}}
\put(1275,2864){\makebox(0,0)[lb]{\smash{{{\SetFigFont{10}{12.0}{\rmdefault}{\mddefault}{\updefault}$f_1\in T\cap U$}}}}}
\put(3225,1964){\makebox(0,0)[lb]{\smash{{{\SetFigFont{10}{12.0}{\rmdefault}{\mddefault}{\updefault}$f\in U\setminus T$}}}}}
\end{picture}
}
\end{center}

Revenons maintenant au cas g\'en\'eral. Nous avons une suite de liens
$f_1, \dots,f_n \in T\cap U$ partant du vertex $v$ suivant le chemin
${\cal P}$ et aboutissant \`a un lien $f\in U\setminus T$.
Alors on effectue la suite de changements \'el\'ementaires sur l'arbre $U$
\'echangeant les liens
$e\lrarr f_1, \dots, f_{n-1}\lrarr f_n$, et ainsi cr\'eant la suite
d'arbres maximaux $U_1,\dots U_n$.
Partant de $v$, tous les liens $e,f_1,\dots, f_{n-1}$
sont \`a la fois dans $T$ et dans $U_n$, $f_n$ est dans $T\setminus U_n$,
$f$ est dans $U_n\setminus T$, et tous les autres liens revenant vers $w$
sont dans $U_n$.
Nous sommes donc de retour dans le cas simple pr\'ec\'edent et nous
pouvons proc\'eder \`a un ultime changement   $f_n\lrarr f$
sur l'arbre maximal $U_n$ cr\'eant ainsi l'arbre (maximal) $\tl{U}$
tel que l'enti\`ere boucle ${\cal P}$ de $v$ \`a $v$
est \`a la fois dans $T$ et $\tl{U}$ except\'e
le lien $f$ qui n'est dans aucun.
En pratique, nous sommes partis d'une boucle avec tous les liens dans $U$
sauf un qui est dans $T$ (c'est le lien initial $e$), et en effectuant
des changements \'el\'emenatires, nous l'avons boug\'e le long du chemin
jusqu'\`a ce qu'il rencontre un lien qui n'est pas dans $T$
et qu'ils ``s'annulent'' ensemble.  

Ceci conclut l'absorbtion du lien $e$:
l'ensemble $T\setminus \tl{U}$ contient
un lien de moins que $T\setminus U$.
Maintenant,nous pouvons r\'ep\'eter la m\^eme proc\'edure
en partant du nouvel arbre $\tl{U}$.

\medskip
A pr\'esent, il est possible de d\'emontrer le r\'esultat suivant:
\begin{theo}
\label{moveind}
Le Jacobien du changement de variables
\Ref{chvar} est 1, de telle sorte que la mesure
$\dmu_T$ ne d\'epend pas du choix de l'arbre $T$.
\end{theo}

Ce th\'eor\`eme assure l'existence d'une
{\bf mesure $\dmu^{(\Gamma)}=\dmu_T$}
ind\'ependante du choix de $T$ et par cons\'equent \'egalement
ind\'ependante des arbitraires de toute la proc\'edure de fixation de jauge.
Elle permet d'int\'egrer les fonctions invariantes de jauge et de d\'efinir
l'espace des fonctions invariantes $L^2$ sur $\Gamma$.
Comme nous le verrons dans le prochain chapitre, c'est
l'espace ${\cal H}_\Gamma$ des r\'eseaux de spin d\'efini sur
le graphe $\Gamma$.

\medskip

La proposition \ref{elmoves} implique qu'il suffit de d\'emontrer le
th\'eor\`eme \ref{moveind} pour des changements \'el\'ementaires.
Consid\'erons donc une changement \'el\'ementaire sur l'arbre $T$
autour du vertex $v$ et d\'efinissant le nouvel arbre
$U=T\cup f \setminus e$.
Pour chaque lien $a\notin U$ sur la fleur r\'eduite d\'efinie par $U$,
on d\'efinit la variable $\gu_a$. Il s'agit de les exprimer en fonction
des variables $\gt_b$. Dans le cas $a\notin U$ et
$\notin T$, $\gu_a$ et $\gt_a$ sont \'egaux \`a une multiplication
pr\`es \`a droite ou \`a gauche par $\gt_f$ ou son inverse.
Le seul autre cas est celui de $a=e$; alors $\gu_e=(\gt_f)^{\pm
1}$. Alors, utilisant les invariances de la mesure 
(th\'eor\`eme \ref{totalinv}),
on peut conclure que le changement de variables ci-dessus a
un Jacobien trivial.

\chapter{D{\'e}finir les r{\'e}seaux de spins} \label{chap:spinnet}

Maintenant que nous avons muni les espaces invariants $A_\Gamma$ d'une
mesure (de Haar), nous pouvons d\'efinir  l'espace des
fonctions invariantes $L^2$ sur un graphe $\Gamma$.
Dans le cas $G=SU(2)$, une base de cet espace
est fournie par les r\'eseaux de spin $SU(2)$. Dans ce chapitre, il s'agit
de g\'en\'eraliser ce r\'esultat au cas d'un groupe de Lie $G$ quelconque,
pour pouvoir en particulier l'appliquer au cas des groupes de Lorentz dans le
cadre d'une approche \`a la \lg invariante par Lorentz. Les r\'eseaux
de spin g\'en\'eralis\'es seront d\'efinis comme les vecteurs propres
d'un ensemble d'op\'erateurs Laplaciens commutants et Hermitiens.
On s'attachera ensuite \`a discuter la g\'en\'eralisation potentielle du
formalisme d'Ashtekar-Lewandowski au cas d'un groupe non-compact, en vue
de d\'efinir un espace d'\'etats quantiques de la connexion/g\'eom\'etrie
en sommant sur tous les graphes $\Gamma$ possibles.
Je soulignerai les diff\'erences entre  le cas d'un groupe compact
et d'un groupe non-compact et j'expliquerai des possibles
constructions alternatives.

\section{Diagonaliser les Laplaciens}

Dans cette section, je vais d\'efinir les op\'erateurs diff\'erentiels
Laplaciens  sur un graphe $\Gamma$. Puis je montrerai que ce sont des
op\'erateurs hermitiens pour le produit scalaire d\'efini par la mesure
$d\mu_\Gamma$, r\'esultat se g\'en\'eralisant \`a tout op\'erateur
diff\'erentiel invariant de jauge. Alors, les r\'eseaux de spin
seront d\'efinis en tant que vecteurs propres g\'en\'eralis\'es
des op\'erateurs Laplaciens. Ce qui permettra de mieux comprendre
la structure de l'espace d'Hilbert $\H_\Gamma$ des fonctions invariantes
$L^2$ sur $\Gamma$.

\subsection{Op{\'e}rateurs Laplacien/Casimir}

Soit un graphe $\Gamma$. Une fonction (invariante de jauge) d\'efinie
sur $\Gamma$ d\'epend de $E$ \'el\'ements du groupe
$g_1,\dots,g_E$.
D\'enotons par $X$ un \'el\'ement de l'alg\`ebre de Lie et par
$\dd_{X}^{R_e}$ (resp. $\dd_{X}^{L_e}$)
la d\'erivation correspondante invariante \`a droite (resp. gauche)
agissant sur le $j$-i\`eme \'el\'ement du groupe associ\'e
au lien $e$:
\beqs
\dd_{X}^{R_1}f(g_{1},\cdots,g_{N}) =f(Xg_{1},\cdots,g_{N}),\\
\dd_{X}^{L_1}f(g_{1},\cdots,g_{N}) =f(g_{1}(-X),\cdots,g_{N}).
\eeqs
L'action du groupe de jauge agit sur ces op\'erateurs diff\'erentiels
par conjugaison au niveau des vertex:
$$
\dd_{X}^{R_e} \rightarrow \dd_{k_{s(e)}Xk_{s(e)}^{-1}}^{R_e}.
$$
Nous nous int\'eressons aux op\'erateurs diff\'erentiels invariants
de jauge. L'alg\`ebre de ces op\'erateurs est g\'en\'er\'ee par les
op\'erateurs Laplaciens  $\Delta^{(i)}_e$ o\`u 
$e$ d\'ecrit les liens du graphe $\Gamma$, $i$ est un entier entre 1 et
$r$ le rang du groupe.
Pour chaque lien $e$, il y a une correspondance un-\`a-un entre
l'ensemble des Laplaciens et l'ensemble des op\'erateurs Casimir
sur l'alg\`ebre de Lie.
Par cons\'equent, l'ensemble de ces Laplaciens forme bien une base
compl\`ete d'op\'erateurs commutants. 
En effet, ils commutent bien entre eux puisque, pour un lien $e$ donn\'e, 
deux Casimirs commutent et que les op\'erateurs
diff\'erentiels associ\'es \`a deux liens diff\'erents commutent \'egalement.

Le but est de d\'efinir les r\'eseaux de spin comme la base
des vecteurs propres de cet ensemble complet d'op\'erateurs diff\'erentiels
invariants de jauge.
Pour arriver l\`a, il faut montrer que ces op\'erateurs sont Hermitiens
pour la mesure $d\mu^{(\Gamma)}$ construite dans le chapitre pr\'ec\'edent.
Je vais donner la preuve dans le cas des op\'erateurs Laplaciens
$\Delta_e=\sum_i\dd_{X_i}^{R_e}\dd_{X_i}^{R_e}$
(not\'e plus simplement $\Delta_e = \dd^{R_e}.\dd^{R_e}$)
o\`u $X_i$ d\'enote une base orthonormale de l'alg\`ebre de  Lie.
Le cas g\'en\'eral est similaire, mais n\'ecessite des notations
plus lourdes.

\medskip

La mesure $\dmu^{(\Gamma)}$ \'etant d\'efinie sur la fleur r\'eduite
correspondant \`a $\Gamma$, il faut suivre la proc\'edure
de fixation de jauge d\'efinissant les variables ``r\'eduites''
$G_i,\dots,G_F$ sur la fleur et exprimer les op\'erteurs
$\Delta_e$ en fonction des d\'eriv\'ees $\tl{\dd}^{L,R}_i$
par rapport \`a ces nouvelles variables.

Pour commencer,  regardons l'exemple suivant: le cas
d'une fleur \`a deux p\'etales issue soit du graphe $\Theta$
soit du graphe \`a lunettes.
Fixons de jauge le graphe $\Theta$:

\ni
\begin{minipage}{8cm}
\begin{center}
\setlength{\unitlength}{0.00041667in}
\begingroup\makeatletter\ifx\SetFigFont\undefined%
\gdef\SetFigFont#1#2#3#4#5{%
  \reset@font\fontsize{#1}{#2pt}%
  \fontfamily{#3}\fontseries{#4}\fontshape{#5}%
  \selectfont}%
\fi\endgroup%
{\renewcommand{\dashlinestretch}{30}
\begin{picture}(2359,2187)(0,-10)
\put(1315.795,1084.682){\arc{1406.508}{4.5441}{6.2714}}
\put(1315.795,1102.318){\arc{1406.507}{0.0118}{1.7391}}
\put(1080.939,1081.743){\arc{1412.058}{3.1575}{4.8790}}
\thicklines
\path(1015.926,1724.624)(1198.000,1778.000)(1020.315,1844.543)
\thinlines
\put(1080.939,1105.257){\arc{1412.057}{1.4042}{3.1256}}
\thicklines
\path(1020.315,342.457)(1198.000,409.000)(1015.926,462.376)
\thinlines
\put(356,1094){\blacken\ellipse{84}{84}}
\put(356,1094){\ellipse{84}{84}}
\put(2006,1094){\blacken\ellipse{84}{84}}
\put(2006,1094){\ellipse{84}{84}}
\path(375,1056)(1200,1056)
\thicklines
\path(1380.000,1116.000)(1200.000,1056.000)(1380.000,996.000)
\path(1200,1056)(2025,1056)
\put(0,981){\makebox(0,0)[lb]{\smash{{{\SetFigFont{9}{10.8}{\rmdefault}{\mddefault}{\updefault}A}}}}}
\put(2175,981){\makebox(0,0)[lb]{\smash{{{\SetFigFont{9}{10.8}{\rmdefault}{\mddefault}{\updefault}B}}}}}
\put(975,1956){\makebox(0,0)[lb]{\smash{{{\SetFigFont{9}{10.8}{\rmdefault}{\mddefault}{\updefault}$g_1$}}}}}
\put(1050,81){\makebox(0,0)[lb]{\smash{{{\SetFigFont{9}{10.8}{\rmdefault}{\mddefault}{\updefault}$g_2$}}}}}
\put(975,1281){\makebox(0,0)[lb]{\smash{{{\SetFigFont{9}{10.8}{\rmdefault}{\mddefault}{\updefault}$g_3$}}}}}
\end{picture}
}
\end{center}
\end{minipage}
\begin{minipage}{5cm}
\setlength{\unitlength}{0.00041667in}
\begingroup\makeatletter\ifx\SetFigFont\undefined%
\gdef\SetFigFont#1#2#3#4#5{%
  \reset@font\fontsize{#1}{#2pt}%
  \fontfamily{#3}\fontseries{#4}\fontshape{#5}%
  \selectfont}%
\fi\endgroup%
{\renewcommand{\dashlinestretch}{30}
\begin{picture}(3319,1119)(0,-10)
\put(937,563){\ellipse{1068}{1068}}
\put(1987,563){\ellipse{1068}{1068}}
\put(1462,563){\blacken\ellipse{108}{108}}
\put(1462,563){\ellipse{108}{108}}
\put(1350,0){\makebox(0,0)[lb]{\smash{{{\SetFigFont{9}{10.8}{\rmdefault}{\mddefault}{\updefault}A}}}}}
\put(0,450){\makebox(0,0)[lb]{\smash{{{\SetFigFont{9}{10.8}{\rmdefault}{\mddefault}{\updefault}$G_1$}}}}}
\put(2550,450){\makebox(0,0)[lb]{\smash{{{\SetFigFont{9}{10.8}{\rmdefault}{\mddefault}{\updefault}$G_2$}}}}}
\end{picture}
}
\end{minipage}

Fixant de jauge \`a partir du point $A$, nous avons
$G_1=g_1g_3^{-1}$ et $G_2=g_2g_3^{-1}$.
Il est facile de v\'erifier que 
$\dd^R_1=\tl{\dd}^R_1$, $\dd^R_2=\tl{\dd}^R_2$
et $\dd^R_3=\tl{\dd}^{L_1}+\tl{\dd}^{L_2}$;
et par cons\'equent que $\Delta_1=\tl{L}_1$, $\Delta_2=\tl{L}_2$
et $\Delta_3=\tl{\Delta}_1+\tl{\Delta}_2+2\tl{\Delta}_{12}$,
avec $\tl{\Delta}_{12}=\tl{\dd}^{L}_1.\tl{\dd}^{L}_2$.

Dans le cas du graphe \`a lunettes:

\begin{center}
\setlength{\unitlength}{0.00041667in}
\begingroup\makeatletter\ifx\SetFigFont\undefined%
\gdef\SetFigFont#1#2#3#4#5{%
  \reset@font\fontsize{#1}{#2pt}%
  \fontfamily{#3}\fontseries{#4}\fontshape{#5}%
  \selectfont}%
\fi\endgroup%
{\renewcommand{\dashlinestretch}{30}
\begin{picture}(4276,1097)(0,-10)
\put(937,541){\ellipse{1068}{1068}}
\put(3037,541){\ellipse{1068}{1068}}
\put(1462,541){\blacken\ellipse{108}{108}}
\put(1462,541){\ellipse{108}{108}}
\put(2512,541){\blacken\ellipse{108}{108}}
\put(2512,541){\ellipse{108}{108}}
\path(2100,503)(2550,503)
\path(1500,503)(2100,503)
\thicklines
\path(1920.000,443.000)(2100.000,503.000)(1920.000,563.000)
\put(1425,203){\makebox(0,0)[lb]{\smash{{{\SetFigFont{9}{10.8}{\rmdefault}{\mddefault}{\updefault}A}}}}}
\put(2325,203){\makebox(0,0)[lb]{\smash{{{\SetFigFont{9}{10.8}{\rmdefault}{\mddefault}{\updefault}B}}}}}
\put(1725,728){\makebox(0,0)[lb]{\smash{{{\SetFigFont{9}{10.8}{\rmdefault}{\mddefault}{\updefault}$g_3$}}}}}
\put(0,503){\makebox(0,0)[lb]{\smash{{{\SetFigFont{9}{10.8}{\rmdefault}{\mddefault}{\updefault}$g_1$}}}}}
\put(3600,503){\makebox(0,0)[lb]{\smash{{{\SetFigFont{9}{10.8}{\rmdefault}{\mddefault}{\updefault}$g_2$}}}}}
\end{picture}
}
\end{center}
Nous avons $G_1=g_1$ et $G_2=g_3g_2g_3^{-1}$. Ce qui donne
$\Delta_1=\tl{\Delta}_1$, $\Delta_2=\tl{\Delta}_2$
(car le Laplacien est invariant sous l'action de $Ad(G)$) et
 $\Delta_3=(\tl{\dd}_2^R-\tl{\dd}_2^L)^2$.

\medskip

Dans le cas g\'en\'eral, pour un lien $e$ fix\'e,
s'il existe un arbre maximal $T$ qui ne contient pas $e$,
alors $e\notin T$ sera aussi sur la fleur r\'eduite
et $\Delta_e$ sera simplement le Laplacien
$\tl{\Delta}_e$ avec les d\'eriv\'ees
par rapport \`a la nouvelle variable $G_e$ sur la fleur.

Ce qui arrive aux liens qui sont dans tous les arbres possibles, tel
le lien au milieu du graphe \`a lunettes, est un peu plus d\'elicat.
Consid\'erons un tel lien $e$ et fixons de jauge \`a partir de son vertex
source $v=s(e)$. Alors  $\dd_e^R$ est \'egale \`a la somme des
 $\tl{\dd}_f^R$ sur les liens $f$
dont la boucle correspondante (liant $v$ au lien $f$) commence par le lien $e$
et $\tl{\dd}_f^L$ pour les liens  dont la boucle se termine par le lien $e$.

Dans tous les cas, les op\'erateurs diff\'erentiels initiaux
$\Delta_e$ s'\'ecrivent comme une somme
d'op\'erateurs $\tl{\Delta}_{ij}^{RR}=\tl{\dd}^{R}_i.\tl{\dd}^{R}_j$ et
$\tl{\Delta}_{ij}^{LR}=\tl{\dd}^{L}_i.\tl{\dd}^{R}_j$
o\`u $i,j =1,\cdots,g$. Ces op\'erateurs sur $G^h$
sont invariants sous l'action diagonale de $Ad(G)$.
Cela peut se montrer facilement en partant du fait que l'invariance
de jauge d'une fonction
 $\phi_\Gamma$ est \'equivalente \`a l'\'equation
\beq
(\sum_{e|s(e)=v} \dd^{R_e} + \sum_{e|t(e)=v} \dd^{L_e}) \phi=0,
\eeq
valable en chaque vertex $v$.
Choisir un arbre (maximal) revient \`a utiliser ces \'equations
pour exprimer toutes les d\'eriv\'ees sur les liens de l'arbre
en fonction des autres d\'eriv\'ees $\tl{\dd}_i^{L,R}$.
Cela nous laisse alors avec une unique relation:
$\sum_{i=1}^h (\tl{\dd}_i^{R}+ \tl{\dd}^{L}_i)\phi =0 $.

Maintenant que nous avons d\'efini les op\'erateurs Laplaciens
et \'etudi\'e leur structure en d\'etails,
nous allons
montrer qu'ils sont Hermitiens et nous allons d\'efinir les r\'eseaux de spin.

\subsection{R{\'e}seaux de Spins comme vecteurs propres et la structure
de $\H_\Gamma$}

\begin{theo}
Les op\'erateur Laplaciens $\tl{\Delta}_{ij}^{RR}$ et
$\tl{\Delta}_{ij}^{LR}$ sont Hermitiens pour le produit scalaire
d\'efini par la mesure
$\dmu_h$, $h\ge2$.
\end{theo}
Je vais donner la d\'emonstration pour
$\tl{\Delta}_{i}\equiv\tl{\Delta}_{ii}^{RR}$. La preuve pour les op\'erateurs
$LR$ se fait exactement de la m\^eme mani\`ere.

Consid\'erons
\beq
\label{sym}
\int_{A_{h}} (\varphi\tl{\Delta}_i \psi -\psi\tl{\Delta}_i \varphi) d\mu_h
\eeq
o\`u $\varphi,\psi$ sont des fonctions invariantes de jauge.
Introduisons une fonction $\phi$, pour fixer la jauge, telle que
\beq\label{gfix}
\int_{G} \,{}^g\phi dg =1,
\eeq
o\`u
${}^g\phi(g_{1},\cdots, g_{N})
= \phi(gg_{1}g^{-1},\cdots, gg_{N}g^{-1})$.
L'int\'egrale (\ref{sym}) se r\'e-\'ecrit:
\beq
\int_{G^{h}}(\varphi\tl{\Delta}_i \psi -\psi\tl{\Delta}_i \varphi)
 \phi \,dg_{1}\cdots dg_{h}.
\eeq
Gr\^ace \`a l'invariance de la mesure de Haar sous multiplication \`a gauche,
nous pouvons int\'egrer par parties les d\'eriv\'ees invariantes
\`a droite. Enlevant les symboles $\tl{}$ pour all\'eger les notations,
on obtient:
\beq
\int_{A_{h}} (\varphi\Delta \psi -\psi\Delta \varphi) d\mu =
\int_{G^{h}} \psi\dd_{X_{j}}^{R_i}\varphi \dd_{X_{j}}^{R_i}\phi
-\varphi\dd_{X_{j}}^{R_i}\psi \dd_{X_{j}}^{R_i}\phi \,dg_{1}\cdots dg_{h}.
\eeq
Nous pouvons factoriser le premier terme:
\beq
\int_{G^{h}} dg_{1}\cdots dg_{h} \,
\psi\dd_{X_{j}}^{R_i}\varphi \dd_{X_{j}}^{R_i}\phi
=
\int_{A_{h}} \psi
\left[
\int_{G}dg\, {}^g\dd_{X_{i}}^{R_k}\varphi {}^g\dd_{X_{i}}^{R_l}\phi
\right]d\mu_h,
\eeq
o\`u nous avons utiliser la d\'efinition de la mesure invariante \Ref{facton}
et $^g\psi=\psi$.
Ensuite, utilisant la propri\'et\'e des d\'eriv\'ees
\beq
{}^g\dd_{X_{i}}\phi = \dd_{Ad(g)^{-1}\cdot X_{i}} {}^g\phi,
\eeq
et l'invariance de l'op\'erateur diff\'erentiel quadratique
\beq
 \sum_{i} \dd_{Ad(g)\cdot X_{i}}^{R} \otimes \dd_{Ad(g)\cdot X_{i}}^{R}
 = \sum_{i} \dd_{X_{i}}^{R}\otimes \dd_{X_{i}}^{R}
\eeq
on obtient:
\beq
\int_{A_{h}}
\psi
\left[
\int_{G}dg  \dd_{X_{i}}{}^g\varphi
\dd_{X_{i}}{}^g\phi \,dg
\right]\dmu_h.
\eeq
Finalement, 
puisque ${}^g\varphi=\varphi$ ($\varphi$ \'etant invariante de jauge) 
et que la condition (\ref{gfix}) implique
$\dd_{X_{i}}\int {}^g\phi dg =0$, nous pouvons conclure que
l'int\'egrale (\ref{sym}) est nulle.

\medskip

\begin{defi}
Les op\'erateurs  $\Delta_e$ formant un ensemble
d'op\'erateurs Hermitiens commutants deux \`a deux sur l'espace d'Hilbert
$L^2(\dmu^{(\Gamma)})$,
nous pouvons les diagonaliser dans une base commune, et leur
vecteurs propres forment une base orthonormale de $L^2(\dmu^{(\Gamma)})$.
On appelle ces vecteurs propres les {\bf r\'eseaux de spin} ou
{\it spin networks}.
\end{defi}

Dans le cas (tr\`es probable) o\`u une des valeurs propres 
fait partie d'un spectre continu,
ces vecteurs propres doivent \^etre consid\'er\'es comme  des vecteurs
g\'en\'eralis\'es $\delta$-normalisable, donc en tant que
distributions invariantes et non des fonctions invariantes.

\medskip

Nous obtenons ainsi une base de fonctions/distributions labell\'ees
par les valeurs propres des op\'erateurs Laplaciens -les Casimirs du groupe-
sur chaque lien du graphe. Autrement dit,
 si on appelle $d\rho(\lambda)$ la mesure
spectrale de l'op\'erateur Laplacien, on peut \'ecrire:
\beq
H_\Gamma
\equiv L^2(\dmu^{(\Gamma)})= \oplus_e \int
d\rho(\lambda_e)\otimes_v I_v(\lambda),
\eeq
o\`u $I_v(\lambda)$ correspond \`a la d\'eg\'en\'er\'escence
des espaces propres labell\'es par la famille de valeurs propres
$\{\lambda_e, e\in\Gamma\}$. $I_v(\lambda)$ est habituellement associ\'e
\`a l'espace des entrelaceurs (vecteurs invariants)
entre les repr\'esentations port\'ees par les liens
se rencontrant \`a un vertex $v$.

On fait correspondre habituellement une repr\'esentation de $G$
\`a  un ensemble de valeurs propres des Laplaciens correspondant \`a un lien
donn\'e.  Ainsi on obtient un vecteur d\'ecrit comme un graphe dont
les liens portent chacun une repr\'esentation de $G$. Cela
fonctionne sans probl\`eme par exemple pour $G=SU(2)$, car 
il y a alors une bijection entre les
valeurs possibles du Casimir et les repr\'esentations unitaires
irr\'eductibles (de dimension finie). Dans le cas g\'en\'eral, cela n'est
plus aussi certain et plusieurs repr\'esentations peuvent avoir les m\^emes
valeurs des Casimir.
Par exemple, dans le cas de $\slr$, la s\'erie des repr\'esentations
discr\`etes donne deux repr\'esentations pour une m\^eme valeur du Casimir.
Par contre, dans le cas de $\slc$, les repr\'esentations
unitaires sont uniquement d\'etermin\'ees par les valeurs des 2 Casimirs.

Maintenant,
pour regarder de plus pr\`es la structure de $I_v(\lambda)$ et pourquoi
la d\'eg\'en\'erescence (mis \`a part celle associ\'ee
\`a la correspondance ``valeurs propres des Casimirs
$\lrarr$ repr\'esentations'')
est associ\'ee aux vertex du graphe, il convient
de d\'eplier les vertex du graphe $\Gamma$. Par cela, j'entends
 remplacer chaque
vertex par un arbre (minimal) trivalent. Pour un vertex donn\'e $v$,
cela est possible en regroupant les liens se rencontrant en $v$
deux par deux, en cr\'eant un vertex trivalent pour chacune de ces paires, puis
en r\'ep\'etant le processus jusqu'a un vertex trivalent final. Cette
proc\'edure permet de d\'efinir un {\it graphe d\'epli\'e} $\Gamma_0$.
Puisque les fleurs correspondants \`a $\Gamma$ et $\Gamma_0$ sont les m\^emes,
on a $L^2(\dmu^{(\Gamma)})=L^2(\dmu^{(\Gamma_0)})$.
On construit alors les r\'eseaux de spin sur $\Gamma_0$ labellant
tous ses liens, donc \`a la fois les liens originaux de $\Gamma$ et les liens
{\it internes} aux vertex de $\Gamma$. On a ainsi acc\`es \`a
la structure interne des vertex des r\'eseaux de spin et tout le probl\`eme
est ramen\'e \`a \'etudier les vertex trivalents i.e. les entrelaceurs
trivalents. Dans le cas de $G=SU(2)$, ces entrelaceurs sont
uniques (\`a normalisation pr\`es): ce sont les coefficients de Clebsh-Gordan.
Malheureusement, d\`es que le rang du groupe grandit un peu,
l'espace des entrelaceurs peut \^etre de dimension infinie (comme pour $sl(3,\R)$).

\medskip
Au final, cette construction  des r\'eseaux de spin pour un
groupe de Lie arbitraire reproduit bien la structure attendue des r\'eseaux
de spin, similaire au cas $SU(2)$ qu'il permet d'ailleurs
de retrouver enti\`erement. Cela nous permet donc de comprendre la structure
de ces observables (fonctions cylindriques de la connexion)
des th\'eories de jauge dans
le cas o\`u le groupe de jauge est non-compact. Le raisonnement met
\'egalement en relief les points d\'elicats qui peuvent se poser lors de
l'application de ce formalisme g\'en\'eral \`a des cas particuliers.

\section{L'espace d'Hilbert des r{\'e}seaux de spins}

\subsection{La diff{\'e}rence compact $\leftrightarrow$ non-compact}

Dans le cas du groupe compact $SU(2)$, il est possible de coller
tous les espaces d'Hilbert des connexions sur des graphes en un seul
espace d'Hilbert d\'ecrivant les \'etats quantiques en tant que fonctionnelles
sur un espace de connexion g\'en\'eralis\'ee. Ceci est obtenu au travers des 
m\'ethodes projectives expliqu\'ees dans
la section \ref{ashlewMeasure}, pouvant \^etre g\'en\'eralis\'ees \`a tous les
groupes compacts \cite{ash-lew}. Cependant elles ne s'appliquent pas
directement au cas des groupes non-compacts, le probl\`eme principal
\'etant que les mesures invariantes $d\mu_\Gamma$ d\'efinies sur les graphes
ne sont pas compatibles avec la structure de projections/injections
d\'efinies entre les espaces de connexions discr\`etes $A_\Gamma$.

\medskip

N\'eanmoins, on peut toujours construire le grand espace d'Hilbert comme somme de
tous les $\H_\Gamma$. Il existe une diff\'erence de taille entre les groupes
compacts et les groupes non-compacts: la repr\'esentation triviale
($j=0$ dans le cas $G=SU(2)$) n'est pas $L^2$ et n'intervient pas dans
les labels des r\'eseaux de spin pour les groupes non-compacts. Ainsi, si on
compare la situation avec le cadre  de la construction GNS d\'ecrite comme
alternative \`a la construction d'Ashtekar-Lewandowski dans la section 
\ref{ashlewMeasure}, nous avons directement construit les espaces
$\Htl_\Gamma$, g\'en\'er\'es par les r\'eseaux de spin labell\'es
par aucune repr\'esentation triviale. De ce point de vue, nous n'avons
pas besoin d'injecter les espaces les uns dans les autres et nous
n'avons pas besoin de quotienter. Nous pouvons donc
construire l'espace d'Hilbert
des \'etats quantiques de la connexion en tant que somme directe:
\beq 
\label{Hfinal}
\H_{\mathrm{inv}}
= \bigoplus_{\Gamma}\H_{\Gamma}.
\eeq

Il reste cependant une ambigu\"\i t\'e de normalisation dans la somme
ci-dessus: nous pouvons {\it a priori} normaliser les espaces $\H_\Gamma$
ind\'ependamment les uns des autres. Cette libert\'e dans la normalisation
relative des espaces  d'Hilbert est due \`a l'ambigu\"\i t\'e de normalisation
de la mesure de Haar. Dans le cas compact, nous pouvons imposer qu'elle
soit une mesure de probabilit\'e \cite{baez:spinnet}
et la normaliser pour que le groupe ait un volume unit\'e. Cela n'est plus
possible dans le cas des groupes non-compacts.

N\'eanmoins, vue la d\'efinition des mesures sur les fleurs, il est naturel
d'imposer que la mesure de Haar utilis\'ee partout (par toutes les fleurs,
tous les graphes) soit toujours normalis\'ee de la m\^eme mani\`ere.
Plus pr\'ecis\'ement, si on consid\`ere une fonction sur $G^{(n+1)}/Ad(G)$,
nous pouvons
int\'egrer un de ses arguments en utilisant la mesure de Haar et
nous obtenons une fonction sur $G^n/Ad(G)$. Il est naturel de
demander \`a ce que les int\'egrales de ces deux fonctions soient identiques.
Cet argument fixe la mesure de Haar \`a une constante pr\`es et si
nous la multiplions par un facteur $\alpha$, alors toutes les mesures 
$\dmu_n$ prendront un facteur $\alpha^{(n-1)}$. On peut consid\'erer
se fixer un facteur $\alpha$
comme un choix d'\'echelle dans notre th\'eorie physique.

Une autre mani\`ere de proc\'eder consiste \`a remarquer que, bien que 
$\H_{\mathrm{inv}}$ ne semble pas \^etre un espace $L^2$, il ressemble
fortement \`a un espace de Fock. Dans ce cadre, les injections/projections
de l'approche d'Ashtekar-Lewandowski seraient remplac\'ees par des
op\'erateurs d'annihilation et de cr\'eation, supprimant et
cr\'eant des boucles, qui agiraient comme des
isom\'etries entre les diff\'erents espaces d'Hilbert 
$\H_{\Gamma}= L^2(G^E/G^V)=L^2(G^{h_\Gamma}/Ad(G))$ et fixeraient
les normalisations relatives.
Plus pr\'ecis\'ement,  consid\'erons un graphe limite
 $\Gamma_\infty$ d'une suite de graphe $(\Gamma_i)_{i\in
\N},\, \Gamma_i \in  \G$ telle que $\Gamma_i \prec \Gamma_{i+1}$
(avec inclusion stricte) alors l'espace
\beq
{\cal{F}}_{\Gamma_\infty} =\bigoplus_{i}\Htl_{\Gamma_i},
\eeq
semble exactement \^etre un espace de Fock o\`u l'addition d'une
boucle/p\'etale serait un op\'erateur de cr\'eation.

La difficult\'e de munir ${\cal{F}}_{\Gamma_\infty}$
d'une structure d'espace de Fock
provient de la
sym\'etrie de jauge r\'esiduelle non-compacte $Ad(G)$. Une fixation
de jauge naturelle serait d'effacer cette sym\'etrie r\'esiduelle
en consid\'erant les fonctions cylindriques invariantes de jauge
\`a tous les vertex sauf en un vertex particulier $A$.
Ainsi, en suivant la proc\'edure de fixation de jauge d\'ecrie
dans le chapitre \ref{chap:fleur}, l'espace des connexions \`a support sur
le graphe serait simplement 
$G^{\otimes h_\Gamma}$ et l'espace d'Hilbert correspondant
$L^2(G^{\otimes h_\Gamma},dg^{\otimes h_\Gamma})$
(la mesure est donn\'ee simplement par la mesure de Haar sur $G$).
Il est alors facile d'empiler ces espaces pour former un espace de Fock
${\cal F}$,
en sommant sur les graphes. Les \'etats (quantiques) de la connexion
seraient, dans ce cadre, des ensembles de boucles de point de d\'epart $A$
(des fleurs autour de $A$).
Les op\'erateurs de cr\'eation et d'annihilation seront les
op\'erateurs usuels permettant de passer de $L^2(G^N)$ \`a $L^2(G^{N\pm 1})$,
cr\'eant ou d\'etruisant une boucle partant de $A$.
De ce point de vue, ${\cal F}$ repr\'esente les fluctuations
de la connexion autour du point $A$. Mais alors comment
restaurer l'invariance de
jauge au point $A$? L'imposer directement sur ${\cal F}$
cause des divergences.
A la place, nous pourrions nous placer au point $A$, ignorer l'invariance
de jauge et imposer que les \'etats consid\'er\'es se transforment selon une
repr\'esentation fix\'ee de $G$. Cela reviendrait \`a introduire une particule
en $A$ ou de mani\`ere \'equivalente un {\bf observateur}. Mais ceci
est une question d'interpr\'etation physique ne permettant pas de r\'esoudre
le probl\`eme math\'ematique de construire un espace de connexion
g\'en\'eralis\'ee pour un groupe non-compact.

\subsection{Un espace de Fock: th\'eorie de jauge sur $G$}
\label{analogieparticule}

Nous voulons coller les espaces 
$L^2(G^n/Ad(G))$ ensemble. Une analogie tr\`es utile
est d'interpr\'eter ces espaces comme espaces d'\'etats de 
particules sur un groupe $G$ \cite{frolich}. En fait,
$L^2(G^n,({\textrm{d}}g)^n)$ - $\textrm{d}g$ est la mesure de Haar
sur $G$ - est l'espace correspondant \`a une particule libre vivant sur
le groupe $G^n$ ou, de mani\`ere \'equivalente,
$n$ particules libres sur le groupe $G$.
Son  action est fonction de la trajectoire
$g(t):\R\rightarrow G^n$:
\beq
S_{\mathrm{free}}=\f{1}{2}\int dt \textrm{Tr}
\left((g^{-1}\partial_tg)^2\right)
\eeq
Cette action est invariante sous multiplication (constante) \`a droite et \`a
gauche dans $G^n$.
On peut proc\'eder \`a l'analyse Hamiltonienne du syst\`eme. L'espace
des phases est le fibr\'e cotangent du groupe $G$.
L'\'equation du mouvement est:
\beq
\dd_t(g^{-1}\dd_tg)=0
\eeq
On choisit $\pi^{(l)}=g^{-1}\dd_tg$ comme moment conjugu\'e
(\`a la place du moment canonique). C'est la charge de Noether
 associ\'ee \`a l'invariance \`a gauche.
Les solutions sont alors param\'etr\'ees comme
\beq
g_{g_0,\pi^{(l)}}(t)=g_0 \exp(\pi^{(l)} t).
\eeq
Ce sont les g\'eod\'esiques sur le groupe.
On peut \'egalement choisir le moment \`a droite
d\'efini par $\pi^{(r)}=-\dd_tg g^{-1}$. Alors les trajectoires solutions
s'\'ecrivent
\beq
g_{g_0,\pi^{(r)}}(t)=\exp(-\pi^{(r)}t)g_0.
\eeq

Le crochet de Poisson est:
\beq
\begin{array}{ccc}
\{g,\hat{g}\} &= & 0 \\
\{g,\pi^{(l)}_X\} &=& Xg\\
\{\pi^{(l)}_X,\pi^{(l)}_Y\} &=& \pi^{(l)}_{[X,Y]}
\end{array}
\eeq
o\'u on note $g,\, \hat{g}$ des \'el\'ements du groupe,
$X,Y$ sont dans l'alg\`ebre de Lie et
$\pi^{(l)}_X=\textrm{Tr}(X \pi^{(l)})$ est la
composante de $\pi^{(l)}$ dans la direction $X$.

\medskip
On peut alors cr\'eer l'espace de Fock des \'etats de particules libres:
\beq
{\cal F}=\bigoplus_{n\ge 0} L^2(G^n),
\eeq
qui prend en compte les \'etats \`a diff\'erents nombres de particules.
On peut \'ecrire explicitement les op\'erateurs de cr\'eation
$a^\dagger_\varphi$ et d'annihilation $a_\varphi$, ajoutant ou retirant
une particule \`a un \'etat (sym\'etris\'e) \`a $n$ particules,
not\'e $\psi$:
\beq
\label{dest}
(a_\varphi\psi)(g_1,g_2,\dots,g_{n-1})=
\int dg_n \psi(g_1,g_2,\dots,g_n)\overline{\varphi}(g_n)
\eeq

\beq \label{crea}
(a^\dagger_\varphi\psi)(g_1,g_2,\dots,g_{n+1})=\sum_i
\psi(g_1,\dots, g_{i-1},g_{i+1},\dots,g_n)\varphi(g_{i}).
\eeq

Il est aussi possible d'\'ecrire une th\'eorie de champs libre
correspondant \`a cet espace de Fock.
Pour cela, d\'efinissons les op\'erateurs champs
 $\Phi(g) = a_{\delta_{g}} $ et
$\Phi^\dagger(g) = a^\dagger_{\delta_{g}} $, o\`u $\delta_{g}$ est 
la fonction delta de Dirac au point $g$.
Alors l'op\'erateur impulsion totale, comme agissant
sur l'espace d'Hilbert \`a $N$ particules,
s'exprime en fonction des op\'erateurs champs:
$ \sum_{i=1}^{N}\pi^{i{(l/r)}}_{X} = 
\int_{G} dg \Phi^\dagger(g)(-i\nabla^{{(l/r)}}_{X})\Phi(g)$,
o\`u  $ \nabla^{{(l/r)}}_{X}$
est l'op\'erateur d\'eriv\'ee invariante \`a gauche/droite
dans la direction $X$.
De la m\^eme mani\`ere, on peut \'ecrire l'op\'erateur Hamiltonien: 
\beq
H= -\int_{G} dg \Phi^\dagger(g)\Delta \Phi(g).
\eeq 
L'action gouvernant la quantification et la dynamique du champ
s'\'ecrit en fonction d'un champ $\Phi(t,g)$ (dans l'``espace-temps''
form\'e par $\R\times G$):
\beq \label{Sfree}
S[\Phi(t,g)]=\int_{\R\times G}dt dg 
\Phi^\dagger(g)(i{\partial \over \partial t} + \Delta)\Phi(g).
\eeq

Il s'agit maintenant de suivre les m\^emes \'etapes dans
le cas de particules jaug\'ees, pour lesquelles on impose
une sym\'etrie globale sous $Ad(G)$.
Ceci peut \^etre pris en compte en introduisant un champs de jauge
 $A$ \`a valeur dans l'alg\`ebre de Lie $\G$. Alors
l'action devient\footnotemark:
\beqs
S_{\mathrm{gauged}} [g(t)\in G^n,A]&= & -\f{1}{2}\int dt
\textrm{Tr}\left((g^{-1}\partial_tg)^2\right) \nonumber \\
&& +\int dt \textrm{Tr}\left(
(g\dd_tg^{-1})A+A(g^{-1}\dd_tg)+gAg^{-1}A-A^2
\right)
\eeqs
\footnotetext{Nous avons modifi\'e l\'eg\`erement l'action propos\'e
dans \cite{frolich}.}
Cette action est invariante sous des transformations
de jauge $Ad(G)$ d\'efinies par des fonctions arbitraires
$h(t)$ \`a valeurs dans $G$:
\beq
\left\{
\begin{array}{ccc}
g & \rightarrow & hgh^{-1}\\
A & \rightarrow & hA h^{-1} + h\dd_th^{-1}
\end{array}
\right.
\eeq
L'espace des \'etats de notre syst\`eme est alors $L^2(G^n/Ad(G))$.
Dans le cas de ces particules jaug\'ees,
nous aimerions \'ecrire, comme dans le cas des particules libres,
des op\'erateurs de cr\'eation et d'annihilation, et construire
ainsi une th\'eorie de champs correspondante.
Le probl\`eme r\'eside dans le changement de la sym\'etrie:
la sym\'etrie $L^2(G^n/Ad(G))$ est une sym\'etrie globale $Ad(G)$ du 
 syst\`eme \`a $n$ particules et il n'est pas ais\'e de trouver
un \'equivalent de cette invariance au niveau des espaces de Fock.
Une analogie possible serait d'\'etudier un syst\`eme de $N$ particules
dans l'espace-temps invariant sous transformation globale de Poincar\'e.
Coller ces espaces \`a  $N$ particules n'est pas chose facile.

La m\'ethode la plus simple pour \'ecrire des op\'erateurs de cr\'eation 
et d'annihilation consisterait \`a fixer de jauge.
Ainsi,  commen\c cant par un graphe $\Gamma$ appliquant la proc\'edure
de fixation de jauge, nous avons montr\'e que l'espace $L^2(A_\Gamma)$
est isomorphe \`a $L^2(G^{h_\Gamma-1})$.
Par cons\'equent, l'espace associ\'e \`a un graphe limite serait l'espace
de Fock somme/limite de ces $L^2(G^{h_\Gamma-1})$.
Dans ce contexte, les op\'erateurs de cr\'eation 
et d'annihilation reviendraient bien \`a ajouter et enlever une boucle
au graphe. Il faudrait sans doute pousser ces id\'ees plus loin
pour comprendre la structure de ces op\'erateurs de champs.
Notons tout de m\^eme que l'action pour la th\'eorie de champs
de ces particules jaug\'ees s'obtient en rajoutant
le terme $$\int dt dg A(t)( \nabla^{{(l)}} - \nabla^{{(r)}})\Phi(g,t)$$
\`a l'action \Ref{Sfree} o\`u $A(t)$ est un champ de jauge.

\subsection{Une Repr{\'e}sentation de l'alg{\`e}bre des holonomies?}
\label{part2:discussion}

Pour conclure la discussion sur l'espace d'Hilbert
des fonctionelles cylindriques de la connexion dans le
cas de groupes non-compacts, il faut mentionner
que, dans l'approche suivie ici, l'accent a \'et\'e mis
sur les op\'erateurs d\'eriv\'ees, qui fournissent une
repr\'esentation des op\'erateurs form\'es \`a partir de la
triade/t\'etrade dans le cadre de la \lqg
(voir partie I pour plus de d\'etails).
Il faut \'egalement discuter si l'espace d'Hilbert construit 
$\H_{\mathrm{inv}}= \bigoplus_{\Gamma\in\G}\H_{\Gamma}$
porte aussi une repr\'esentation des observables de la connexion,
tel l'alg\`ebre des fonctionnelles cylindriques -ou r\'eseaux 
de spin- elles-m\^emes en tant qu'op\'erateurs Hermitiens
(pour les observables r\'eelles).

A priori on s'attend \`a ce qu'elles soient repr\'esent\'ees
par des op\'erateurs multiplication sur l'espace
d'Hilbert des r\'eseaux de spin.
Mais, comme d'habitude, dans le cas des
groupes non-compacts, nous sommes confront\'es
\`a des probl\`emes de divergence.
En effet, d\^u au caract\`ere distributionnel des fonctionnelles
r\'eseaux de spins dans le cas de groupes non-compacts, il est d\'elicat
de multiplier deux paquets d'onde ou fonctionnelles cylindriques $L^2$
dans la base des r\'eseaux de spin.
Une autre approche est d\'ecrit dans \cite{karim} pour le cas de $\slc$
(et les r\'esultats s'appliquent \'egalement au cas de $\slr$).
Les r\'eseaux de spin $\slc$ sont labell\'es par des repr\'esentations
unitaires de $\slc$, qui sont de dimension infinie et labell\'es par
un param\`etre (valeur propre du Laplacien) continu, que l'on note $\rho$. 
A la place de repr\'esenter ces quantit\'es, on peut s'int\'eresser
\`a repr\'esenter des r\'eseaux de spin labell\'es par les
repr\'esentations irr\'eductibles de dimension finie. Ils sont labell\'es
par des param\`etres discrets, ici not\'es $n$, et ne sont bien s\^ur
plus des fonctionnelles cylindriques $L^2$. N\'eanmoins, on peut les faire
agir sur $\H_{\mathrm{inv}}$. Faire agir un r\'eseau de spin $n$ sur
un paquet d'onde $f(\rho)$ holomorphe (cela correspond plus ou moins
\`a une fonctionnelle cylindrique \`a support compact sur l'espace
des orbites)  revient \`a shifter $f$ \`a $f(\rho+in)$.
Il est alors possible de v\'erifier que cet op\'erateur est bien d\'efini
et est Hermitien.
Il serait bien de pouvoir \'etendre cette approche au cas d'un
groupe $G$ g\'en\'erique et montrer que l'espace invariant de jauge
$\H_{\mathrm{inv}}$ porte bien une repr\'esentation des observables
r\'eelles de la connexion (r\'eseaux de spin) sous la forme
d'op\'erateurs Hermitiens.

\bigskip

Finalement, nous avons d\'ecrit la structure de l'espace des fonctionnelles
cylindriques de la connexion, pour un groupe de Lie $G$ ``quelconque''.
Cette espace correspondant aux observables d'une th\'eorie de jauge, l'espace
des fonctionnelles $L^2$ (pour la mesure invariante construite ici) forment
l'espace des fonctions d'onde ou \'etats quantiques de la th\'eorie. Et nous
avons exhib\'e la base des r\'eseaux de spin, qui diagonalise les
op\'erateurs Laplaciens/Casimir du groupe $G$. 
Maintenant, dans le dernier chapitre de cette partie, je vais illustrer
le formalisme g\'en\'eral d\'evelopp\'e jusqu'ici dans les cas particuliers
les plus simples, mais \'egalement les plus utiles pour la quantification de
la relativit\'e g\'en\'erale.

\chapter{Des exemples utiles} \label{chap:exemples}

Le sujet de ce chapitre est l'application des techniques
d\'evelopp\'ees pr\'ec\'edemment
aux th\'eories de la gravit\'e. Par cons\'equent, je vais d\'evelopper
les exemples des fonctions cylindriques pour les groupes $\slc$,
$\slr$ et $\su$.
En effet, $\slc$ correspond au groupe de Lorentz en $3+1$ dimensions,
$\slr$ est le groupe de Lorentz pour la gravit\'e en  $2+1$ dimensions
et $\su$ est \`a la fois le groupe de jauge de la \lqg (en $3+1$ dimensions)
et le groupe de sym\'etrie pour la gravit\'e Euclidienne en $3$ dimensions.

La m\'ethode est de d\'eriver tout d'abord les r\'esultats
dans le cadre de $\slc$, dont la structure est relativement simple
(un unique sous-groupe de Cartan), puis de d\'eriver les cas de $\su$
(pour v\'erifier que l'on retrouve bien les r\'esultats usuels) et de $\slr$
(qui est un peu plus complexe \`a cause de ses deux sous-groupes de Cartan).
Je d\'ecrirai le cas \`a une boucle, qui revient en fait \`a \'etudier
les caract\`eres des repr\'esentations des groupes, et le cas
de la fleur \`a deux p\'etales, puisqu'il est \`a la base de la construction
pour toutes les fleurs. 

\section{La Boucle}

Le cas \`a une boucle revient \`a \'etudier l'espace $G/Ad(G)$, qui
est singuli\`erement diff\'erent des autres espaces
$G^h/Ad(G)\sim G^{h-1}$ pour $h\ge2$. Il est n\'eanmoins tr\`es int\'eressant
puisque c'est l'espace des fonctions cylindriques
invariantes de jauge le plus simple et que nous sommes suppos\'es 
retrouver les caract\`eres des repr\'esentations unitaires comme
base de l'espace $L^2(G/Ad(G))$.

\subsection{Les Caract{\`e}res de  $\slc$}

Consid\'erons le cas du groupe $SL(2,\C)$. Il s'agit de
d\'ecrire le quotient 
$(SL(2,\C))//Ad(SL(2,\C))$ tel que d\'efini par la g\'eom\'etrie
alg\'ebrique dans la sous-section \ref{geomalg}.

Nous noterons $g=\mat{a}{b}{c}{d} \in \slc$. 
L'alg\`ebre $\C[\slc]^{\slc}$ des polyn\^omes invariants sous l'action
adjointe est g\'en\'er\'ee par
$$X(g)=\f{1}{2}tr(g),$$
la trace \'etant prise dans la repr\'esentation fondamentale
i.e. $tr(g)=a+d$.
En effet, les  polyn\^omes invariants
sont des combinaisons lin\'eaires des $tr(g^n)$. De plus
$tr(g^n)$ s'exprime comme un polyn\^ome en $X$
gr\^ace \`a la relation $g^{2}-tr(g) g +1=0$.
Plus pr\'ecis\'ement $tr(g^n) =T_{n}(X)$ o\`u $T_{n}$ sont
les polyn\^omes de  Chebichev du premier type.
Par cons\'equent, le spectre de l'alg\`ebre affine des
polyn\^omes invariants est simplement $\C$ et
le morphisme vers le quotient est la trace.

De plus  $X^{-1}(x), x \neq \pm 1$ est exactement une orbite
d'un \'el\'ement (strictement) r\'egulier. Et
 $$
X^{-1}(\pm 1)= \{\pm Id\} \cup
\left\{\pm\mat{1}{z}{0}{1}| z\in\C\right\} \cup \left\{\pm\mat{1}{0}{z}{1}|
z\in\C\right\}$$
contient plusieurs orbites, mais seules celles de plus ou moins l'identit\'e
$\pm Id$ sont ferm\'ees.
Notant $G_1=\{g|X(g)\ne \pm1\}$,
on peut donc consid\'erer $\slc//Ad(\slc)$ comme le quotient g\'eom\'etrique
de $G_{1}\cup \pm Id$ par $Ad(G)$.
La mesure $Ad(G)$-invariante (la mesure de Weyl) est induite
par la mesure de Haar sur $\slc$ et s'\'ecrit:
\beq \label{dhslc}
\mu(f)=\int_{\C}|X^{2}-1|f(X) dX
\eeq
o\`u l'int\'egration se fait sur le plan complexe priv\'e
de l'intervalle r\'eel $[-1,+1]$ avec la mesure de Lebesgue
habituelle sur $\C$.
Plus explicitement, $\slc$ a un unique sous-groupe de Cartan $H$
consistant en l'ensemble des matrices diagonales
$\{diag(\lambda,\lambda^{-1}),\lambda\in \C^*\}$.
Le groupe de Weyl est $Z_2$ et $diag(\lambda,\lambda^{-1})$ est
conjugu\'e \`a $diag(\lambda^{-1},\lambda)$.
La formula d'int\'egration de Weyl s'\'ecrit:
\beq
 \int_{\slc} f(g) dg =
 \int_{H} dh\,\left[\int_{\slc/H} f(xhx^{-1}) dx \right]
 |\Delta(h)|^{2}.
\eeq
La mesure invarainte est obtenue en enlevant l'int\'egration redondante
sur  $\slc/H$ et en int\'egrant seulement sur $H$. Alors on retrouve
la mesure \Ref{dhslc}.

\medskip

La s\'erie principale de repr\'esentations unitaires de
$\slc$ est une famille de repr\'esentations unitaires
irr\'eductibles de $\slc$ index\'es par des couples
$(j,\rho)$ avec $j\in \Z/2$ et $\rho \in \R$.
Elles sont r\'ealis\'ees dans $L^{2}(\C)$ et l'action 
$R_{j,\rho}$ de $\slc$ est:
\beq
R_{j,\rho}\mat{a}{b}{c}{d} f(z) = |bz+d|^{-2-2i\rho}\left(\f{bz+d}
{|bz+d|}\right)^{2j} f\left(\f{az+c}{bz+d}\right),
\eeq
avec $z\in \C$ et $f\in L^{2}(\C)$.
Les caract\`eres sont les traces des \'el\'ements du groupe
dans ces repr\'esentations et sont donn\'es par:
\beq
\chi_{j,\rho}\mat{e^{x+i\theta}}{0}{0}{e^{-x-i\theta}} =
\f{e^{i\rho x}e^{i j\theta}+e^{-i\rho x}e^{-i j\theta}}{|e^{x+i\theta} - e^{-x-i\theta}|^2}.
\eeq
En utilisant la mesure \Ref{dhslc} avec $X=(e^{ x + i \theta}+e^{-x-i
\theta})/2$ et effectuant le changement de variables vers $x,\theta$, il
est rapide de v\'erifier que
\beq
\mu\left(\chi_{j_1,\rho_1}\chi_{j_2,\rho_2}\right)=
\delta_{j_1j_2}\delta(\rho_1-\rho_2).
\eeq
Donc les caract\`eres de ces repr\'esentations forment bien
une base\footnotemark orthonormale (en tant que distributions) de l'espace d'Hilbert
$L^2(A_1)=L^2(\slc//Ad(\slc))$.

\footnotetext{En fait, les representations $(j,\rho)$ et $(-j,-\rho)$ d\'efinissant
les m\^emes caract\`eres, il faut se restreindre par exemple \`a $j\ge 0$ pour obtenir
une base.}

\subsection{Les Caract{\`e}res de $\su$}

Le cas de $SU(2)$ est tr\`es similaire au cas de $\slc$. C'est 
une certaine section r\'eelle de $\slc$.
Il y a
un unique sous-groupe de Cartan $H$ consistant en les matrices diagonales
$$
\left\{h_\theta=\mat{e^{i\theta}}{0}{0}{e^{-i\theta}},
\theta\in[-\pi,\pi]\right\}.
$$
Le groupe de Weyl est $Z_2$: $h_\theta$ est conjugu\'e \`a son inverse
$h_{-\theta}$.
La mesure invariante par $SU(2)$ est:
\beq
\mu_{SU(2)}(f)=\f{2}{\pi}\int_{-1}^{+1}dX \sqrt{1-X^{2}}f(X)
=\f{2}{\pi}\int_{0}^{\pi}d\theta\sin^2\theta f(\theta)
\eeq
avec $X=1/2tr(g)=\cos\theta$. C'est-\`a-dire que l'on restreint le cas $\slc$
aux configurations $X\in[-1,+1]$.

Une base orthonormale de $L^2(SU(2)/Ad(SU(2)))$ est fournie par les
caract\`eres des repr\'esentations irr\'eductibles unitaires (de
dimension finie) de $SU(2)$:
\beq
\chi_{j}(h_\theta) =
\f{\sin(j+1)\theta}{\sin(\theta) }
\eeq
o\`u  $j$ d\'ecrit $\N$ (l'entier est le double du {\it spin} usuel).
Effectuant un changement de variables de
 $X$ \`a $\theta$, il est facile de v\'erifier la propri\'et\'e
d'orthonormalit\'e:
\beq
\mu_{SU(2)}(\chi_j\chi_k)=\delta_{jk}.
\eeq

\subsection{Les Caract{\`e}res de $\slr$}

$\slr$ est le groupe de Lorentz en $2+1$ dimensions et peut
s'obtenir aussi comme une section r\'eelle de $\slc$.
Cette fois-ci, l'\'etude est plus compliqu\'ee car nous avons
maintenant 
deux sous-groupes de Cartan. Le premier est compact et correspond
aux rotations d'espace (rotations dans le plan):
\beq
H_0=\left\{k_\theta
= \mat{\cos\theta}{\sin\theta}{-\sin\theta}{\cos\theta}, \,
0\le\theta\le 2\pi\right\}.
\eeq
Le second est non-compact et correspond aux boosts:
\beq
H_1=\pm\left\{a_t=\mat{e^t}{0}{0}{e^{-t}},\,t\in R\right\}.
\eeq
Le groupe de Weyl $W(H_0)$ est trivial mais
on a toujours $W(H_1)=Z_2$, $a_t$ \'etant conjugu\'e \`a
$a_{-t}$. Un \'el\'ement r\'egulier de $\slr$ est soit conjugu\'e
\`a $H_0$ soit \`a $H_1$. Une fonction invariante sous $Ad(SL(2,\R))$
 $f$ sera donc d\'ecrite par son action sur les deux sous-groupes de
Cartan i.e. par deux fonctions
$f_0(\theta)$ et $f^\pm_1(t),t\ge0$.

\medskip

Pour construire la mesure, il s'agit de diviser par les volumes
de $G/H_0$ et de $G/H_1$. Ces deux volumes sont en effet infinis.
Malheureusement,  le probl\`eme est compliqu\'e par le fait
que le rapport de ces deux volumes est lui aussi infini
(d\^u au fait que $H_0$ est compact alors que $H_1$ est non-compact).
Cela r\'esulte en une ambigu\"\i t\'e et nous obtenons une famille
\`a un param\`etre de possibles mesures invariantes sous $Ad(\slr)$:
\beq
\label{dmusu}
\mu_{SL(2,\R)}(f)=\alpha_0\int_0^{2\pi}d\theta\sin^2\theta f_0(\theta)
+\alpha_1\int_{0}^{+\infty}dt\sh^2t f^\pm_1(t).
\eeq
La propri\'et\'e formelle que nous voulons que la mesure
satisfasse est 
\beq
\label{1loopinv}
\mu(^Gf)=\int_G dg \,f(g)
\eeq
o\`u  $^Gf$ d\'enote la fonction
moyenn\'ee par l'action $Ad(\slr)$
d'une fonction $f$ \`a support compact.
Le point subtil est que le centralisateur pour l'action $Ad(\slr)$
d'un \'el\'ement du groupe g\'en\'erique 
est (conjugu\'e) soit \`a $H_0$ soit \`a $H_1$
d\'ependant sur l'\'el\'ement du groupe en question.
Par cons\'equent, la proc\'edure de moyennage devrait
prendre en compte ce fait.
Ainsi $f_0$ (resp. $f_1$) \'etant \`a support sur l'espace
 $G_{(0)}$ (resp. $G_{(1)}$) des \'el\'ements du  groupe
conjugu\'es \`a $H_0$ (resp. $H_1$), nous d\'efinissons:
\beq
^Gf_i(g=xh_ix^{-1})=\int_{G/H_i}dx_i\,f(x_ih_ix_i^{-1}).
\eeq
Dans ce cas, il est direct de voir, en utilisant
\Ref{Weyl}, que \Ref{1loopinv} est satisfait par l'unique choix
$\alpha_i=1$.

\medskip

Mais cela n'est pas tout.
En effet, le moyen le plus simple de construire une mesure invariante
est de choisir un cutoff
 $\lambda$ et des sous-ensembles compacts $G_\lambda$ de $G$,
tel que la suite des $G_\lambda$ soit un suite
croissante, ``remplissant'' $G$ quand $\lambda$ tend
vers l'infini:
$$
lim_{\lambda\arr\infty}G_\lambda = G.
$$
Alors on obtient une mesure invariante dans la limite infinie i.e.
\beq
\mu(f)=\lim_{\lambda \rightarrow \infty}
\f{\int_{G_\lambda }f(g) dg}
{\int_{G_\lambda} dg},
\eeq
pour une fonction $f$ invariante par $G$. Cette mesure limite
correspond au choix $(\alpha_0=1, \alpha_1=0)$: cette mesure
donne un poids nul aux fonctions \`a support sur $G_{(1)}$.

\medskip

La mani\`ere de r\'econcilier ces deux points de vue est d'accepter 
les diff\'erences entre $H_0$ et $H_1$.
On d\'efinit  ainsi deux espaces d'Hilbert diff\'erents:
l'espace ${\cal H}_0$ des fonctions \`a support sur $G_{(0)}$
et d\'efini par la mesure $(\alpha_0 =1, \alpha_1=0)$
et l'espace ${\cal H}_1$ des fonctions \`a support sur
$G_{(1)}$ et d\'efini par la mesure $(\alpha_0 =0$, $\alpha_1=1)$.
Ceci prend en compte le fait que l'espace quotient par $Ad(\slr)$
n'est pas connexe avec des volumes de centraliseurs incommensurable.
En pratique, cela signifie que les deux secteurs ne peuvent pas communiquer
i.e. qu'il n'existe pas d'op\'erateurs physiques reliant
les \'etats physiques des deux secteurs. 
Ceci a \'et\'e montr\'e rigoureusement par
 Gomberoff et Marolf dans \cite{gomb} dans un contexte similaire
en utilisant le vocabulaire de {\it group averaging} et de {\it rigging maps}.

\medskip

$\slr$ a trois s\'eries principales de repr\'esentations unitaires: 
la s\'erie continue ${\cal C}_s$ labell\'ee par un nombre r\'eel positif
 $s$ et deux s\'eries discr\`etes (conjugu\'ee complexe l'une de l'autre)
 ${\cal D}^\pm_n$
toutes deux d\'efinies par un entier $n\ge1$.
Les caract\`eres de la s\'erie continue sont:
\beq
\chi_s(k_\theta) =  0
\eeq
\beq
\chi_s(\pm a_t) = \f{\cos s t}{|\sh t|}
\eeq
et les caract\`eres des s\'eries discr\`etes ${\cal D}^\pm_n$ sont
\beq
\chi^\pm_n(k_\theta) = \mp \f{e^{\pm i(n-1)\theta}}{2i\sin\theta}
\eeq
\beq
\chi^\pm_n(a_t) = \f{e^{-(n-1)|t|}}{2|\sh t|}
\textrm{ avec un facteur $(-1)^n$ pour } -a_t.
\eeq
Vue la formule pour les mesures, il est
clair que les caract\`eres des s\'eries discr\`etes
(resp. de la s\'erie continue) sont orthonormaux
dans l'espace d'Hilbert ${\cal H}_0$ (resp. ${\cal H}_1$).
De plus, tous ces caract\`eres sont des vecteurs propres du Laplacien.
En effet, le Laplacien s'\'ecrit:
\beq
\Delta = \f{1}{\sin\theta}\f{\dd^2}{\dd\theta^2}\sin\theta +\f{1}{4}
\textrm{ sur } H_0 \quad,\quad
\Delta = - \f{1}{\sh t}\f{\dd^2}{\dd t^2}\sh t +\f{1}{4}
\textrm{ sur } H_1 \textrm{ pour }t\ge 0.
\eeq
Ainsi la valeur propre correspondant \`a
 $\chi_s$ est $s^2+1/4$ et celle correspondant \`a 
 $\chi^{\pm}_n$ est $m(1-m)$ o\`u l'on note
$m=n-1/2$.
De plus, on remarque que, pour une mesure g\'en\'erique \Ref{dmusu}
pour des $(\alpha_0,\alpha_1)$ arbitraires, les caract\`eres
 $\chi_n^\pm$ des s\'eries discr\`etes ne sont pas orthonormaux,
ce qui serait en contradiction avec le fait que la Laplacien est Hermitien.
La seule mesure coh\'erente est alors le choix $(\alpha_0=1,\alpha_1=0)$,
qui donne l'espace d'Hilbert ${\cal H}_0$,
qui est donc la seule mesure possible quand on consid\`ere
les s\'eries discr\`etes.

\medskip

En fait, le probl\`eme est plus  profond et la question est \`a propos
du domaine de d\'efinition du Laplacien $\Delta$. On \'etudie
son action sur les espaces ${\cal H}_0$ et ${\cal H}_1$ pour tenir
compte du moyennage par $Ad(G)$, mais le probl\`eme des valeurs/vecteurs
propres n'est bien d\'efini qu'en se pla\c cant sur les distributions sur
tout le groupe $G$ (prenant en compte les rotations nulles, qui sont des
\'el\'ements non-r\'eguliers). L'\'etude
des distributions invariantes sur tout le groupe $G$
(au lieu de $G_{(0)} \coprod G_{(1)}$) est \`a la base des travaux,
lanc\'es par Harish-Chandra, sur l'analyse harmonique des groupes non-compacts
\cite{vara}. Ainsi la ``queue'' des $\chi_n$ sur $G_{(1)}$
est due aux conditions \`a la fronti\`ere (r\'egularit\'e sur
les rotations nulles)
dans le probl\`eme aux valeurs propres.

\medskip

Finalement, le cas de la boucle -le graphe \`a une p\'etale- est assez
compliqu\'e d\^u au fait que l'espace $A_1$ n'est pas connexe,
puisque nous avons exclus les rotations nulles du groupe.
Pour les fleurs \`a plus de p\'etales, la situation sera
heureusement plus simple car le centraliseur d'un point g\'en\'erique de
$G^h$ sera $G$. Ainsi la question de ``recoller'' les rotations nulles
(\'el\'ements non-r\'eguliers) pour le probl\`eme aux valeurs propres
sera plus facile \`a \'etudier, comme nous allons le voir dans le
cas de la fleur \`a deux p\'etales.

\section{Les fleurs {\`a} 2 p{\'e}tales}

L'\'etude du cas de la fleur \`a deux p\'etales est essentielle
\'etant donn\'ee la construction de la mesure d\'ecrite dans le chapitre
\ref{chap:mesure}. En effet, la mesure de la fleur \`a deux p\'etales
permet de construire directement la mesure sur toutes les fleurs. Etudier
ce cas particulier permet ainsi de comprendre la structure g\'en\'erale
des mesures quotients.

\subsection{Le cas $\slc$}

Consid\'erons le groupe $G=\slc$.
Nous voulons d\'ecrire le quotient $(\slc)^{2}
//Ad(\slc)$, comme d\'efini dans la sous-section \ref{geomalg}.

Pour  $(g_{1},g_{2}) \in \slc^{2}$, on d\'efinit $X_{1}(g_{1},g_{2})=
(1/2) tr(g_{1})$, $X_{2}(g_{1},g_{2})= (1/2) tr(g_{2})$ et
$X_{3}(g_{1},g_{2}^{-1})= (1/2) tr(g_{1}g_{2})$.
Ceci cr\'ee un morphisme  $\pi: \slc^{2}\rightarrow \C^{3}$
invariant sous $Ad(\slc)$.
Il a la propri\'et\'e suivante:
\begin{prop}
   $\pi$ fournit un isomorphisme entre l'alg\`ebre des
polyn\^omes invariants $\C[\slc^{2}]^{\slc}$ et les
polyn\^omes complexes \`a trois variables $\C[X_{1},X_{2},X_{3}]$.
\end{prop}
D\'emontrons cette proposition.
Notant $[,]_G$ le commutateur sur le groupe, on d\'efinit
l'ensemble
$$
G_{2}(\slc)=\left\{(g_{1},g_{2})\in \slc | tr(g_{1})^{2}\neq 4
\textrm{ ou }
tr(g_{2})^{2}\neq 4,{\mathrm et}\,tr([g_{1},g_{2}]_G) \neq 2 \right\}.
$$
L'image de cet ensemble par $\pi$ est $\C^3\setminus\Delta$,
o\`u on note  $\Delta$ l'ensemble ferm\'e dans $\C^{3}$ 
sur lequel les polyn\^omes $X_{1}^{2} -1$ ou $X_{2}^{2}- 1$, et 
$\Theta(X_{1},X_{2},X_{3})\equiv (X_{3}-X_{1}X_{2})^{2}
-(X_{1}^{2}-1)(X_{2}^{2} -1)$ sont nuls.
Ceci d\'ecoule du fait que
 $ tr([g_{1},g_{2}]_G) -2 = 4\Theta(X_{1},X_{2},X_{3})$.
Le point cl\'e est que ceci fournit un isomorphisme entre
$G_{2}(\slc)/\slc$ et $ \C^{3}\setminus\Delta$.
Construisons donc explicitement l'application inverse.
Pour cela, introduisons les matrices 
$s(\vec{X})=(s_{1},s_{2})$:
\beqs
 s_{1}(\vec{X}) &=& \mat{ X_{1}+\sqrt{X_{1}^{2}-1} }{0}
 {0}{ X_{1}-\sqrt{X_{1}^{2}-1} }, \label{s1(X)}\\
 s_{2}(\vec{X}) &=&
\mat{X_{2} -\f{ X_{1}X_{2}-X_3}{\sqrt{X_{1}^{2}-1}}}
 {1}
 {- \f{\Theta(\vec{X})}{X_{1}^{2}-1} }
 {X_{2} + \f{ X_{1}X_{2}-X_3}{\sqrt{X_{1}^{2}-1}} }.
\label{s2(X)}
\eeqs
Il faut faire attention \`a la d\'efinition de $s$ car elle
fait intervenir une racine carr\'ee.
Heureusement, un changement (de signe) dans le choix de la racine carr\'ee
peut s'impl\'ementer par une transformation de jauge. Par cons\'equent
$s$ est une fonction bien d\'efinie \`a valeur dans 
le quotient $G_{2}(\slc)/\slc$. Plus pr\'ecis\'ement,
notons $\tilde{s}(\vec{X})$ l'application correspondant \`a une
autre d\'etermination de la racine carr\'ee i.e.  obtenue \`a partir de 
$s$ en rempla\c cant  $\sqrt{X_{1}^{2}-1}\arr -\sqrt{X_{1}^{2}-1}$. Alors,
on peut v\'erfier que:
 \beq
 \tilde{s}(\vec{X}) = Ad\mat{0}{i}{i}{0}\cdot s(\vec{X}).
 \eeq
Maintenant, il est facile de voir que
 $\pi \circ s$ est l'identit\'e sur
$\C^{3}\setminus\Delta$.

Il est \'egalement vrai que $s\circ \pi$ est l'identit\'e sur
 $G_{2}(\slc)/\slc$.
En effet, consid\'erons un couple $(g_{1},g_{2})\in  G_{2}(\slc)$,
nous pouvons tout d'abord diagonaliser
$g_{1}$ puisque c'est un \'el\'ement r\'egulier.
Ensuite, cela ne fixe pas compl\'etement l'action du groupe de jauge puisqu'on
peut encore agir par une transformation de jauge diagonale et par une
transformation de Weyl (i.e. $g_{1}\rightarrow
g_{1}^{-1}$).
Une transformation de jauge diagonale $diag(\lambda,\lambda^{-1})$ agit
alors comme
$g_{2}=\mat{a}{b}{c}{d}\rightarrow
\mat{a}{\lambda^{2}b}{\lambda^{-2}c}{d}$. Et puisque
$tr(g_{1}g_{2}g_1^{-1}g_2^{-1}) -2 = -(\lambda -\lambda^{-1})^2 bc$,
la condition
$\Theta \neq 0$ se traduit par $bc\neq 0$. Ainsi, on peut fixer
l'action r\'esiduelle en imposant $b=1$. De cette mani\`ere, nous avons
effectivement ramener le couple $(g_1,g_2)$ sous la forme $(s_1,s_2)$
donn\'ee par les \'equations \Ref{s1(X)} et \Ref{s2(X)}.
Par cons\'equent, $s\circ \pi$ est bien l'identit\'e.

\begin{prop}
La mesure invariante $\mu$ d\'efinie dans la d\'efinition \ref{defmeasure}
est simplement la mesure de Lebesgue sur $\C^{3}$ traduite par les
applications $X_{1},X_{2},X_{3}$.
Plus pr\'ecis\'ement, soit $F$ une fonction sur $\C^{3}$,
$\pi^{*}F$ est alors une fonction invariante et
     \beq
     \int_{(\slc)^{2}//\slc}\pi^{*}F(g_{1},g_{2}) d\mu(g_{1},g_{2})
     =\int_{\C^{3}} F(\vec{X}) d^{2}X_{1}d^{2}X_{2}d^{2}X_{3}.
     \eeq
\end{prop}

Afin d\'emontrer cette proposition, rappelons la d\'efinition de la mesure
de Haar sur $\slc$:
$$
dg = d^{2}a d^{2}b d^{2}c d^{2}d \delta^{2}(ad-bc-1)
\qquad \textrm{pour} \qquad g=\mat{a}{b}{c}{d}.$$
Notons
\beq
y= \mat{a}{1}{c}{d},\,
h=\mat{\lambda}{0}{0}{\lambda^{-1}},\,
g=yhy^{-1}.
\eeq
La mesure  $A_{2}$ est d\'efinie par $d\mu =dg$. Elle s'\'ecrit donc
\beq
dg = |\lambda-\lambda^{-1}|^{2} d^{2}(\lambda+\lambda^{-1}) d^{2}a 
d^{2}d.
\eeq
De plus $X_{1}= \lambda+\lambda^{-1}$, $X_{2}=a+d$ et
$X_{3}=\lambda a +\lambda^{-1}d$. D'o\`u, apr\`es un petit calcul,
$dg =  d^{2}X_{1}d^{2}X_{2}d^{2}X_{3}$.

\medskip

Regardons maintenant les fonctionelles invariantes
\beq
\Phi_{\vec{j},\vec{\rho}}(g_{1},g_{2}) =
\chi_{j_{1},\rho_{1}}(g_{1})
 \chi_{j_{2},\rho_{2}}(g_{2})
 \chi_{j_{3},\rho_{3}}(g_{1}g_{2}).
\eeq
Un calcul explicite donne alors:
\beq
\int \Phi_{\vec{j},\vec{\rho}}  d\mu(g_{1},g_{2})
= \prod_{i=1}^{3} \delta(\rho_{i}) \delta_{j_{i}}.
\eeq

\subsection{D{\'e}duire $\su$ de  $\slc$}

Il est possible de retrouver le cas $G=SU(2)$ dans le formalisme
pr\'ec\'edent.
Ceci en imposant la contrainte
$\vec{X}(g_{1},g_{2}) \in I_{3}$ avec
\beq
I_{3} \equiv \left\{ X_{i} \in ]-1,1[, \,
\Theta(\vec{X})=(X_{3}-X_{1}X_{2})^{2}
-(X_{1}^{2}-1)(X_{2}^{2} -1)<0 \right\}.
\label{sl2tosu2}
\eeq
La mesure invariante est alors:
\beq
\label{su2mes}
\int_{I_3}d\vec{X}.
\eeq

En posant $X=\cos\theta$,
nous pouvons r\'e-exprimer la contrainte en fonction des angles
$\theta_{1,2,3}\in[0,\pi]$:
\beq
\cos(\theta_1+\theta_2)\le\cos\theta_3\le\cos(\theta_1-\theta_2).
\eeq
On retrouve la contrainte habituelle appara\^\i ssant
quand on multiplie deux \'el\'ements de $SU(2)$ ou, de mani\`ere
\'equivalente, quand on additionne deux vecteurs dans un espace sph\'erique.
En effet, deux \'el\'ements  $g_1,g_2$ de
$SU(2) \sim S^3$
d\'eterminent  un triangle dans $S^3$ avec comme vertex $1,g_1$ et $g_2$.
La g\'eom\'etrie invariante/intrins\`eque de ce triangle
est donn\'ee par les longueurs des trois c\^ot\'es,
qui sont  alors $\theta_1, \theta_2, \theta_3$.
Ecrites dans ces variables, la mesure se lit
$\sin\theta_1 \sin\theta_2 \sin\theta_3 d\theta_1d\theta_2d\theta_3$
et le domaine d'int\'egration est
\beq
\left|
\begin{array}{cc}
\theta_1 +\theta_2 \leq \theta_3\,&\textrm{plus permutations cycliques.} \\
\theta_1 +\theta_2 + \theta_3\leq 2\pi.&
\end{array}
\right.
\eeq
Nous pouvons \'egalement d\'ecrire la g\'eom\'etrie du triangle en fonction
de deux de ses c\^ot\'es: leur longueurs $\theta_1, \theta_2$ et
l'angle $\tl\theta_3$ qu'ils forment.
Cet angle est d\'etermin\'e par les longueurs des trois c\^ot\'es
par la formule $\cos\theta_3=cos\theta_1\cos\theta_2
-\sin\theta_1\sin\theta_2 \cos\tl\theta_3$.
Dans ces nouvelles variables g\'eom\'etriques, la contrainte \Ref{sl2tosu2}
s'\'ecrit
$\Theta=-\sin^2\theta_1\sin^2\theta_2\sin^2\tl\theta_3 \neq 0$, ce qui
signifie simplement que nous excluons les triangles d\'eg\'en\'er\'es.
La mesure est alors tr\`es simplement
$\sin^2\theta_1 \sin^2\theta_2 d\theta_1d\theta_2
\sin\tl\theta_3 d\tl\theta_3$.
Ce qui nous permet de v\'erifier facilement que
\beq
\int_{I_{3}} \chi_{j_{1}}(X_{1})
\chi_{j_{2}}(X_{2})\chi_{j_{3}}(X_3) d\vec{X}=
\delta_{j_{1},j_{2}} \delta_{j_{2},j_{3}} \f{1}{d_{j_{3}}}
\eeq
comme attendu!

\subsection{$\slr$: les triangles de $AdS_3$}

Dans le cas de $\slr$, la section de $\slc$ consid\'er\'ee est
donn\'ee par la  contrainte
$\vec{X}(g_{1},g_{2}) \in J_{3}$ avec
\beq
\label{j3}
J_{3} \equiv \left\{ X_{i}\in \R, \,
\Theta(\vec{X}) \neq 0,
(X_{1}^{2}\neq 1 \textrm{ or } X_{2}^2 \neq 1)  \} \textrm{ and } 
\vec{X} \neq I_{3} \right\}.
\eeq
La mesure invariante est alors:
\beq\label{slmes}
\int_{J_{3}}d\vec{X}.
\eeq
On note que  $J_{3}$ correspond \`a une section r\'eelle de 
 $\C^{3}$, qui est compl\'ementaire \`a $I_{3}$ d\'ecrivant $\su$.
Et similairement au cas
$\su$, il est possible de donner une interpr\'etation g\'eom\'etrique
\`a l'espace des configurations
 $J_3$ et \`a la condition de non-d\'eg\'en\'erescence.
Cela permettra de comprendre d'une mani\`ere intuitive
les singularit\'es de l'espace quotient.

\medskip

Il est bien connu que $\slr$ est isomorphe \`a
l'espace Anti-de-Sitter $AdS_3$ en 3 dimensions, qui
 peut \^etre d\'ecrit comme une hyperbolo\"ide
dans l'espace plat \`a 4 dimensions
$AdS_3=\{-(X_0)^2+(X_1)^2+(X_2)^2-(X_3)^2 =-1\}$.
Explicitement, cet isomorphisme se lit
\beq \label{gx}
 g(X)=
\mat{X_0 +X_1} {X_2 +X_3}{X_2-X_3}{X_0-X_1}.
\eeq
$AdS_3$ est un espace Lorentzien
et son groupe d'isom\'etrie est $SO(2,2)$.
Ainsi, l'espace des couples d'\'el\'ements 
$(g_1,g_2)$ de $\slr$
correspond \`a l'espace des triangles g\'eod\'esiques dans $AdS_3$ dont
l'un des vertex est fix\'e \`a l'identit\'e.
L'action adjointe de $\slr$ sur
$(g_1,g_2)$ se traduit alors par l'action du sous-groupe de
$SO(2,2)$ qui fixe l'identit\'e,
c'est-\`a-dire par l'action du groupe de Lorentz $SO(2,1)$
qui tourne les triangles.
Par cons\'equent, l'espace des orbites
est l'espace qui d\'ecrit la g\'eom\'etrie intrins\`eque des triangles
dans Anti-de-Sitter.

Les triangles sont de 4 types:
ils peuvent \^etre de genre espace, de genre temps,
de genre nul ou d\'eg\'en\'er\'e
(dans le cas o\`u les trois vertex du triangle sont sur une
m\^eme g\'eod\'esique), selon qu'ils sont dans un plan
de genre espace, temps ou nul.
Les c\^ot\'es des triangles peuvent eux-aussi \^etre de 4 types,
le cas d\'eg\'en\'er\'e correspondant au cas o\`u les deux vertex du c\^ot\'e
co\"incident.
Contrairement au cas $G=SU(2)$,
la g\'eom\'etrie invariante n'est pas enti\`erement
d\'etermin\'ee par la longueur des c\^ot\'es (le {\it carr\'e}
de la longueur en fait) puisque, par exemple,
la longueur d'un c\^ot\'e est nulle \`a la fois pour un c\^ot\'e de
genre nul ou un c\^ot\'e d\'eg\'en\'er\'e.

N\'eanmoins si nous nous restraignons \`a l'espace des triangles
satisfaisant $\Theta\neq 0$, alors la g\'eom\'etrie
du triangle est d\'etermin\'ee de mani\`ere unique par la longueur de
ces c\^ot\'es. Plus pr\'ecis\'ement:
\begin{prop}
\label{tetriang}
$\Theta(g_1,g_2)= 0$  \'equivaut \`a demander 
que le triangle AdS $(1,g_1,g_2)$ est d\'eg\'en\'er\'e ou de genre nul.
De plus $\Theta(g_1,g_2) < 0$ (resp. $\Theta(g_1,g_2) > 0$) ssi le
triangle AdS $(1,g_1,g_2)$ est de genre espace (resp. temps).
\end{prop}
La d\'emonstration de cette proposition n\'ecessite un peu de g\'eom\'etrie
dans l'espace $AdS$ et le lecteur int\'eress\'e peut trouver la
preuve dans \cite{spinnet}.

\medskip

Cette proposition nous dit que l'espace $J_3$ est l'espace des triangles
de genre non-nul et non-d\'eg\'en\'er\'es.
Cet espace est clairement non-connexe. Ses deux parties connexes
correspondent aux r\'egions pour lesquelles la normale
au triangle est de genre temps ou de genre espace\footnotemark.
\footnotetext{Il n'y a pas des distinction entre pass\'e et futur
dans $AdS_3$ puisque cet espace est p\'eriodique dans le temps. En particulier,
une g\'eod\'esique de genre temps retraversera la surface normale initiale.
On ne distingue donc pas les normales de genre temps
orient\'ees vers le pass\'e
de celles orient\'ees vers le futur.}
Ainsi, similairement, au cas de la boucle,
la mesure \Ref{slmes} permet de d\'efinir le produit scalaire de
deux espaces d'Hilbert diff\'erents, un pour les fonctions
\`a support sur les triangles de genre espace et un pour les
fonctions \`a support sur les triangles de genre temps.
Cependant, la situation est essentiellement diff\'erente du
cas \`a une p\'etale. En effet, dans le cas \`a une boucle,
l'espace quotient \'etait non-connexe et le centraliseur \'etait
fondamentalement diff\'erent dans les deux parties connexes.
A pr\'esent, dans le cas \`a deux boucles, le centralisateur (le groupe
laissant invariant un triangle donn\'e) est trivial dans les deux secteurs.
Cela implique qu'il n'existe aucune r\`egle de supers\'election, qui
interdit les op\'erateurs invariants envoyant un secteur sur l'autre. Et
on \'evite les probl\`emes du cas \`a une boucle, comme d\'ecrit dans
la section pr\'ec\'edente et dans \cite{gomb}.

En fait, il est possible d'\'etendre l'espace $J_3$ \`a l'espace
 $\tilde{J}_3 $ de tous les triangles non-d\'eg\'en\'er\'es
 (incluant les triangles de genre nul),
le centralisateur d'un triangle de $\tilde{J}_3 $ \'etant encore trivial.
Par cons\'equent, on peut \'etendre
la d\'efinition de la mesure \`a 
$\tilde{J}_3 $ qui est {\bf connexe}, et par cons\'equent
il existe une {\bf unique  mesure invariante} sur cet espace!
Plus techniquement, si $\phi(g_1,g_2)$ d\'enote une fonction \`a
support compact sur $\tilde{J}_3 $ alors
$^G\phi(g_1,g_2)=\int \phi(gg_1g^{-1},gg_2g^{-1})dg$ est encore bien
d\'efini pour tout $(g_1,g_2)\in \tilde{J}_3 $.
Cela permet d'\'etendre directement les distributions invariantes
sur $J_3$ (obtenu par moyennage par $Ad(\slr)$) \`a des
distributions invariantes sur $\tilde{J}_3$.
Les fonctionelles r\'eseaux de spin seront d\'efinies en tant que telles 
et la structure de l'espace d'Hilbert est fix\'e de mani\`ere unique.
Ce r\'esultat se g\'en\'eralise \`a tous les espaces correspondant aux fleurs
avec plus de 2 p\'etales, \'etant donn\'e que la mesure
est d\'efinie directement
\`a partir de la mesure du cas \`a deux p\'etales. Ainsi, seul le cas
d'une simple boucle est singulier.


\part{Invariance de Lorentz en Quantification Canonique}

Dans cette partie, je vais d\'ecrire les formalismes de \lg
en $2+1$ dimensions et $3+1$ dimensions utilisant le groupe de Lorentz,
respectivement $\slr$ et $\slc$.

La gravit\'e $2+1$ est ``triviale'' dans le sens que c'est une
th\'eorie topologique sans degr\'e de libert\'e locaux et que
nous savons plus ou moins comment la quantifier exactement.
Etudier le formalisme de la \lqg et
d\'evelopper les techniques de r\'eseaux de spin
dans ce cas est ainsi un travail pr\'eliminaire
pour ``tester'' les m\'ethodes dans un cadre plus simple
que la relativit\'e g\'en\'erale (en $3+1$).
Cela permettra de voir les probl\`emes principaux, et il
s'agira \'egalement de v\'erifier dans quelle mesure on retrouve bien
les m\^emes r\'esultats que les autres approches \`a la th\'eorie quantique
en $2+1$.
Je m'int\'eresserai particuli\`erement aux op\'erateurs g\'eom\'etriques
-spectre des longueurs et des aires- et t\^acherai de souligner les liens
reliant le pr\'esent formalisme aux formalismes usuels, dont le lecteur
peut trouver une description assez exhaustive (plus ou moins d\'etaill\'ee)
dans \cite{carlip}.

Puis je d\'ecrirai comment  d\'evelopper une \lqg en $3+1$ dimensions
bas\'ee sur le groupe de Lorentz (donc sans fixation de jauge
du type {\it time gauge}). Le formalisme canonique de base utilis\'e
a \'et\'e principalement \'etudi\'e par Alexandrov
\cite{sergei1,sergei2,sergei3} avec qui j'ai travaill\'e sur les
th\'eories quantiques correspondantes \cite{3+1}.
La motivation principale de ce travail est de trouver un lien
explicite entre les structures canoniques de la \lg et les mod\`eles
covariants des mousses de spin (espace-temps form\'e par des histoires
de r\'eseaux de spin) que je d\'ecrirai dans la partie IV.
J'ai explicit\'e la probl\'ematique de cette \'etude dans \cite{psn}:
les mousses de spin sont des structures invariantes par le groupe de Lorentz
$\slc$ alors que la \lqg usuelle est fond\'ee sur des quantit\'es invariantes
de jauge $\su$. Dans cette recherche, la th\'eorie en $2+1$ dimensions
pourra servir de guide 
car les deux formalismes (\lqg et mousse de spin) utilisent alors tous deux
le groupe de Lorentz $\slr$.
A travers la quantification de la th\'eorie canonique ``covariante''
(utilisant le groupe de Lorentz)
\cite{3+1}, je montrerai qu'il est possible
de retrouver d'une part le formalisme usuel de la \lqg $\su$ et d'autre part
de d\'eriver les structures cin\'ematiques des mod\`eles
de mousse de spin.
L'outil principal sera des r\'eseaux de spins ``projet\'es'' \cite{psn}
qui permettront de prendre en compte la mani\`ere dont
l'hypersurface canonique est plong\'e dans l'espace-temps.
En fait, l'hypersurface sera d\'ecrite (localement)
au niveau quantique \`a travers son
plongement dans l'espace-temps.

\chapter{La th\'eorie $2+1$}
\label{lqg3d}

La relativit\'e g\'en\'erale en $2+1$ dimensions peut \^etre
formul\'ee en fonction  d'une connexion (1-forme)
$SO(2,1)$ (groupe de Lorentz dans ce cadre),
not\'ee $A^i_\mu(x)$, et d'une triade (base orthonorm\'ee)
$e^i_\mu(x)$. Ici, $\mu=0,1,2$ d\'enote un indice d'espace-temps,
labellant une base du fibr\'e (co-)tangent, et $i=0,1,2$ est un indice
interne, labellant une base de l'alg\`ebre $so(2,1)$.
Je me placerai dans un espace-temps de signature $(-++)$;
ainsi les indices internes pourront \^etre mont\'es et descendus
\`a l'aide de la m\'etrique plate $\eta_{ij}={\rm diag}[-++]$.
L'action est d\'efinie \`a l'aide de la triade et de la courbure
de la connexion $F^{i}_{\mu\nu}=
\partial_\mu A^{i}_{\nu} -\partial_\nu A^{i}_{\mu}
+\eta^{ij}\epsilon_{jkl} A^{k}_{\mu} A^{l}_{\nu}$ 
($\epsilon_{ijk}$ est le tenseur compl\`etement anti-sym\'etrique):
\begin{equation}
    S(A,e)
=\f{1}{G}\int \mathrm{tr}( e\wedge F)
=\frac{1}{G}\int d^3x \ \eta_{ij}\ \epsilon^{abc}\ e^i_a\
    F^{j}_{bc},
\label{action2+1}
\end{equation}
o\`u la trace est prise sur $so(2,1)$, le produit $\wedge$ d\'efini
entre formes et $G$ la constante de Newton. Je travaille sans constante
cosmologique ($\Lambda=0$).

Il est facile de voir que les \'equations du mouvement imposent
que la connexion soit {\bf plate} ($F=0$) et que la triade soit compatible
avec la connexion (${\rm d}_Ae=0$). Il n'y a donc pas de perturbation
du type onde gravitationnelle en $2+1$ dimensions et la th\'eorie se r\'esume
\`a \'etudier l'espace des connexions plates.

\medskip

Un point crucial dans l'\'etude de la th\'eorie quantique est
la reformulation de la gravit\'e $2+1$ sous la forme d'une
th\'eorie de Chern-Simons sur le groupe de Poincar\'e $ISO(2,1)$
\cite{witten:2+1}. En effet, on peut former une $iso(2,1)$-connexion $\om$
en regroupant la connexion $A$
et la triade en un seul objet
$$
\om_\mu=A^i_\mu J_i + e^i_\mu P_i \in iso(2,1),
$$
o\`u les $J_i$ sont les g\'en\'erateurs du groupe de Lorentz $SO(2,1)$
et les $P_i$ les g\'en\'erateurs des translations:
$$
[J_i,J_j]=\epsilon_{ij}^kJ_k \qquad
[J_i,P_j]=\epsilon_{ij}^kP_k \qquad
[P_i,P_j]=0.
$$
L'action \Ref{action2+1} se r\'e-\'ecrit alors:
\beq
S(A,e)=S(\om)=\f{1}{G}\int \la \om \wedge d\om\ra +\f{2}{3}
\la \om\wedge\om\wedge\om\ra,
\eeq
o\`u $\la,\ra$ d\'enote une forme quadratique invariante
(non-d\'eg\'en\'er\'ee) sur $iso(2,1)$:
$$
\la J_i,P_j\ra=\eta_{ij} \qquad
\la J_i,J_j\ra=\la P_i,P_j\ra=0.
$$
Dans le cas d'une constante cosmologique $\Lambda>0$
(resp. $\Lambda<0$), on
obtient le groupe de De Sitter $SO(3,1)$ (resp. le groupe
anti-De Sitter  $SO(2,2)$) au lieu du groupe de Poincar\'e.
Ainsi, on reformule la gravit\'e en $2+1$ dimensions comme
une th\'eorie {\bf topologique} de jauge  $ISO(2,1)$.
Les \'equations du mouvement se r\'eduisent \`a $F(\om)=0$,
c'est-\`a-dire aux connexions $\om$ de courbure nulle.
On peut pousser cette approche assez loin et quantifier exactement
la th\'eorie  en fournissant une repr\'esentation de l'alg\`ebre
des holonomies de la connexion $\om$ 
(voir par exemple \cite{karim,carlip}).

\medskip

Ce n'est pas cette approche que je vais suivre ici. Je vais
d\'ecrire l'approche \lg o\`u on utilise comme fonctions d'onde
des fonctionnelles de la connexion de Lorentz $A$, en gardant
\`a l'esprit l'application de ce formalisme \`a la gravit\'e en
$3+1$ dimensions. Je vais donc commencer par l'analyse de la structure
canonique de l'action \Ref{action2+1}, puis proc\'eder \`a
la quantification en utilisant les fonctionnelles cylindriques
en la connexion $A$. Pour cela, j'exploiterai les r\'esultats
de la partie II sur les r\'eseaux de spin
pour $\slr\sim SO(2,1)$ et je d\'eriverai le spectre d'op\'erateurs
longueurs et aires sur ces \'etats quantiques de la g\'eom\'etrie.

\section{Le formalisme de la \lg en $2+1$}

Le formalisme de \lg en $2+1$ que je vais expliquer dans cette section
a \'et\'e \'etudi\'e dans \cite{2+1bis,2+1}, en collaboration
avec Carlo Rovelli et Laurent Freidel. Il est diff\'erent du traitement
de \cite{2+1toymodel} o\`u les auteurs proc\`edent \`a la repr\'esentation
des boucles de Wilson ($T_0$) et de leur ``moments'' associ\'es ($T^1$).
Je m'attacherais ici \`a utiliser les fonctionnelles cylindriques
et les r\'eseaux de spin, suivant la m\^eme approche en $3+1$. Il existe
\'egalement un autre formalisme de \lqg en $2+1$  se fondant sur une
rotation de Wick et utilisant ainsi un groupe de jauge compact $\su$
\cite{jacek}. Mais conserver le groupe de Lorentz (non-compact)
$SO(2,1)$ permet une comparaison directe avec les mod\`eles de mousse
de spin (de type Ponzano-Regge) d\'efinis en $2+1$ dimensions
\cite{davids,laurent:2+1}.

\medskip

Effectuons l'analyse Hamiltonienne habituelle sur l'action \Ref{action2+1},
en choisissant la coordonn\'ee
$x^0$ comme le param\`etre de l'\'evolution temporelle
et $x^a=(x^1,x^2)$ des coordonn\'ees sur la surface initiale (canonique)
 $\Sigma$, que nous prenons ferm\'ee et orientable.
Alors, on peut r\'e-\'ecrire l'action:
\beqs
    S&=& \frac{1}{G}\int dt \int_\Sigma dx^a \ \left( \eta_{ij}\,
    \epsilon^{ab}\ e^i_a\left( \partial_0 A^j_{b} -\partial_b A^j_{0}
    +\eta^{jk}\epsilon_{klm} A^l_{b} A^m_{0}\right) +
    \eta_{ij}\epsilon^{ab} e^i_0\ F^j_{ab} \right)\nonumber \\
&=& \f{1}{G}\int dt \int_\Sigma dx^a \left(
\epsilon^{ab}\eta_{ij} e^i_a\partial_0 A^j_{b}+
A_0^i\epsilon^{ab}(\eta_{ij}\partial_be_a^j+
\epsilon_{ijk}e^j_aA^k_b)
+e^i_0\eta_{ij}\epsilon^{ab}  F^j_{ab}
\right),
    \label{eq:deco}
\eeqs
avec $\epsilon^{ab}=\epsilon^{0ab}$. Cette expression nous donne
directement les variables canoniques  $A^i_{a}(x)$ 
et leur moments conjugu\'es $\pi^{a}_{i}(x) = \frac{1}{G} \eta_{ij}
\epsilon^{ab} e^j_b(x)$. 
Le crochet de Poisson fondamental est alors:
\begin{equation}
    \{ A^i_a(x), e^j_b(y)\}=G\ \epsilon_{ab}\, \eta^{ij}\,
    \delta^{(2)}(x,y).
    \label{eq:pb}
\end{equation}
Les multiplicateurs de Lagrange  $A_0^i$ et $e^i_0$  imposent
respectivement les contraintes
$\epsilon^{ab}{\cal D}_a e^i_b=0$ et $F^i_{ab}=0$.
La premi\`ere  -la {\it loi de Gauss}-
g\'en\`ere les transformations de jauge $SO(2,1)$, agissant sur
la connexion et la triade.
La seconde contrainte $F^i=0$
(puisque la connexion est anti-sym\'etrique, $F^i_{ab}$ est enti\`erement
donn\'ee par $F^i=\epsilon^{ab} F^i_{ab}$)
impose que le courbure soit plate. Elle 
g\'en\`ere une variation de la triade seule:
\beq
\left|
\begin{array}{ccl}
\delta e^i_a &=& {\cal D}_a \lambda^i \\
\delta A^i_a &=& 0.
\end{array}
\right.
\eeq
Quand le rep\`ere mobile $e$ est non-d\'eg\'en\'er\'e,
cette deuxi\`eme contrainte
peut se d\'ecomposer en une contrainte vectorielle imposant une
invariance sous diff\'eomorphismes spatiaux (de la tranche 2-dimensionnelle)
et en une contrainte scalaire (ou contrainte Hamiltonienne)
\cite{thiemann:2+1}.
La structure de la th\'eorie $2+1$ est alors tr\`es similaire \`a la
structure canonique de la \lg, telle que d\'ecrie dans la partie I.
Plus explicitement, introduisons la (co-)triade conjugu\'ee \`a la
connexion:  
\beq
E^a_i=\epsilon^{ab}\eta_{ij}e^j_b,
\eeq
et le vecteur normal densit\'e:
\beq
n^i=\f{1}{2}\epsilon^{ijk}\epsilon_{ab}E^a_jE^b_k
=\vec{e_1}\wedge\vec{e_2},
\label{normaldensity}
\eeq
o\`u j'utilise la notation vectorielle pour les indices internes.
Utilisant que
 $|\vec{n}|^2=\eta_{ij}n^in^j=-\textrm{det}({}^2g)$
est le d\'eterminant de la 2-m\'etrique (le signe moins \'etant d\^u \`a la
signature Lorentzienne de la m\'etrique $\eta$)
et que $E^a_in^i=0$, on peut d\'ecomposer
la contrainte de platitude en
\cite{thiemann:2+1}
\beq
N^iF^i=N^aV_a+NH,
\eeq
$N^a$ et $N$ \'etant respectivement le {\it Shift} et le {\it Lapse}
d\'efinis comme:
\beq
N^i=N^a\epsilon_{ab}E^b_i+N\f{n^i}{\sqrt{\textrm{det}({}^2g)}}
\qquad\Leftrightarrow\qquad
N^a=\epsilon_{ijk}\f{n^iE^a_j}{\textrm{det}({}^2g)}N^k
\textrm{ et }
N=\f{N^in^i}{\sqrt{\textrm{det}({}^2g)}}.
\eeq
$V_a$ est la contrainte vectorielle imposant l'invariance sous diff\'eomorphismes spatiaux
et  $H$ est la contrainte Hamiltonienne cens\'ee dicter
l'\'evolution dans le temps:
\beq
\left\{
\begin{array}{ccccc}
V_a &=&F^i_{ab}E^b_i &=& \vec{e_a}\cdot \vec{F}\\
H &=&\f{1}{2}\epsilon_{ijk}F^i_{ab}\f{E^a_jE^b_k}{\sqrt{\textrm{det}({}^2g)}}
&=& \f{\vec{n}}{|\vec{n}|}\cdot\vec{F}
\end{array}
\right.
\eeq

\medskip

Je vais d\'ecrire la structure cin\'ematique de la th\'eorie quantique.
Plus pr\'ecis\'ement, je vais m'occuper principalement des quantit\'es
invariantes de jauge: les fonctionnelles cylindriques invariantes de jauge.
Je vais ainsi utiliser les r\'eseaux de spin $SO(2,1)\sim\slr$
pour quantifier la th\'eorie. Pour prendre en compte les contraintes
vectorielles, il suffira de consid\'erer les classes d'\'equivalence sous
diff\'eomorphismes de graphes sur $\Sigma$. Ce sera l'espace
``cin\'ematique'' de la \lqg.
L'\'etape suivante (que je ne vais pas \'etudier ici) serait de prendre
en compte la contrainte Hamiltonienne pour faire \'evoluer dans le temps
les r\'eseaux de spin. Cette contrainte Hamiltonienne  agirait sur
les r\'eseaux de spin en modifiant \`a la fois le graphe-support et les
repr\'esentations du r\'eseau de spin. Dans le
cas de la gravit\'e Euclidienne en $3$
dimensions, on est cens\'e retrouver alors l'amplitude du mod\`ele
de Ponzano-Regge (avec $SU(2)$) d\'efini par les symboles $\{6j\}_{(\su)}$,
qui  permet bien au final de r\'eduire l'espace cin\'ematique
\`a l'espace physique des connexions plates \cite{ooguri}. Dans notre
cadre Lorentzien, on s'attendrait donc \`a reconstruire
\`a partir de la contrainte scalaire $H$ un mod\`ele de
Ponzano-Regge Lorentzien \cite{laurent:2+1}.

\subsection{Quantification et l'Op\'erateur Longueur}

Dans cette sous-section, je d\'efinis le cadre de la \lqg en
$2+1$ dimensions, fond\'ee sur l'utilisation des
fonctionnelles cylindriques de la connexion comme fonctions d'onde.
Je d\'efinis l'espace d'Hilbert des \'etats de la th\'eorie quantique,
dont une base  est donn\'ee par les r\'eseaux de spin $SO(2,1)$. Et 
je d\'efinis l'op\'erateur longueur d'une courbe, qui n'est en fait
rien d'autre que l'op\'erateur Laplacien (sa racine carr\'e plus
pr\'ecis\'ement). Il sera donc diagonal dans la
base des r\'eseaux de spin et ses valeurs propres seront donn\'ees par 
la valeur du Casimir de $SO(2,1)$. Le spectre explicite sera donn\'e
dans la sous-section suivante.

\medskip

En \lqg, les fonctions d'onde de la th\'eorie sont des fonctionnelles
de la connexion, contrairement \`a l'approche de la th\'eorie de
Chern-Simons o\`u on consid\`ere des fonctionnelles de la connexion $\om$
donc \`a la fois de $A$ et de $e$.
Plus pr\'ecis\'ement, on consid\`ere des
fonctionnelles cylindriques de la connexion $A$, d\'efinies par 
un graphe-support $\Gamma$ (avec $V$ vertex et $E$ liens)
et une fonction $\psi$ sur $(SO(2,1))^E$:
\beq
\Psi_{\Gamma,\psi}(A)=\psi(U_{1}(A),\ldots,U_E(A)),
\eeq
o\`u
$$
U_{\gamma}(A)={\cal P}\exp\left({\int_\gamma ds \,\dot{\alpha}^a(s)\,
A^i_a(\alpha(s))\ X_i}\right),
$$
est l'holonomie de la connexion $A$ sur d'un lien $\gamma$ du graphe.
Ici  $X_i$ d\'enote les 3  g\'en\'erateurs de l'alg\`ebre de
Lie $so(2,1)$.

Suivant la partie II, on d\'efinit pour un graphe $\Gamma$ fix\'e
une mesure invariante sur l'espace invariant de jauge.
Cela fournit l'espace d'Hilbert $\H_\Gamma$ des fonctionnelles $L^2$
\`a support sur le graphe, dont une base est donn\'ee par les r\'eseaux
de spin $SO(2,1)$.
Sur cet espace, deux sortes d'op\'erateurs sont d\'efinies.
Tout d'abord, les op\'erateurs de type boucle de Wilson (trace
de l'holonomie de la connexion $A$ le long d'une boucle).
Ces op\'erateurs agissent par multiplication. Il faut faire
particuli\`erement attention au domaine de d\'efinition d'un
tel op\'erateur (voir la discussion dans la sous-section
\ref{part2:discussion}).
Les autres op\'erateurs sont les op\'erateurs d\'eriv\'ees, correspondant
au champ $e^i_a(x)$:
\begin{equation}
\widehat{e^i_a(x)} =
-i\hbar G\epsilon_{ab}\eta^{ij}\f{\delta}{\delta A^j_b(x)},
\label{op:e}
\end{equation}
o\`u $\hbar G$ est la longueur de Planck $l_P$ en 3 dimensions.
Rappelons que
\begin{equation}
\f{\delta}{\delta A^j_b(x)}U_\gamma(A)=
\int_\gamma ds \, \frac{d\gamma^b(s)}{ds} \delta^{(2)}(\gamma(s),x)\
U_{\gamma_1(s)}(A) [X^i U_{\gamma_2(s)}](A),
\label{holoderivee}
\end{equation}
o\`u $\gamma_1(s)$ and $\gamma_2(s)$ sont les deux parties du lien $\gamma$
s\'epar\'ees par le point $x$ et $X^i$ le g\'en\'erateur
de l'action \`a gauche de $SO(2,1)$ sur les fonctions sur le groupe.
L'alg\`ebre quantique de ces op\'erateurs fournit une quantification
de leur alg\`ebre de Poisson classique.

\medskip

Ainsi, le produit scalaire de notre th\'eorie est d\'efini par:
\beq
\la \Psi_{\Gamma\psi}|\Psi_{\Gamma\psi'}\ra
=\int d\mu_\Gamma(U_1...U_E) \,\overline{\psi(U_1...U_E)}\ \psi'(U_1...U_E)
\eeq
Il faut alors faire attention aux conditions de r\'ealit\'e:
les op\'erateurs $\what{A^i_a}$ et $\what{e^i_a}$ doivent
\^etre des op\'erateurs Hermitiens.  $\what{A^i_a}$  agissant
par multiplication, on a trivialement (\`a quelques ennuis
de domaine de d\'efinition pr\`es) impl\'ement\'e
$\what{A^i_a}^\dagger=\what{A^i_a}$. De plus, comme montr\'e dans
la partie II, tout op\'erateur diff\'erentiel invariant de jauge
est Hermitien sur l'espace $\H_\Gamma$ pour la mesure $d\mu_\Gamma$,
et donc bien sur tout l'espace
$\H_{\rm inv}=\bigoplus_{\Gamma}\H_\Gamma$. 

\medskip

En utilisant la d\'ecomposition de  Plancherel des fonctions $L^2$ 
sur $SO(2,1)$ (pour la mesure de Haar) ou, de mani\`ere \'equivalente, la 
proc\'edure de la partie II,
une base orthonorm\'ee de
$\H_\Gamma$ est fournit par les r\'eseaux de spin labell\'es par les
repr\'esentations unitaires (de dimension infinie) de $SO(2,1)$.
Rappellons en effet que les fonctions $L^2$ sur $SO(2,1)$
peuvent \^etre d\'ecompos\'ees sur une base orthonorm\'ee
d\'efinie par des repr\'esentations unitaires irr\'eductibles.
Dans cette d\'ecomposition dite de Plancherel, les repr\'esentations
appartiennent \`a deux s\'eries de repr\'esentations: la s\'erie
principale et la s\'erie discr\`ete (en fait, la s\'erie discr\`ete
positive et la s\'erie discr\`ete n\'egative). Je donnerai des d\'etails
sur cela dans la prochaine sous-section.

Pour construire cette base des r\'eseaux de spin, on commence par
se fixer un graphe $\Gamma$.
On choisit une repr\'esentation unitaire ${\cal I}_e$ de 
$SO(2,1)$ pour chaque lien $e$ du graphe.
On contracte les matrices de repr\'esentation des
 holonomies $U_e^{{\cal I}_e}$ \`a l'aide d'un entrelaceur $SO(2,1)$ 
\`a chaque vertex (entrela\c cant les repr\'esentations associ\'ees
aux liens se rencontrant \`a ce vertex).
La fonctionnelle r\'esultante est un r\'eseau de spin $SO(2,1)$.
Elle depend du graphe, des repr\'esentations associ\'ees aux liens 
et des entrelaceurs associ\'es aux noeuds.
L'ensemble de ces fonctionelles forme  une base orthonormale
(au sens g\'en\'eralis\'e des distributions) compl\`ete des
fonctionnelles cylindriques invariantes de jauge.

\medskip

Dans ce contexte de \lg en $2+1$ dimensions, on s'int\'eresse
\`a l'op\'erateur longueur, qui est l'\'equivalent de l'op\'erateur aire
en $3+1$ dimensions.
Tout d'abord, la longueur d'une courbe lisse
$c:\tau\in [0,1]\rightarrow c(\tau)\in\Sigma$ est donn\'ee par
\begin{equation}
    L_c=\int_{[0,1]} d\tau\ \sqrt{\dot{c}^a(\tau)\, \dot{c}^b(\tau)\ g_{ab}(c(\tau))}
    =\int_{[0,1]} d\tau\ \sqrt{\dot{c}^a(\tau)\, \dot{c}^b(\tau)\ \eta_{ij}\,
    e^i_a(c(\tau))\, e^j_b(c(\tau))}.
    \label{length}
\end{equation}
Pour simplifier les notations, introduisons le vecteur
$$
e^i(c(\tau))=e^i_a(c(\tau))\dot{c}^a(\tau).
$$
Je me restreindrai \`a l'\'etude du cas
o\`u la norme  $\eta_{ij}e^ie^j$ du vecteur $e^i$
ne change pas de signe le long de la courbe.
Cela revient  \`a demander que la courbe soit enti\`erement
de genre temps ou de genre espace.
Remarquons que dans le cas d'une courbe de genre temps, i.e.
 $\eta_{ij}e^ie^j<0$,
il existe une autre quantit\'e invariante de jauge mis \`a part la norme
de  $e^i$:
le signe de $e^0$. Ce signe est en effet
invariant sous $SO(2,1)$ et  correspond \`a l'orientation dans le temps,
pass\'e ou futur,
de la courbe.

La longueur d'une courbe de genre espace
 ($\eta_{ij}e^ie^j>0$) est bien d\'efinie
par \Ref{length} comme un nombre r\'eel.
Dans le cas d'une courbe de genre temps ($\eta_{ij}e^ie^j<0$), il est
pr\'ef\'erable de d\'efinir un intervalle temporel r\'eel orient\'e
\beq
T_c=\textrm{sign}(e^0)\int_{[0,1]} dt\ \sqrt{|\eta_{ij}e^ie^j|}.
\label{time}
\eeq
L'op\'erateur longueur quantique repr\'esentant la longueur classique
est obtenue
en rempla\c cant l'op\'erateur triade $e^i_a(x)$ par l'op\'erateur
quantique \Ref{op:e} dans ces expressions.
Regardons maintenant l'action de cet op\'erateur longueur
sur les fonctionnelles r\'eseaux de spin, suivant l'exemple
de l'op\'erateur aire en \lqg en $3+1$ dimensions.

Consid\'erons une courbe  $c$ et un \'etat r\'eseau de spin
tel que la courbe et le graphe-support  s'intersecte en un seul point (et
pas en un vertex du graphe) et regardons l'action de
l'op\'erateur longueur de la courbe $c$ sur cet \'etat.
La proc\'edure peut \^etre facilement \'etendue aux
cas d'intersections multiples et d'intersections au niveau des noeuds.
Appelons $\gamma$ le lien du graphe intersect\'e par la courbe $c$ et
notons ${\cal I}$  la repr\'esentation irr\'eductible  associ\'ee to
$\gamma$.

L'action de $e^i(x)$ sur l'\'etat r\'eseau de spin ins\`ere la matrice
$X^i_{\cal I}$ du g\'en\'erateur $X^i$ dans la
repr\'esentation ${\cal I}$ au milieu de l'holonomie $U_\gamma$
le long du lien $\gamma$. On obtient alors:
\begin{equation}
\what{L_c}\Psi^{({\cal I})}=\hbar G \left[\int_c d\tau\,\sqrt{
    \left(\int_\gamma ds\,\epsilon_{ab}\dot{c}^a(\tau)
    \frac{d\gamma^b}{ds}\delta^{(2)}(\gamma(s),c(\tau))
    \right)^2
    \left(-\eta_{jk}X^j_{({\cal I})}X^k_{({\cal I})}\right)
    }\ \right]\ \ \Psi^{({\cal I})}.
    \label{eq:l}
\end{equation}
L'int\'egrale
$$
\int_c d\tau\,\int_\gamma ds\,\left|\epsilon_{ab}\dot{c}^a(\tau)
    \frac{d\gamma^b}{ds}\delta^{(2)}(\gamma(s),c(\tau))\right|
$$
compte le nombre d'intersections entre la courbe $c$ et
le lien $\gamma$, ici 1,
car $\epsilon_{ab}\dot{c}^a(\tau)\gamma'^b(s)$ est le Jacobien
de la transformation entre les coordonn\'ees orthonormales
$(x_1,x_2)$ et les coordonn\'ees locales $(\tau,s)$
\footnotemark.
\footnotetext{Cela est vrai que dans le cas o\`u la courbe $c$ et le
lien $\gamma$ ne sont pas tangents au point d'intersection.
Dans ce cas particulier, l'action de $\what {L_c}$ est nulle.}
Cela montre que les r\'eseaux de spin sont des vecteurs
propres de l'op\'erateur longueur, qui n'est autre que
l'op\'erateur Laplacien:
\begin{equation}
\what{L_c}\ \Psi^{({\cal I})}= \hbar G\
    \sqrt{-\eta_{jk}X^j_{({\cal I})}X^k_{({\cal I})}}\ \ \Psi^{({\cal I})}= \hbar G\
    \sqrt{q^{({\cal I})}}\ \Psi^{({\cal I})}.
    \label{eq:eigenvalue}
\end{equation}
o\`u $q^{({\cal I})}$ est la valeur de
l'op\'erateur Casimir $Q=-\eta_{ij}X^i X^j$
de $SO(2,1)$ dans la repr\'esentation $\cal I$.
Cela fournit le spectre des longueurs de la gravit\'e
$2+1$ dans la quantification \`a boucle.

D\'ependant du signe de $q^{(\cal I)}$,
on obtient soit une longueur de genre espace ou de genre temps ou de genre nul.
Dans les cas de genre temps ou de genre nul,
il faut \'egalement sp\'ecifier l'orientation temporelle i.e. le signe
de l'observable $\textrm{sign}(e^0)$.
Tout comme la longueur, cette information devrait \^etre
sp\'ecifi\'ee par la repr\'esentation ${\cal I}$:
dans les repr\'esentations de genre temps/nul, l'op\'erateur
 $\what{e^0}$ devrait avoir une spectre compl\`etement
positif ou compl\`ement n\'egatif.

Pour finir, l'action de l'op\'erateur longueur d'une courbe agissant sur un
r\'eseau de spin dont elle intersecte le graphe-support en plusieurs points
est donn\'ee par une contribution pour chaque lien intersect\'e:
\beq
{L_c}\ \Psi=l_P\sum_{e|e\cap c\ne\emptyset}
\sqrt{q^{({\cal I}_e)}}\, \Psi.
\eeq

\medskip
Maintenant que nous avons d\'efini l'espace d'Hilbert des \'etats quantiques
et l'action de l'op\'erateur longueur, regardons plus en d\'etails
les r\'eseaux de spin $SO(2,1)$, pour en d\'eduire le spectre exact
de l'op\'erateur longueur. Pour cela, il faut d\'ecrire les repr\'esentations
de $SO(2,1)$.

\subsection{Repr\'esentations de $SO(2,1)$ et spectre des longueurs}

L'alg\`ebre de Lie $so(2,1)$ est de dimension 3 et g\'en\'er\'ee
dans sa repr\'esentation fondamentale\footnote{En fait,
c'est la repr\'esentation fondamentale de
 $SU(1,1)$, qui est le double recouvrement de $SO(2,1)$.
Je me bornerai ici \`a n'utiliser que le groupe $SO(2,1)$
car sa th\'eorie des repr\'esentations est plus simple.
N\'eanmoins, tous les r\'esultats peuvent \^etre g\'en\'eralis\'es
sans aucun probl\`eme au cas $SU(1,1)$, et le lecteur int\'eress\'e
peut trouver les d\'etails dans l'appendice de \cite{2+1}.} par
les trois matrices suivantes:
\begin{equation}
\tau_0=\frac{i}{2}\matt{cc}{1 & 0 \\ 0 & -1}, \quad
\tau_1=\frac{1}{2}\matt{cc}{0 & 1 \\ 1 & 0}, \quad
\tau_2=\frac{i}{2}\matt{cc}{0 & -1 \\ 1 & 0}.
\label{tau}
\end{equation}
Les relations de commutations entre ces g\'en\'erateurs sont:
$$
[\tau_0,\tau_1]=-\tau_2 \qquad [\tau_1,\tau_2]=\tau_0 \qquad [\tau_2,\tau_0]=-\tau_1.
$$
On peut v\'erifier que ce sont les bons signes pour le groupe de sym\'etrie
$SO(2,1)$ d'un espace Lorentzien $(-,+,+)$.
Soient $X_i$ les g\'en\'erateurs d'une repr\'esentation lin\'eaire du groupe.
Ils satisfont eux-aussi:
$$
[X_0,X_1]=-X_2 \qquad [X_1,X_2]=X_0 \qquad [X_2,X_0]=-X_1.
$$
Il est important de ne pas confondre les propri\'et\'es d'Hermicit\'e
des matrices  $\tau_i$ et des op\'erateurs $X_i$.
Comme signal\'e pr\'ec\'edemment, les repr\'esentations
intervenant en gravit\'e quantique sont celles apparaissant
dans le formule de Plancherel.
Ces repr\'esentations sont unitaires i.e. les op\'erateurs $iX_i$
sont Hermitiens. En fait, ce sont les quantit\'es
correspondant aux op\'erateurs triades et leur Hermicit\'e refl\`ete
que la triade est r\'eelle!

Il est utile d'\'etudier l'alg\`ebre en introduisant les op\'erateurs
 $H$ and $J_\pm$:
\beq
H=-iX_0 \qquad J_\pm = \pm X_1 +i X_2.
\eeq
Ils satisfont les relations de commutation:
\beq
[H,J_\pm]=\pm J_\pm \qquad [J_+,J_-]=-2H.
\label{comm}
\eeq
La diff\'erence avec l'alg\`ebre r\'eelle $so(3)$ est le signe moins
dans la seconde relation de commutation. L'op\'erateur Casimir s'\'ecrit:
\beq
Q=(X_0)^2-(X_1)^2-(X_2)^2
=-H^2+\frac{1}{2}(J_+J_-+J_-J_+).
\label{casimir}
\eeq
Les conditions de r\'ealit\'e exprimant que les op\'erateurs
 $iX_i$ sont  Hermitiens se lisent:
\beq
H^\dagger=H \qquad
(J_\pm)^\dagger=J_\mp.
\label{reality}
\eeq

\medskip

On peut alors \'etudier les
 repr\'esentations de $SO(2,1)$ dans le m\^eme type de base que pour $SO(3)$.
Et il est facile de v\'erifier que
\begin{equation}\label{rep}
    \left\{
    \begin{array}{ccc}
    H|m\ra&=& m|m\ra, \\
    J_+|m\ra&=& (q +m(m+1))^{1/2}|m+1\ra, \\
    J_-|m\ra&=& (q +m(m-1))^{1/2}|m-1\ra
    \end{array}
    \right.
\end{equation}
fournit une repr\'esentation de $SO(2,1)$ sur l'espace g\'en\'er\'e
par les vecteurs $\{|m\ra , \ m\in\Z\}$\footnotemark.
Le param\`etre $q$ donne la valeur de l'op\'erateur Casimir. 
\footnotetext{Si on g\'en\'eralise $m$ pour prendre en compte des valeurs
demi-enti\`eres
$m+1/2$ partout dans \Ref{rep}, on obtient une repr\'esentation de
$SU(1,1)$. Et si on remplace $m$ par des valeurs $m+\alpha$,
avec $0<\alpha<1$, on obtient une repr\'esentation
du recouvrement universel de $SO(2,1)$.}

Les repr\'esentations unitaires de $SO(2,1)$ sont de dimension infinie
car $SO(2,1)$ est non-compact. De plus, leur op\'erateur de
Casimir est Hermitien $Q^\dagger=Q$, donc $q$ est r\'eel.
Regardons donc les diverses repr\'esentations obtenues pour des valeurs
r\'eels du param\`etre $q$. Commen\c cons par les valeurs n\'egatives
 $q\le0$.
Pour des valeurs g\'en\'eriques,
les repr\'esentation obtenues sont irr\'eductibles. Cependant
$(q+m(m+1))$ prend des valeurs n\'egatives, ce qui contredit la condition
de r\'ealit\'e $(J_\pm)^\dagger=J_\mp$.
Par contre, pour les valeurs sp\'eciales $q=-n(n-1)\le0$, avec $n\in \N^*$,
$(q+m(m+1))$ s'annule pour $m=n-1$ and $m=-n$. Par cons\'equent,
la repr\'esentation n'est pas irr\'eductible.
En fait, elle se d\'ecompose en 3 repr\'esentations:
\begin{itemize}
\item Il y a la repr\'esentation interm\'ediaire, not\'ee $V^n$.
Elle est de dimension finie et g\'en\'er\'ee par les vecteurs
 $\{|m\ra\ , \ -(n-1)\le m\le (n-1)\}$.
C'est la m\^eme repr\'esentation
que la repr\'esentation irr\'eductible de $SO(3)$ de spin $j=n-1$.
Mais $(q+m(m+1))<0$ et on a $(J_\pm)^\dagger=-J_\mp$, ce qui viole
la condition de r\'ealit\'e: ce n'est pas une repr\'esentation unitaire.
\item La repr\'esentation sup\'erieure ${\cal D}^+_n$ est une repr\'esentation
de dimension infinie de plus bas poids g\'en\'er\'ee par les valeurs
$m\in n+\N$. C'est une repr\'esentation unitaire.
\item La repr\'esentation inf\'erieure ${\cal D}^-_n$ est une repr\'esentation
de dimension infinie de plus haut poids g\'en\'er\'ee par les valeurs
$m\in -(n+\N)$. C'est aussi une repr\'esentation unitaire.
\end{itemize}
Pour une valeur positive du Casimir $q>0$,
$(q+m(m+1))=q-1/4 +(m+ 1/2)^{2}$ reste toujours positif
et on obtient une repr\'esentation de dimension infinie unitaire
g\'en\'er\'ee par les vecteurs $m\in{\mathbf Z}$.
$0<q<1/4$ correspond \`a la s\'erie dite exceptionnelle ou compl\'ementaire.
Et $q>1/4$ \`a la s\'erie dite principale.
Ces derni\`eres sont not\'ees ${\cal C}_s$, avec $q=s^2+1/4$.

Les repr\'esentations unitaires irr\'eductibles
${\cal D}^+_n$, ${\cal D}^-_n$ et
${\cal C}_s$ sont celles intervenant dans la {\it d\'ecomposition de
Plancherel} d'une fonction $L^2$ $f$ sur le groupe $SO(2,1)$:
\beqs
f(g)&=&
\sum_{n\ge 1}\ (2n-1)\ \textrm{Tr}(f^+_nR_{{\cal D}^+_n}(g))
+\sum_{n\ge 1}\ (2n-1)\ \textrm{Tr}(f^-_nR_{{\cal D}^-_n}(g))
\nonumber \\
&&+\int_{s>0} ds\ \frac{\coth(\pi s)}{4\pi s}\
\textrm{Tr}(f_sR_{{\cal C}_s}(g)),
\eeqs
Il est important de noter que les repr\'esentations continues
commencent \`a $q=1/4$ au lieu de $q=0$.

\medskip

Les valeurs propres de l'op\'erateur longueur
associ\'e \`a une courbe fix\'ee sont donn\'ees par la racine carr\'e
de l'op\'erateur Casimir des repr\'esentations port\'ees par les liens
du graphe-support que la courbe intersecte.
Une repr\'esentation continue
${\cal C}_s$ a un Casimir {\em positif} et correspond donc \`a une
{\it longueur spatiale} avec valeur:
\beq
L_s=\sqrt{s^2+1/4}.
\eeq
Notons le ``trou'' initial $1/2$. Cela signifie qu'il existe une longueur
spatiale minimale m\^eme si le spectre est {\it continu}.
Ce trou correpond \`a la s\'erie compl\'ementaire de
repr\'esentations unitaires. Elles ont une mesure de Plancherel nulle
i.e. elles n'interviennent pas dans la d\'ecomposition de Plancherel.

Une repr\'esentation discr\`ete ${\cal D}^\epsilon_n$
($\epsilon=\pm$ et $n\in\N$)
a un Casimir {\em n\'egatif} et correspond donc \`a une courbe
de {\it genre temps}.
Son orientation pass\'e ou futur $\textrm{sgn}(e^0)$
est donn\'ee par $\epsilon$. En effet, $\epsilon$ est le signe du
g\'en\'erateur $H$ (de ses valeurs propres), qui est l'op\'erateur
quantification de $e^0$. Ainsi le spectre des longueurs
d'une courbe de genre temps sera {\it discret}. Et les valeurs propres
sont:
\beq
T_{\epsilon,n}=\epsilon\sqrt{n(n-1)}.
\eeq
Remarquons que les valeurs ne sont pas \'egalement espac\'ees.
Notons aussi que les premi\`eres repr\'esentations discr\`etes
${\cal D}^\pm_{n=1}$ ont un Casimir nul et par cons\'equent
correspondent \`a un courbe de {\it genre nul}, le signe $\pm$
refl\'etant l'orientation pass\'e ou futur de la courbe.

\begin{figure}[t]
\begin{center}
\psfrag{p1}{$1^+$}
\psfrag{m1}{$1^-$}
\psfrag{nm2}{$n=2^-$}
\psfrag{np2}{$n=2^+$}
\psfrag{np3}{$n=3^+$}
\psfrag{np4}{$n=4^+$}
\psfrag{r2}{$T=i\sqrt{2}\sim 1.41$}
\psfrag{r6}{$T=i\sqrt{6}\sim 2.45$}
\psfrag{r12}{$T=i\sqrt{12}\sim 3.46$}
\psfrag{s0}{$s=0,L=\f{1}{2}$}
\includegraphics[width=6.5cm]{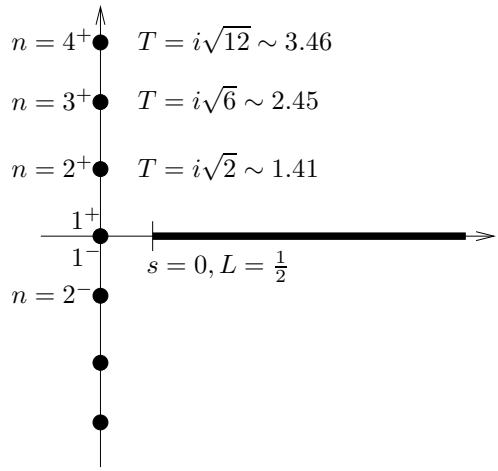}
\end{center}
\caption{Le spectre de l'op\'erateur longueur.}
\end{figure}

\begin{figure}[t]
\begin{center}
\psfrag{pm0}{$(1/2^\pm)$}
\psfrag{p1}{$1^+$}
\psfrag{m1}{$1^-$}
\psfrag{nm2}{$n=2^-$}
\psfrag{np2}{$n=2^+$}
\psfrag{np3}{$n=3^+$}
\psfrag{l1}{$L=i\f{1}{2}$}
\psfrag{l2}{$L=i\f{3}{2}$}
\psfrag{l3}{$L=i\f{5}{2}$}
\includegraphics[width=6.5cm]{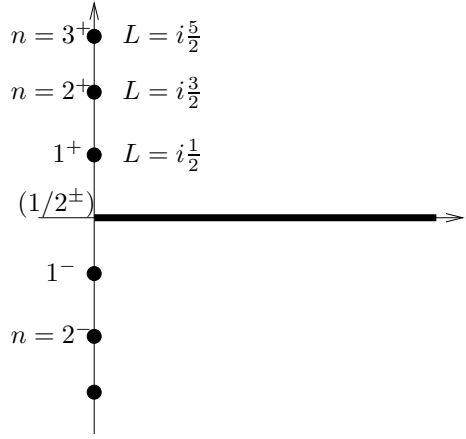}
\caption{Le spectre de l'op\'erateur longueur sym\'etrique.}
\end{center}
\end{figure}

Il existe une alternative \`a ce spectre, due \`a une ambig\"uit\'e  dans
la quantification (au niveau de la r\'egularisation) \cite{alekseev}.
En effet, nous pouvons utiliser la proc\'edure de quantification
sym\'etrique  pour quantifier $\what{e^i}$
comme  d\'erivation sur  l'alg\`ebre de Lie (en tant qu'espace vectoriel) 
au lieu de  d\'erivation sur le groupe de Lie\footnotemark.
\footnotetext{Cette ambig\"uit\'e peut \^etre retrouv\'ee
dans l'analogie avec les mod\`eles de particules sur le groupe
d\'ecrit dans \ref{analogieparticule}.
L'action pour une particule libre sur le groupe est
$$
S=\f{1}{2}\int dt\, \textrm{Tr}((g^{-1}\dd_tg)^2).
$$
Comme variable conjugu\'ee \`a  $g$, on peut choisir
soit le moment canonique $p=\dd_tg^{-1}$, qui est commutatif, soit
la charge de Noether non-commutative $\Pi=g\dd_tg^{-1}$. Cette derni\`ere
g\'en\`ere la multiplication \`a gauche et satisfait
 $\{\Pi_X,\Pi_Y\}=\Pi_{[X,Y]}$,
o\`u $\Pi_X=\textrm{Tr}(X\Pi)$ est la composante en $X$ de $\Pi$.
}
Alors l'op\'erateur Casimir $q$ est translat\'e \`a $q-1/4$
et le spectre de la longueur devient:
\beq
\left|
\begin{array}{ccccc}
L_s&=&s & \textrm{for space-like } &{\cal C}_{s>0}\\
T_{\epsilon,n}&=&\epsilon\left(n-\f{1}{2}\right) & \textrm{for time-like }&
{\cal D}^{\epsilon=\pm}_{n\ge1}
\end{array}
\right.
\label{lengthspectrum}
\eeq
Il y a maintenant un intervalle temporel minimal et aucune repr\'esentation
de genre nul donn\'e par les s\'eries discr\`etes.
De plus le spectre des longueurs temporelles devient \'egalement espac\'e. 
De l'autre c\^ot\'e, le trou initial pour le spectre des longueurs spatiales
dispara\^\i t et on a la possibilit\'e d'une courbe de genre nul dans
la limite $s\rightarrow 0$.
Ce second spectre est compatible avec les r\'esultats obtenus
dans les mod\`eles de mousse de spin en $2+1$ dimensions (mod\`ele
de Ponzano-Regge Lorentzien) \cite{davids,laurent:2+1, laurent:asymp}.

\subsection{Spectre de l'aire}

L'aire d'une surface ${\cal S}$
plong\'ee dans la surface canonique $\Sigma$ est donn\'ee par
\beq
\A_{\cal S}=\int_{\cal S} ds^2 \sqrt{\textrm{det}({}^2g)}
\eeq
o\`u $g_{ab}=e^i_ae^j_b\eta_{ij}$ est la 2-m\'etrique sur $\Sigma$.
On regarde donc l'op\'erateur quantification de cette quantit\'e.

Le  d\'eterminant de la m\'etrique s'\'ecrit
$\textrm{det}({}^2g)=-\eta^{ij}n_in_{j}$
en fonction du vecteur normal densit\'e
$n_i(x)=\f{1}{2}\epsilon_{ijk}\epsilon^{ab}e_a^j(x)e_b^k(x)$.
La triade $e_a^i(x)$ agissant sur un r\'eseau de spin
a une action non nulle ssi  $x$ appartient au graphe-support.
Et quand $x$ est au milieu d'un lien (pas un noeud) cette action
est proportionelle au vecteur tangent $\dot{\gamma}_a(s)X^i$ au lien.
Alors, puisque $\epsilon^{ab}\dot{\gamma}_a\dot{\gamma}_b=0$,
l'op\'erateur  $\what{n_i(x)}$ a une action nulle sur le r\'eseau de spin.
Par cons\'equent, les seuls points o\`u $\what{n_i(x)}$
a une actio non-nulle est au niveau des noeuds/vertex du graphe-support:
 {\it l'op\'erateur aire est \`a support sur seulement les noeuds
du graphe-support}.

Pour calculer l'action de l'op\'erateur aire d'une surface ${\cal S}$
sur un \'etat r\'eseau de spin,
on d\'ecoupe la surface en petits morceaux,
chacun d'eux contenant au plus un noeud du graphe-support.
On peut donc se restreindre au cas d'une surface \'el\'ementaire
ne contenant qu'un seul vertex $v$ du r\'eseau de spin sur lequel
l'op\'erateur aire agit.
Pour simplifier les notations, je me contente d'expliciter
le cas d'un noeud tri-valent.
Pour d\'efinir l'aire,
il faut choisir une {\it orientation} pour $\Sigma$, m\^eme si le r\'esultat
final sera ind\'ependant de ce choix. Cela revient \`a se fixer un ordre
coh\'erent des 3 liens se rencontrant \`a chaque vertex du graphe. 

Le vertex $v$ a trois liens incidents
$e=1,2,3$ (ordonn\'e suivant l'orientation) labell\'e
par les  repr\'esentations ${\cal I}_e$ de $SO(2,1)$.
Commen\c cons par regarder l'action de l'op\'erateur
 $\what{n_i(x)}$ sur la fonctionnelle r\'eseau de spin
au niveau du noeud $v$:
\beq
\what{n_i}(x) \Psi_{v}^{{\cal I}_1{\cal I}_2{\cal I}_3}=
\alpha(x,v)\ {\wtl{n_i}} \Psi_{v}^{{\cal I}_1{\cal I}_2{\cal I}_3},
\eeq
avec le facteur g\'eom\'etrique
\beq
\alpha(x,v)=
\sum_{e,e'}\ \int ds dt\ \delta^{2}(x,\gamma_{e}(s))\
\delta^{2}(x,\gamma_{e'}(t))\
|\epsilon_{ab} \gamma^a_{e}(s) \gamma^b_{e'}(t)|,
\eeq
et l'op\'erateur ${\wtl{n_i}}$ ins\'erant des  $X$ au vertex $v$ 
\beq
 {\wtl{n_i}} \Psi^{{\cal I}_1{\cal I}_2{\cal I}_3}=
 -\frac{l_P^2}{2} \epsilon_{ee'}\epsilon_{ijk}
X^j_{{\cal I}_{e}}X^k_{{\cal I}_{e'}}
\Psi^{{\cal I}_1{\cal I}_2{\cal I}_3},
\eeq
o\`u $e,e'$ sont deux liens arbitraires se rencontrant en $v$
et $\epsilon_{ee'}$ refl\`ete  l'orientation relative des deux liens
autour du vertex.
Notons qu'en utilisant
$\vec{X}_{{\cal I}_1}+\vec{X}_{{\cal I}_2}+\vec{X}_{{\cal I}_3}=0$,
on peut obtenir une expression plus sym\'etrique de ${\wtl{n_i}}$
en sommant l'expression pr\'ec\'edente sur les couples de liens
$(e,e')$ (avec un facteur $1/3$).

On peut r\'egulariser le facteur g\'eom\'etrique et on voit qu'il 
est proportionel \`a
$\delta^{2}(x,v)$.
Par cons\'equent, l'action de l'op\'erateur aire
sur le r\'eseau de spin est:
\beqs
\A_{\cal S}\,\Psi^{{\cal I}_1{\cal I}_2{\cal I}_3} &=&
\sqrt{-\eta^{ii'}{\wtl{E_i}}{\wtl{E_i'}}}
\,\Psi^{{\cal I}_1{\cal I}_2{\cal I}_3} \\
&=&
l_P^2
\sqrt{-\f{1}{4}
\eta^{ii'}\epsilon_{ijk}\epsilon_{i'j'k'}
X^j_{{\cal I}_1}X^k_{{\cal I}_2}
X^{j'}_{{\cal I}_1}X^{k'}_{{\cal I}_2}}
\ \Psi^{{\cal I}_1{\cal I}_2{\cal I}_3} \nonumber\\
&=&
l_P^2\f{1}{2}
\sqrt{\left(
(\vec{X}_{{\cal I}_1})^2(\vec{X}_{{\cal I}_2})^2
-(\vec{X}_{{\cal I}_1}.\vec{X}_{{\cal I}_2})^2
\right)}
\Psi^{{\cal I}_1{\cal I}_2{\cal I}_3} \nonumber \\
&=&
l_P^2\f{1}{2}\sqrt{
|\vec{X}_{{\cal I}_1}\wedge\vec{X}_{{\cal I}_2}|^2
}
\Psi^{{\cal I}_1{\cal I}_2{\cal I}_3}.
\eeqs
En utilisant que
$\vec{X}_{{\cal I}_1}+\vec{X}_{{\cal I}_2}+\vec{X}_{{\cal I}_3}=0$,
nous pouvons r\'e-exprimer le facteur pr\'ec\'edent en fonction
des op\'erateurs Casimir
$q^{{\cal I}_\alpha}$ des 3 repr\'esentations.
Ainsi l'op\'erateur aire est diagonal dans la base des r\'eseaux de spin
avec comme valeurs propres:
\beq
\A_{\cal S} =
l_P^2
\f{1}{2}\sqrt{
q^{{\cal I}_1}q^{{\cal I}_2}-
\f{1}{4}(q^{{\cal I}_3}-q^{{\cal I}_1}-q^{{\cal I}_2})^2
}.
\eeq
Cela correspond bien avec la d\'efinition de l'aire d'un triangle
g\'eom\'etrique d\'efini par les longueurs de ces 3 c\^ot\'es
$L_\alpha=\sqrt{q^{{\cal I}_\alpha}}$.
Plus explicitement, la formule ci-dessus se r\'e-\'ecrit:
\beq
\A_{\cal S} =
l_P^2 \f{1}{4}
\sqrt{
(L_1+L_2+L_3)
(-L_1+L_2+L_3)
(L_1-L_2+L_3)
(L_1+L_2-L_3)
}.
\eeq
On obtient donc un tout coh\'erent puisque 
$L_\alpha=\sqrt{q^{{\cal I}_\alpha}}$ est pr\'ecis\'ement
la valeur propre de l'op\'erateur longueur agissant sur un lien
$\alpha$ du r\'eseau de spin.

\medskip

Pour trouver le spectre de l'op\'erateur aire explicitement, il faut
caract\'eriser les noeuds admissibles i.e. les triplets de
repr\'esentations $({\cal I}_1,{\cal I}_2,{\cal I}_3)$
admettant un entrelaceur non nul. Pour cela, regardons
la d\'ecomposition du produit tensoriel de deux repr\'esentations
de $SO(2,1)$.
La d\'ecomposition des produits tensoriels est\footnotemark
\footnotetext{Le lecteur pourra trouver des expressions explicites
des coefficients de Clebsh-Gordan dans \cite{davids}.}:
\begin{equation}
    {\cal D}^\pm_{n_1}\otimes{\cal D}^\pm_{n_2}=
    \bigoplus_{n\ge n_1+n_2}{\cal D}^\pm_n,
\label{n+}
\end{equation}
\begin{equation}
    {\cal D}^+_{n_1}\otimes{\cal D}^-_{n_2}=
    \bigoplus_{n=1}^{n_1-n_2}{\cal D}^+_{n} \oplus
    \bigoplus_{n=1}^{n_2-n_1}{\cal D}^-_{n} \oplus
    \int ds \ {\cal C}_s,
\end{equation}
\begin{equation}
    {\cal D}^\pm_{n_1} \otimes {\cal C}_{s_2} =
    \bigoplus_{n\ge 1} {\cal D}^\pm_{n} \oplus
    \int ds \ {\cal C}_s,
\end{equation}
\begin{equation}
    {\cal C}_{s_1} \otimes {\cal C}_{s_2}=
    \bigoplus_{n\ge 1} {\cal D}^+_{n} \oplus
    \bigoplus_{n\ge 1} {\cal D}^-_{n} \oplus
    2\int ds\  {\cal C}_s.
\label{ccc}
\end{equation}
Ces d\'ecompositions peuvent \^etre interpr\'et\'ees comme des r\`egles
de sommation de vecteurs (dans l'espace de Minkovski $M^{2+1}$).
Intuitivement, une representation correspond \`a la classe d'\'equivalence
d'un vecteur sous l'action du groupe de Lorentz, et on distingue ainsi
des vecteurs de genre espace et des vecteurs de genre temps pass\'e et futur.
Le Casimir de la repr\'esentation correspond \`a la norme de ces vecteurs.
Consid\'erer le produit tensoriel de deux repr\'esentations revient
\`a additionner deux vecteurs. Ainsi,
similairement \`a l'\'etude
des mod\'eles de mousse de spin en 3 dimensions,
on regarde les classes d'\'equivalence
de {\it triangles} (sous l'action du groupe de Lorentz),
comme remarqu\'e dans le chapitre \ref{chap:exemples} de la partie II.

Ainsi on identifie les repr\'esentations ${\cal D}^\pm_n$ 
\`a des vecteurs de genre temps pass\'e et futur
de norme $L_n=n-1/2$ et les repr\'esentations ${\cal C}_s$ \`a des vecteurs
de genre espace de norme $L_s=s$.
Remarquons que ceci correspond exactement au spectre de l'op\'erateur de
longueur (sym\'etrique).
Alors, l'\'equation \Ref{n+} correspond au fait que la somme de
deux vecteurs de genre temps de m\^eme orientation
est de nouveau un vecteur de genre temps de la m\^eme orientation.
De plus, le produit tensoriel refl\`ete \'egalement l'in\'egalit\'e
anti-triangulaire $L_n\ge L_{n_1}+L_{n_2}$.
Similairement, l'\'equation \Ref{ccc}
correspond \`a sommer deux vecteurs de genre espace. Le r\'esultat
peut \^etre soit un vecteur de genre espace, soit un vecteur de genre temps,
et il n'y a aucune in\'egalit\'e (anti-)triangulaire.
Notons que cela implique que restreindre la th\'eorie
\`a la seule s\'erie principale continue de repr\'esentations
pour ne consid\'erer que les liens et des surfaces de genre espace ne
fonctionne pas: il faudrait en plus imposer \`a la main les in\'egalit\'es
triangulaires sur les produits tensoriels, ce qui ne semble pas tr\`es
naturel \`a la vue de la th\'eorie des repr\'esentations.

Finalement, les valeurs propres de l'op\'erateur aire
reproduisent pr\'ecis\'ement les diff\'erents types de triangles obtenus
en sommant deux vecteurs comme d\'ecrit par les r\`egles
de d\'ecomposition des produits tensoriels.

\subsection*{Conclusion: un espace continu et un temps discret}

En conclusion,
il est possible de quantifier la gravit\'e en $2+1$ dimensions
en suivant la proc\'edure de la \lqg. Je n'ai pas d\'ecrit comment
\'etudier la dynamique de la th\'eorie et je me suis attach\'e \`a
l'interpr\'etation g\'eom\'etrique des \'etats r\'eseaux de spin, qui sont
les \'etats quantiques de la g\'eom\'etrie de la surface canonique.
Ainsi un r\'eseau de spin (graphe habill\'e par des repr\'esentations
de $SO(2,1)$) a une interpr\'etation naturelle d'une vari\'et\'e 2d
discr\`ete/triangul\'ee: les faces ont une aire finie et sont duales aux
vertex du graphe-support, ces faces sont s\'epar\'ees par des c\^ot\'es
de longueur finie duaux aux liens du graphe.
De plus, le spectre des longueurs est fix\'e et a \'egalement une
interpr\'etation g\'eom\'etrique: 
les longueurs des c\^ot\'es de genre espace ont un spectre {\bf continu}
et les c\^ot\'es de genre temps ont un spectre des longueurs {\bf discret}!

\section{Variables \`a la 't Hooft}

Une autre approche dans laquelle il est possible de retrouver le m\^eme
r\'esultat de ``espace continu versus temps discret'' est la quantification
de la gravit\'e $2+1$ par 't Hooft \cite{hooft}. Il s'agit de travailler
sur un espace triangul\'e (en fait, d\'ecomposition cellulaire arbitraire)
et d'\'etudier comment la triangulation \'evolue dans le temps.
Dans ce cadre, 't Hooft d\'erive assez simplement une dynamique des tranches
spatiales.

Je propose ici de d\'eriver directement ce formalisme
\`a partir du  cadre canonique.
Cela nous permettra  d'avoir un nouveau point de vue
sur la dynamique en \lqg.
En particulier d'obtenir des
relations de commutation canonique entre des variables de
longueur et des variables d'angle. Le calcul de base est donc
le crochet de Poisson de la longueur d'une courbe $c$
et de (la trace de) l'holonomie de la connexion $A$ le long
d'une courbe $\gamma$. On consid\`ere le cas d'une seule
intersection entre $c$ et $\gamma$.
\begin{figure}[t]
\begin{center}
\psfrag{g}{$\gamma$}
\psfrag{g1}{$\gamma_1$}
\psfrag{g2}{$\gamma_2$}
\psfrag{c}{$c$}
\psfrag{P}{$P$}
\psfrag{n}{$\vec{n}$}
\psfrag{e}{$\vec{e}(P)$}
\includegraphics[width=8cm]{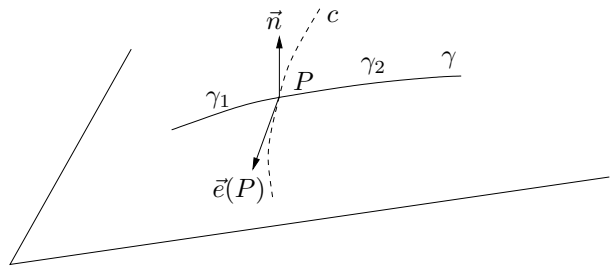}
\end{center}
\caption{La courbe $c$ dont on regarde la longueur et
la courbe $\gamma$ le long de laquelle on calcule l'holonomie,
se recontrant au point $P$ qui partage $\gamma$
en $\gamma_1$ et $\gamma_2$.}
\end{figure}

\medskip

Plus pr\'ecis\'ement, la longueur d'une courbe  param\'etr\'ee
par $c:\tau\in[0,1]\arr c(\tau)\in\Sigma $ est fonction de
la seule triade:
\beq
l_c=\int_cd\tau\,\sqrt{\eta_{ij}e^i(\tau)e^j(\tau)}
\quad{\rm avec}\quad e^i(\tau)=e^i_a\f{dc^a(\tau)}{d\tau}.
\eeq
D'autre part, notons $U_\gamma(A)\in SO(2,1)\sim SU(1,1)$
l'holonomie de la connexion $A$ le long de $\gamma$.
Une matrice
$$
U=\mat{\alpha+i\alpha_0}{\alpha_1-i\alpha_2}
{\alpha_1+i\alpha_2}{\alpha-i\alpha_0}\in SU(1,1)
$$
avec $\alpha^2+\alpha_0^2-\alpha_1^2-\alpha_2^2=1$
peut \^etre de deux types:
\begin{itemize}
\item soit c'est un {\it boost} si
$\beta^2=-(\alpha_0^2-\alpha_1^2-\alpha_2^2)>0$.
Dans ce cas, on d\'efinit l'angle du boost
$\eta>0$ tel que $\alpha^2=\ch^2\eta$ et $\beta^2=\sh^2\eta$.
Ainsi, en notant $X_i$ les matrices de Pauli g\'en\'erant $SU(1,1)$,
on \'ecrit
$$
U=\pm\ch\eta \,{\rm Id} + \sh\eta \,\vec{u}\cdot\vec{X}
$$
o\`u $\vec{u}$ est un vecteur unit\'e de genre espace dans l'espace
de Minkovski $M^{(-++)}$, simplement l'axe du boost.
Remarquons que changer $\vec{u}$ en $-\vec{u}$ revient \`a changer
$\eta$ en $-\eta$, d'o\`u la restriction $\eta>0$.
Notons finalement que
$$
\f{1}{2}{\rm Tr}(U)=\alpha=\ch\eta \qquad
\f{1}{2}{\rm Tr}(U\vec{X})=\vec{u}\sh\eta.
$$
\item soit c'est une {\it rotation} quand
$\beta^2=(\alpha_0^2-\alpha_1^2-\alpha_2^2)>0$.
Dans ce cas, on d\'efinit l'angle de la rotation
$\theta\in[-\pi,\pi[$ tel que $\alpha^2=\cos^2\theta$ et $\beta^2=\sin^2\theta$.
Ainsi, on obtient
$$
U=\cos\theta \,{\rm Id} + \sin\theta \,\vec{u}\cdot\vec{X}
$$
o\`u $\vec{u}$ est un vecteur unit\'e de genre temps, l'axe de la rotation.
Remarquons que changer $\vec{u}$ en $-\vec{u}$ revient \`a changer
$\theta$ en $-\theta$. Ainsi on peut se restreindre \`a  $\theta\in[-\pi,\pi[$
et $\vec{u}$ orient\'e futur (sur l'hyperbolo\"ide sup\'erieure).
\end{itemize}
On d\'efinit alors la trace de l'holonomie
\beq
t_\gamma=\f{1}{2}{\rm Tr}\,U_\gamma(A)=
\left|
\begin{array}{ccc}
\pm\ch\eta &\quad& {\rm (boost)} \\
\cos\theta &\quad& {\rm (rotation)}
\end{array}
\right.
\eeq
Alors le crochet de Poisson de $l_c$ et $t_\gamma$ 
consiste en leur d\'eriv\'ees respectives  par rapport \`a $e$ et \`a $A$:
$$
\left\{l_c,t_\gamma\right\}=
G\epsilon_{ab}\eta^{ij}\int_\Sigma d^2x\,
\f{\delta l_c}{\delta e^i_a(x)}
\f{\delta t_\gamma}{\delta A^j_b(x)}.
$$
D\'eriver $t_\gamma$ en fonction de $A^j(x)$ revient
\`a ins\'erer le g\'en\'erateur $X^j$ au point $x=\gamma(s)$ au
milieu de $U_\gamma$, c'est-\`a-dire on d\'ecoupe $\gamma$
en deux morceaux $\gamma_1(s)$ et $\gamma_2(s)$ s\'epar\'es par $x$
et on consid\`ere la quantit\'e
$$
\f{1}{2}{\rm Tr}\,U_{\gamma_1(s)}X^jU_{\gamma_2(s)}=
u^j(s)\times\left|
\begin{array}{c}
\sh\eta \\
\sin\theta
\end{array}
\right.
$$
avec
$\vec{u}(s)=U_{\gamma_1(s)}\vec{u}=U_{\gamma_2(s)}^{-1}\vec{u}$
o\`u $\vec{u}$ est l'axe de l'holonomie $U_\gamma$
(i.e. $\vec{u}(s)$ est simplement le transport\'e parall\`ele
de $\vec{u}$ le long de la courbe $\gamma$).
Alors, un calcul direct donne
\beq
\left\{l_c,t_\gamma=
\left|
\begin{array}{c}
\pm\ch\eta \\
\cos\theta
\end{array}
\right.
\right\}=
G\left|
\begin{array}{c}
\sh\eta \\
\sin\theta
\end{array}
\right. \times
\int_cd\tau\int_\gamma ds\,
\left(
\epsilon_{ab}
\f{dc^a(\tau)}{d\tau}
\f{d\gamma^b(s)}{ds}
\delta^{(2)}(c(\tau),\gamma(s))
\right)
\vec{u}(s).\f{\vec{e}(\tau)}{|\vec{e}(\tau)|}
\label{lt:poisson1}
\eeq
L'int\'egrande est compos\'e de deux facteurs. Le premier est
simplement la fonction delta de Dirac (avec le Jacobien
du changement de coordonn\'ees $(x^1,x^2)\arr(\tau,s)$)
nous pla\c cant au point d'intersection de $c$ et $\gamma$.
Le second est le produit scalaire (un invariant de jauge)
de l'axe de la rotation/boost $\vec{u}(s)$ avec le vecteur unit\'e
tangent \`a la courbe $c$.

\medskip

Ici, j'ai utilis\'e des quantit\'es invariantes de jauge $SO(2,1)$.
Mais il reste \`a impl\'ementer
l'autre contrainte de jauge $\vec{F}=0$. Je vais les
utiliser pour ``tourner'' la triade $e$ et donc le vecteur $\vec{e}(\tau)$
pour l'aligner sur $\vec{u}(s)$ et simplifier l'expression du
crochet de Poisson \Ref{lt:poisson1}. Plus pr\'ecis\'ement, je vais fixer
la jauge. La condition de fixation de jauge
\footnotemark\footnotetext
{Il est facile de v\'erifier que son crochet de Poisson avec les contraintes
$\vec{F}$ est non-nul et par cons\'equent
que c'est une fixation de jauge l\'egitime.
Plus pr\'ecis\'ement, il est possible d'atteindre la surface contrainte
$\vec{C}=0$ par des transformations de jauge ``topologiques'' g\'en\'er\'ees
par $\vec{F}=0$.}
que je choisis est
\beq
\vec{{\cal C}}=
\left|
\begin{array}{c}
\sh\eta \\
\sin\theta
\end{array}
\right.\vec{u}(P)\wedge\vec{e}(P)=0
\label{jaugehooft}
\eeq
au point d'intersection $P=c(\tau)=\gamma(s)$.
Une propri\'et\'e tr\`es int\'eressante de ce choix est que
$$
\{\vec{{\cal C}},l_c\}=\{\vec{{\cal C}},t_\gamma\}=0,
$$
par cons\'equent le crochet de Dirac de $l$ et $t$ sera identique
\`a leur crochet de Poisson!
De plus,
$\vec{u}(P)$ et $\vec{e}(P)$ sont colin\'eaires et le produit scalaire
dans \Ref{lt:poisson1} est $\pm 1$ suivant l'orientation relative de $c$
et $\gamma$. Ainsi
\beq
\left\{l_c,t_\gamma=
\left|
\begin{array}{c}
\pm\ch\eta \\
\cos\theta
\end{array}
\right.
\right\}_D=
G\left|
\begin{array}{c}
\sh\eta \\
\sin\theta
\end{array}
\right. 
\label{dirachooft}
\eeq
d'o\`u les crochets canoniques (dans la fixation de jauge\footnotemark):
\begin{center}
\framebox{\makebox{
$ \displaystyle
\{l_c,\eta\}_D=\pm 1 \quad{\rm et}\quad
\{l_c,\theta\}_D= 1
$}}
\end{center}

\footnotetext{Il existe une possibilit\'e de fixation de jauge alternative
\`a \Ref{jaugehooft} aboutissant aux m\^emes crochets canoniques. C'est la
contrainte scalaire:
$$
{\cal C}=\left|
\begin{array}{c}
\sh\eta \\
\sin\theta
\end{array}
\right.\vec{u}(P)\cdot\vec{n}(P)
=0
$$
o\`u $\vec{n}=\vec{e_1}\wedge \vec{e_2}=
\epsilon^{ab}\epsilon^i_{jk}e_a^je_b^k$ est le vecteur densit\'e normal.
Cette contrainte commute avec $l_c$ et $t_\gamma$. Cependant elle ne fixe pas
directement le crochet  de Dirac \Ref{dirachooft}. N\'eanmoins, une fois
$\vec{u}(P)$ orthogonale au plan tangent, il
est possible de choisir une courbe $c$
telle que $\vec{e}(P)$ soit colin\'eaire \`a $\vec{u}$ et par cons\'equent
satisfaisant le crochet canonique \Ref{dirachooft}.
}

L'angle $\pm\eta$ ou $\theta$ est conjugu\'e \`a la longueur $l_c$ d\'ependant
du secteur de l'espace des phases o\`u nous sommes. Ainsi lors
de la quantification, comme $\eta$ n'est pas born\'e, le spectre de $l_c$ sera
continu et, comme $\theta\in[0,2\pi]$, le spectre de $l_c$ sera discret!
Ceci confirme les r\'esultats de l'approche \lqg d\'ecrite dans
la section pr\'ec\'edente.
De plus, ce sont les relations de commutation canonique utilis\'ees par 't Hooft
dans \cite{hooft} pour quantifier son mod\`ele de
gravit\'e quantique en $2+1$ dimensions. On a donc bien d\'eriv\'e le m\^eme
cadre th\'eorique.

\medskip

Je ne vais pas d\'ecrire la dynamique de la th\'eorie. Mais en fait,
la fixation de jauge $\vec{{\cal C}}$ ne fixe pas enti\`erement les
transformations de jauge g\'en\'er\'ees par les $\vec{F}$. Il reste la
composante $\vec{F}$ g\'en\'erant des transformations de $\vec{e}(P)$
colin\'eaires \`a $\vec{u}(P)$ et donc \`a $\vec{e}(P)$: ce sont
des variations de la norme de $\vec{e}(P)$. Cette contrainte suppl\'ementaire
va donc s'assimiler \`a la
contrainte Hamiltonienne g\'en\'erant un flot de courbes $c$ (ou $\gamma$)
d\'ecrivant leur \'evolution dans le temps. On remarque qu'elle g\'en\`ere
des variations de la longueur $l_c$ (sans toucher aux holonomies) et qu'elle
s'apparente ainsi aux transformations conformes.
Tout le probl\`eme est alors
de {\it choisir un temps} c'est-\`a-dire une variable conjugu\'ee
\`a cette derni\`ere contrainte. Il s'agit de d\'efinir ce que l'on
veut dire par une tranche \`a temps constant. Sans doute que l'inclusion
de particules dans le formalisme pourrait aider \`a d\'efinir un
temps pr\'ef\'erentiel.

\medskip

Pour retrouver le cadre utilis\'e par 't Hooft, on peut appliquer
les calculs pr\'ec\'edents \`a une triangulation de la surface canonique
et consid\'erer comme variables canoniques les longueurs des liens de la
triangulation. Les variables conjugu\'ees seront des angles/boosts
associ\'es \`a chaque lien d\'ecrivant le changement de rep\`ere d'un c\^ot\'e
\`a l'autre du lien. La dynamique modifiera alors les longueurs (et angles) et, quand
des longueurs s'annulent, modifiera la triangulation elle-m\^eme \cite{hooft}.
On peut \'egalement penser appliquer ce formalisme \`a la \lqg. En effet,
on peut prendre un graphe et consid\'erer comme variables canoniques les angles/boosts
trace des holonomies le long des liens du graphe. Alors les variables conjugu\'ees
seront les longueurs de courbes traversant les liens du graphe. Et la dynamique
agira sur les angles/boosts et longueurs et modifiera s\^urement le graphe initial.
Cela permettrait d'avoir un autre point de vue sur la dynamique de la \lg en $2+1$ dimensions.

Finalement, pour avoir une id\'ee de la dynamique dans ce formalisme,
on pourrait regarder  le cas du tore. Ainsi, choisissons deux courbes/cercles
correspondant aux deux cycles du tore. Alors l'angle du cercle 1 est conjugu\'e
\`a la longueur du cercle 2 et r\'eciproquement. 
Ceci est tr\`es similaire  au formalisme
issu  de la reformulation de la gravit\'e $2+1$ sous forme
de th\'eorie de Chern-Simons, dont
on conna\^\i t tr\`es bien les d\'etails. Il s'agirait
de comparer les deux formalismes pour en tirer
des informations pr\'ecises sur la structure du Hamiltonien dans notre situation.

\section*{Conclusion: G\'eom\'etrie Quantique en $2+1$ dimensions}

J'ai effectu\'e l'analyse canonique de la gravit\'e (Lorentzienne) en $2+1$
dimensions dans un formalisme de \lqg, c'est-\`a-dire utilisant le
couple de variables canoniques conjugu\'ees connexion $SO(2,1)$ et diade
\`a valeur dans $so(2,1)$ (couple de 3-vecteurs formant une base orthonorm\'ee
de la surface canonique). La quantification m\`ene \`a des \'etats quantiques
de la surface canonique d\'ecrits par des r\'eseaux de spin $SO(2,1)$:
ce sont des graphes dont les liens sont labell\'es par des repr\'esentations
unitaires de $SO(2,1)$ et on peut interpr\'eter leur dual en tant  que vari\'et\'e
simpliciale 2d dont les longueurs des c\^ot\'es sont donn\'es par
 le Casimir (sa racine carr\'ee) des repr\'esentations.

On peut alors distinguer
deux types de repr\'esentations: celles \`a Casimir positif, correspondant
\`a des c\^ot\'es de genre espace et celles \`a Casimir n\'egatif correspondant \`a
des c\^ot\'es de genre temps. Dans ce contexte, on obtient que les \'etats de
g\'eom\'etrie de la surface canonique contiennent des c\^ot\'es des deux genres et
qu'ils n'y a pas de crit\`ere alg\'ebrique pour se restreindre de mani\`ere naturelle
\`a des triangulations uniquement de genre espace.
De plus, r\'esultat important, on obtient:
\begin{enumerate}
\item un spectre continu des longueurs de genre espace,
\item un spectre discret des intervalles de genre temps.
\end{enumerate}

On peut retrouver cette m\^eme conclusion en identifiant une variable
conjugu\'ee \`a la longueur d'une courbe $c$ dessin\'ee sur la surface:
la trace de l'holonomie le long d'une courbe $\gamma$ intersectant $c$.
On remarque alors qu'il y a deux secteurs de l'espace des phases suivant
que l'holonomie est conjugu\'ee \`a une rotation ou un boost. Dans le cas d'un
boost, la courbe $c$ est de genre espace et, l'angle param\`etre du boost d\'ecrivant
l'intervalle non-compact $\R_+$, le spectre de la longueur ne sera pas
quantif\'e. Alors que dans le cas d'une rotation, le courbe $c$ est de genre temps
et l'angle param\`etre d\'ecrivant $[0,2\pi]$, la longueur de $c$ sera quantifi\'ee!
Cette analyse canonique permet de d\'eriver le mod\`ele de gravit\'e
quantique discret  en $2+1$ dimensions de 't Hooft, qui lui avait permis
de trouver les m\^emes r\'esultats ``{\bf espace continu \& temps discret}''.

\chapter{Un nouveau formalisme canonique en $3+1$ dimensions}

Je vais maintenant expliquer comment formuler la \lg 
sous une forme invariante sous le groupe de Lorentz $\slc$ et non pas
seulement sous le groupe des rotations 3d $\su$.
L'approche est fond\'ee sur l'analyse canonique r\'ealis\'ee par Alexandrov
\cite{sergei1,sergei2,sergei3}. Comme dans le cas de la th\'eorie
$2+1$ expos\'ee pr\'ec\'edemment, la quantification utilisera des r\'eseaux
de spin sur le groupe de Lorentz. Il en r\'esultera que le spectre
des aires (\'equivalent en $3+1$ dimensions
de la longueur en $2+1$ dimensions) de genre espace sera
en g\'en\'eral {\it continu}.

Plus pr\'ecis\'ement, on trouvera une ambigu\"\i t\'e dans la quantification.
Et il sera d'une part possible de retrouver le cadre de la \lqg usuelle
(d\'ecrite dans la partie I) avec un spectre de l'aire discret.
D'autre part, on d\'erivera une th\'eorie quantique compatible avec
les mod\`eles de mousse de spin Lorentziens d\'ecrits dans la partie IV.
Dans ce second cas, le spectre de l'aire sera bien entendu continu.

La motivation du choix de quantification permet alors un nouveau point de vue
sur la \lqg et la question d'invariance sous Lorentz. En effet, on retrouvera
que la connexion de Lorentz utilis\'ee pour d\'eriver la \lg usuelle
ne se transforme pas normalement sous l'action des
diff\'eomorphismes d'espace-temps: y aurait-il un probl\`eme au niveau
de la contrainte Hamiltonienne en \lqg $\su$?
D'un autre c\^ot\'e, la connexion utilis\'ee pour faire le lien avec
les mod\`eles de mousse de spin est l'unique connexion de Lorentz se
comportant bien sous l'action des diff\'eomorphismes d'espace-temps: 
cela simplifierait-il l'analyse de la dynamique de la th\'eorie?
Tout cela r\'ev\`ele une autre question: est-il possible
de trouver un lien explicite entre la \lqg $\su$ et les mod\`eles de mousse
de spin covariants?

\section{Analyse Canonique de la th\'eorie $3+1$}

Rappellons qu'il s'agit d'analyser et de quantifier
l'action de Palatini g\'en\'eralis\'ee \Ref{palatini}:
\beq
S[\om,e]=\f{1}{2}\int_{\cal M}
\epsilon_{IJKL}e^I\w e^J \w F^{KL}(\om)
-\f{1}{\imm}\int_{\cal M} e^I\w e^J \w F_{IJ}(\om).
\eeq
o\`u les $I,J,\dots$ sont des indices \`a valeur dans l'espace
de Minkovski $M^{(3+1)}$ allant de 0 \`a 3.
Je vais suivre l'analyse canonique d'Alexandrov \cite{sergei1}.
On commence par effectuer un splitting $3+1$ de l'espace ${\cal M}$
en une vari\'et\'e $\R\times \Sigma$  en distinguant une
direction temporelle des dimensions d'espace. Cela revient \`a d\'ecomposer
la t\'etrade $e^I$ (en tant que 1-forme) en:
\beqs
e^0&=&N{\rm d}t+\chi_i E^i_a {\rm d}x^a  \nonumber \\
e^i&=&E^i_aN^a{\rm d}t+E^i_a{\rm d}x^a 
\eeqs
o\`u $i$ est un indice interne allant de 1 \`a 3 (restriction de $I$
aux composantes ``espace'') et $a$ est l'indice d'espace labellant
les coordonn\'ees $x^a$. $N$ et $N^a$ sont comme la {\it lapse} et
le {\it shift}. $\chi^i$ donne la d\'eviation de la normale \`a l'hypersurface
canonique par rapport \`a la direction temps.
On peut donc d\'efinir une (direction) normale (de genre temps)
par le 4-vecteur
$$
\chi=\f{1}{\sqrt{1-|\vec{\chi}|^2}}\left(1,\chi_i\right)
\in\H_+
$$
vivant sur l'hyperbolo\"\i de sup\'erieure dans $M^{(3+1)}$ des vecteurs
unit\'e de genre temps (orient\'es vers le futur).
D\'enotant par $X,Y$ des indices dans l'alg\`ebre de Lie $sl(2,\C)$
(couple anti-sym\'etrique $[IJ]$) allant de 1 \`a 6, on d\'efinit
des nouvelles variables de connexion/triade \`a valeur dans $sl(2,\C)$
(au lieu de $\su$ comme dans la partie I). Commen\c cons par la connexion
$A^X_a$:
\beq
A^X=(\f{1}{2}\om^{0i},\f{1}{2}\epsilon^i_{jk}\om^{jk}).
\eeq
Puis, on peut d\'efinir une triade ``rotation''
\beq
R^a_X=(-\epsilon_i^{jk}E^i_a\chi_k,E^i_a)
\eeq
et une triade ''boost''
\beq
B^a_X=(\star R^a)_X=(E^i_a,\epsilon_i^{jk}E^i_a\chi_k),
\eeq
o\`u $\star$ d\'enote l'op\'erateur de Hodge sur $sl(2,\C)$ \'echangeant
les parties rotation et boost dans l'alg\`ebre.
On peut en fait \'ecrire des projecteurs sur les parties rotation et boost
de $sl(2,\C)$ en prenant les ``carr\'es'' de $R$ et $B$. En effet:
\beq
(P_R)^X_Y=R_a^XR^a_Y=\left(
\begin{array}{cc}
-\frac{\delta_a^b\chi^2-\chi_a\chi^b}{1-\chi^2} &
-\frac{ {\eps_a}^{bc}\chi_c}{1-\chi^2} \\
-\frac{ {\eps_a}^{bc}\chi_c}{1-\chi^2} &
\frac{\delta_a^b-\chi_a\chi^b}{1-\chi^2}
\end{array} \right)
\eeq
est le projecteur de $sl(2,\C)$ sur le sous-espace $su(2)_\chi$ g\'en\'erant
les rotations laissant le vecteur (de genre temps) $\chi$ invariant.
Egalement
\beq
(P_B)^X_Y=B_a^XB^a_Y=\left(
\begin{array}{cc}
\frac{\delta_a^b-\chi_a\chi^b}{1-\chi^2} &
\frac{ {\eps_a}^{bc}\chi_c}{1-\chi^2} \\
\frac{ {\eps_a}^{bc}\chi_c}{1-\chi^2} &
-\frac{\delta_a^b\chi^2-\chi_a\chi^b}{1-\chi^2}
\end{array} \right)
\eeq
d\'efinit le projecteur sur le sous-espace des boosts, suppl\'ementaire
\`a $su(2)_\chi$.
L'action exprim\'ee dans ces nouvelles variables se lit:
\beq
S=\int dtd^3x\,\left(
\left(B^a_X-\f{1}{\imm}R^a_X\right)\dd_tA^X_a
+\Lambda^X\G_X+{\cal N}^a\H_a+{\cal N}\H
\right).
\eeq
Le couple de variables canoniques est donc:
\beq
\left\{A^X_a(x),\left(B^b_Y-\f{1}{\imm}R^b_Y\right)(y)\right\}=
\delta^X_Y\delta^b_a\delta^{(3)}(x,y).
\eeq
Puis on a les contraintes de {\it premi\`ere classe}:
\beqs
\G_X&=&{\cal D}_A \left(B_X-\f{1}{\imm}R_X\right), \nonumber\\
\H_a&=& -\left(B^b_X-\f{1}{\imm}R^b_X\right)F_{ab}^X(A),\nonumber\\
\H&=&\f{1}{1+\frac{1}{\imm^2}}
\left(B-\f{1}{\imm}R\right)\left(B-\f{1}{\imm}R\right)F(A).
\eeqs
$\G$ g\'en\`ere les transformations de jauge $\slc$. $\H_a$ est
une contrainte vectorielle imposant (\`a travers une combinaison
lin\'eaire de $\G$ et de $\H$) l'invariance sous diff\'eomorphisme
spatiaux de $\Sigma$. Et finalement, $\H$ est la contrainte scalaire
ou Hamiltonienne, dictant l'\'evolution de nos variables canoniques.
Le lecteur pourra trouver des expressions explicites dans
\cite{sergei1,sergei2}.
En plus de ces contraintes de premi\`ere classe, s'ajoute des
contraintes de {\it seconde classe}:
\beqs
\phi^{ab}&=&(\star R^a)^XR^b_X=0\\
\psi^{ab}&\approx& RR{\cal D}_AR.
\eeqs
La contrainte $\phi=0$ est la contrainte dite de {\it simplicit\'e}
et est similaire \`a la contrainte \Ref{lqg:2nde}. La contrainte
$\psi=0$ est issue du crochet de Poisson $\{\H,\phi\}$ et revient \`a imposer
que la contrainte $\phi=0$ est pr\'eserv\'ee au cours de l'\'evolution
dans le temps. Il est int\'eressant de noter la ressemblance de $\psi$ avec
la contrainte de r\'ealit\'e \Ref{Ereality} du formalisme self-dual.
Le crochet de Poisson des contraintes $\varphi=(\phi,\psi)$ est de la forme
\cite{sergei1,sergei2}:
$$
\Delta_{rs}=\{\varphi_r,\varphi_s\}=
\mat{0}{D_1}{-D_1}{D_2}
$$
et son inverse est
$$
\Delta^{-1}_{rs}=\mat{D_1^{-1}D_2D_1^{-1}}{-D_1^{-1}}{D_1^{-1}}{0},
$$
ce qui nous permet de calculer le crochet de Dirac:
$$
\{f,g\}_D=\{f,g\}-\{f,\varphi_r\}\Delta^{-1}_{rs}\{\varphi_s,g\}.
$$
L'alg\`ebre des contraintes de premi\`ere classe n'est pas modifi\'ee,
et en d\'efinissant
\begin{eqnarray}
&&{\cal G}(\Lambda)=\int d^3x\, \Lambda^X{\cal G}_X, \qquad
\H(N )=\int d^3x\, N \H,
\nonumber \\
&&{\cal D}(\vec N)=\int d^3x\, N^a(\H_a+A_a^X{\cal G}_X), \label{smcon}
\end{eqnarray}
les contraintes  ont des crochets de Poisson similaires au cas de la
\lg explicit\'es dans la partie I:
\begin{eqnarray}
&&\left\{ {\cal G}(\Lambda_1) ,{\cal G}(\Lambda_2) \right\}_D=
{\cal G}([\Lambda_1, \Lambda_2]),
\nonumber \\
&&\left\{ {\cal D}(\vec N) ,{\cal D}(\vec M) \right\}_D=
-{\cal D}([\vec N ,\vec M ]),\nonumber \\
&&\left\{ {\cal D}(\vec N) ,{\cal G}(\Lambda) \right\}_D=-
{\cal G}( N^a\partial_a\Lambda), \nonumber \\
&&\Bigl\{ \H(N ) ,{\cal G}(\Lambda) \Bigr\}_D =0, \label{algA} \\
&&\Bigl\{ {\cal D}(\vec N) ,\H(N ) \Bigr\}_D=
-\H({\cal L}_{\vec N}N ), \nonumber \\
&&\Bigl\{ \H(N ),\H(M ) \Bigr\}_D =
{\cal D}(\vec K)-{\cal G}(K^bA_b), \nonumber
 \end{eqnarray}
avec
\beqs
[\Lambda_1, \Lambda_2]^X=f^X_{YZ}\Lambda_1^Y\Lambda_2^Z & \quad&
[\vec N ,\vec M ]^a=
N^b\partial_bM^a-M^b\partial_bN^a \nonumber \\
{\cal L}_{\vec N}N =
N^a\partial_a N-N\partial_aN^a& \quad&
K^b=(N\partial_aM-M\partial_aN)R^a_XR^b_Y g^{XY}
\eeqs
o\`u $f^X_{YZ}$ sont les constantes de structure de l'alg\`ebre
$sl(2,\C)$ (les $A\in\{1,2,3\}$ sont des indices de boosts et
les $B\in\{4,5,6\}\sim\{1,2,3\}$ sont des indices de rotations):
$$
\begin{array}{ccc}
f_{A_1 A_2}^{A_3}=0,&
f_{A_1 B_2}^{A_3}=-\eps^{A_1 B_2 A_3},&
f_{B_1 B_2}^{A_3}=0, \\
f_{B_1 B_2}^{B_3}=-\eps^{B_1 B_2 B_3},&
f_{A_1 B_2}^{B_3}=0,&
f_{A_1 A_2}^{B_3}=\eps^{A_1 A_2 B_3}.
\end{array}
$$
Ainsi, comme dans la section \ref{loopgravity}, $\G$ g\'en\`ere les
transformations de jauge $\slc$ (au lieu de $\su$), ${\cal D}$
g\'en\`ere les diff\'eomorphismes spatiaux sur $\Sigma$ et $\H$ induit
les diff\'eomorphismes dans la direction temps
i.e. les reparam\'erisations du temps.

Cependant les relations de commutation entre la
connexion $A$ et la triade $R,B$
changent. Plus pr\'ecis\'ement, la triade commutent toujours avec elle-m\^eme.
Mais la connexion n'est plus conjugu\'ee \`a la triade. De plus,
la connexion ne commute plus avec elle-m\^eme et l'expression de son crochet
de Dirac avec elle-m\^eme est  assez lourde (pour les
formules explicites, le lecteur peut se r\'ef\'erer \`a
\cite{3+1,sergei2,sergei3}). Bien entendu, puisque nous utilisons le crochet
de Dirac, les variables canoniques initiales perdent leur r\^ole
de variables pr\'ef\'er\'ees et nous sommes libres de choisir d'autres
variables mieux adapt\'ees pour d\'ecrire la th\'eorie. Suivant
\cite{sergei3}, nous ne modifions pas la triade mais cherchons
une nouvelle connexion ${\cal A}$
satisfaisant les crit\`eres -naturels et n\'ecessaires pour assurer
une signification physique/g\'eom\'etrique \`a ${\cal A}$- suivants:
\begin{itemize}
\item[$\bullet$]
${\cal A}$ doit \^etre une connexion de Lorentz i.e. se transformer
correctement sous la loi de Gauss $\G$:
\beq
\{\G(\Lambda),{\cal A}^X_a\}_D=
\dd_a\Lambda^X-[\Lambda,A_a]^X
=\dd_a\Lambda^X-
f^X_{YZ}\Lambda^YA^Z_a.
\eeq
\item[$\bullet$]
${\cal A}$ doit \^etre une 1-forme et donc se transformer correctement
sous diff\'eomorphisme spatiaux (sur $\Sigma$):
\beq
\{{\cal D}(\vec{N}),{\cal A}\}_D=
{\cal A}^X_b\dd_aN^b-
N^b\dd_a{\cal A}^X_b.
\eeq
\item[$\bullet$]
On aimerait que ${\cal A}$ soit ``conjugu\'ee'' \`a la triade, cela
permettrait  \'egalement que les op\'erateurs aires
$aire_{\cal S}\sim\int_{\cal S} d^2x\,
\sqrt{n_an_bR^a_XR^{bX}}$ (o\`u $n_a$ est la normale
\`a la surface ${\cal S}$)
soient diagonalisables. Cette condition s'\'ecrit:
\beq
\{{\cal A}^X_a(x),R^b_Y(y)\}_D\propto
\delta_a^b
\delta^{(3)}(x,y).
\eeq
\end{itemize}

On obtient une famille \`a deux param\`etres de possibles
connexion ${\cal A}(\lambda,\mu)$ satisfaisant ces conditions
\cite{sergei3}. Leur expression en fonction de $A$ et de $R$ est:
\beqs
{\cal A}_a^X(\lambda,\mu) &=&
A_a^X+\frac{1}{2}\left(\left(1+\frac{\lambda}{\imm}\right)
- \frac{1}{\imm}(1-\mu) \star\right)
P_R\f{1}{1+\frac{1}{\imm^2}}\left(1-\f{1}{\imm}\star\right)
[B_a,\G]^X
\nonumber \\
&&+
(\lambda + \mu \star)
\left( P_R\star A_a^X + \Theta_a^X(R)\right),
\eeqs
avec
\beq
\Theta_a^X(R)=\Theta_a^X(\chi)=
\left(- \frac{\eps^{ijk}\chi_j\dd_a\chi_k}{1-\chi^2},
\frac{\dd_a\chi^i}{1-\chi^2} \right).
\eeq
Les propri\'et\'es de commutation de ces connexions avec
la(les) triade(s) sont simples:
\beq
\{ {\cal A}^X_a(\lambda,\mu),B_Y^b\}_D=\delta_a^b \left[\left(
(1-\mu)-\lambda\star \right) P_B\right]^X_Y
\eeq
\beq
\{ {\cal A}^X_a(\lambda,\mu),P_B\}_D=
\{ {\cal A}^X_a(\lambda,\mu),\chi\}_D=0.
\eeq
Malgr\'e cela, le crochet de Dirac de ${\cal A}$ avec elle-m\^eme reste
g\'en\'eriquement assez compliqu\'e.

A partir de l\`a, on obtiendrait, en utilisant comme observables
de base les boucles de Wilson/r\'eseaux de spin de la connexion
${\cal A}(\lambda,\mu)$, une famille de quantifications
non-\'equivalentes. 
Ainsi, on peut calculer l'action d'un op\'erateur aire comme dans le
cas de la \lqg (partie I) sur une boucle de Wilson de
${\cal A}(\lambda,\mu)$ dans la repr\'esentation $(n\in\N,\rho\ge0)$
et on trouverait \cite{3+1,sergei3}:
\beq
aire_{\cal S}\sim
l_P^2 \sqrt{(\lambda^2 + (1-\mu)^2) C(su(2)_\chi) -(1-\mu)^2 C_1(sl(2,\C))
+\lambda(1-\mu) C_2(sl(2,\C))}
\eeq
o\`u $C(su(2)_\chi)=\vec{J}\cdot\vec{J}=j(j+1)$ est le
Casimir du groupe $\su_\chi$ (laissant
le vecteur $\chi$ invariant),
$C_1(sl(2,\C))=T^XT_X=\vec{J}^2-\vec{K}^2=n^2-\rho^2$
et $C_2(sl(2,\C))=(\star T)^XT_Y=\vec{J}\cdot\vec{K}=n\rho$ sont les deux 
Casimirs du groupe $\slc$ (on note $T^X$ les g\'en\'erateurs
de $\slc$ et le lecteur peut trouver des d\'etails sur les repr\'esentations
de $\slc$ en appendice). Remarquons que cette formule d'aire est toujours
bien d\'efinie gr\^ace \`a la condition $j\ge n$ sur le
d\'ecomposition des repr\'esentations
unitaires de $\slc$ en repr\'esentations de $\su$.

Il s'agit par cons\'equent de choisir une (ou des)
quantifications ``sp\'eciales'' distingu\'ees par des propri\'et\'es
suppl\'ementaires impos\'ees sur ${\cal A}(\lambda,\mu)$
qui seraient ``naturelles''.
Deux cas particuliers se distinguent:\\ 
\begin{itemize}
\item[$\bullet$]
D'une part, il existe une {\bf unique connexion commutative},
qui correspond au choix $(\lambda,\mu)=(-\imm,1)$ i.e.
$$
\{{\cal A}(-\imm,1),{\cal A}(-\imm,1)\}_D=0.
$$ 
Notons cette connexion $\bA$. Alors
\beq
\{\bA^X_a,R^b_Y\}_D=\imm\delta_a^b(P_R)^X_Y,
\eeq 
c'est-\`a-dire que seule la partie ``rotation'' de la connexion
importe et est conjugu\'ee \`a la triade ``rotation'' $R$.
Le spectre de l'aire r\'esultant 
ne prend en compte que la partie ``rotation'' du Casimir de $\slc$
-la partie $\su_\chi$- et reproduit exactement la formule
de l'aire de la \lqg {\it d\'ependant explicitement du param\`etre
d'Immirzi}:
\beq
{\rm aire}_{\cal S}\sim
l_P^2 \imm\sqrt{C(su(2)_\chi)}.
\label{airebA}
\eeq
De plus,
on peut expliciter les parties rotations et boosts de $\bA$:
\beq
\left|
\begin{array}{ccl}
P_R\bA &=&P_R(1-\imm\star)A-\imm\Theta\\
P_B\bA &=& \star\Theta(\chi)=\star(\chi\wedge\dd\chi)
\end{array}
\right.
\label{contraintesbA}
\eeq
En particulier, la partie boost $P_B\bA$
n'est pas une variable ind\'ependante et est fonction du
champ de normale $\chi$ (donc fonction de la triade).
De plus \'etant donn\'e que $\Theta$ s'annule
quand le champ $\chi$ est constant\footnotemark, 
dans la {\it time gauge} $\chi=\chi_0$,
$\bA$ se r\'eduit simplement \`a la connexion
$\su$ d'Ashtekar-Barbero \`a la base du formalisme
r\'eel de la \lg $\su$.

\footnotetext{Notons que $P_B\Theta=0$.}

Plus pr\'ecis\'ement, $\bA$ est l'unique connexion $\slc$
extension de la connexion $\su$ $P_R(1-\imm\star)A$
(le lecteur peut trouver la d\'emonstration dans les appendices de
\cite{3+1}). On a donc une pure connexion ``rotation''
ou {\it connexion d'espace}.
\item[$\bullet$]
D'autre part, il existe une {\bf unique connexion d'espace-temps},
qui correspond au choix $(\lambda,\mu)=(0,0)$ i.e. qui se transforme bien
sous l'action des diff\'eomorphismes d'espace-temps ($\H$
engendre bien les diff\'eomorphismes dans la direction temps). En effet,
c'est l'unique connexion qui co\"\i ncide avec la connexion originale
$A$ sur la surface des contraintes $\G^X=\H_a=\H=0$.
Notons cette connexion simplement $\cA$.
Le crochet de Dirac avec la triade est:
\beq
\{\cA^X_a,B^b_Y\}_D=\delta_a^b(P_B)^X_Y,
\eeq 
c'est-\`a-dire que seule la partie ``boost'' de la connexion
importe et est conjugu\'ee \`a la triade ``boost'' $B$.
Le spectre de l'aire r\'esultant {\it
ne d\'epend pas du param\`etre d'Immirzi}
et ne prend en compte que la partie {\it boost} du Casimir de $\slc$:
\beq
aire_{\cal S}\sim
l_P^2 \sqrt{C(su(2)_\chi)-C_1(sl(2,\C))}.
\label{airecA}
\eeq
De plus, on peut expliciter la partie rotation boost de $\cA$, qui
est simplement \'egale \`a la spin-connexion fonction de la triade\footnotemark:
\beq
P_R\cA^X_a =\Gamma(R)^X_a=
\frac12 f^{W}_{YZ}(P_R)^{XY} R_a^Z \dd_c R^c_W
+\frac12 f^{ZW}_Y\left(R^T_aR_{Tb}(P_R)^{XY}
+R_b^XR_a^Y -R_a^XR_b^Y \right) R^c_Z \dd_c R^b_W.
\eeq
Cette contrainte  est \`a comparer avec la contrainte de r\'ealit\'e
\Ref{Areality} et \Ref{opAreality} du formalisme self-dual
de la \lg qui impose que la partie r\'eelle de la connexion complexe
est \'egale \`a la spin-connexion.
Dans ce cas, on voit que nous travaillons avec une {\it pure connexion
de boost}.
\end{itemize}

\footnotetext{En fait, pour obtenir une expression de $P_R\cA$ reli\'ee \`a
la connexion de spin, il est n\'ecessaire d'utiliser les contraintes de seconde classe, que
l'on peut ajouter \`a n'importe quelle fonction des variables sans modifier les
crochets de Dirac.}

\medskip

Maintenant, je vais proc\'eder \`a la quantification de ces
deux formalismes, qui se distinguent naturellement du cas g\'en\'eral
$(\lambda,\mu)$ quelconque. Mais avant je tiens \`a remarquer
que la connexion commutative $\bA$ ne se comporte pas normalement
sous les diff\'eomorphismes
d'espace-temps (sous la contrainte Hamiltonienne $\H$), ainsi il risque
d'avoir des probl\`emes d'interpr\'etation g\'eom\'etrique de l'action
de $\H$ au niveau quantique (risque d'anomalie)!
D'un autre c\^ot\'e,
la connexion d'espace-temps $\cA$ n'est
pas commutative (pour le crochet de Dirac) donc il
sera plus compliqu\'e de quantifier la th\'eorie dans ce cas!

Les boucles de Wilson $\slc$ et les r\'eseaux de spin correspondants sont
toujours des observables de la th\'eorie. N\'eanmoins, l'aire d\'epend
de la normale $\chi$ comme on peut le voir sur les formules de ``spectre''
que j'ai donn\'ees ci-dessus. Ainsi si nous voulons quantifier la th\'eorie
et obtenir des \'etats quantiques de g\'eom\'etrie vecteurs
propres des op\'erateurs aire comme dans le cadre usuel de
la \lqg, il faudra consid\'erer comme
fonctions d'onde des fonctionnelles invariantes d\'ependant
\`a la fois de la connexion (de Lorentz),
$\bA$ ou $\cA$, et du champ de normales $\chi$. J'y consacrerai
 la section suivante, avant de proc\'eder explicitement
\`a la quantification des deux th\'eories (classiquement \'equivalentes)
ci-dessus qui nous m\`enera aux r\'esultats principaux de cette partie:
\begin{itemize}
\item[$\bullet$]
d'une part, \`a partir de $\bA$, on re-d\'erivera la \lqg $\su$
et on en donnera une formulation
invariante sous le groupe de Lorentz,
\item[$\bullet$]
d'autre part, en utilisant $\cA$,
on obtiendra le cadre des mod\`eles de mousse de spin (voir partie IV) avec
des r\'eseaux de spin {\it simples} d\'ecrivant la g\'eom\'etrie
de l'hypersurface canonique $\Sigma$\
\`a travers son immersion dans l'espace-temps.
\end{itemize}

\section{Outils: r\'eseaux de spins projet\'es}

On s'int\'eresse donc \`a des fonctions d'une connexion de
Lorentz $A\in sl(2,\C)$ {\bf et} du champ de normales
$\chi\in \slc/\su$ ($\chi$ peut \^etre g\'eom\'etriquement/intuitivement
compris comme la normale \`a
l'hypersurface canonique $\Sigma$).
J'ai introduit le mat\'eriel que je d\'ecris dans
cette section dans l'article \cite{psn}, o\`u le lecteur
pourra trouver des preuves de toutes les affirmations
non-d\'emontr\'ees ici.
Dans ce cadre, une transformation de jauge $\slc$ agit \`a la fois
sur $A$ et sur $\chi$ et
l'invariance de jauge d'une fonction $f(A,\chi)$ se traduit par:
\beq
\forall g\in\slc\,f(A,\chi)=f({}^gA=gAg^{-1}+g\dd g^{-1},g.\chi).
\eeq
A l'aide d'une transformation de jauge, on remarque que l'on peut
toujours tourner les vecteurs $\chi(x)$ jusqu'\`a atteindre
un m\^eme vecteur fix\'e $\chi_0$ (par exemple $(1,0,0,0)$).
Ainsi, une fonction invariante est enti\`erement d\'etermin\'ee par 
sa section \`a $\chi=\chi_0$ constant (sur $\Sigma$)
$f_{\chi_0}(A)=f(A,\chi=\chi_0)$:
\beq
f(A,\chi)=f_{\chi_0}({}^gA) \quad \textrm{pour tout } g
\textrm{ tel que } g.\chi=\chi_0.
\eeq
Remarquons que $f_{\chi_0}(A)$ a une invariance r\'esiduelle
sous $\su_{\chi_0}$. Ainsi, nous sommes en train d'\'etudier des fonctions
d'une connexion de Lorentz, non pas invariantes (d'une mani\`ere effective)
sous $\slc$ mais simplement sous le groupe {\it compact} $\su$!
Cette fixation de jauge est simplement la {\it time gauge} utilis\'ee
en \lg.

Dans ce cadre, je vais introduire des fonctions cylindriques, ne d\'ependant
de la connexion et de $\chi$ qu'\`a travers un nombre fini de variables.
Tout comme les fonctionnelles cylindriques introduites pr\'ecedemment, elles
seront \`a support sur un graphe.
Je munirai les fonctions \`a support sur un graphe donn\'e
d'une mesure et d\'ecrirai l'espace des fonctionnelles $L^2$.
Une base sera donn\'ee par des r\'eseaux de spin ``projet\'es''.
Puis j'\'etudierai l'espace total des fonctions cylindriques en le
d\'ecrivant comme un espace de Fock et en discutant la question de raffiner
le graphe-support. Enfin, j'expliquerai comment \'etendre le
pr\'esent formalisme \`a n'importe quel groupe de Lie non-compact.

\subsection{Fonctions Cylindriques et R\'eseaux de Spin Projet\'es}

Fixons-nous un graphe orient\'e $\Gamma$ dans cette section, avec
$E$ liens et $V$ vertex.
Une fonction cylindrique de $A$ et de $\chi$ d\'ependra
des holonomies $U_1,\dots,U_E$ de $A$ le long des liens de $\Gamma$
et des valeurs $\chi_1,\dots, \chi_V$ de $\chi$ aux vertex du graphe.
Une transformation de jauge agissant sur les holonomies
\`a leurs extr\'emit\'es, l'invariance de jauge d'une fonction cylindrique
s'\'ecrit:
\beq
\label{newsym}
\forall \{k_v\}\in \slc^{\otimes V}, \,
\phi (U_1,U_2,\dots,U_E,\chi_1,\dots,\chi_V)=
\phi(k_{s(1)}U_1k_{t(1)}^{-1},\dots,k_{s(E)}U_Ek_{t(E)}^{-1},
k_1.\chi_1,\dots, k_V.\chi_V).
\eeq
Gr\^ace aux nouvelles variables $\chi_v$, bien que la fonctionelle
$\phi$ soit invariante sous $\slc^V$, la sym\'etrie de jauge
effective est compacte. En effet, \`a $\chi$ fix\'e, $\phi$ n'est invariante
que sous $\times_{v=1}^{V}\su_{\chi_i}$.
Plus pr\'ecis\'ement, nous pouvons ``fixer de jauge'' en tournant
$\chi$ sur $\chi_1,\dots,\chi_V=\chi_0$, comme fait sur une
fonction g\'en\'erique. Cela d\'efinit la fonction
$$
\phi_{\chi_0}(U_1,U_2,\dots,U_E)=
\phi (U_1,U_2,\dots,U_E,\chi_1=\chi_0,\dots,\chi_V=\chi_0)
$$
dont l'invariance de jauge est r\'eduite \`a  $(\su_{\chi_0})^V$.

Physiquement, le plongement de $\Sigma$ dans l'espace-temps ${\cal M}$
est d\'efini par le champ $\chi$. Ici, nous avons retenu qu'un nombre
 fini de valeurs $\chi_1,\dots,\chi_V$ et par cons\'equent, du point de vue
de la fonction cylindrique, le plongement de $\Sigma$ n'est d\'efini
qu'en ce nombre fini de points et est laiss\'e flou partout ailleurs:
$\Sigma$, vu par la fonction cylindrique, n'est d\'efini qu'en un nombre fini
de points... En ces points, la normale $\chi_v$ est fix\'ee et donc la
sym\'etrie r\'eduite de $\slc$ \`a $\su_{\chi_v}$.

\medskip

Puisque la sym\'etrie de jauge effective est compacte,
une mesure naturelle
pour int\'egrer les fonctions cylindriques invariantes est
la mesure de Haar sur $\slc$.
Plus pr\'ecis\'ement, on d\'efinit la mesure:
\beq
\mu(\phi)=\int_{\slc^E}\prod_e\tr{d}g_e
\phi(g_1,\dots,g_E,\chi_1,\chi_2,\dots,\chi_V)=
\int_{\slc^E}\prod_e\tr{d}g_e
\phi_{\chi_0}(g_1,\dots,g_E),
\eeq
et le produit scalaire:
\beqs
\langle \phi | \psi \rangle&=&
\int_{\slc^E} \prod_i\tr{d}g_i
\bar{\phi}(g_1,\dots,g_E,\chi_1,\chi_2,\dots,\chi_V)
\psi(g_1,\dots,g_E,\chi_1,\chi_2,\dots,\chi_V)\nonumber\\
&=&
\int_{\slc^E} \prod_i\tr{d}g_i
\bar{\phi}_{\chi_0}(g_1,\dots,g_E)
\psi_{\chi_0}(g_1,\dots,g_E).
\label{newprod}
\eeqs
Il est facile de v\'erifier que ces quantit\'es sont bien d\'efinies
pour des fonctionnelles invariantes de jauge.
On peut alors consid\'erer l'espace des fonctions $L^2$ pour cette mesure.

Avant d'\'etudier la structure de cet espace, remarquons qu'il
sera toujours possible de regarder des observables invariantes de jauge
et fonction uniquement de la connexion  sur cet espace d'Hilbert. En effet,
pour ${\cal O}(U_1,\dots,U_E)$
invariantes de jauge (fonctions cylindriques usuelles
de la connexion), on peut consid\'erer les \'el\'ements de matrices
\beq
\langle\phi|{\cal O}|\psi\rangle=
\int_{SO(3,1)^E} \prod_i\tr{d}g_i
\bar{\phi}(g_1,\dots,g_E,\chi_1,\dots,\chi_V)
{\cal O}(g_1,\dots,g_E)
\psi(g_1,\dots,g_E,\chi_1,\dots,\chi_V),
\eeq
qui ne d\'ependent pas des $\chi$ et sont bien d\'efinis pour ${\cal O}$
born\'e.

\medskip

Il est facile d'exhiber une base orthonormale de l'espace $L^2$,
que nous noterons $H_\Gamma$. Elle sera donn\'ee par
des r\'eseaux de spin ``projet\'es''. Leur structure est
tr\`es similaire aux r\'eseaux de spin d\'ecrit pr\'ec\'edemment.
En effet, puisque la sym\'etrie de jauge (effective) est r\'eduite \`a
$\su$, je vais attacher des entrelaceurs $\su$ aux vertex au
lieu d'entrelaceurs $\slc$ comme on s'attendrait
pour des r\'eseaux de spin $\slc$. D'o\`u l'appelation ``projet\'e''
puisque la sym\'etrie de
jauge est projet\'ee de $\slc$ \`a $\su_\chi$.

Explicitement, le proc\'edure normale pour construire un r\'eseau de spin
est d'assigner une repr\'esentation ${\cal I}_i$ de $SL(2,\C)$ \`a
chaque lien  $e_i$ du graphe $\Gamma$, puis de choisir
un $\slc$-entrelaceur $I_v$ pour chaque vertex $v$ du graphe et
de construire la fonctionnelle r\'eseau de spin en contractant les
holonomies $U_i[A]$ (le long des liens $e_i$)
dans la repr\'esentation ${\cal I}_i$ avec les
entrelaceurs pour obtenir un scalaire:
\beq
\phi (A)= \bigotimes_v I_v \bigotimes_i D^{{\cal I}_i} (U_i).
\eeq
Dans notre cas, on choisit en plus une repr\'esentation $j_i^{(v)}$
de $\su$ pour chaque lien $e_i$ du graphe -plus pr\'ecis\'ement
deux par lien, un pour chaque vertex aux extr\'emit\'es-
et un $SU(2)$-entrelaceur $i_v$ entrela\c cant les repr\'esentations
autour de chaque vertex $v$.
Pour coller les holonomies en utilisant ces entrelaceurs,
on va les projeter au niveau des vertex.

Pour \^etre plus explicite, notons  $R^{\cal I}$ l'espace
d'Hilbert de la repr\'esentation ${\cal I}$ de $SL(2,\C)$
et $V^j$ l'espace d'Hilbert de la repr\'esentation $j$ de $SU(2)$.
Les repr\'esentations unitaires irr\'eductibles de $\slc$
sont de dimension infinie et labell\'ee par un couple
$(n\in\N,\rho\ge0)$ (voir en appendice pour plus de d\'etails).
Alors on peut d\'ecomposer $R^{(n,\rho)}$ en repr\'esentations de $SU(2)$.
Pour cela, choisissons un sous-groupe $\su$ particulier de $\slc$
 i.e. choisissons une normale (temps)
$x \in SL(2,C)/SU(2)$ et regardons le sous-groupe $SU(2)_x$ des
transformations laissant le vecteur $x$ invariant. Alors:
\beq
R^{(n,\rho)}=\bigoplus_{j\ge n} V^j_{(x)}.
\eeq
Appelons $P_{(x)}^j$ le projecteur orthogonal de
$R^{(n,\rho)}$ sur $V^j_{(x)}$.
Il s'exprime
\beq
P_{(x)}^j=\Delta_j\int_{SU(2)_x}\tr{d}g\,\overline{\zeta}^j(g)D^{(n,\rho)}(g),
\eeq
o\`u $\Delta_j=(2j+1)$ est la dimension de  repr\'esentation $j$,
l'int\'egration sur le sous-groupe $SU(2)_x$,
$D^{(n,\rho)}(g)$ la matrice de $g$ dans la repr\'esentation $(n,\rho)$ 
et $\zeta^j$ le caract\`ere
de la repr\'esentation $j$.

Pour construire le r\'eseau de spin projet\'e, on ins\`ere ce projecteur
aux deux extr\'emit\'es de chacun des liens
et la fonctionnelle a la structure suivante autour d'un certain
noeud $v$ (trivalent pour simplifier les notations):
\beq
\phi(U_1,U_2,U_3,\dots, \chi_v,\dots)=
i_v^{j_1j_2j_3}\prod_{i=1}^3|{\cal I}_i\chi_vj_i^{(v)}m_i\rangle
\langle{\cal I}_i\chi_vj_i^{(v)}m_i|D^{{\cal I}_i}(U_i) \dots\tr{autres vertex},
\eeq
o\`u $|{\cal I} xjm\rangle$ est la base de $V^j_{(x)}\harr R^{\cal I}$,
avec $m$ allant de $-j$ \`a $j$, et o\`u on somme sur les vecteurs
 $m_i$. En quelques mots, on trace sur $V^j_{(\chi)}$ au lieu
de  $R^{\cal I}$ au niveau des vertex.

\begin{figure}[t]
\begin{center}
\psfrag{u1}{$U_1$}
\psfrag{u2}{$U_2$}
\psfrag{u3}{$U_3$}
\psfrag{c1}{$\chi_{A}$}
\psfrag{c2}{$\chi_{B}$}
\psfrag{c3}{$\chi_{C}$}
\psfrag{i1}{$i_{A}$}
\psfrag{i2}{$i_{B}$}
\psfrag{i3}{$i_{C}$}
\psfrag{r1}{$({\cal I}_1,j_1)$}
\psfrag{r2}{$({\cal I}_2,j_2)$}
\psfrag{r3}{$({\cal I}_3,j_3)$}
\includegraphics[width=8cm]{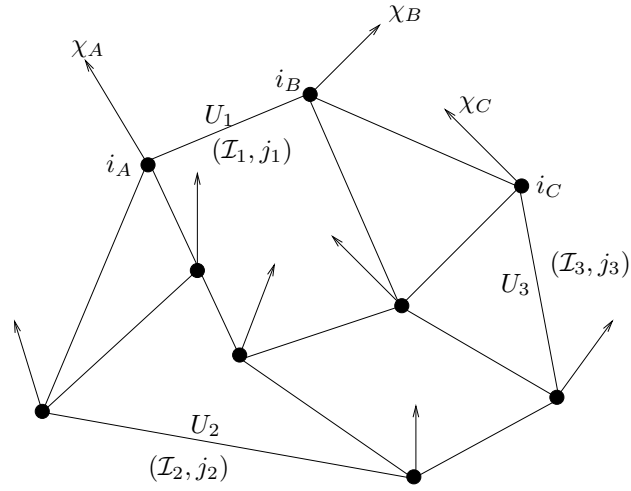}
\end{center}
\caption{Un r\'eseau de spin projet\'e, fonctionnelle
des holonomies $U_i$ et des normales $\chi$:
les liens sont labell\'es
par les couples de repr\'esentations $({\cal I},j)$
de $\slc$ et $\su$ et les vertex par des entrelaceurs $\su$.}
\end{figure}

Il est facile de v\'erifier que ces nouveaux r\'eseaux de spin 
satisfont l'invariance de jauge \Ref{newsym}.
Et un calcul simple permet de montrer qu'une fois choisie
une base orthonormale d'entrelaceurs de $\su$,
les r\'eseaux de spin projet\'es r\'esultants forment une base
de l'espace $H_\Gamma$
des fonctionnelles cylindriques $L^2$ de $A$ et $\chi$ \`a support sur
le graphe $\Gamma$ i.e. le produit scalaire de deux r\'eseaux de spin
est une fonction delta en les repr\'esentations ${\cal I}$ et $j$ et
en les $\su$-entrelaceurs.

Insistons sur le fait qu'a priori il y a deux repr\'esentations
de $\su$ diff\'erentes pour chaque lien, li\'ees \`a
ses deux extr\'emit\'es. Quand on s'occupera de raffiner les
r\'eseaux de spin projet\'es, on sera r\'eduit naturellement
\`a consid\'erer que les cas o\`u les deux repr\'esentations
co\"\i ncident et donc que chaque lien n'est labell\'e que par une
seule repr\'esentation de $\su$.

\medskip

Il est int\'eressant de remarquer que les r\'eseaux de spin projet\'es
sont intimement li\'es aux r\'eseaux de spin usuels $\su$.
En effet, consid\'erons la restriction d'un r\'eseau de spin projet\'e, dans la fixation
de jauge $\forall v,\, \chi_v=\chi_0$, au sous-groupe $SU(2)^E$ (au lieu de $\slc^E$).
Alors l'invariance de jauge r\'esiduelle signifie que cette restriction est justement
une fonctionnelle cylindrique sur $\su$ invariante sous des transformations $\su$.
Et on obtient en fait un r\'eseau de spin $\su$, avec des liens labell\'es
par les repr\'esentations de $\su$ donn\'ees par le r\'eseau de spin projet\'e
(en fait, il faut que les deux repr\'esentations $\su$ associ\'ees \`a chaque lien 
co\"\i ncident sinon la fonctionnelle restreinte est simplement nulle).
Ce r\'eseau de spin $\su$ ne d\'epend plus des repr\'esentations 
${\cal I}_e=(n_e,\rho_e)$ de $\slc$ attach\'ees aux liens,
mais seulement des repr\'esentations $j_e$ de $\su$.
De ce point de vue, les $j_e$ suffisent \`a d\'ecrire la g\'eom\'etrie intrins\`eque 3d
de l'hypersurface. Et les $(n_e,\rho_e)$ d\'ecrivent
la mani\`ere dont l'hypersurface change sous des transformations
de Lorentz infinit\'esimal i.e.
le plongement de l'hypersurface dans l'espace-temps environnant.
En fait, dans les chapitres suivants, nous verrons que d\^ues aux conditions impos\'ees
sur $\bA$ et sur $\cA$, nous obtiendrons en fait deux formulations compl\'ementaires:
avec $\bA$, les \'etats d\'ecriront la g\'eom\'etrie
intrins\`eque de $\Sigma$ tandis qu'avec
$\cA$, les \'etats d\'ecriront le plongement de $\Sigma$ dans l'espace-temps.

\subsection{Raffiner les fonctions cylindriques}

Tout comme pour l'\'etude des r\'eseaux de spin expos\'ee dans la partie II,
dans l'\'etude des r\'eseaux de spin projet\'es, j'ai commenc\'e
par la construction des espaces
d'Hilbert $H_\Gamma$ des fonctions cylindriques $L^2$ \`a support
sur un graphe donn\'e $\Gamma$.
Maintenant, je vais m'int\'eresser \`a la structure de l'espace total
des fonctionnelles cylindriques c'est-\`a-dire \`a la somme
sur les graphes de tous les $H_\Gamma$.

\medskip

Tout d'abord, nous utilisons simplement la mesure de
Haar sur $\slc$ et l'espace $H_\Gamma$ est
l'espace des fonctions $L^2$ sur $\slc^E$ invariantes sous le groupe compact $\su^V$.
Ainsi, comme le groupe d'invariance est compact, il
suffit de prendre des fonctions quelconques sur $\slc^E$
et d'impl\'ementer l'invariance sous $\su$ en int\'egrant partiellement les fonctions.
Il est alors tr\`es facile de munir l'espace total
$H=\oplus_\Gamma H_\Gamma$ d'une structure
d'{\it espace de Fock} en \'ecrivant des op\'erateurs de cr\'eation et d'annihilation.

Par exemple, consid\'erons deux graphes $\Gamma_1\subset \Gamma$ et notons
$\Gamma'=\Gamma\setminus\Gamma_1$.
Choisissons maintenant une fonction $\varphi \in H_{\Gamma_1}$.
On d\'efinit l'op\'erateur
d'annihilation agissant sur $H_\Gamma$
\beq
\forall f\in H_\Gamma\, (a_\varphi f)(g_{f\in\Gamma'},x_{w\in\Gamma'})=
\int \prod_{e\in \Gamma_1}\tr{d}g_e
\bar{\varphi}(g_{e\in\Gamma_1},x_{v\in \Gamma_1})
f(g_{e\in\Gamma_1},g_{f\notin\Gamma_1},x_{v\in\Gamma})
\eeq
o\`u on prend les m\^emes variables $x_{v\in \Gamma_1}$ pour $\varphi$ et $f$.
$a_\varphi$ envoie bien $H_\Gamma$ sur $H_{\Gamma'}$.

De m\^eme, on peut rajouter des liens \`a un graphe
$\Gamma$ en multipliant les fonctions de
$H_\Gamma$ par des fonctions cylindriques $L^2$ \`a support
seulement sur les liens \`a rajouter.
On pourrait \'egalement coller deux graphes avec des
liens en commun si les fonctions cylindriques
d\'efinies sur les deux graphes sont \`a support compact.

\medskip

Un autre point int\'eressant est la possibilit\'e de
{\it raffiner} les fonctionelles cylindriques
projet\'ees. En effet, physiquement, la fonctionelle cylindrique
ne conna\^\i t qu'une nombre fini
de points de $\Sigma$: les vertex du graphe, o\`u on conna\^\i t explicitement la normale
\`a l'hypersurface. On peut donc se poser la question de donner
plus d'informations \`a la fonctionnelle
cylindrique pour qu'elle en connaisse plus sur la g\'eom\'etrie de $\Sigma$. En fait,
lors de la quantification, les fonctionnelles cylindriques deviendront les fonctions d'onde
de la g\'eom\'etrie de $\Sigma$ et l'\'etat quantique de
$\Sigma$ ne sera d\'efini qu'\`a travers
l'information lue \`a partir de la fonctionnelle cylindrique.

Rajouter l'information de la normale \`a un point $x$
de $\Sigma$ revient \`a projeter la fonctionnelle
\`a ce point de $\slc$ sur $\su_{\chi(x)}$. On s'int\'eresse
donc aux vertex bivalents: pour raffiner
une fonctionnelle cylindrique, on rajoute des vertex bivalents
le long des liens. Dans la limite,
on aura projet\'e la fonctionnelle en tout point. Pour
d\'efinir une telle limite, on aimerait
d\'efinir des suites de fonctionnelles cylindriques coh\'erentes \`a support
sur des graphes de plus en plus fins.

Commen\c cons donc par d\'efinir la notion de ``coh\'erent'',
c'est-\`a-dire comment rajouter/enlever
un vertex bivalent \`a une fonctionnelle cylindrique. Pour cela,
concentrons-nous tout d'abord
sur la relation entre l'espace des fonctions ne d\'ependant que
d'un lien et l'espace de celles o\`u
on rajoute un vertex au milieu du lien:
\beq
\begin{array}{cccc}
f(g,x,y)&=&&f(kgh^{-1},kx,hy) \\
&  \updownarrow &? & \\
\phi(g_1,g_2,x,y,z) &= &&
\phi(ag_1c^{-1},cg_2b^{-1},ax,by,cz).
\end{array}
\eeq
L'op\'eration la plus naturelle entre ces deux ensembles 
est de commencer avec une fonction $f$ et de d\'efinir une fonction $\phi$
en contractant les deux \'el\'ements du groupe:
\beq
f \rightarrow \phi(g_1,g_2,x,y,z)=
f(g=g_1g_2,x,y).
\label{contract}
\eeq
On peut alors d\'efinir une op\'eration inverse pour aller de
$\phi$ \`a $f$ en int\'egrant sur la
valeur $z$ de la normale au point suppl\'ementaire. Plus pr\'ecis\'ement,
$\int dz \,\phi(g_1,g_2,x,y,z)$
ne d\'epend de $g_1$ et $g_2$ qu'\`a travers le produit $g=g_1g_2$.
Ainsi, on peut exprimer la m\^eme int\'egrale par un produit de convolution:
\beq
\phi \rightarrow
f(g,x,y)=\int_{SL(2,C)} \tr{d}\tl{g}\,
\phi(\tl{g},\tl{g}^{-1}g,x,y,z)
=\int_{{\cal H}_+} \tr{d}z \,
\phi(g_1,g_2,x,y,z).
\label{inverse}
\eeq

N\'eanmoins, cela n'est pas vraiment ce que nous recherchons, car la fonction $\phi$
d\'efinie \`a partir de $f$ ne d\'epend pas de la
variable normale $z$. Ce que nous voulons
c'est rajouter l'information $z$ donc nous cherchons
une nouvelle fonction $\phi$ d\'ependant de $z$.

Pour avoir une id\'ee plus pr\'ecise de la proc\'edure, il est plus simple de regarder
les r\'eseaux de spin projet\'es. Concentrons-nous de nouveau sur un lien particulier
du r\'eseau.
Trois variables vivent sur ce lien: un \'el\'ement du groupe $g$ et deux normales $x$ et $y$
aux deux vertex extr\'emit\'es.
Le r\'eseau de spin d\'epend alors d'une repr\'esentation ${\cal I}$ de $SL(2,\C)$
et de deux repr\'esentations $j_1,j_2$ de $\su$ et la
fonctionnelle correspondante s'\'ecrit:
\beq
f(g,x,y,\dots)=\langle{\cal I} x j_1 m_1|
D^{\cal I}(g)|{\cal I} y j_2 m_2\rangle \dots
\eeq
On peut {\it cr\'eer un vertex bivalent} en ins\'erant au milieu de ce lien l'identit\'e
\beq
\forall j,\,\tr{Id}_{R^{\cal I}}=\f{1}{\Delta_j}\int_{{\cal H}_+}
\tr{d}z P^j_{(z)}.
\eeq
Cela n'est possible que si $j_1=j_2=j$. En effet, on veut imposer une invariance par $\su$ au niveau
du nouveau vertex et, pour cela, il fautque cela soit la m\^eme repr\'esentation de $\su$ de
chaque c\^ot\'e du vertex bivalent. Ainsi, nous sommes naturellement r\'eduit \`a
une seule repr\'esentation $j$ de $\su$ pour chaque lien du graphe!
Maintenant, on consid\`ere le nouvelle fonctionnelle
\beq
\phi(g_1,g_2,x,y,z,\dots)=
\f{1}{\Delta_j}
\langle{\cal I} x j_1 m_1|D^{\cal I}(g_1)
P^j_{(z)}D^{\cal I}(g_2)|{\cal I} y j_2 m_2\rangle
\dots
\eeq
o\`u on a ins\'er\'e le projecteur $P^j_{(z)}$ sur la repr\'esentation $j$ de $\su_z$.
Si nous appliquons l'application \Ref{inverse} et que nou
 int\'egrons $\phi$ sur $z$, on retrouve la fonction $f$ de d\'epart.

Le ``probl\`eme'' avec cette proc\'edure est que cette {\it projection} n'est pas compatible
avec le produit scalaire, c'est-\`a-dire que les normes $L^2$ de $f$ et de $\phi$ ne
sont pas \'egales. En fait, pour une fonction $\phi$ $L^2$ arbitraire,
$f$ n'est pas forc\'ement $L^2$... N\'eanmoins, si $\phi$
\'etait (grosso modo) $L^4$, alors
$f$ est $L^2$. On est ainsi amen\'e \`a travailler avec des fonctions dans
$L^2\cap L^4 \cap L^6 \cap \dots$, par exemple des fonctions \`a support compact
(en fait \`a support compact dans l'espace de Fourier, o\`u on exprime les fonctionnelles
comme fonctions des repr\'esentations ${\cal I}=(n\in\N,\rho\in\R)$ de $\slc$
au lieu de
fonctions d'\'el\'ements du groupe). Il existe alors
des {\it suites coh\'erentes} de fonctionnelles
cylindriques de plus en plus raffin\'ees! Malheureusement,
je ne sais pas si l'ensemble de ces suites
peut \^etre muni d'une structure d'espace d'Hilbert. Le lecteur pourra cependant trouver
une discussion plus d\'etaill\'ee de cette question dans \cite{psn}.

\subsection{G\'en\'eralisation \`a un groupe quelconque}

Commen\c cons par rappeller quelques faits de th\'eorie des groupes et introduisons
le sous-groupe compact maximal d'un groupe donn\'e $G$ et rappelons ses propri\'et\'es.
 On consid\`ere donc un groupe connexe lin\'eaire r\'eductif $G$
i.e. un groupe connexe ferm\'e de matrices r\'eelles ou complexes stable sous transconjugaison
On note ${\cal G}$ son alg\`ebre de Lie.
On d\'efinit l'automorphisme $\Theta$ de $G$  consistant \`a prendre 
l'inverse du transconjugu\'e. On a $\Theta^2=1$ et on appelle $\Theta$
{\it l'involution de Cartan}.
Alors
\beq
K=\{g\in G | \Theta g=g\}
\eeq
est connexe et compact et est un sous-groupe compact maximal de $G$.
Par exemple, c'est  $SO(3)$ dans le cas de $G=SO(3,1)$ et
c'est $\su$ dans le cas de $G=SL(2,\C)$.
La diff\'erentielle $\theta$ de $\Theta$ \`a l'identit\'e $1$ est un
automorphisme de ${\cal G}$, donn\'ee par l'oppos\'e de la transconjugaison.
Puisque $\theta^2=1$, on peut d\'efinir la
{\it d\'ecomposition de Cartan} de ${\cal G}$
\beq
{\cal G}=p\oplus m\, ,
\eeq
o\`u $p$ and $m$ sont les espaces propres de correspondant aux valeurs propres
1 et -1 de $\theta$. $p$ consiste des \'el\'ements anti-Hermitiens de ${\cal G}$ et
$m$ des \'el\'ements Hermitiens.
Et on obtient facilement que:
\beq
\begin{array}{ccc}
\left[p,p\right] & \subset & p \\ 
\left[p,m\right] & \subset & m \\ 
\left[m,m\right] & \subset & p
\end{array}
\eeq
La d\'ecomposition de Cartan de $G$ est donn\'ee par l'application:
\beq
\begin{array}{ccc}
K\times m &\rightarrow & G \\
(k,u) & \rightarrow & k \,\exp(u)
\end{array}.
\eeq
C'est un diff\'eomorphisme, qui identifie le quotient
$X=G/K$ du groupe $G$ par l'action \`a gauche de  $K$ \`a l'espace $m$.

\medskip

Une {\it  fonction cylindrique projet\'ee} est d\'efinie \`a support sur un graphe
 $\Gamma$ -avec $E$ liens orient\'es et  $V$ vertex-
et d\'epend d'un \'el\'ement du groupe pour chaque lien et d'une variable
 $x\in G/K$ pour chaque vertex. De plus, on demande l'invariance de jauge sous $G$:
\beq
\forall k_i\in G, \,
\phi (g_1,g_2,\dots,g_E,x_1,\dots,x_V)=
\phi(k_{s(1)}g_1k_{t(1)}^{-1},\dots,k_{s(E)}g_Ek_{t(E)}^{-1},
k_1.x_1,\dots, k_V.x_V).
\eeq
Une telle fonction peut \^etre construite \`a partir d'une fonction quelconque
sur $G^{\otimes E}$ en int\'egrant (\`a gauche) sur la partie $K$
des $E$ \'el\'ements du groupe.
Une fixation de jauge correspond \`a fixer toutes les variables $x$
\`a des valeurs arbitraires et le cas particulier
$\chi=\chi_0$ revient \`a fixer tous les $x$ \`a la classe d'\'equivalence
 $[\textrm{Id}]$ de l'identit\'e
i.e. le  sous-groupe $K$ lui-m\^eme. 
La mesure que j'introduis est le mesure de Haar $dg^{\otimes E}$.
L'espace d'Hilbert r\'esultant est l'espace des fonctions invariantes $L^2$.
Ainsi, nous avons ``\'evit\'e'' les probl\`emes auxquels nous sommes
confront\'es dans l'\'etude des espaces des orbites pour les groupes non-compacts
de la partie II.

\medskip

On g\'en\'eralise la construction des r\'eseaux de spin projet\'es.
Pour cela, regardons comment une repr\'esentation de $G$
se d\'ecompose en repr\'esentations de $K$.
Soit donc une repr\'esentation $\pi$ de $G$ d\'efinie sur un espace d'Hilbert $V$.
 Quand $K$ agit par des op\'erateurs unitaires, alors
on peut d\'ecomposer $\pi$ en repr\'esentations irr\'eductibles de $K$
(voir chapitre 8 de \cite{knapp}):
\beq
\pi \, = \, \sum_{\tau\in\hat{K}} n_\tau \tau
\eeq
o\`u  $\hat{K}$ est l'ensemble des classes d'\'equivalence (unitaire) de
repr\'esentations irr\'eductibles de $K$
et $n_\tau\in\N\cup\{+\infty\}$ est la
multiplicit\'e de chaque repr\'esentation.
De plus, si $\pi$ est une repr\'esentation irr\'eductible unitaire, alors
toutes les multiplicit\'ees $n_\tau$ sont finies et satisfont
$n_\tau \le \textrm{dim}\tau$.

On peut \'egalement introduire la notion de vecteurs $K$-finis qui sont les vecteurs
 $v$ tels que $\pi(K)v$ g\'en\`ere un espace de dimension finie.
Pour des r\'epr\'esentations unitaires (ou plus g\'en\'eralement admissibles),
tous les vecteurs $K$-finis sont des vecteurs  $C^\infty$, l'espace
des vecteurs $K$-finis est stable sous
$\pi({\cal G})$ et tous les \'el\'ements de matrices
$g\rightarrow (\pi(g)u,v)$, avec $u$ $K$-finis,
sont des fonctions analytiques sur $G$.

On peut alors introduire les r\'eseaux de spin projet\'es.
On choisit $E$ \'el\'ements du groupe $g_1,\dots, g_E$ sur les liens du graphe
et  $V$ \'el\'ements $x_1,\dots,x_V$ sur le quotient
$G/K$. On choisit aussi $E$ repr\'esentations ${\cal I}$
(unitaires irr\'eductibles)
de $G$, et 2$E$ repr\'esentations de $K$. On suit exactement
la m\^eme construction que dans le cas $\slc$ choisissant
 $V$ $K$-entrelaceurs et les contractant avec les \'el\'ements de matrice
de $\pi^{{\cal I}_i}(g_i)$ \`a l'aide d'une base
de l'espace d'Hilbert de la repr\'esentation du sous-groupe $K$ (en fait du sous-groupe
$K_{(x)}$ stabilisateur de $x$).
Finalement, gr\^ace \`a la $K$-finitude des vecteurs utilis\'es dans la construction, la fonctionnelle
r\'esultante est parfaitement bien d\'efinie.

\section{Quantification: retrouver la \lg $\su$}

On aimerait quantifier canoniquement la relativit\'e g\'en\'erale en utilisant
les boucles de Wilson de la connexion de Lorentz $\bA$,
qui semble n'\^etre qu'une simple extension \`a
$\slc$ de la connexion $\su$ d'Ashtekar-Barbero.
Cela sera confirm\'e au niveau de la th\'eorie quantique.
En effet, je vais montrer que l'on retrouve
exactement comme espace des \'etats quantiques
le m\^eme espace d'Hilbert des r\'eseaux de spin $\su$
qu'en \lqg. En fait, on aura une extension \`a $\slc$ de
ces structures i.e. on pourra agir sur
les r\'eseaux de spin $\su$ avec des boosts et donc
explorer la th\'eorie en dehors de la fixation
de jauge $\chi(x)=\chi_0=(1,0,0,0)$ ({\it time gauge}). Egalement, on retrouvera le
m\^eme spectre de l'aire discret et d\'ependant du param\`etre d'Immirzi.
On aura ainsi compl\`etement d\'eriv\'e
le formalisme usuel de la \lqg \`a partir du nouveau
formalisme canonique covariant.  Cela permet d'avoir
un nouveau point de vue ``covariant'' 
sur la \lqg, et dans ce cadre, il est utile de se rappeller
que la connexion $\bA$  ne se
transforme pas normalement sous les diff\'eomorphismes
d'espace-temps et que cela peut induire
un probl\`eme d'interpr\'etation g\'eom\'etrique (dans l'espace-temps) de la
contrainte Hamiltonienne $\H$.

Ce probl\`eme semble \^etre reli\'e  \`a la
remarque de Samuel \cite{samuel} comme quoi
les boucles de Wilson de la connexion d'Ashtekar-Barbero n'ont pas d'interpr\'etation
dans l'espace-temps. Ainsi, modifier le plongement de l'hypersurface dans l'espace-temps
change la valeur de cette boucle de Wilson.  N\'eanmoins, cette difficult\'e dispara\^\i t
compl\`etement dans le nouveau formalisme o\`u nous tenons compte de plongement de
l'hypersurface dans l'espace-temps gr\^ace au champ
de normales $\chi$. Ainsi la connexion $\bA$
d\'epend explicitement du plongement de $\Sigma$ dans ${\cal M}$ \`a cause de la contrainte
\Ref{contraintesbA} sur $P_R\bA$. Par contre, les boucles de Wilson de $\bA$ sont des
objets (d'espace-temps) bien d\'efinis invariants de jauge
$\slc$ donc invariants sous changement
de $\chi$. Le probl\`eme r\'eel se situe alors au niveau de la dynamique de $\bA$ dans
l'espace-temps i.e. au niveau de la contrainte Hamiltonienne $\H$. Reste que la th\'eorie
est toujours invariante sous diff\'eomorphismes
de l'espace-temps car l'alg\`ebre des contraintes
n'est pas modifi\'ee par le choix de la connexion! Par contre, choisir une connexion
particuli\`ere revient \`a choisir une repr\'esentation
particuli\`ere des diff\'eomorphismes.
Peut-\^etre que certaines repr\'esentations posent
des probl\`emes lors de la quantification.
Egalement on pr\'ef\`ere g\'en\'eralement la repr\'esentation
naturelle des diff\'eomorphismes
(de l'espace-temps), ce qui permet de conserver
l'interpr\'etation g\'eom\'etrique des quantit\'es.
Ainsi, il est naturel de pr\'ef\'erer la th\'eorie quantique bas\'ee sur la connexion
d'espace-temps $\cA$, que je d\'ecrirai dans la prochaine section, \`a la pr\'esente
quantification reproduisant la \lqg usuelle.

\medskip

Consid\'erons une boucle de Wilson de la connexion de Lorentz $\bA$
\beq
U_{\alpha}[\bA]={\cal P}\exp\left(\int_\alpha dx^i \bA_i^X T_X\right),
\eeq
o\`u $\alpha$ est une boucle orient\'ee (et $T_X$ les g\'en\'erateurs de $\slc$).
C'est l'objet invariant de jauge de base.
On regarde habituellement les repr\'esentations
principales unitaires irr\'eductibles $R^{(n,\rho)}$
de $\slc$, car ce sont celles intervenant dans la formule de Plancherel
(voir en appendice pour plus de d\'etails).
Mais comme remarqu\'e pr\'ec\'edemment, ces observables ne
nous suffisent pas dans notre cas car ces fonctionnelles
ne sont pas des vecteurs propres de l'op\'erateur aire
et par cons\'equent n'ont pas d'interpr\'etation
physique/g\'eom\'etrique simple en tant que
fonctions d'onde de la g\'eom\'etrie.
En effet, consid\'erons une petite surface intersectant
la boucle en un point $x$, pour calculer l'op\'erateur
aire, il faut d\'ecomposer la repr\'esentation $R^{(n,\rho)}$
de $\slc$ en les repr\'esentations $V^j_{\chi(x)}$ du sous-groupe
$\su_{\chi(x)}$ (laissant le vecteur $\chi(x)$ invariant).
D'apr\`es la formule \Ref{airebA} donnant l'action de l'op\'erateur aire,
chaque sous-espace $V^j$ donnera une contribution $\imm\sqrt{j(j+1)}$
et l'op\'erateur aire total ne sera pas une simple multiplication:
l'op\'erateur aire n'est pas diagonal sur les boucles de Wilson
(ou r\'eseaux de spin $\slc$).

Pour obtenir des vecteurs propres, nous avons besoin de pouvoir s\'electionner
un sous-espace particulier $V^j_{\chi(x)}$.
Puisque ce sous-espace d\'epend de $\chi$, nous devons consid\'erer
des fonctionnelles invariantes d\'ependant \`a la fois de $\bA$ et du champ $\chi$.
Cela justifie l'\'etude de la section pr\'ec\'edente. En effet, les r\'eseaux
de spin projet\'es construits projetent bien la repr\'esentation $R^{(n,\rho)}$
de $\slc$ sur des repr\'esentations $V^j_{\chi(x)}$ de $\su$:
un r\'eseau de spin donn\'e est un vecteur propre
des op\'erateur aires d'une surface intersectant son graphe-support \`a un/des vertex.
Si l'on veut consid\'erer une surface n'intersectant le
graphe-support \`a un vertex, il faut raffiner le r\'eseau de spin en introduisant
un vertex bivalent sur le graphe au niveau du point d'intersection.
Dans la limite d'une fonctionnelle infiniement raffin\'ee, on aurait un vecteur propre
de tous les op\'erateurs aire.

Dans ce cadre, on a des vecteurs propres de l'aire et
on peut l\'egitimement \'ecrire le spectre des aires:
\beq
aire_{\cal S}\sim\imm\sqrt{C(\su_\chi)}\sim\imm\sqrt{j(j+1)},
\eeq
qui reproduit bien le spectre de l'aire de la \lqg $\su$!

Les repr\'esentations $(n,\rho)$ de $\slc$ n'interviennent pas dans cette formule.
On aimerait dire qu'elles d\'ecrivent simplement la mani\`ere dont
les r\'eseaux de spin projet\'es se comportent sous des boosts de Lorentz
et donc d\'ecrivent le plongement de l'hypersurface $\Sigma$ dans l'espace-temps.

\medskip

A ce niveau, on aurait tendance \`a penser que le travail est fini.
Cependant, il faut d\'ecrire la structure d'espace d'Hilbert de l'espace
des \'etats (ici: les r\'eseaux de spin projet\'es). On aimerait prendre
simplement la mesure de Haar naturelle comme utilis\'ee dans la section
pr\'ec\'edente. Mais, en fait, il faut tenir compte des contraintes
de seconde classe, qui se manisfestent par les contraintes sur la connexion $\bA$.
En effet, $P_B\bA$ est fonction de $\chi$, et cela r\'eduit de 18 \`a 9 le nombres
de composantes ind\'ependantes de la connexion $\bA$. Cette contrainte
devra \^etre pris en compte dans la mesure: on ne veut int\'egrer que sur les configurations
des champs $\bA$ et $\chi$ satisfaisant cette contrainte!

\medskip

Je rappelle l'expression exacte de la contrainte liant $\bA$ et $\chi$:
$$
P_B\bA = \star\Theta(\chi)=\star(\chi\wedge\dd\chi).
$$
Comme nous l'avons vu, une fonction $f$, de $\bA$ et de $\chi$,
invariante de jauge est enti\`erement  d\'etermin\'ee
par sa valeur $f_{\chi_0}$ dans la fixation de jauge $\chi=\chi_0$
(qui n'est qu'autre que la {\it time gauge}).
Commen\c cons par analyser ce qui se passe dans cette jauge.
Dans ce cas, $P_B\bA =0$: $\bA$ est r\'eduit \`a sa
partie $\su_{\chi_0}$ et devient simplement une connexion
$\su$. Plus pr\'ecis\'ement, les holonomies
de $\bA$ sont des \'el\'ements de $\su$.
Par cons\'equent, pour impl\'ementer cette contrainte, on peut
se restreindre aux fonctions $f_{\chi_0}$ d\'efinies sur $\slc^E$
\`a support seulement sur $\su_{\chi_0}^E$
ou, de mani\`ere \'equivalente, il suffit de consid\'erer
le produit scalaire:
\beq
\langle f | g\rangle=
\int_{[SU_{\chi_0}(2)]^E} dU_e \,\overline{f_{\chi_0}(U_e)}
g_{\chi_0}(U_e).
\label{su2prod}
\eeq
Cette mesure est bien invariante sous $\slc$:
je l'ai d\'efinie sur la section $\chi=\chi_0$, il suffit de
faire une transformation de jauge pour sortir de cette jauge
particuli\`ere.
Par contre, si on regarde ce produit scalaire sur les r\'eseaux de
spin projet\'es, il y a une \'enorme d\'eg\'en\'erescence. Et tout comme
dans le cadre de la construction GNS, il faut enlever tous
les \'etats dont la norme est nulle.

Pour un r\'eseau de spin projet\'e d\'efini par
les repr\'esentations $(n_e,\rho_e)$ de $\slc$
et les repr\'esentations $j_e$ de $\su$, qui satisfont
$j_e\ge n_e$ (condition pour trouver la repr\'esentation
$j$ dans la d\'ecomposition de la repr\'esentation $(n,\rho)$),
nous avons d\'ej\`a vu que la restriction de la fonctionnelle,
dans la jauge $\chi=\chi_0$, \`a $\su^E$ est simplement le
r\'eseau de spin $\su$ labell\'e par les repr\'esentations
$j_e$. Ainsi, tous les r\'eseaux de spin partageant
les m\^emes labels $j_e$ sont \'equivalents du point de vue
du produit \Ref{su2prod} quelque soient leurs repr\'esentations
$(n_e,\rho_e)$. On retrouve ainsi exactement le m\^eme
espace d'Hilbert des \'etats quantiques que pour la \lqg
usuelle!

\medskip

De plus, on peut maintenant \'egalement d\'ecrire les r\'eseaux
de spin en dehors de la {\it time gauge} i.e. d\'ecrire les solutions
aux contraintes de seconde classe en toute
g\'en\'eralit\'e et ne pas seulement les d\'ecrire
dans la jauge $\chi=\chi_0$.
Pour cela, prenons une solution des contraintes dans la jauge
$\chi=\chi_0$, puis agissons dessus par des transformations de jauge.
Par exemple, prenons un r\'eseau de spin $\su$. Comme nous venons
de le voir, c'est une solution des contraintes. Agissons dessus par
une transformation de Lorentz. Il s'agit de faire tourner le champ
$\chi(x)$ et donc les sous-espaces $\su_\chi$, en tout point de l'espace,
en particulier en tout point du graphe-support. Pour cela, on est amen\'e
\`a consid\'erer des r\'eseaux de spin projet\'es en tout point, ou
des r\'eseaux de spin projet\'es infiniment raffin\'es, ou encore
des r\'eseaux de spin {\it totalement projet\'es}.

Pour \^etre plus explicite, regardons le cas d'une simple
holonomie $U_\alpha^{(n,\rho)}[\bA]$ le long d'un lien $\alpha$
dans la repr\'esentation $(n,\rho)$.
On choisit une repr\'esentation $j$ de $\su$ associ\'e au lien $\alpha$.
On peut d\'efinir une
``holonomie'' totalement projet\'ee en choisissant une partition
de $\alpha=\coprod_{k=1}^N [a_k,a_{k+1}]$,
projetant aux $N$ points $a_k$ de cette partition, puis raffinant
la partition $N\arr\infty$:
\beq
{\cal U}^{(n,\rho,j)}_{\alpha}[\bA,\chi]= \lim\limits_{N \rightarrow \infty}
{\cal P}\left\{ \prod\limits_{n=1}^{N}
P^{j}_{\chi_{_{v_{n+1}}}}
U^{(n,\rho)}_{\alpha_n}[\bA] P^{j}_{\chi_{_{v_{n}}}} \right\},
\eeq
o\`u $P^j_{(x)}$ d\'enote comme pr\'ec\'edemment le projecteur
$R^{(n,\rho)}\arr V^j_{(x)}$.
On aimerait qu'une telle boucle de Wilson ou qu'un r\'eseau de spin
d\'efini similairement soit solution des contraintes de seconde classe
i.e. ne d\'ependent  que de la partie $\su_\chi$ de la connexion $\bA$
en tout point.
Dans la limite d'un raffinage infini de la partition, \`a un point
donn\'e $x\in\alpha$, on est amen\'e \`a consid\'erer
$P^{j}_{(\chi)}\bA(x)_XT^XP^{j}_{(\chi)}$. On veut  que la partie $T^X=K$
de cette expression soit nulle. En regardant les repr\'esentations
de $\slc$ (voir en annexe), on aper\c coit que:
$$
P^{j}_{(\chi)}\vec{K} P^{j}_{(\chi)}=\beta_{(j)}
P^{j}_{(\chi)}\vec{J}P^{j}_{(\chi)} 
\qquad \textrm{avec }\beta_{(j)}=\frac{n\rho}{j(j+1)}.
$$
Par cons\'equent la partie boost dispara\^\i t \`a condition que
$\beta_{(j)}=0$ i.e. si et seulement si $n\rho=0$. Ceci s\'electionne
les repr\'esentations {\it simples} de $\slc$ avec soit $n=0$
(s\'erie continue) soit $\rho=0$ (s\'erie discr\`ete). Dans ces cas,
les r\'eseaux de spin totalement projet\'es, labell\'es
par des repr\'esentations simples de $\slc$, sont bien solutions
des contraintes de seconde classe (liant $\bA$ et $\chi$).
Leurs sections \`a $\chi=\chi_0$ sont toujours bien les
r\'eseaux de spin $\su$.

\medskip

Ceci conclut l'\'etude des \'etats quantiques de la th\'eorie fond\'ee
sur la connexion d'Ashtekar-Barbero g\'en\'eralis\'ee $\bA$.
Ce sont des r\'eseaux de spin totalement projet\'es labell\'es
par des repr\'esentations simples ($(0,\rho)$ ou $(n,0)$) de $\slc$.
Ils se r\'eduisent dans la jauge $\chi=\chi_0$ \`a de simples
r\'eseaux de spin $\su$. Le produit scalaire peut \^etre donn\'e
dans cette jauge particuli\`ere: c'est la m\^eme mesure
qu'en \lqg, d\'efinie simplement \`a partir de la mesure de Haar $\su$.

\section{L'alternative: une \lg covariante}

On s'int\'eresse maintenant \`a une quantification
utilisant des fonctionnelles de la connexion $\cA$.
Rappelons la particularit\'e de cette connexion: c'est l'unique
connexion se transformant correctement sous les diff\'eomorphismes
d'espace-temps dans le formalisme canonique.

La difficult\'e principale lors de la quantification dans ce cas est
la {\it non-commutativit\'e} (classique) de $\cA$. Et il faut prendre
en compte ce crochet de Dirac non-nul de $\cA$ avec elle-m\^eme
au niveau des op\'erateurs correspondant $\what{\cA}$.
Dans ce cadre, je n'ai pas quantifi\'e totalement la th\'eorie, et je ne
vais donner qu'une repr\'esentation des crochets de Dirac en
un {\it nombre fini de points} de $\Sigma$. N\'eanmoins, dans cette
quantification partielle, on trouve exactement les m\^emes structures
cin\'ematiques que dans le mod\`ele de mousse de spin de Barrett-Crane,
que je d\'ecrirai dans la partie IV. Cette quantification permet ainsi
d'obtenir un lien explicite entre le formalisme canonique et les
th\'eories covariantes de mousse de spin!

\medskip

Commen\c cons par \'etudier le cas limite $\imm\arr\infty$. Alors
l'action de Palatini g\'en\'eralis\'ee se r\'eduit \`a l'action
de Palatini usuelle avec un unique terme $\star(e\wedge e)\wedge F(\om)$.
Les crochets de Dirac s'\'ecrivent:
\beqs
\{B,B\}_D&=&0,\nonumber \\
\{P_B\cA,P_B\cA\}_D&=&0,\nonumber \\
\{P_B\cA^X_a,B^b_Y\}_D&=&\delta_a^b(P_B)^X_Y,\nonumber \\
\{P_R\cA,P_R\cA\}_D&=&0,
\eeqs
et la non-commutativit\'e de $\cA$ est entre sa partie rotation
$P_R\cA$ et sa partie boost $P_B\cA$ et r\'eside dans la contrainte
$$
P_R\cA=\Gamma(R),
$$
exprimant $P_R\cA$ comme une fonction (la connexion de spin)
de la triade $R$ ou $B$.
Puisque la partie $P_R\cA$ n'est pas une variable ind\'ependante
et n'intervient pas dans les crochets de Poisson de $\cA$ avec la triade,
on aimerait l'ignorer. Plus pr\'ecis\'ement, on aimerait
utiliser des fonctions d'onde (toujours invariantes
sous $\slc$) ne d\'ependant pas de $P_R\cA$.

Si on se r\'eduit \`a chercher des fonctions satisfaisant ce
crit\`ere \`a un nombre fini de points, une solution est
fournie directement par les r\'eseaux de spin projet\'es
labell\'es par des repr\'esentations triviales $j=0$ de $\su$.
De telles fonctionnelles sont nomm\'ees {\it r\'eseaux de spin
simples}.

\begin{figure}[t]
\begin{center}
\psfrag{r}{$\rho$}
\psfrag{r1}{$\rho_1$}
\psfrag{r2}{$\rho_2$}
\psfrag{r3}{$\rho_3$}
\psfrag{r4}{$\rho_4$}
\psfrag{r5}{$\rho_5$}
\psfrag{aire}{${\cal A}\equiv l_P^2\sqrt{\rho^2+1}$}
\psfrag{c1}{$\chi_{A}$}
\psfrag{c2}{$\chi_{B}$}
\psfrag{c3}{$\chi_{C}$}
\includegraphics[width=8cm]{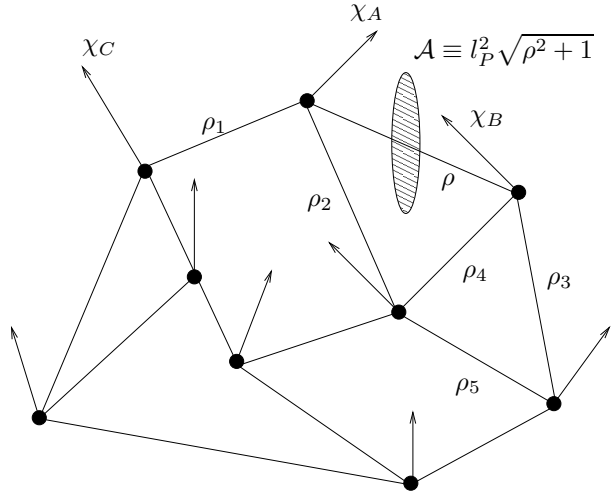}
\end{center}
\caption{Un r\'eseau de spin simple labell\'e par les repr\'esentations
simples $\rho_i$, qui d\'efinissent l'aire des surface duales aux liens.
Les $\chi$ donnent la normale \`a l'hypersurface au niveau des liens.}
\end{figure}

Regardons cela plus pr\'ecis\'ement. Choisissons un nombre fini de points
sur $\Sigma$ et consid\'erons des graphes dont les vertex sont ces points
particuliers. Sur un tel graphe, on labelle les liens $e$
avec des repr\'esentations $(n_e,\rho_e)$. On consid\`ere les holonomies
$U_e[\cA]$ le long de ces liens et on regarde leurs \'el\'ements de matrice
sur les vecteurs invariants sous $\su$. Remarquons que l'espace
$R^{(n,\rho)}$ ne contient un vecteur invariant sous $\su$ que si
$n=0$. On est donc restreint naturellement aux repr\'esentations simples
$(0,\rho_e)$. La fonctionnelle s'\'ecrit alors:
\beq
s^{\{\rho_e\}}(U_e,\chi_v)=
\prod_e \langle\rho_e\,\chi_{s(e)}\,j=0|U_e
|\rho_e\,\chi_{t(e)}\,j=0\rangle,
\eeq
o\`u $\chi_v$ est la valeur du champ $\chi$ au vertex $v$ et
$|\rho\chi_vj=0\rangle$ le vecteur de la repr\'esentation $\rho$ qui est
 invariant sous $SU(2)_{\chi_v}$. Ce sont des fonctionnelles cylindriques
ne d\'ependant pas de $P_R\cA$ aux vertex!
Une expression explicite de $s^{\{\rho_e\}}(U_e,\chi_v)$
est fournie par les noyaux $K_\rho$
d\'ecrits en annexe:
$$
s^{\{\rho_e\}}(U_e,\chi_v)=\prod_e
 K_{\rho_e}(\chi_{s(e)}^{-1}U_e\chi_{t(e)}).
$$

Alors, le crochet de Dirac de deux telles fonctionnelles,
\`a support sur des graphes dont les vertex sont les
points choisis au d\'ebut, est nul. D'un certain de vue, les points fix\'es
au d\'ebut sont les seuls points de $\Sigma$ existants. En effet,
ce sont les seuls points o\`u on conna\^\i t explicitement
la normale \`a $\Sigma$. Les autres points \'etant rendus
{\it flous} dans l'espace-temps. Dans ce cadre, si on consid\`ere deux
r\'eseaux de spin simples s'intersectant \`a un autre point
que leur vertex, il nous faut ``d\'efinir'' ce point et pr\'eciser
la normale \`a ce point: il faut projeter les r\'eseaux de spin
en ce point i.e. raffiner les fonctionnelles en rajoutant un vertex
bivalent. De cette mani\`ere, on pourrait construire une tour
de fonctionnelles d\'ependant de $\cA$ qui commuteraient les
unes avec les autres. On pourrait \'egalement consid\'erer la limite
de r\'eseaux de spin simples infiniment raffin\'es.

De toute mani\`ere, on peut regarder l'action de l'op\'erateur
aire sur un r\'eseau de spin simple pour une
surface intersectant le graphe-support au niveau d'un vertex.
Le r\'eseau de spin est un vecteur avec comme valeur propre:
\beq
{\rm aire}_{{\cal S}}\sim\sqrt{-C_1(\slc)}=\sqrt{\rho_e^2+1},
\eeq
en appliquant la formule \Ref{airecA} pour $j=0$.
On trouve ainsi un spectre des aires {\bf continu}. Le point
particuli\`erement int\'eressant est que ces r\'eseaux de spin simples
repr\'esentent aussi les \'etats cin\'ematiques des mousses de spin
(voir partie IV pour plus de d\'etails).

Pour terminer la quantification, il faut pr\'eciser l'impl\'ementation
des op\'erateurs $\what{\cA}$ et $\what{B}$:
\beqs
\what{P_B\cA}&=&P_B\cA\times, \nonumber \\
\what{B}&=&\f{\delta}{\delta P_B\cA},\nonumber \\
\what{R}&=&\what{\star B}\nonumber, \\
\what{P_R\cA}&=& \Gamma(\what{R}).
\eeqs

\medskip

Toute la proc\'edure peut se g\'en\'eraliser au cas g\'en\'erique
$\imm$ quelconque.
En effet, les crochets de Dirac sont alors:
\beqs
\{B,B\}_D&=&0,\nonumber \\
\left\{P_B\left(1+\f{1}{\imm}\star\right)\cA,
P_B\left(1+\f{1}{\imm}\star\right)\cA\right\}_D &=&0,\nonumber \\
\{\cA^X_a,B^b_Y\}_D&=&\delta_a^b(P_B)^X_Y,\nonumber \\
\{P_R\cA,P_R\cA\}_D&=&0.
\eeqs
On remarque que comme $P_R\cA$ commute avec $B$, on peut r\'e-\'ecrire
la relation de commutation entre $\cA$ et $B$ telle que:
\beq
\left\{P_B\left(1+\f{1}{\imm}\star\right)\cA^X_a,B^b_Y\right\}_D=\delta_a^b(P_B)^X_Y,
\eeq
\'etant donn\'e que $P_B\star =\star P_R$.
On a ainsi la m\^eme structure que pr\'ec\'edemment en rempla\c cant
$$P_B\cA\lrarr P_B\left(1+\f{1}{\imm}\star\right)\cA.$$
Par cons\'equent, on peut garder les m\^emes fonctionnelles simples
que dans le cas $\imm\arr\infty$ en changeant l\'eg\`erement
la repr\'esentation des op\'erateurs:
on impl\'emente maintenant $P_B(1+1/\imm\star)\cA$ par la multiplication
par $P_B\cA$. D'o\`u la quantification:
\beqs
\what{P_B\left(1+\f{1}{\imm}\star\right)\cA}&=&P_B\cA\times, \nonumber \\
\what{B}&=&\f{\delta}{\delta P_B\cA},\nonumber \\
\what{P_R\cA}&=& \Gamma(\what{R}),\nonumber \\
\what{P_B\cA}&=&P_B\cA\times +\f{1}{\imm}\what{P_R\cA}.
\eeqs

\medskip
On a donc quantifi\'e la th\'eorie en un nombre fini de points.
Je ne sais pas explicitement comment la quantifier enti\`erement,
bien qu'il semble possible d'utiliser pour cela l'alg\`ebre des 
fonctionnelles cylindriques d\'ecrit ci-dessus ou la limite de
fonctionnelles cylindriques infiniement raffin\'ees.
N\'eanmoins, on obtient des r\'eseaux de spin simples qui diagonalise
l'op\'erateur aire d'une surface et le spectre est continu. Ces
m\^emes r\'eseaux sont \'egalement les \'etats cin\'ematiques
des mod\`eles de mousse de spin et forment donc un pont entre
le formalisme canonique et le formalisme covariant des mousses de spin.
Finalement, on obtient une th\'eorie quantique (et un spectre
de l'aire) {\it ind\'ependant du param\`etre d'Immirzi $\imm$}.
Ce r\'esultat est compatible avec l'impl\'ementation de
l'int\'egrale de chemin r\'ealis\'e dans \cite{sergei:pathintegral},
qui montre qu'elle est ind\'ependante du param\`etre d'Immirzi.

\section*{Conclusion}

Le formalisme canonique invariant sous Lorentz (sans fixation de jauge)
n\'ecessite l'utilisation du crochet de Dirac tenant compte
des contraintes de seconde classe. Les relations de commutation
de la connexion (avec elle-m\^eme) en ressortent bien plus compliqu\'ees.
N\'eanmoins, on peut changer les variables canoniques de base
et il est possible de trouver deux connexions de Lorentz aux
propri\'et\'es bien particuli\`eres.

La premi\`ere est l'unique
connexion de Lorentz commutative (sous les crochet de Dirac) et
impl\'emente en fait une extension Lorentzienne de la connexion
d'Ashtekar-Barbero. Elle permet de retrouver la \lqg usuelle \`a partir
de notre contexte covariant. De plus, on peut maintenant consid\'erer
l'action
de boosts de Lorentz sur les r\'eseaux de spin $\su$ (\'etats de la
g\'eom\'etrie en \lg) Le travail effectu\'e
ouvre donc les portes vers l'\'etude de changement des \'etats
sous modification du plongement de l'hypersurface canonique
et ainsi de l'action des boosts de Lorentz sur les op\'erateurs aire
de la \lqg $\su$. Notons que cette th\'eorie quantique
d\'epend fortement du param\`etre d'Immirzi $\imm$.

La seconde connexion est l'unique connexion se transformant correctement
sous diff\'eomorphismes de l'espace-temps. Le fait qu'elle soit
non-commutative complique la proc\'edure de quantification. N\'eanmoins,
il est possible de quantifier la th\'eorie (de mani\`ere partielle,
car s'\'etant restreint \`a un nombre fini de points ou, en d'autres termes,
quantifiant une classe restreinte d'observables) et on obtient
comme \'etats quantiques de g\'eom\'etrie les r\'eseaux de spin simples
 repr\'esentant \'egalement les \'etats cin\'ematiques
des mod\`eles de mousse de spin.
De plus, on d\'erive un spectre de l'aire continu par opposition au 
spectre discret de la \lg. Notons que cette th\'eorie quantique
ne d\'epend pas du tout du param\`etre d'Immirzi.

Pour le moment, il semble donc avoir un choix entre un spectre discret
d\'ependant de $\imm$ et un spectre continu ind\'ependant de $\imm$.
Reste toujours l'\'etude de la dynamique exacte de ces th\'eories.
L'usage de la connexion d'espace-temps sera peut-\^etre alors
plus avantageux que celui de la connexion d'Ashtekar-Barbero
(g\'en\'eralis\'ee): le comportement ``bizarre'' de cette derni\`ere
sous diff\'eomorphismes de l'espace-temps pourrait causer des probl\`emes
lors de l'impl\'ementation de la contrainte Hamiltonienne
au niveau quantique\dots

\part{L'Espace-Temps en Mousse de Spins}

Les mod\`eles de Mousses de Spin, ou {\it Spin Foams},
sont des mod\`eles d'int\'egrale de chemin de la gravit\'e.
Ils appara\^\i ssent au carrefour de diverses approches \`a la
question de la gravit\'e quantique: \lqg, th\'eories topologiques
({\it State Sum Models}),
approches discr\`etes \`a la Regge, int\'egrale de chemin
(somme sur les g\'eom\'etries) \`a la Hartle-Hawking,
approches cat\'egoriques.
Pour le lecteur int\'eress\'e, il existe deux revues
compl\`etes sur le sujet \cite{dan:review, alej:review}.

Le point de vue de la gravit\'e \`a boucles nous
pr\'esente les mousses de spin comme des histoires de r\'eseaux de spin
\cite{carlo&reis}, et c'est de l\`a qu'est issu le nom de {\it Spin Foam}
\cite{baez:sf}. Ainsi une mousse de spin est un espace-temps
repr\'esentant l'\'evolution d'un r\'eseau de spin dans le temps
(sous l'action d'un Hamiltonien ou d'une contrainte Hamiltonienne):
les faces de la mousse de spin sont labell\'ees par des
repr\'esentations d'un groupe ($\su$ si l'on \'etudie la \lg ou
$\slc$ dans une approche invariante sous Lorentz), aux c\^ot\'es
sont associ\'es des entrel\^aceurs (entre les repr\'esentations
des faces adjacentes au lien) et les vertex/points repr\'esentent
les ``\'ev\`enements'' dans l'espace-temps i.e les lieux de
la dynamique.
De cette mani\`ere, si on consid\`ere une tranche hypersurface
de genre espace d'une mousse de spin, on obtient bien un graphe
dont les liens sont labell\'es par des repr\'esentations i.e. un
r\'eseau de spin!
Alors un mod\`ele de mousse de spin particulier d\'efinit une amplitude
\`a chaque histoire de r\'eseau de spin. La fonction de partition
du mod\`ele est alors la somme de ces amplitudes sur toutes
les histoires possibles et en cela impl\'emente une int\'egrale de chemin.

\medskip

Le mod\`ele de mousse de spin le plus prometteur est le mod\`ele
de Barrett-Crane, cens\'e d\'efinir la gravit\'e quantique.
Il en existe une version Euclidienne \cite{bc1} et une version Lorentzienne
\cite{bc2}. Il est issu des techniques de quantification des th\'eories
topologiques. Plus particuli\`erement, la relativit\'e
g\'en\'erale peut \^etre reformul\'ee comme une th\'eorie BF contrainte
(voir par exemple \cite{bf}).
Or la th\'eorie BF est une th\'eorie
topologique que nous savons quantifier. De plus, son caract\`ere
``topologique''
permet de l'\'etudier sur une triangulation (ou d\'ecomposition
cellulaire g\'en\'erique) de l'espace-temps sans en alt\'erer le contenu.
Par cons\'equent, on peut discr\'etiser la th\'eorie BF et d\'ecouper
l'espace-temps en 4-simplex, quantifier la th\'eorie BF discr\'etis\'ee, puis
imposer les contraintes au niveau quantique.
Dans ce contexte, le mod\`ele de Barrett-Crane associe des repr\'esentations
du groupe de Lorentz $\slc$ aux faces de la triangulation et d\'efinit
une amplitude pour chaque 4-simplex comme fonction des repr\'esentations
qui le labellent. Les contraintes sur la th\'eorie BF se refl\`etent
par des contraintes sur les repr\'esentations admissibles, qui
sont restreintes \`a \^etre simples.
La mousse de spin est alors le dual (topologique) de la
triangulation de l'espace-temps et l'amplitude d'une histoire/triangulation
est le produit des amplitudes des 4-simplex la composant.
Une tranche 3d de cette mousse de spin d\'efinit un r\'eseau de spin simple!
Et on retrouve ainsi les \'etats cin\'ematiques de la \lg covariante
construits dans la partie pr\'ec\'edente.
On peut aussi voir ce mod\`ele comme une quantification g\'eom\'etrique
des 4-simplex \cite{bc1,bb}, ce qui fournit une interpr\'etation
g\'eom\'etrique directe de la th\'eorie. Ainsi on peut le construire
\`a l'aide de bivecteurs quantiques se combinant pour former des 4-simplex
quantiques. On obtient un espace-temps fait de patchs plats (les 4-simplex
quantiques), la courbure de l'espace-temps r\'esidant dans
la mani\`ere dont ces patchs sont reli\'es \cite{causal1}.

\medskip

Il existe des analogues en basse dimension. Ce sont des mod\`eles
discrets quantifiant la th\'eorie topologique BF.
En deux dimensions, l'approche
mousse de spin rejoint l'\'etude des mod\`eles de matrices (en ``0 dimension'')
de la th\'eorie des cordes. On peut comparer le contenu
du mod\`ele de mousse de spin avec la th\'eorie BF
classique et sa quantification canonique et v\'erifier que tout
est bien coh\'erent \cite{2d}.
En trois dimensions, c'est le mod\`ele
de Ponzano-Regge. Il quantifie la gravit\'e Euclidienne
en trois dimensions et permet
de comprendre le contexte des mousses de spin, m\^eme si son caract\`ere
topologique le distingue irr\'em\'ediablement du cas de la relativit\'e
g\'en\'erale. Il est construit en d\'ecoupant l'espace-temps en t\'etra\`edre
(3-simplex): on attache des repr\'esentations de $\su$
aux liens de la triangulation puis on d\'efinit des amplitudes invariantes
sous $\su$ associ\'ees \`a chaque t\'etra\`edre. Cette proc\'edure
produit des amplitudes infinies et il faut les r\'egulariser. Il existe
diverses m\'ethodes. On peut imposer un cut-off sur les repr\'esentations.
Ou on peut utiliser une d\'eformation quantique du groupe $\su$:
on obtient le mod\`ele de Turaev-Viro qui est bien d\'efini et cens\'e
quantifier la gravit\'e 3d avec une constante cosmologique \cite{ooguri}.
On peut aussi remarquer que ces infinis correspondent \`a une invariance
r\'esiduelle par diff\'eomorphismes et fixer de jauge cette sym\'etrie
\cite{diffeo}. Au final, on associe une amplitude finie \`a
chaque triangulation. Le dual de ces triangulations d\'efinit une mousse
de spin, dont chaque tranche 2d donne un r\'eseau de spin $\su$:
le mod\`ele d\'efinit une dynamique des r\'eseaux de spin
(voir par exemple \cite{fotini:sf}).
Il existe \'egalement des versions Lorentziennes de ce mod\`ele
\cite{davids,laurent:2+1} utilisant le groupe $\slr$ au lieu de $\su$.

Le lien avec la gravit\'e 3d classique est souvent effectu\'e
\`a l'aide d'asymptotiques de l'amplitude des t\'etra\`edres. On retrouve
en effet l'exponentielle de l'action de Regge. La m\^eme
proc\'edure peut \^etre suivie avec le mod\`ele de Barrett-Crane et
on retrouve une nouvelle fois l'action de Regge
\cite{asymp:ruth&barrett}.
Mais cela ne fonctionne pas aussi simplement. Tout d'abord, on ne trouve
pas exactement l'exponentielle, mais le cosinus. Il faut orienter la mousse
de spin et r\'etablir la causalit\'e pour reconstruire l'exponentielle
\cite{causal1,causal2}. Egalement, il semble que ce soient des configurations
d\'eg\'en\'er\'ees qui dominent la fonction de
partition \cite{laurent:asymp, asymp:baez,asymp:barrett}, donc il reste du
travail pour v\'erifier que le mod\`ele correspond bien \`a une th\'eorie
de la gravit\'e quantique.


\medskip

Un autre point tr\`es int\'eressant est la reformulation des mod\`eles
sous la forme d'une th\'eorie des champs sur un groupe. En effet, le mod\`ele
de Ponzano-Regge peut \^etre \'ecrit sous la forme d'une th\'eorie
des champs sur $\su^{\times n}$, dite de Boulatov-Ooguri \cite{ooguri:gft}.
Dans ce cadre,
la triangulation/mousse de spin est reconstruite comme un diagramme
de Feynman de la th\'eorie des champs. Ainsi la th\'eorie g\'en\`ere
toutes les triangulations de l'espace-temps et son int\'egrale de chemin
d\'efinit la fonction de partition des mousses de spin.
Pour Ponzano-Regge, il existe une version convergente
de l'int\'egrale de chemin \cite{summation}, ce qui permet de lui donner
un sens physique intrins\`eque au lieu de se contenter de son r\^ole
de fonction g\'en\'eratrice.

La m\^eme construction est possible en quatre dimensions. On peut ainsi
g\'en\'erer le mod\`ele de Barrett-Crane \`a partir  d'une th\'eorie de champs
sur $Spin(4)$ pour la version Euclidienne \cite{gft,alej:e}
et sur $\slc$ pour la version Lorentzienne \cite{alej:l}.
Plus g\'en\'eralement, il est possible de faire de m\^eme 
pour n'importe quel mod\`ele de mousse de spin \cite{carlo:gft}.
Dans le cas de Barrett-Crane, il existe aussi
des r\'esultats de finitudes des amplitudes
\cite{finite:alej,finite:bb,finite}
et on a donc bien un mod\`ele bien d\'efini. On peut aussi
se servir de l'int\'egrale de chemin pour d\'efinir
les observables de la th\'eorie \cite{observables} \`a travers
des corr\'elations de la th\'eorie de champs.

\medskip

L'avantage des mod\`eles de mousse de spin est que ce sont
des structures covariantes et construisant l'espace-temps m\^eme:
les mousses de spin sont ``{\it background independent}'', elles ne sont
pas d\'efinies sur une vari\'et\'e mais la d\'efinissent elles-m\^emes.
En cela, ils pr\'esentent un v\'eritable espoir de gravit\'e quantique.
Et m\^eme si les mod\`eles actuels en quatre dimensions se r\'ev\`elent
 incorrects
(on ne retrouve pas la relativit\'e g\'en\'erale dans un
r\'egime semi-classique), ils semblent fournir un exemple bien d\'efini
d'une th\'eorie d'int\'egrale de chemin sur les classes de
g\'eom\'etrie de l'espace-temps.

\medskip

Dans un premier chapitre, je commencerai par d\'ecrire les mousses de spin
2d. Durant ma th\`ese, j'ai \'etudi\'e leur relation avec la th\'eorie BF
2d avec Carlo Rovelli et Alejandro Perez \cite{2d}.
Ici, je pr\'esente directement le mod\`ele \`a travers sa th\'eorie de champs
g\'en\'eratrice. Je rappelle comment la fonction de partition et corr\'elations
de la th\'eorie g\'en\`erent une int\'egrale de chemin sur les g\'eom\'etries 2d.
Et dans ce cadre, j'ai tent\'e au cours de ma th\`ese  de comprendre
 la signification non-perturbative et le r\^ole de cette
 th\'eorie de champs \`a travers l'effet non-perturbatif le plus simple:
les solutions classiques de la th\'eorie de champs.

Puis je d\'ecris   le mod\`ele de Ponzano-Regge quantifiant la gravit\'e 3d.
Cela permet de pr\'esenter la g\'eom\'etrie
des mousses de spin et de comprendre la logique de l'utilisation de
ces mod\`eles.
J'aborde dans ce contexte la question
de l'\'evolution et de la causalit\'e.
A c\^ot\'e de ces consid\'erations g\'eom\'etriques, je fournis \'egalement
une interpr\'etation des solutions classiques de la th\'eorie de champs.

Dans un second chapitre, je d\'ecris le mod\`ele de gravit\'e quantique
d\'efini par les mousses de spin de Barrett-Crane.
Dans un premier temps, j'ai \'etudi\'e la quantification de la
relativit\'e g\'en\'erale aboutissant au mod\`ele de mousse de spin
\cite{bf1,bf2}, en concentrant mon attention \`a la possibilit\'e
d'une ambigu\"\i t\'e de type Immirzi.
Puis je me suis attach\'e \`a comprendre la g\'eom\'etrie quantique
du mod\`ele et le r\^ole de la causalit\'e \cite{causal1,causal2}.

\chapter{Mousse de Spin en Basse Dimension}

Regarder les mod\`eles de mousse de spin en basse dimension
permet de comprendre la structure de la th\'eorie BF et les outils
math\'ematiques n\'ecessaires \`a son \'etude. Il est alors possible
de faire des calculs explicites et de comparer les mod\`eles
de mousse de spin aux autres approches \`a la quantification
de la th\'eorie BF. De plus, nous pouvons \'egalement tester
nos interpr\'etations g\'eom\'etriques et le sens physique que
nous attribuons aux diverses structures de la th\'eorie.
En esp\'erant que cela nous m\`ene \`a une meilleure compr\'ehension
des mousses de spin en quatre dimensions, cens\'ees quantifier
la relativit\'e g\'en\'erale. En effet, les structures de base
sont similaires m\^eme si malheureusement ces th\'eories sont beaucoup
trop simples pour pouvoir servir de v\'eritable guide dans l'\'etude
de la gravit\'e quantique.

\medskip

Je commence par d\'ecrire la th\'eorie BF en deux dimensions. Je construis les
mousses de spin et compare l'approche \`a la quantification canonique.
Je montre aussi comment d\'eriver le mod\`ele \`a partir d'une
th\'eorie de champs. Dans ce cadre, j'explore la signification
non-perturbative de cette th\'eorie de champs \`a travers l'effet le plus
simple: les solutions classiques de la th\'eorie.

Je commence par d\'ecrire le mod\`ele de mousse de spin en deux dimensions
quantifiant la th\'eorie BF. Je montre comment la th\'eorie de champs auxiliaire
permet de g\'en\'erer l'espace-temps \`a partir de rien et j'\'etudie
la possibilit\'e d'utiliser
les solutions classiques comme {g\'eom\'etrie de \it background}.

Puis je d\'ecris 
le mod\`ele de Ponzano-Regge, explicite  la g\'eom\'etrie
du mod\`ele et son interpr\'etation en terme d'histoire de r\'eseau de spin.
Cela permettra de mieux comprendre la logique du mod\`ele de Barrett-Crane.
Finalement, je regarde les solutions de la th\'eorie de champs g\'en\'erant
le mod\`ele de Ponzano-Regge et discute leur interpr\'etation dans le cadre
de la quantification de la gravit\'e 3d.

\section{Th\'eorie BF en 2d}

\subsection{Th\'eorie de champs g\'en\'eratrice}

Le mod\`ele de mousse de spin en deux dimensions correspond
\`a la quantification de la th\'eorie BF:
\beq
S(A,B)=\int_{\cal M}{\rm Tr} BF(A)=\int_{\cal M}B^iF^i(A)
\eeq
o\`u ${\cal M}$ est une vari\'et\'e 2d, $B$ un champ scalaire sur ${\cal M}$
\`a valeur dans $su(2)$,
$A$ une 1-forme \`a valeur dans $su(2)$ et $F$ sa courbure.
Classiquement, les \'equations du mouvement imposent que la connexion
$A$ est plate. La th\'eorie est invariante sous $\su$
et invariante sous diff\'eomorphisme. De plus, 
c'est une th\'eorie topologique. On peut la discr\'etiser et la quantifier.
La proc\'edure est r\'ealis\'ee explicitement dans \cite{2d} o\`u
on construit les mousses de spin \`a partir de la th\'eorie classique et o\`u
on compare ses r\'esultats avec la quantification canonique. Ici, je ne d\'ecrirai pas
ce travail, et je m'attacherai \`a pr\'esenter les structures des mousses de spin 2d
dans leur globalit\'e.

\medskip

Construisons les mousses de spin 2d
directement \`a partir de la th\'eorie de champs g\'en\'eratrice.
Soit donc un champ $\varphi:\su^2\arr\C$ et l'action:
\beq
S[\varphi]\equiv\f{1}{2}\int
dg_1dg_2\,|\varphi(g_1,g_2)|^2-\f{\lambda}{3!}\int
dg_1dg_2dg_3\,\varphi(g_1,g_2)\varphi(g_2,g_3)\varphi(g_3,g_1).
\eeq
On impose une invariance de $\varphi$ sous $\su$:
\beq
\forall g\in\su,\, \varphi(g_1,g_2)=\varphi(g_1g,g_2g).
\eeq
On peut donc exprimer la th\'eorie avec le champ $\phi$ d\'efini par
$$
\phi(g_1g_2^{-1})\equiv\varphi(g_1,g_2).
$$
L'action devient alors:
\beq
S[\phi]=\f{1}{2}\int_{\su} dg\,|\phi(g)|^2-
\f{\lambda}{3!}\int dg d\tl{g}
\phi(g)\phi(\tl{g})\phi(\tl{g}^{-1}g^{-1}).
\eeq
On demande \'egalement que $\varphi$ soit r\'eel et sym\'etrique
$g_1\lrarr g_2$. Cela revient \`a imposer $\phi$ r\'eel
et invariant sous passage
\`a l'inverse $g\lrarr g^{-1}$. Afin  de mieux saisir le contenu de la
th\'eorie, on passe \`a la transform\'ee de Fourier d\'ecomposant
$\phi$ sur les repr\'esentations de $\su$:
$$
\phi(g)=\sum_{j\in\N/2}\Delta_j\phi^j_{ab}D^j_{ab}(g)
$$
o\`u $\Delta_j=(2j+1)$ est la dimension de la repr\'esentation
$j$, $D^j_{ab}(g)$ la matrice repr\'esentant l'\'el\'ement $g\in\su$ et
$\phi^j_{ab}$ la matrice $\Delta_j\times\Delta_j$ transform\'ee de Fourier
de $\phi$. Imposer que $\phi$ soit sym\'etrique se traduit par
$\epsilon{}^t\phi^j=\phi^j\epsilon$,
o\`u $\epsilon$ est l'op\'erateur de conjugaison/parit\'e\footnotemark
sur les repr\'esentations de $\su$.
\footnotetext{$\epsilon^{(j)}$ est tel que
$\forall g\in\su,\,\overline{D^j(g)}=\epsilon D^j(g)\epsilon^{-1}$.
On a alors $^t\epsilon^{(j)}=(-)^{2j}\epsilon^{(j)}$.}
Alors, demander $\phi$ r\'eel rend la matrice $\phi^j$ hermitienne.
Un calcul rapide fournit
\beq
S[\phi]=\sum_j\Delta_j\left(
\f{1}{2}{\rm Tr}\phi_j^2 -\f{\lambda}{3!}{\rm Tr}\phi_j^3
\right).
\eeq
Les secteurs \`a $j$ diff\'erents de la th\'eorie ne communiquent pas
et, pour chaque $j$ donn\'e, on reconna\^\i t un mod\`ele
de matrice simple, correspondant aux mod\`eles de matrices de la th\'eorie
des cordes z\'ero-dimensionnelle.

\medskip

Concentrons-nous donc sur une composante $j$ fix\'ee et \'etudions le mod\`ele
de matrice:
\beq
S_N[M]=N\left(\f{1}{2}{\rm Tr}M^2-\f{\lambda}{3!}{\rm Tr}M^3\right),
\label{matrice}
\eeq
o\`u $N$ est la taille des matrices $M$.
\begin{figure}[t]
\begin{center}
\psfrag{w1}{$\f{1}{N}$}
\psfrag{w2}{$N\times\lambda$}
\includegraphics[width=6cm]{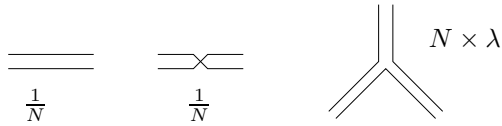}
\end{center}
\caption{Propagateurs et Vertex pour les diagrammes de Feynman du
mod\`ele de matrices \Ref{matrice} de taille $N$.}
\label{2drule}
\end{figure}
On peut calculer la fonction de partition (et les corr\'elations) de la th\'eorie
de mani\`ere perturbative en d\'eveloppant ses diagrammes de Feynman. Les r\`egles
de Feynman sont d\'efinies par un propagateur trivial ($1/N$ fois l'identit\'e)
et un vertex trivalent de poids $N\times\lambda$.
En fait, il y a deux composantes du propagateur dues aux sym\'etries de la matrice $M\equiv \phi_j$:
l'identit\'e et un twist (voir figure \ref{2drule}.).

Alors les diagrammes de Feynman ferm\'es,
dont la somme permet de calculer la fonction de partition, g\'en\`erent des triangulations
de vari\'et\'e 2d. En effet, un diagramme de Feynman donn\'e est un graphe trivalent ferm\'e
dont les noeuds sont les vertex d'interactions, les liens les propagateurs. Mais c'est m\^eme un 2-complexe
dont les faces peuvent \^etre reconstruites comme les boucles du diagramme de Feynman, prenant en compte
qu'il est construit \`a partir de lignes doubles.
On peut construire le dual topologique du diagramme de Feynman
en consid\'erant les {\it triangles duaux} aux vertex ``d'interaction''
(voir \cite{2dsurface} pour une approche syst\'ematique du probl\`eme de la reconstruction
d'une surface 2d \`a partir du graphe).
A ce point, notons que le propagateur twist est n\'ecessaire pour retrouver
toute les topologies possibles de vari\'et\'e 2d \`a partir du d\'eveloppement perturbatif.
Alors il est facile de calculer l'amplitude d'un graphe donn\'e en fonction
de son nombre de vertex $V$, de c\^ot\'es $E$ et de faces $F$:
$$
{\cal A}^{(E,V,F)}=N^{V-E+F}\lambda^V.
$$
Elle  d\'epend de la caract\'eristique d'Euler $\chi({\cal M})=V-E+F$ de la surface
2d ${\cal M}$ duale au graphe consid\'er\'e.
$\chi({\cal M})$ est une fonction du genre $g$ de de la vari\'et\'e ${\cal M}$ et de son orientabilit\'e:
$\chi({\cal M})=2-2g({\cal M})$ pour une vari\'et\'e orientable et
$\chi({\cal M})=2-g({\cal M})$ pour une vari\'et\'e non-orientable.
Ainsi, le cas $\chi=2,g=0$ correspondant \`a une sph\`ere, $\chi=0,g=1$ \`a un tore,
$\chi=1,g=1$ \`a l'espace projectif $P^2\R$, $\chi=0,g=2$ \`a la bouteille de Klein,
et ainsi de suite. La fonction de partition est alors
obtenue en sommant sur tous les graphes possibles:
\beq
Z_N=\sum_{g,\epsilon}\sum_V {\cal N}(g,\epsilon,V)\lambda^VN^{2-(1+\epsilon)g},
\eeq
o\`u ${\cal N}(g,\epsilon,V)$ compte le nombre de triangulations
de genre $g$ avec $V$ triangles orientable pour $\epsilon=1$ et non-orientable
pour $\epsilon=0$.
Ces nombres ne sont malheureusement pas connus explicitement. N\'eanmoins, on conna\^\i t des
asymptotiques permettant de d\'eduire le comportement convergeant/divergeant de $Z$
\cite{ooguri:gft}:
\beqs
{\cal N}(g=0,V)&\sim& V^{-\f{7}{2}}e^{\beta_c V} \quad\textrm{avec}\quad
e^{\beta_c}=\f{2^4}{3^{3/2}},\nonumber \\
{\cal N}(g\ge 1,V)&\sim& V^{-\f{5}{2}g+\f{3}{2}}e^{\beta_c V}.
\eeqs
Le cas {\it topologique} correspond au cas o\`u l'amplitude d'un graphe ne d\'epend plus que
de sa topologie i.e. de son genre $g$ (et de son orientabilit\'e).
Il est atteint quand $\lambda=1$. Ce cas correspond
\`a la th\'eorie BF. Cependant, la somme des graphes de Feynman
est alors clairement divergente,
ce qui rend l'\'etude de ce cas particulier difficile dans le contexte des mousses de spin.

\medskip

On peut introduire des fronti\`eres aux triangulations en calculant des corr\'elations.
En effet, on obtiendra alors des diagrammes de Feynman ouverts (avec bord).
Les observables de la th\'eorie (invariants sous $\su$ donc sous conjugaison dans le langague
des matrices) sont g\'en\'er\'ees par les ${\rm Tr}M^n$. Une telle observable ${\rm Tr}M^n$
correspond \`a une fronti\`ere d\'efinie par un cercle de longueur $n$ unit\'es. Plus
pr\'ecis\'ement, la corr\'elation
$$
W=\int[dM]\,{\rm Tr}M^ne^{-S[M]}
$$
se d\'eveloppe en des diagrammes de Feynman duaux \`a des surfaces triangul\'ees avec bord
un cercle ``triangul\'e'' par $n$ segments.
On peut g\'en\'eraliser cela et d\'efinir les corr\'elations
$W(c=\{c_0,c_1,\dots\})$ correspondant \`a un bord avec $c_n$ cercles de longueur $n$.

On d\'efinit alors des {\it \'etats fronti\`eres} $|c\ra=|c_0,c_1,\dots\ra$
et les corr\'elations $W$ d\'efinissent  des {\it amplitudes de transition} entre eux.
Plus pr\'ecis\'ement, sur ces \'etats orthogonaux par d\'efinition, on peut d\'efinir
un produit scalaire ``physique'' par
\beq
\la c'|c\ra_{{\rm phys}}=W(c\arr c')\equiv W(c\cup c').
\eeq
Puis pour obtenir un vrai espace d'Hilbert, il faudrait enlever les configurations
non-physiques donn\'ees par les \'etats d\'eg\'en\'er\'es (dont la norme est nulle).
Math\'ematiquement, pour cela, on peut utiliser la construction GNS: on consid\`ere
la $C^*$-alg\`ebre ${\cal F}$ 
des ``\'etats fronti\`eres'' $c$ et la forme lin\'eaire positive d\'efinie sur ${\cal F}$
par $W$, alors on construit l'id\'eal de Gelfand ${\cal I}=\{c|W(c*c)=0\}$
et l'espace d'Hilbert physique est le compl\'et\'e de ${\cal F}/{\cal I}$.

Ceci est la proposition de \cite{observables} d\'ecrivant les mousses de spin
comme une impl\'ementation  propre de l'int\'egrale de chemin sur les g\'eom\'etries.
Cependant, pour pouvoir mener ce projet \`a terme, il faudrait conna\^\i tre la fonction
de partition et les corr\'elations de mani\`ere explicite, un calcul perturbatif
ne suffisant pas. Dans ce cadre, il est l\'egitime  de poser la question
``cette th\'eorie de champs g\'en\'erant l'espace-temps \`a partir de rien (vide total),
a-t-elle une v\'eritable signification non-perturbative?''
Dans ce cadre, le plus simple effet non-perturbatif est fourni par les solutions
classiques de la th\'eorie, que je propose maintenant d'\'etudier.

\subsection{Solutions ``Classiques'': Instantons de la th\'eorie}

Regardons les solutions classiques de l'action $S[M]$. La th\'eorie n'ayant pas de
param\`etre $\hbar$, il n'y a pas de sens \'evident dans lequel ces instantons
dominent la fonction de partition/les corr\'elations. Le param\`etre math\'ematiquement
le plus similaire reste la taille $N$ des matrices et il semble que ces configurations
``classiques'' dominent l'int\'egrale dans la limite $N\arr\infty$.
Une autre utilisation de ces solutions classiques est de les consid\'erer comme
de nouveaux {\it backgrounds} et d'\'etudier la th\'eorie perturbativement autour de ces
configurations.

\medskip

Commen\c cons par calculer ces solutions. Elles sont solutions de l'\'equation
\beq
M_0-\f{\lambda}{2}M_0^2=0,
\eeq
donc elles sont diagonalisables avec valeurs propres 0 et $-2/\lambda$:
\`a un facteur pr\`es, ce sont des projecteurs.
L'action \'evalu\'ee sur une solution $M_0$ d\'epend de la dimension de projecteur $k$
(dimension de l'espace propre associ\'e \`a $-2/\lambda$):
$$
S[M_0]=\f{2}{3}\f{k}{\lambda^2}.
$$
L'action minimale est atteinte pour $k=0$ i.e. la solution nulle.

On peut faire de m\^eme pour la th\'eorie de champs sur le groupe. 
L'\'equation du mouvement s'\'ecrit
\beq
\varphi(g_1,g_2)=\f{\lambda}{2}\int dg_3\,\varphi(g_2,g_3)\varphi(g_3,g_1)
\quad\textrm{ou bien}\quad
\phi(g)=\f{\lambda}{2}\int d\tl{g}\,\phi(\tl{g})\phi(\tl{g}^{-1}g^{-1}),
\eeq
ce qui signifie que $\phi=\lambda/2\phi\star\phi$ o\`u $\star$ d\'enote le produit
de convolution\footnotemark  sur $\su$. D\'efinissant
l'application
$$
P_\phi:f\in L^2(\su)\arr\f{\lambda}{2}\phi\star f,
$$
$\phi$ est une solution classique si et seulement si 
$P_\phi$ est un projecteur dans l'espace d'Hilbert $L^2(\su)$.
Reste qu'il est plus simple pour les calculs de travailler sur les composantes
de Fourier $\phi_j$, dans quel cas on peut se restreindre sans probl\`eme \`a
l'\'etude du mod\`ele de matrice.

\footnotetext{Le produit de convolution sur $\su$ est d\'efini par
$a\star b(g)=\int d\tl{g}\,a(g\tl{g}^{-1})b(\tl{g})$.}

\medskip

On peut comprendre une solution classique $M_0$ comme une configuration 
``auto-raffinante'' du champs: similairement \`a la logique des triangulations
explicit\'ee ci-dessus, $M_0$ peut se penser comme un segment sur le bord, ce segment $M_0$
\'etant ``\'equivalent'' \`a deux segments $M_0^2$ et ainsi de suite.
Plus pr\'ecis\'ement, regardons la fonctionnelle bord ``cercle de longueur n''
$$
{\rm Tr}M_0^n=\left(\f{2}{\lambda}\right)^{n-1}{\rm Tr}M_0,
$$
qui est \'egale (sur les $M_0$) \`a la fonctionnelle bord ``cercle de longueur 1''
(\`a un facteur pr\`es).
D'une certaine mani\`ere, $M_0$
ressemble \`a un {\it \'etat coh\'erent du segment} (quantifi\'e)!
On peut m\^eme aller plus loin. On regarde les fonctions (invariantes de jauge)
d'une matrice:
ce sont des fonctions du ``segment quantifi\'e'', c'est-\`a-dire les \'etats
de la th\'eorie.
On consid\`ere alors la relation d'\'equivalence suivante:
\beq
f(M)\sim g(M) \darr \forall M_0\,\textrm{solution classique,}\,
f(M_0)=g(M_0),
\eeq
d\'efinie par l'\'egalit\'e des fonctions sur l'ensemble des solutions classiques\footnotemark.
\footnotetext{Cette construction me fut sugg\'er\'ee par Laurent Freidel.}
L'ensemble des fonctions invariantes de jauge \'etant g\'en\'er\'ee par les
${\rm Tr}M^n$, il est facile de voir qu'aux yeux de la relation $\sim$ tous les ${\rm Tr}M^n$
pour tous $n$ sont \'equivalents
$$
\forall n,m\ge1, \,\left(\f{\lambda}{2}\right)^{n}{\rm Tr}M^n\sim
\left(\f{\lambda}{2}\right)^{m}{\rm Tr}M^m
$$
et que
le quotient par $\sim$ est g\'en\'er\'e simplement par des \'etats d\'ependant
du nombre de cercles et plus de leur longueur! De cette mani\`ere, on retrouve bien les
m\^emes r\'esultats du formalisme canonique rendant compte de l'invariance par
diff\'eomorphismes et du caract\`ere topologique du mod\`ele.
Les solutions classiques d\'efinissent ainsi les \'etats quantiques
de la th\'eorie BF originale.

\medskip
Maintenant que j'ai discut\'e la possible d'interpr\'etation physique de ces solutions
classiques, regardons \`a quoi ressemble les perturbations autour de ces configurations.
Pour cela, d\'efinissons la nouvelle action
\beq
S_{M_0}[M]=S[M_0+M]-S[M_0]=
N\left(\f{1}{2}{\rm Tr}M^2-\f{\lambda}{3!}{\rm Tr}M^3
-\f{\lambda}{2}{\rm Tr}M^2M_0\right).
\eeq
Dans un contexte non-perturbatif, l'int\'egrale de chemin de cette th\'eorie
devrait co\"\i ncider avec l'originale.
Dans le contexte perturbatif des diagrammes de Feynman, le terme suppl\'ementaire d\'ependant
de $M_0$ introduit une modification au propagateur de la th\'eorie.
En effet, le terme quadratique en $M$ s'\'ecrit $1/2{\rm Tr}M^2(1-\lambda M_0)$
et le propagateur devient $(1-\lambda M_0)^{-1}=1-\lambda M_0$.
Le vertex (d'interaction) de la th\'eorie lui demeure inchang\'e.
C'est la contribution des faces dans les diagrammes de Feynman qui va changer,
n'\'etant plus simplement $N={\rm Tr}Id$ mais ${\rm Tr}(1-\lambda M_0)^p$
avec $p$ nombre de c\^ot\'es/propagateurs autour de la face consid\'er\'ee.
Ce nombre d\'epend de $M_0$ et vaut bien $N$ quand $p$ est pair mais vaut $N-2k$ quand
$p$ est impair. Cela casse l'invariance topologique de la th\'eorie et permet d'introduire
donc un {\it background} non-trivial dans la th\'eorie\footnotemark.
\footnotetext{Remarquons la solution classique
$\displaystyle \phi_j=\f{2}{\lambda}Id_j$ correspondant \`a la distribution
$\displaystyle \phi(g)=\f{2}{\lambda}\delta(g)$,
autour de laquelle il serait tr\`es int\'eressant de perturber. Dans ce cas-l\`a,
la contribution des faces est soit $N$ soit $-N$ et on n'introduit que des signes
suppl\'ementaires par rapport \`a la th\'eorie topologique.}

Dans ce contexte, il est \'egalement naturel de modifier l'alg\`ebre
des ``\'etats fronti\`eres'' pour prendre en compte la solution classique.
En effet, on pourrait regarder le terme ${\rm Tr}M^2M_0$ comme un nouveau
vertex trivalent et non comme une correction au propogateur. De ce point de vue,
on obtient sur la fronti\`ere des termes en $M$ mais aussi en $M_0$.
Math\'ematiquement, on consid\`ere des observables du type ${\rm Tr}M_0^iM^j$
et les corr\'elations correspondantes
$$
W_{M_0}=\int[dM] {\rm Tr}M_0^iM^j e^{-S_{M_0}[M]}.
$$
Ces \'etats fronti\`eres o\`u se m\^elent le champs/matrice $M$ et la solution
classique $M_0$ peuvent s'interpr\'eter comme des perturbations de la g\'eom\'etrie
autour de $M_0$. Ce point de vue se g\'en\'eralise aux mod\`eles en dimension sup\'erieure.
Par exemple, en dimension 3, notant $\phi$ le champ g\'en\'erateur, la th\'eorie
autour de $\phi_0$ contient des termes en ${\rm Tr}\phi_0\phi^3$
qui se peuvent s'interpr\'eter comme des
corrections au propagateur mais uniquement en termes de nouveaux vertex m\^elant $\phi$
et la solution classique $\phi_0$.

Reste \`a savoir si tout cela permettra d'avoir des informations suppl\'ementaires
sur les int\'egrales de chemin (son comportement asymptotique par exemple)
et si on pourra utiliser les $M_0$ pour construire
des g\'eom\'etries de fond int\'eressantes\dots

\section{G\'eom\'etrie du Mod\`ele de Ponzano-Regge}

Le mod\`ele de Ponzano-Regge est un mod\`ele de mousse de spin quantifiant
la gravit\'e Euclidienne en 3d, version $\su$ (au lieu de $\slr$) de la th\'eorie
que j'ai d\'ecrite dans le chapitre \ref{lqg3d}.
On peut le d\'efinir \`a la fois comme une quantification g\'eom\'etrique
d'une triangulation de l'espace-temps ou comme une discr\'etisation de la
gravit\'e 3d (qui est une th\'eorie topologique de type BF!). On
peut aussi le voir comme une version dynamique de la \lqg en trois dimensions,
d\'ecrivant la structure de l'espace-temps et d\'efinissant l'\'evolution
des r\'eseaux de spin. Dans ce contexte, le mod\`ele d\'efinit un projecteur sur
les \'etats physiques solutions de toutes les contraintes (y compris
la contrainte Hamiltonienne). Enfin, on peut le d\'eriver \'egalement d'une th\'eorie
de champs g\'en\'eratrice dont j'\'etudierai les solutions classiques et
tenterai de leur donner une interpr\'etation physique.

\subsection{D\'efinition du mod\`ele}

Le mod\`ele de Ponzano-Regge est un mod\`ele de g\'eom\'etrie simplicial:
on construit l'espace-temps  (tri-dimensionel) en collant des t\'etra\`edres ensemble
et on associe une amplitude \`a chaque configuration.
Plus pr\'ecis\'ement,  une repr\'esentation $V^j$ de $\su$ peut \^etre
consid\'er\'ee comme un vecteur quantifi\'e de norme $j$ \footnotemark
(voir par exemple \cite{bb}),
ou intuitivement comme la classe d'\'equivalence d'un vecteur de norme $j$ sous
transformation $\su$ (donc sous les rotations 3d).
Dans le contexte de la triangulation, il est par cons\'equent naturel d'associer
\`a chaque c\^ot\'e $e$ de la triangulation une repr\'esentation $V^{j_e}$ de $\su$,
$j_e$ correspondant \`a sa longueur (tout comme dans le cadre de la \lqg 3d).
Il s'agit ensuite d'associer une amplitude \`a chaque triangulation d\'efinie
par l'ensemble de ces repr\'esentations $\{j_e\}$. Pour cela,
on d\'efinit le mod\`ele localement et on attribue des poids \`a chaque
c\^ot\'e/lien, triangle/face et t\'etra\`edre.

Le poids associ\'e \`a un c\^ot\'e est simplement
la dimension de la repr\'esentation associ\'ee $\Delta_j=(2j+1)$.
$\Delta_j$  correspond aux nombres d'\'etats $|m\rangle \in V^j$ possibles
du c\^ot\'e/vecteur ayant fix\'e sa longueur $j$.
Le poids d'un triangle $t$
est une fonction des trois repr\'esentations  $j_1,j_2,j_3$ attach\'ees
\`a ses c\^ot\'es.
Si il existe un entrelaceur entre ces trois repr\'esentations
(i.e. le coefficient de Clebsch-Gordan est non nul), alors ce poids
est 1, sinon il est simplement 0. Ceci correspond au nombre
de triangles possibles (\`a rotation pr\`es) \'etant donn\'ee la longueur
des trois c\^ot\'es. Math\'ematiquement, ce poids peut \^etre d\'efini
par l'\'evaluation du  graphe $\Theta$ labell\'e par les
trois repr\'esentations  $j_1,j_2,j_3$. C'est le produit
$\Cm{1}{2}{3}\Cm{1}{2}{3}$
de deux coefficients de Clebsch-Gordan.
Finalement, on associe \`a chaque t\'etra\`edre $tet$ le symbole $\{6j\}$
d\'ependant des six repr\'esentations des c\^ot\'es de $tet$.
Ce symbole est d\'efini par le produit  des entrelaceurs
(coefficients de Clebsch-Gordan) associ\'es aux 4 triangles de $tet$:
\beq
\{6j\}=\sum_{m_i}\Cm{1}{2}{3}
\Cm{3}{4}{5} \Cm{5}{2}{6} \Cm{6}{4}{1}.
\eeq

\footnotetext{Il y a en fait une ambigu\"\i t\'e dans la norme exacte, certains
pr\'econisent  $j$, $j+1/2$ ou $\sqrt{j(j+1)}$\dots}

En multipliant tous ces poids, on obtient alors une amplitude pour la configuration
de la triangulation (ensemble des t\'etra\`edres recoll\'es) d\'efinie par les $j_e$.
Et la fonction de partition pour une triangulation $T$ (sans bord)
donn\'ee est:
\beq
Z(T)\,=\,\mathcal{N}_T\left(\prod_e\sum_{j_{e}}\Delta_{j_e}\right)\prod_{t}
\Theta_t(j)\prod_{tet}\{6j\}_{tet} \label{eq:PR}
\eeq
o\`u $\mathcal{N}_T$ est un  facteur (combinatoire) ne d\'ependant que
de la triangulation (et non des labels $j_e$).

\medskip

On peut voir ce mod\`ele comme une discr\'etisation de la th\'eorie BF topologique,
d\'efinie sur les variables connexion $\om$ et multiplicateur de Lagrange $B$.
Cette th\'eorie BF est exactement la gravit\'e tri-dimensionnelle avec le champ $B$
\'etant simplement la triade $E$
(voir partie III pour plus de d\'etails).
Au niveau de l'int\'egrale de chemin (pour la fonction de partition par exemple),
en int\'egrant sur le champ $B$, il ne reste que l'int\'egration sur les configurations
de $\om$ contrainte \`a \^etre de courbure nulle $F=0$:
$$
Z=\int [dB]\int[d\om]\, e^{i\int{\rm Tr}B\w F}=
\int[d\om]\, \delta(F[\om]).
$$
Il est alors facile de discr\'etiser cette int\'egrale\footnote{On peut \'egalement
discr\'etiser directement la th\'eorie BF, puis effectuer l'int\'egrale sur le champ
$B$ seulement au niveau discret. On retrouve le m\^eme r\'esultat final, mais
cela permet de mieux suivre ce qui advient des sym\'etries continues (comme
l'invariance sous diff\'eomorphismes) au niveau discret \cite{diffeo}.}.
Pour cela pla\c cons-nous sur le {\it dual de la triangulation}.
On fait donc correspondre \`a chaque t\'etra\`edre un vertex (repr\'esentant ainsi
un \'ev\`enement dans l'espace-temps), \`a chaque triangle correspond un lien (1d)
reliant les deux t\'etra\`edres auxquels il appartient,
\`a chaque c\^ot\'e une surface appel\'ee {\it plaquette} d\'elimit\'ee par une boucle
reliant les t\'etra\`edres partageant ce c\^ot\'e, et enfin \`a chaque point un volume
3d ou {\it bulle}. Ce dual est la mousse de spin elle-m\^eme: l'espace-temps
est fait de bulles dont les surfaces (faites des plaquettes) sont labell\'ees
par des repr\'esentations de $\su$ (en effet, les $V^j$ sont attach\'ees aux liens, donc
aux plaquettes duales).
\begin{figure}
\begin{center}
\includegraphics[width=8cm]{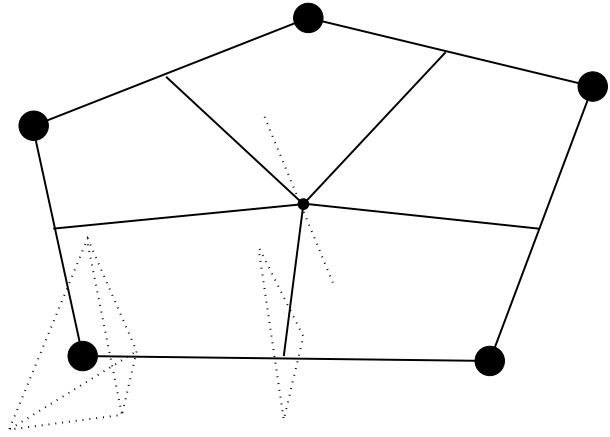}
\caption{Une {\it plaquette} (dual d'un c\^ot\'e de la vari\'et\'e simpliciale)
divis\'ee en {\it wedges}, avec ses liens limites (duaux aux triangles)
et ses vertex limites (duaux aux tetra\`edres).}
\end{center}
\end{figure}
On commence par discr\'etiser la connexion $\om$. On la remplace par
une connexion ``discr\`ete'': des \'el\'ements $g_l$ de $\su$ associ\'es
\`a chaque lien dual $l$. Ils sont ainsi rattach\'es chacun \`a un triangle
et d\'ecrivent le transport parall\`ele d'un t\'etra\`edre \`a un autre \`a travers
le triangle (d\'ecrivant donc le changement de rep\`ere
d'un t\'etra\`edre \`a l'autre).
Il est alors normal de discr\'etiser la 2-forme courbure $F[\om]$ sur des surface, ici
plaquettes. En effet, on peut remplacer $F[\om]$ sur une plaquette/face
$f$
par l'holonomie de la connexion autour de cette plaquette
multipliant les \'el\'ements de groupe $g_l$ des liens sur la fronti\`ere
de la plaquette. Alors la condition $F=0$ se traduit par contraindre
l'holonomie \`a \^etre l'identit\'e sur $\su$. Ainsi on discr\'etise l'int\'egrale
de chemin de la th\'eorie BF:
\beq
Z=\left(\prod_l\int dg_l\right)\,\prod_f
\delta\left(\overrightarrow{\prod_{l\in \partial f}} g_l\right).
\label{PRconnect}
\eeq
On peut alors ``d\'evelopper'' les fonctions delta en utilisant la formule
de Peter-Weyl sur $\su$:
$$
\delta(g)=\sum_j\Delta_j\chi_j(g),
$$
o\`u $\chi_j$ est le caract\`ere de la repr\'esentation $j$ (la trace sur $V^j$).
Ceci associe une repr\'esentation \`a chaque face (duale) $f$ i.e. \`a chaque
c\^ot\'e $e$ de la triangulation. Un calcul rapide permet alors de retrouver
la m\^eme fonction de partition que pr\'ec\'edemment.

\subsection{Ponzano-Regge comme un projecteur}

Etudions maintenant les propri\'et\'es de ce mod\`ele et regardons comment
s'en servir pour calculer des amplitudes de transition en gravit\'e quantique (3d).
Tout d'abord, la propri\'et\'e fondamentale du mod\`ele de Ponzano-Regge est que
c'est une {\it th\'eorie topologique}\footnote{En fait, les amplitudes d\'efinies
sont infinies et le mod\`ele n'est pas bien d\'efini en g\'en\'eral. N\'eanmoins,
on peut le r\'egulariser soit par cutoff, soit en utilisant un groupe quantique
$\su_q$ (mod\`ele de Turaev-Viro), dans ces cas, on obtient un mod\`ele topologique
bien d\'efini.}: les amplitudes calcul\'ees ne d\'ependent pas des d\'etails de
la triangulation fix\'ee mais de sa topologie uniquement.

Plus pr\'ecis\'ement, il existe des \'egalit\'es ``pentagonales'' exprimant le produit
de 2 symboles $\{6j\}$ en fonction de la somme de produits de 3 symboles $\{6j\}$,
et similairement un seul symbole $\{6j\}$ en fonction de la somme de produits de
4 symboles $\{6j\}$. Ces identit\'es peuvent \^etre traduites en terme d'invariance
de la fonction de partition sous les mouvements de Pachner $2\lrarr 3$ et $1\lrarr 4$,
rempla\c cant respectivement 2 t\'etra\`edres par 3 t\'etra\`edres (gardant la m\^eme 
fronti\`ere) et 1 t\'etra\`edre par 4 t\'etra\`edres, et r\'eciproquement.

Cette propri\'et\'e exprime l'invariance du mod\`ele sous raffinement
de la triangulation, ce qui refl\`ete que le mod\`ele discret d\'ecrit
exactement la th\'eorie (quantique) continue et non pas seulement une approximation.

\medskip

Introduisons des bords/fronti\`eres dans le mod\`ele. Du point de vue de la
triangulation, on peut d\'efinir le bord d'une vari\'et\'e simpliciale tri-dimensionnelle
par une vari\'et\'e simpliciale bi-dimensionnelle constitu\'ee de triangles coll\'es
les uns aux autres. Dans le mod\`ele de Ponzano-Regge, les c\^ot\'es des triangles
de la fronti\`ere sont labell\'es par des spins $j$.
Du point de vue dual, si on prend une tranche 2d de la mousse de spin
(duale \`a la triangulation), on obtient un r\'eseau de spin labell\'e par
des repr\'esentations $j$ et avec des vertex trivalents. Ces vertex trivalents peuvent 
\^etre consid\'er\'es comme duaux \`a des triangles, alors le r\'eseau de
spin s'interpr\`ete comme dual \`a une vari\'et\'e simpliciale 2d et tout est
bien coh\'erent. La fonction de partition d'une vari\'et\'e avec bords
est donn\'ee par exactement la m\^eme formule que pr\'ec\'edemment sauf
que nous ne sommons pas sur les spins/repr\'esentations $j$ externes.

Ayant introduit des fronti\`eres, on peut alors interpr\'eter le mod\`ele
de Ponzano-Regge comme nous fournissant des amplitudes de transition entre vari\'et\'es
bi-dimensionnelles.
En fait, il d\'efinit un projecteur, projetant les \'etats cin\'ematiques
(r\'eseaux de spin labell\'ees par des repr\'esentations arbitraires) sur des \'etats
physiques solutions de toutes les contraintes de la gravit\'e 3d
(voir par exemple \cite{ooguri}).
Plus pr\'ecis\'ement, 
consid\'erons une 3-vari\'et\'e $M$ et d\'ecomposons-la en trois parties
$M_1$, $M_2$ et $N$, avec $N$ ayant la topologie d'un cylindre
$\Sigma\times[0,1]$ ($\Sigma$ compact) et les bords $\partial M_1$ et $\partial
M_2$ \'etant  isomorphes \`a $\Sigma$. La fonction de partition de Ponzano-Regge
peut s'\'ecrire:
\beq
Z_M\,=\,\mathcal{N}_T\,\sum_{j_e\in\Delta_1,j_{\tilde{e}}\in\Delta_2}\,Z_{M_1,\Delta_1}(j_e)\,P_{\Delta_1,\Delta_2}(j_e,j_{\tilde{e}})\,Z_{M_2,\Delta_2}(j_{\tilde{e}}),
\eeq
en choisissant une triangulation de $M$ telle qu'aucun t\'etra\`edre
ne soit partag\'e par deux des r\'egions $M_1$, $M_2$ ou $N$.
Ici $\Delta_i$ d\'enote une triangulation des fronti\`eres.
La somme sur les spins attach\'es aux liens internes \`a
$M_1$, $M_2$ et $N$ est implicite dans la d\'efinition des fonctions
$Z_{M_i,\Delta_i}$ et $P_{\Delta_1,\Delta_2}$.

On a alors la relation suivante  grace \`a l'invariance topologique du mod\`ele:
\beq
\mathcal{L}_{\Delta_2}\sum_{j_{\tilde{e}}\in\Delta_2}\,P_{\Delta_1,\Delta_2}(j_e,j_{\tilde{e}})\,P_{\Delta_2,\Delta_3}(j_{\tilde{e}},j'_e)\,=\,P_{\Delta_1,\Delta_3}(j_e,j'_e),
\eeq
o\`u $\mathcal{L}$ est une constante de normalisation d\'ependant uniquement
de la triangulation $\Delta_2$.
Cela permet de d\'efinir un op\'erateur de projection agissant sur
les \'etats de r\'eseaux de spin $\phi_\Delta$ d\'efinis sur la fronti\`ere
d'une triangulation:
\beq
\mathcal{P}[\phi_\Delta](j)\,=\,\mathcal{L}_\Delta\,\sum_{j'}\,P_{\Delta,\Delta}(j,j')\,\phi_\Delta(j').
\eeq
Il est facile de v\'erifier que
$\mathcal{P}\cdot\mathcal{P}=\mathcal{P}$, et nous pouvons \'ecrire la fonction
de partition comme:
\beq
Z_M\,=\,\mathcal{N}_T\,\sum_{j_e\in\Delta_1,j_{\tilde{e}}\in\Delta_2}\,\mathcal{P}[Z_{M_1,\Delta_1}](j_e)\,P_{\Delta_1,\Delta_2}(j_e,j_{\tilde{e}})\,\mathcal{P}[Z_{M_2,\Delta_2}](j_{\tilde{e}}).
\eeq
Cela nous permet de d\'efinir les {\it \'etats physiques} de la th\'eorie
comme ceux satisfaisant:
\beq
\phi_\Delta\,=\,\mathcal{P}[\phi_\Delta] \label{eq:proj}
\eeq
qui est l'analogue de l'\'equation de Wheeler-DeWitt.
On en d\'eduit le produit scalaire physique:
\beq
\langle \phi_\Delta \mid
\phi'_\Delta\rangle_{\rm phys}\,=\,\sum_{j,j'\in\Delta}\,\phi_\Delta(j)\,P_{\Delta,\Delta}(j,j')\,\phi'_\Delta.
\eeq
Ainsi, les fonctions $Z_{M_1}$ et $Z_{M_2}$ sont solutions de
l'\'equation \Ref{eq:proj} et la fonction de partition de $M$
donne leur produit scalaire! Dans ce contexte, il est possible
de v\'erifier que le projecteur projete bien les r\'eseaux de spin
sur le secteur $F=0$ et que donc le mod\`ele de Ponzano-Regge
impl\'emente bien la gravit\'e quantique Euclidienne en trois dimensions.

\medskip

Il existe une autre mani\`ere de comprendre la projection r\'ealis\'ee par le mod\`ele
en utilisant la construction GNS.
Tout d'abord, dans un formalisme connexion, les \'etats fronti\`eres sont d\'efinis
par des fonctionnelles cylindriques. Plus pr\'ecis\'ement, on d\'efinit les donn\'ees
g\'eom\'etriques du bord par une connexion (discr\`ete) sur le graphe fronti\`ere.
Pour \^etre plus concret, consid\'erons un graphe fronti\`ere donn\'e $\Gamma$ et
num\'erotons ses liens par $i=1,..,N$ et ses vertex par $v$. Un \'etat fronti\`ere,
not\'e $\phi$, est
une fonction cylindrique d\'ependant d'\'el\'ements de groupe $g_i$
et invariante sous $\su$ aux vertex:
$$
\forall k_v\in\su,\,
\phi(g_i)=\phi(k_{s(i)}g_ik^{-1}_{c(i)}),
$$
avec les notations \'evidentes pour les vertex source et cible.
On d\'efinit la mesure de Haar $\mu$ sur $\su^N$
pour int\'egrer ces fonctions.

Ces \'etats fronti\`eres forment une alg\`ebre ${\cal A}$. 
Pour appliquer la construction GNS pour les projeter sur les ``\'etats physiques'',
consid\'erons la fonctionnelle lin\'eaire positive:
\beq
w(\phi)=\mu(a\phi)=\int_{SU(2)^N}\prod_i{dg_i}\,
z(g_i)\phi(g_i)
\eeq
o\`u $z(g_i)$ est la fonction de partition de la triangulation/mousse de spin
avec fronti\`ere d\'efinie par les $g_i$. Elle est d\'efinie \`a partir de la formule
\Ref{PRconnect} valable pour une vari\'et\'e sans fronti\`ere en refermant
les faces/plaquettes ouvertes \`a l'aide de la connexion sur la fronti\`ere:
\beq
z(g_e)=\left(\prod_l\int dg_l\right)\,
\prod_{f\textrm{ interne}} \delta\left(\overrightarrow{\prod_{l\in f}} g_l\right)
\prod_{f\textrm{ ouverte}}
\delta\left(\overrightarrow{\prod_{l\in f}} g_lg_{i\in\dd f}\right).
\eeq
Alors, quotientant par l'id\'eal de Gelfand:
\beq {\cal I}=\{\phi \,|\,
w(\bar{\phi}\phi)=0 \}, \eeq
$w$ d\'efinit un produit scalaire sur ${\cal A}/{\cal I}$
et l'espace d'Hilbert sera le compl\'et\'e de ${\cal A}/{\cal I}$ pour la
norme d\'efinie par $w$.

\begin{figure}
\begin{center}
\includegraphics[width=7cm]{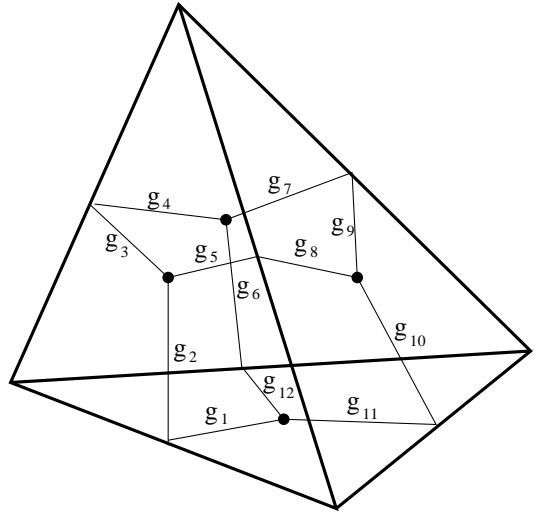}
\caption{Un t\'etra\`edre avec le dual de la fronti\`ere
et les variables de connexion sur la fronti\`ere.}
\end{center}
\end{figure}

Dans le cas du t\'etra\`edre, le graphe fronti\`ere est le graphe trivalent \`a 4
noeuds et 6 liens. Les \'etats fronti\`eres sont des fonctions invariantes
de six \'el\'ements $g_1,..,g_6$. Un calcul rapide donne:
\beq
w(\phi)=\phi(g_i=Id),
\eeq
et donc les \'etats physiques sont donn\'es par les connexions plates 
$g_i=Id$ comme pr\'evu.

\bigskip

Un point important dans la construction d'un projecteur \`a partir du
mod\`ele de Ponzano-Regge est que, tout comme dans les int\'egrales de chemins
formelles pour la gravit\'e, nous avons int\'egr\'e sur les deux signes
du temps propre. Ceci tue la ``causalit\'e''.
On peut voir cela dans l'asymptotique des symboles $\{6j\}$ qui, dans la limite
de grands $j$, tend vers $e^{+iS}+e^{-iS}$ o\`u $S$ est l'action de Regge -discr\'etisation
de la gravit\'e. Les deux termes correspondent \`a des signes de {\it lapse} oppos\'es
et donc \`a des configurations \`a direction temps oppos\'ee.
A un niveau plus profond, cette sym\'etrie se lit dans l'invariance 
de la fonction de partition sous changement d'orientation $g\lrarr g^{-1}$
des \'el\'ements de groupe, qui est due \`a l'invariance des caract\`eres $\chi_j$.
Math\'ematiquement, cela correspond au groupe de Weyl $\Z_2$ de $\su$.

Int\'egrer sur les deux signes du temps propre est 
tout-\`a-fait normal si on veut obtenir un projecteur comme on peut le voir
sur l'exemple de la particule relativiste (voir par exemple \cite{causal1}).
Par contre, si on veut r\'etablir le temps, une structure causale et d\'ecrire
une \'evolution (unitaire) des structures dans le temps, il faut
briser cette sym\'etrie $\Z_2$.

Ce probl\`eme est li\'e directement au domaine d'int\'egration de l'int\'egrale
de chemin. En effet, l'action
$$
S[A,E]=\int{\rm Tr}E\w F(A)
$$
de la gravit\'e 3d est anti-sym\'etrique en $E$. Classiquement, on retrouve la
relativit\'e g\'en\'erale 3d dans le secteur $det(E)>0$. Cependant, la th\'eorie
BF quantique est d\'efinie par l'int\'egrale de chemin sur les deux signes de $E$
et donc de $det(E)$:
$$
Z_{BF}=\int[dE]\,e^{i\int{\rm Tr}E\w F(A)}=
\int_{det(E)>0}[dE]\,\left(e^{i\int E\w F(A)}
+e^{-i\int E\w F(A)}\right)+
\int_{det(E)=0}[dE]\,e^{i\int E\w F(A)}.
$$
N\'egligeant le secteur d\'eg\'en\'er\'e, on voit bien que cette fonction
de partition est diff\'erente de la fonction de partition de la
relativit\'e g\'en\'erale o\`u on int\'egrerait seulement sur les configurations
$det(E)>0$ \cite{laurent:vol3d}.
Int\'egrer sur les deux signes de $det(E)$ effacent la structure
causale de l'espace-temps g\'en\'er\'e car les deux signes de $det(E)$ correspondent
\`a des directions temps oppos\'ees. Et c'est ce m\^eme ph\'enom\`ene qui se refl\`ete
dans le mod\`ele discr\'etis\'e d\'efini par Ponzano-Regge.

En principe, il est possible de distinguer les deux images mirroirs (en coupant
les caract\`eres en deux par exemple), mais il n'y a pas vraiment de motivations
pour imposer une fl\`eche temps dans un mod\`ele Euclidien. Par contre, regardons le
mod\`ele de Ponzano-Regge Lorentzien \cite{davids,laurent:2+1}.
Un tel mod\`ele est fond\'e sur les repr\'esentations unitaires de $SO(2,1)$,
explicit\'ees dans les parties pr\'ec\'edentes. Il existe alors un mod\`ele
topologique utilisant la s\'erie positive discr\`ete de repr\'esentations. Il est
int\'eressant de noter que la sym\'etrie $\Z_2$ est alors bris\'ee et que
les caract\`eres sont de simples exponentielles: $SO(2,1)$ distingue naturellement
deux directions du temps et le mod\`ele Lorentzien h\'erite une fl\`eche temporelle
intrins\`eque. Dans ce sens, il semble donc que le mod\`ele Lorentzien est
une bonne quantification directe de la gravit\'e (tenant compte seulement
d'un secteur du d\'eterminant).

Dans le cas Lorentzien quadri-dimensionnel, le
groupe de base sera $\slc$, et son groupe de Weyl  $\Z_2$, qui ne permet pas de distinguer
naturellement pass\'e et futur. Il n'existe pas encore d'extension naturelle utilisant
un autre groupe permettant une telle extension, et pour r\'etablir la causalit\'e
du mod\`ele, on est oblig\'e de couper le mod\`ele en deux \`a la main, se laissant guider
par l'intuition du mod\`ele de Ponzano-Regge Lorentzien.
Cela sera \'etudi\'e
en d\'etails dans le prochain chapitre.

\subsection{Th\'eorie de champs g\'en\'eratrice}

Il existe une th\'eorie de champs sur $\su$ g\'en\'erant le mod\`ele de
Ponzano-Regge \cite{boulatov}. Tout comme dans le cas 2d, ses diagrammes
de Feynman reproduisent des vari\'et\'es simpliciales\footnote{En fait, les diagrammes
de Feynman donnent  des 2-complexes qui ne sont duaux \`a des triangulations
que localement. En g\'en\'eral, cela fournit des vari\'et\'es \`a singularit\'e
conique. Plus intuitivement, on remarque que les bulles (duaux aux points) ne sont
plus clairement s\'epar\'ees mais s'interp\'en\`etrent, ce qui rend floue
la d\'efinition des points de la triangulation.} habill\'es par des
repr\'esentations de $\su$.

On d\'efinit un champ r\'eel $\phi$ sur $\su^3$ 
invariant sous permutation de ses arguments et invariant sous $\su$:
\beq
\forall g\in\su,\, \phi(g_1,g_2,g_3)=\phi(g_1g,g_2g,g_3g).
\eeq
Intuitivement, c'est un champ sur les triangles de $\R^3$ (triangles
intrins\`eques i.e. classes d'\'equivalence de triangles sous les isom\'etries
de $\R^3$). Passant \`a la transform\'ee de Fourier, $\phi$ s'\'ecrit:
\beq
\phi(g_1,g_2,g_3)=\phi^{j_1j_2j_3}_{a_1a_2a_3}
\prod_{i=1}^3D^{(j_i)}_{a_ib_i}(g_i)
\Cg{1}{2}{3}{b}.
\eeq

L'action de la th\'eorie g\'en\'eratrice est alors:
\beqs
S[\phi]&=&\f{1}{2}\int_{\su^3}dg_1dg_2dg_3\,
|\phi(g_1,g_2,g_3)|^2 \nonumber \\&&-\f{\lambda}{4!}
\int_{\su^6} \prod_1^6 dg_i\,
\phi(g_1,g_2,g_3)\phi(g_3,g_4,g_5)\phi(g_5,g_2,g_6)\phi(g_6,g_4,g_1).
\eeqs
Les r\`egles de Feynman  d\'efinissent un vertex d'interaction dual
\`a un t\'etra\`edre. Et on peut coller ces vertex d'interaction \`a l'aide
de propagateurs, qui repr\'esentent les triangles.
On retrouve alors la fonction de partition d\'ecrit pr\'ec\'edemment quand
$\lambda=1$. Pour $\lambda$ arbitraire, la th\'eorie d\'efinit n'est plus
exactement topologique et on obtient des facteurs de $\lambda$ exposant
le nombre de t\'etra\`edres de la triangulation. De ce point de vue,
$\lambda$ permet de contr\^oler le volume de la triangulation: par exemple les
petites triangulations sont favoris\'ees quand $\lambda$ est petit.

\medskip

Comme dans le cas des mousses de spin 2d, je vais m'int\'eresser aux solutions
classiques de la th\'eorie. L'\'equation du mouvement s'\'ecrit:
\beq
\phi_0(g_1,g_2,g_3)=\f{\lambda}{3!}
\int dg_4dg_5dg_6\,
\phi_0(g_3,g_4,g_5)\phi_0(g_5,g_2,g_6)\phi_0(g_6,g_4,g_1).
\eeq
Passant \`a la transform\'ee de Fourier, cette \'equation s'\'ecrit:
\beq
\phi^{j_1j_2j_3}_{a_1a_2a_3}=\f{\lambda}{3!}
\f{\{6j\}}{\Delta_{j_4}\Delta_{j_5}\Delta_{j_6}}
\phi^{j_3j_4j_5}_{a_3a_4a_5}\phi^{j_5j_2j_6}_{a_5a_2a_6}\phi^{j_6j_4j_1}_{a_6a_4a_1}.
\eeq
D'o\`u la solution \'evidente pour $\phi$ donn\'ee par l'inverse des
Clebsh-Gordan. Mais il existe certainement d'autres solutions. L'action \'evalu\'ee
sur une solution vaut:

$$
S[\phi_0]=\f{1}{4}\int dg_1dg_2dg_3\, \phi_0(g_1,g_2,g_3)^2.
$$
Comme dans le cas des mousses
de spin 2d, on pourra les interpr\'eter comme des \'etats coh\'erents de triangles.
En effet, un tel \'etat $\phi_0$  invariant sous mouvement de
Pachner $1\lrarr3$ est un triangle
\'equivalent \`a trois triangles, et ainsi de suite: c'est
une configuration du champ auto-raffinante.

On peut consid\'erer comme \'etat de la th\'eorie les fonctionnelles du champ $\phi$.
Tout comme dans le cas 2d, on peut introduire la relation d'\'equivalence
d\'efinie par l'\'egalit\'e sur toutes les solutions classiques. Ce seront
des fonctions invariantes sous le mouvement de
Pachner $1\lrarr3$ mais pas forc\'ement sous le mouvement $2\lrarr2$.
Par similarit\'e avec le cadre des mousses de spin 2d, on pourrait consid\'erer ces
classes d'\'equivalence comme les solutions de la th\'eorie quantique canonique.
Mais dans ce cas, puisque nous n'avons pas invariance sous le mouvement $2\lrarr2$,
cela ne projete pas sur les configurations \`a courbure nulle $F=0$. De plus,
physiquement, ces ``\'etats'' ne comportent pas de corr\'elations \`a longues
distances puisqu'elles ne sont possibles que gr\^ace aux mouvements $2\lrarr2$
(non-ultra-locaux comme les mouvements $1\lrarr3$).
On peut trouver dans \cite{pullin} des consid\'erations similaires sur
la dynamique d'une th\'eorie de la gravit\'e 3d ou 4d discr\'etis\'ee.

On peut aussi regarder les perturbations de la th\'eorie autour de ces configurations.
Elles sont d\'efinies par l'action:
\beq
S_{\phi_0}[\phi]=S[\phi_0+\phi]-S[\phi]=
\f{1}{2}\int\phi^2-\f{\lambda}{2}\int\phi^2\phi_0^2
-\f{\lambda}{4!}\int\phi^4-\f{\lambda}{3!}\int\phi^3\phi_0
\eeq
avec une modification \`a la fois du propagateur et du vertex d'interaction.
Cela permet d'\'etudier la th\'eorie autour d'un ``vide'' non-trivial
(une g\'eom\'etrie de fond non nulle).

\subsection{Conclusion: extension \`a la th\'eorie BF en dimension 4}

Les construction pr\'ec\'edentes peuvent \^etre g\'en\'eralis\'ees
\`a toute dimension et r\'esultent en des mod\`eles
de mousse de spin quantifiant la th\'eorie BF topologique.

Examinons la situation en quatre dimensions.
On consid\`ere une vari\'et\'e simpliciale 4d consistant en des 4-simplex
coll\'es ensemble. La mousse de spin est le 2-complexe dual.
Chaque 4-simplex est fait de 10 triangles group\'es
en 5 t\'etra\`edres.
On associe \`a chaque triangle une repr\'esentation de $Spin(4)$
dans le cas Euclidien ou de $\slc$ dans le cas Lorentzien.
Chaque t\'etra\`edre est d\'efini par un entrelaceur 4-valent entre
les 4 repr\'esentations de ses triangles.
$Spin(4)$ se d\'ecompose en $SU(2)\times SU(2)$ et ses repr\'esentations
 irr\'eductibles sont labell\'es par deux demi-entiers
 $J=(j,k)$ correspondant aux deux sous-groupes $\su$.
Une base d'une telle repr\'esentation est fournie par la base usuelle
des repr\'esentations de $\su$ et est labell\'ee par un couple de demi-entiers
$M=(m,n)$. Il n'y a pas un unique entrelaceur 4-valent
\'etant donn\'ees 4 repr\'esentations, mais les entrelaceurs forment un espace
d'Hilbert. Une base est donn\'ee par la d\'ecomposition 
de l'entrelaceur en deux entrelaceurs 3-valents.
Comme les entrelaceurs 3-valents sont uniques
(coefficients de Clebsh-Gordan \`a une normalisation),
la base est labell\'ee par une repr\'esentation de $Spin(4)$ interm\'ediaire.
On note ces entrelaceurs (orthonorm\'es) $\Cq{1}{2}{3}{4}{M}{J}$ avec $J$
la repr\'esentation interm\'ediaire.

Finalement, un 4-simplex est fait de 10 triangles labell\'es par 10 repr\'esentations
et de 5 t\'etra\`edres labell\'es chacun par un entrelaceur.
On associe alors \`a chaque 4-simplex l'unique amplitude invariante
fabriquable \`a partir de ces 15 repr\'esentations. C'est
le symbole $\{15j\}$ de $Spin(4)$, obtenu en contractant
les 10 repr\'esentations avec les 5 entrelaceurs.

Regardons plus pr\'ecis\'ement ce qui se passe au niveau des t\'etra\`edres,
on leur associe un diagramme {\it oeil} labell\'e par les 4 repr\'esentations
$J_1,..,J_4$ des triangles et par deux repr\'esentations ``internes'' $J$ et $K$
d\'ecrivant les entrelaceurs attribu\'es au t\'etra\`edre vu des deux 4-simplex
qui le partagent. Contractant les deux entrelaceurs, l'\'evaluation du diagramme
{\it oeil} vaut 1 si $J=K$ et 0 sinon, ce qui correspond au nombre d'\'etats quantiques
du t\'etra\`edre. On remarque que le t\'etra\`edre est le m\^eme vu des deux 4-simplex.

Comme en 3d, on pourra reconstruire tout cela par une discr\'etisation
de la fonction de partition de la th\'eorie BF et on pourra
\'egalement suivre la m\^eme construction d'un projecteur que dans le cas 3d.

\medskip

A ce point, il est n\'ecessaire de souligner que le 4-simplex construit
tel quel \`a partir de la th\'eorie BF n'est pas {\it g\'eom\'etrique},
dans le sens qu'il n'y a aucune relation entre le champ $B$ de la th\'eorie
et la m\'etrique. Il est par contre possible d'imposer une telle relation
et de construire un 4-simplex g\'eom\'etrique en contraignant les repr\'esentations
qui le labellent \cite{bc1}. Cela m\`ene au mod\`ele de Barrett-Crane
que je discute dans la prochaine partie.
Alors, la repr\'esentation interne d'un t\'etra\`edre sera diff\'erente vu
des deux 4-simplex et l'\'evaluation du diagramme {\it oeil} ne sera plus simplement
0 ou 1 mais on obtiendra un espace d'\'etats quantiques du t\'etra\`edre beaucoup
plus riche \cite{bb}.

\chapter{La g\'eom\'etrie quantique selon Barrett-Crane}

Le mod\`ele de Barrett-Crane est un mod\`ele de mousse de spin 
quantifiant la relativit\'e g\'en\'erale. D'une part, on peut le
voir comme une quantification g\'eom\'etrique de 4-simplex coll\'es par la suite
pour construire un espace-temps, tout comme le mod\`ele de Ponzano-Regge
pouvait \^etre consid\'er\'e comme la th\'eorie d'un t\'etra\`edre quantique.
D'ailleurs ce point de vue se refl\`ete dans la possibilit\'e d'\'ecrire
une th\'eorie de champs g\'en\'eratrice pour le mod\`ele. D'autre part,
on peut d\'eriver le mod\`ele \`a partir de la th\'eorie BF. En effet, il
existe une reformulation de la relativit\'e g\'en\'erale sous la forme
d'une th\'eorie BF contrainte (les contraintes sont d'ailleurs les
contraintes de seconde classe \Ref{lqg:2nde}). Dans ce contexte,
la quantification de la th\'eorie BF en n'importe quelle dimension est connue et
s'effectue d'une mani\`ere parall\`ele au cas tri-dimensionel. Ensuite, il
s'agit d'imposer les contraintes au niveau quantique pour obtenir un mod\`ele
quantifiant l\'egitimement la relativit\'e g\'en\'erale.

\medskip

Dans un premier temps, je vais rappeler en introduction la quantification d'un 4-simplex
effectu\'ee par Barrett et Crane dans les cas Euclidien et Lorentzien \cite{bc1,bc2},
ce qui permettra de comprendre les objections possibles \`a la pertinence du mod\`ele
en tant que gravit\'e quantique. 

Puis dans une premi\`ere section, j'expliciterai la d\'erivation de la relativit\'e
g\'en\'erale (action de Palatini g\'en\'eralis\'ee)
\`a partir de la th\'eorie BF et je montrerai comment traduire
les contraintes au niveau quantique. Dans ce contexte, je soulignerai le r\^ole
du param\`etre d'Immirzi.

Ensuite, je d\'ecrirai le mod\`ele de Barrett-Crane lui-m\^eme et la
g\'eom\'etrie quantique qu'il d\'efinit. En effet, le mod\`ele n'est plus
topologique et il s'agit de comprendre quels sont vraiment les degr\'es de libert\'e
de la th\'eorie.

Enfin, je terminerai ce chapitre par le probl\`eme de la fl\`eche temps: le mod\`ele
est-il causal et peut-on le comprendre en terme d'op\'erateur \'evolution?

\section{Rappels sur la quantification d'un 4-simplex}

Dans notre cadre, la quantification du 4-simplex se fait \`a partir
de sa description g\'eom\'etrique en fonction de {\it bivecteurs}.
A partir d'un 4-simplex g\'eom\'etrique
on peut construire les 10 bivecteurs $b(t)$ correspondant \`a ses 10 triangles $t$
(on prend le bivecteur $e\w f$ pour un triangle dont les vecteurs c\^ot\'es
sont $e,f,g=-(e+f)$).
R\'eciproquement, un 4-simplex g\'eom\'etrique est uniquement d\'etermin\'e
par ses 10 bivecteurs.
Plus pr\'ecis\'ement,
soit un 4-simplex combinatoire avec 10 bivecteurs correspondant \`a ses 10 triangles
et satisfaisant les conditions suivantes \cite{bc1}:
\makeatletter
\renewcommand{\theenumi}{\arabic{enumi}}
\renewcommand{\labelenumi}{(\theenumi)}
\begin{enumerate}
\item Le bivecteur $b(t)$ change de signe si l'orientation
du triangle $t$ change.
\item Chaque bivecteur $b(t)$ est simple i.e. de la forme $e\w f$.
Cela \'equivaut \`a la condition $\la b,*b\ra=0$.
\item Si deux triangles ont un c\^ot\'e en commun, alors la somme
des deux bivecteurs est encore simple.
\item La somme des quatre bivecteurs  correspondant aux faces d'un t\'etra\`edre
est nulle, prenant comme orientation des triangles celle de la fronti\`ere du
t\'etra\`edre.
\item La configuration des bivecteurs est non-d\'eg\'en\'er\'ee: les bivecteurs
de six triangles partageant un m\^eme vertex sont lin\'eairement ind\'ependant.
\item Dans le cas Euclidien, les bivecteurs peuvent \^etre consid\'er\'es
comme des op\'erateurs lin\'eaires en utilisant la m\'etrique Euclidienne.
Alors pour n'importe quels 3 triangles sur un m\^eme t\'etra\`edre, on suppoe que
${\rm Tr}b_1[b_2,b_3]>0$, l'ordre $1,2,3$ \'etant d\'etermin\'e par l'orientation
de la fronti\`ere  du t\'etra\`edre.
Dans le cas Lorentzien, il faut prendre en compte le genre temps ou
espace du t\'etra\`edre.
Par exemple, dans le cas de t\'etra\`edre de genre espace, on impose la m\^eme
condition que dans le cas Euclidien.
\end{enumerate}
Alors les 10 bivecteurs d\'eterminent un unique 4-simplex g\'eom\'etrique \`a translation
pr\`es (et inversion par rapport \`a l'origine pr\`es).
Remarquons que la condition $(6)$ permet de distinguer une configuration $b(t)$
de son oppos\'ee $-b(t)$ (le signe de la trace change) et de son dual $*b(t)$
(on obtient un 4-simplex d\'eg\'en\'er\'e avec ${\rm Tr}b_1[b_2,b_3]=0$).

Notons que dans le cas Euclidien, la contrainte de simplicit\'e d'un bivecteur $b$
s'exprime facilement en utilisant la d\'ecomposition  de $b$ en parties selfduale
et antiselfduale $b=b^++b^-$ avec $*b^\pm=\pm b^\pm$: un bivecteur est simple
ssi $|b^+|^2=|b^-|^2$.

\medskip

Il s'agit de quantifier le 4-simplex en respectant les conditions
$(1)-(6)$\cite{bc1,bc2,bb}. Commen\c cons par le cas Euclidien.
On effectue une quantification g\'eom\'etrique utilisant
une correspondance entre l'ensemble de bivecteurs
$\Lambda^2 \R^4$ et l'alg\`ebre (son dual) $so(4)^*$:
\begin{equation}
\begin{array}{ccccl}
\theta &:& \Lambda^2 \R^4  & \rightarrow & so(4)^*  \\
 & & e \w f & \rightarrow & \theta(e\w f):l\in so(4) \rightarrow \eta(l e,f)\in\R
\end{array}
\label{correspondance}
\end{equation}
o\`u $\eta$ est la m\'etrique Euclidienne (plate).
Il est important pour la suite de noter une ambigu\"\i t\'e dans la d\'efinition de
cette correspondance. En effet, on peut utiliser l'isomorphisme
$\theta\circ *$ \cite{bb}
et m\^eme plus g\'en\'eralement
$\theta\circ (\alpha*+\beta)$ \cite{bf2}.

Identifiant
$so(4)*$ et $so(4)$, on peut remplacer les bivecteurs par des g\'en\'erateurs
du groupe $SO(4)\sim Spin(4)$ et on se retrouve avec  un g\'en\'erateur $J\in so(4)$
sur chaque triangle. Quantifiant g\'eom\'etriquement \cite{bb}, on se retrouve
avec une repr\'esentation de $Spin(4)$ sur chaque triangle et il s'agit de traduire
les conditions $(1)-(6)$ en des contraintes sur les repr\'esentations.

Les repr\'esentations de $Spin(4)\sim \su\times\su$ sont des couples
de repr\'esentations de $\su$ et sont donc labell\'ees
par des couples de demi-entiers $(j_+,j_-)$.
Tenant compte qu'un bivecteur est simple ssi $\la b,*b\ra=0$, on peut quantifier
la condition de simplicit\'e en $\la J,*J\ra=0$. Ceci est le second Casimir
de $spin(4)$ et se traduit donc en:
\beq
j_+(j_++1)-j_-(j_-+1)=0 \quad\textrm{i.e.}\quad j_+=j_-.
\eeq
Ainsi les {\it repr\'esentations simples} de $Spin(4)$ sont labell\'ees
par un unique demi-entier $j=j_-=j_+$.
Ce sont les seules repr\'esentations de $Spin(4)$ contenant un vecteur
invariant sous $\su$. Nous noterons ainsi $|(j,j)x0\ra$, en reprenant
les notations de la partie pr\'ec\'edente, le vecteur de la repr\'esentation
$(j,j)$ invariant sous le groupe $\su_x$ stabilisateur du vecteur
$x\in{\cal S}^3\sim SO(4)/SO(3)$.

Alors, interpr\'etant un t\'etra\`edre comme un tenseur sur les 4 repr\'esentations
attach\'ees \`a ses faces,
la traduction des conditions $(1)-(4)$
est imm\'ediate:
\begin{enumerate}
\item Changer l'orientation d'un triangle change sa repr\'esentation
en son dual.
\item Chaque repr\'esentation est simple i.e. du type $(j,j)$.
\item La somme de deux bivecteurs se traduit par le produit tensoriel
des repr\'esentations.
Pour deux triangles d'un t\'etra\`edre, le produit tensoriel
des deux repr\'esentations se d\'eveloppe sur les repr\'esentations $(j_-,j_+)$
de $spin(4)$. On impose que le tenseur du t\'etra\`edre soit non-nul
seulement sur les composantes simples $(j,j)$ de la d\'ecomposition.
\item  Le tenseur associ\'e \`a un t\'etra\`edre est un tenseur invariant i.e.
un entrelaceur entre les quatre repr\'esentations attach\'ees \`a ses faces.
\end{enumerate}
Ces conditions d\'eterminent de mani\`ere unique l'entrelaceur
attach\'e \`a un t\'etra\`edre en fonction des 4 repr\'esentations
le labellant \cite{reis:bc}. On l'appelle l'{\it entrelaceur de Barrett-Crane}.
C'est l'unique entrelaceur invariant sous $SO(3)^4$ et peut \^etre consid\'er\'e
comme un entrelaceur sur l'espace $SO(4)/SO(3)\sim{\cal S}^3$, fait qui peut se
g\'en\'eraliser \`a toute dimension\cite{puzio}.
Cette invariance au niveau des entrelaceurs peut \^etre interpr\'et\'ee comme
impl\'ementant l'invariance de la g\'eom\'etrie d'un t\'etra\`edre sous le groupe $SO(3)$.

D\'ecomposant l'entrelaceur sur la base d'entrelaceurs trivalents, il se d\'ecompose
uniquement sur les repr\'esentations simples avec avec coefficients leur mesure de
Plancherel (la dimension des repr\'esentations):
\beq
I^{j_1,j_2,j_3,j_4}_{BC}\equiv \sum_j (2j+1)^2
\doubleY{(j_1,j_1)}{(j_2,j_2)}{(j_3,j_3)}{(j_4,j_4)}{(j,j)}
\eeq
o\`u on peut permuter le r\^ole des repr\'esentations $1,2,3,4$ dans la d\'ecomposition
de l'entrelaceur (il est ``isotrope'').
Il peut \'egalement \^etre \'ecrit comme une int\'egrale sur l'espace homog\`ene
${\cal S}^3$:
\beq
I^{j_1,j_2,j_3,j_4}_{BC}\equiv
\int_{{\cal S}^3} dx\,
\prod_{i=1}^4
\la (j_i,j_i)x0|.
\label{intertwinerBC}
\eeq

Il n'y a pas de traduction de la condition $(5)$.  Mais il est possible
d'\'etudier la condition $(6)$. Cette contrainte est importante car
elle permet de distinguer les secteurs
$b\lrarr -b$ et $b\lrarr *b$. Pour cela, il faut introduire des op\'erateurs
volumes chiraux $U^\pm$:
$$
U^\pm=\pm{\rm Tr}b^\pm_1[b^\pm_2,b^\pm_3],
$$
o\`u $b^\pm$ sont les composantes self-duales et anti-self duales de $b$
dans $spin(4)$. Il y a 4 possibilit\'es au niveau classique:
\makeatletter
\renewcommand{\theenumi}{\alph{enumi}}
\renewcommand{\labelenumi}{\theenumi.}
\begin{enumerate}
\item ce sont les bivecteurs
 $b$ eux-m\^emes qui forment le 4-simplex, alors
$U_+>0$ et $U^-=-U^+$.
\item ce sont les bivecteurs
 $-b$ qui forment le 4-simplex, alors
$U_+<0$ et $U^-=-U^+$.
\item ce sont les bivecteurs
 $*b$ qui forment le 4-simplex, alors
$U_+>0$ et $U^-=U^+$.
\item ce sont les bivecteurs
 $-*b$ qui forment le 4-simplex, alors
$U_+<0$ et $U^-=U^+$.
\end{enumerate}
Au niveau quantique, il est possible d'imposer la contrainte
de chiralit\'e $U_++U_-=0$ s\'electionnant le secteur $\pm b$
(et \'eliminant le secteur $\pm *b$). Cela choisit la correspondance
\`a utiliser entre les bivecteurs et l'alg\`ebre (en fait, cela contraint
la structure de Poisson sur les g\'en\'erateurs de l'alg\`ebre)
et il faut que cela soit $\theta\circ *$ et non pas $\theta$ \cite{bb}!
Ainsi l'ambigu\"\i t\'e au niveau de la correspondance est totalement fix\'ee.
N\'eanmoins, il n'est alors pas \'evident du tout d'imposer la contrainte $U^+>0$
au niveau quantique, que l'on laisse de c\^ot\'e pour le moment.
Le lecteur trouvera plus de d\'etails sur la proc\'edure de quantification en appendice.

\medskip

L'extension au cas Lorentzien a \'et\'e \'egalement r\'ealis\'ee par Barrett et Crane
\cite{bc2}. La logique de l'\'etude est la m\^eme que dans le cas Euclidien.
Cette fois-ci, on travaille avec le groupe de Lorentz $SO(3,1)\sim\slc$.
Les repr\'esentations unitaires sont labell\'ees par des couples
$(n\in\N/2,\rho\in\R)$. La simplicit\'e des repr\'esentations
correspond \`a la nullit\'e du second Casimir $n\rho=0$, ce qui restreint
aux repr\'esentations $(0,\rho)$ et $(n,0)$.
Il y a cette fois-ci trois entrelaceurs simples possibles: un correspond
au quotient $\slc/\su$ isomorphe \`a l'hyperbolo\"\i de sup\'erieure ${\cal H}_+$
dans l'espace de Minkowski, un correspond au quotient
$\slc/SU(1,1)$ isomorphe \`a l'hyperbolo\"\i de \`a une nappe ${\cal H}_0$,
et un correspond  au c\^one de lumi\`ere.
Dans les trois cas, si on regarde les fonctions $L^2$ sur ces espaces,
on peut les d\'evelopper en transform\'ee de Fourier (en fait, transform\'ee
de Gelfand-Graev) sur des repr\'esentations simples.

Concentrons-nous sur le cas du quotient $\slc/\su\sim {\cal H}_+$.
Les fonctions $L^2$ sur l'hyperbolo\"\i de se d\'ecompose sur les repr\'esentations
simples du type $(0,\rho)$:
$$
L^2({\cal H}_+)=
\int \rho^2d\rho\, R^{(0,\rho)}.
$$
L'entrelaceur de Barrett-Crane s'\'ecrit dans ce cas comme une int\'egrale sur
${\cal H}_+$:
\beq
\begin{array}{cccl}
I^{\rho_1\rho_2\rho_3\rho_4}_{BC}:&\bigotimes_{i=1}^4 R^{(0,\rho_i)}&\arr& \C \\
&  f_1\otimes..\otimes f_4 &\arr&
\f{1}{2\pi^2}\int_{{\cal H}_+}dx\,f_1(x)..f_4(x)
\end{array}
\eeq
en consid\'erant les \'el\'ements des  $R^{(0,\rho)}$ comme des fonctions sur
l'hyperbolo\"\i de. On peut \'egalement d\'ecomposer cet entrelaceur
en des entrelaceurs trivalents ce qui donne:
\beq
I^{\rho_1\rho_2\rho_3\rho_4}_{BC}=
\int\rho^2d\rho\,\doubleY{\rho_1}{\rho_2}{\rho_3}{\rho_4}{\rho},
\eeq
ou l'\'ecrire en tant qu'int\'egrale sur ${\cal H}_+$ des projections
sur les vecteurs de $R^{(0,\rho_i)}$ invariants sous $\su$ comme
dans le cas Euclidien \Ref{intertwinerBC}.

\medskip

Regardons donc maintenaint le 4-simplex quantique que nous avons construit.
Il est d\'efini par  10 repr\'esentations simples d\'ecrivant ses 10 triangles et
par 5 entrelaceurs simples d\'efinissant les t\'etra\`edres.
Etant donn\'ee la correspondance entre $b\lrarr *J$ d\'efinie par $\theta\circ*$,
le Casimir $\la J,J\ra$ des repr\'esentations simples correspond 
\`a la quantit\'e classique $\la *b,*b\ra=\pm\la b,b\ra$ suivant que
nous sommes dans le cas Euclidien ou Lorentzien.
On obtient ainsi  l'aire (au carr\'e)
du triangle\footnotemark: un triangle labell\'e par $(j,j)$ (cas Euclidien)
aura comme aire $\sqrt{2j(j+1)}$ et un triangle
labell\'e par $(0,\rho)$ (cas Lorentzien) aura comme aire
$\sqrt{\rho^2+1}>0$.
\footnotetext{Il y a toujours des ambigu\"\i t\'es li\'ees \`a la quantification.
Ainsi un autre choix de conventions donnent les spectres $\sqrt{2}(j+1/2)$
et $\rho\ge 0$.}
On remarque que le cas Lorentzien nous fournit le m\^eme spectre de l'aire
qu'obtenu avec l'approche canonique covariante d\'ecrit dans la partie III.
Ensuite les t\'etra\`edres sont d\'efinis par les entrelaceurs simples, qui
s'expriment comme des int\'egrales sur les espaces homog\`enes ${\cal S}^3$
et ${\cal H}_+$. Les vecteurs sur ${\cal S}^3$
et ${\cal H}_+$ peuvent alors s'interpr\'eter comme la normale au t\'etra\`edre.
Dans le cas Lorentzien, cela nous restreint donc \`a des t\'etra\`edres
de genre espace! Ce qui permet une comparaison avec un formalisme canonique
d\'ecrivant l'\'evolution dans le temps de tranches de genre espace.

\medskip

Au final, nous avons quantifi\'e un 4-simplex. Il faudrait maintenant
construire une vari\'et\'e simpliciale en recollant ses 4-simplex
quantiques. Reste que de la mani\`ere de recoller d\'epend 
 la propagation des degr\'es de libert\'e de la th\'eorie.
Pour cela, nous allons partir d'une th\'eorie continue, la relativit\'e
g\'en\'erale (car on veut un mod\`ele de gravit\'e quantique!),
et on va la discr\'etiser. Il se trouve qu'il existe une reformulation
de la gravit\'e en tant qu'une th\'eorie BF contrainte qui permet de red\'eriver
les m\^emes 4-simplex quantiques. Mais de plus, elle permet
de d\'ecrire le couplage entre ces 4-simplex et donc potentiellement
de cr\'eer des corr\'elations. Et c'est ce que je vais d\'ecrire dans
les sections suivantes.

\section{La gravit\'e comme une th\'eorie BF modifi\'ee}
\subsection{Contraindre la th\'eorie BF}

Introduisons l'action de Plebanski \cite{pleb}, qui est une action de type
BF avec des contraintes quadratiques sur le champ $B$:
\beq
S\,=\,S(\omega,B,\phi)\,=\,\int_{\mathcal{M}}\left[
B^{IJ}\,\wedge\,F_{IJ}(\omega)\,-\frac{1}{2}\phi_{IJKL}\,B^{KL}\,\wedge\,B^{IJ}\right]
\eeq
o\`u $\omega$ est une  1-forme \`a  valeur dans $so(4)$ ($so(3,1)$),
$\omega=\omega_{\alpha}^{IJ}J_{IJ}dx^{\alpha}$, $J_{IJ}$ \'etant les g\'en\'erateurs de
$so(4)$ ($so(3,1)$, $F=d_\omega\omega$ est la 2-forme courbure de $\om$,
$B$ est aussi une  2-forme \`a  valeur dans $so(4)$ ($so(3,1)$),
$B=B_{\alpha\beta}^{IJ}J_{IJ}dx^{\alpha}\wedge dx^{\beta}$, et $\phi_{IJKL}$ est un
multiplicateur de Lagrange  satisfaisant $\phi_{IJKL}\epsilon^{IJKL}=0$
(de plus $\phi_{IJKL}$ est suppos\'e sym\'etrique sous \'echange de $(IJ)$ et $(KL)$
et antisym\'etrique sous $I\leftrightarrow J$ et $K\leftrightarrow L$).
Dans la suite, $\alpha,\beta,..$ sont des indices d'espace-temps et
$I,J,K,...$ d\'enotent des indices internes.
Les \'equations du mouvement sont:
\beq
d B\,+\,[\omega,B]=0 \;\;\;\;\;\;\;\;
F^{IJ}(\omega)\,=\,\phi^{IJKL}B_{KL}\;\;\;\;\;\;\;\; 
B^{IJ}\,\wedge\,B^{KL}\,=\,e\,\epsilon^{IJKL} \label{eq:constrB}
\eeq
avec $e=\frac{1}{4!}\epsilon_{IJKL}B^{IJ}\wedge B^{KL}$.
Quand $e\ne0$, les contraintes \Ref{eq:constrB}
sont \'equivalentes \`a
$\epsilon_{IJKL}B^{IJ}_{ab}B^{KL}_{cd}=\epsilon_{abcd}e$
\cite{bf}, ce qui signifie que
$\epsilon_{IJKL}B^{IJ}_{ab}B^{KLab}=0$
i.e. $B_{ab}$ est un bivecteur simple.
De plus, \Ref{eq:constrB}
est satisfaite si et seulement si
 il existe un champ de t\'etrade r\'eel
$e^{I}=e^{I}_{a}dx^{a}$ tel qu'une des \'equations suivantes soit vraie:
\beqs &I_\pm&\;\;\;\;\;\;\;\;\;\;B^{IJ}\,=\,\pm\,e^{I}\,\wedge\,e^{J} \\ 
&II_\pm&\;\;\;\;\;\;\;\;\;\;B^{IJ}\,=\,\pm\,
\frac{1}{2}\,\epsilon^{IJ}\,_{KL}e^{K}\,\wedge\,e^{L}
.
\label{secteurs}  
\eeqs
Restreignant le champ $B$ \`a toujours \^etre dans le secteur
$II_{+}$ (ce qui est parfaitement possible classiquement),
l'action se r\'e-\'ecrit:
\beq
S\,=\,\int_{\mathcal{M}}\,\epsilon_{IJKL}\,e^{I}\,\wedge\,e^{J}\,\wedge\,F^{KL}
\eeq
qui est bien l'action  de Palatini pour le formalisme du premier ordre de
la relativit\'e g\'en\'erale.

\medskip
Il est possible de g\'en\'eraliser les contraintes pour pouvoir d\'eriver
l'action de Palatini g\'en\'eralis\'ee et introduire le param\`etre d'Immirzi.
Ainsi \cite{prieto} introduit l'action:
\beq
S=\int B^{IJ}\w F_{IJ} -\f{1}{2}\phi_{IJKL}B^{IJ}\w B^{KL}+\mu H
\label{actionbf}
\eeq
o\`u $H=a_1\phi_{IJ}\,^{IJ}+a_2\phi_{IJKL}\epsilon^{IJKL}$ avec $a_{1}$
et $a_{2}$ des constantes arbitraires.
$\phi$ (scalaire) et $\mu$ (4-forme) sont des multiplicateurs de Lagrange,
avec $\phi$ ayant comme sym\'etries
$\phi_{IJKL}=-\phi_{JIKL}=-\phi_{IJLK}=\phi_{KLIJ}$.
$\phi$ impose la contrainte sur le champ $B$, tandis que $\mu$ impose
la condition $H(\phi)=0$ sur $\phi$.
L'op\'erateur $*$ agit sur les indices internes de telle sorte que
$*B_{IJ}=1/2\,\epsilon_{IJKL}B^{KL}$. Ainsi $*^2=\epsilon$, avec $\epsilon=1$
dans le cas Euclidien et $\epsilon=-1$ dans le cas Lorentzien.

Ceci est l'action la plus g\'en\'erale  du type BF avec des contraintes quadratiques sur
le champ $B$: la contrainte scalaire $H=0$ est la plus g\'en\'erale possible
\'etant donn\'ees les sym\'etries du champ $\phi$. Alors les \'equations du mouvement
sont les m\^emes que pour l'action de Plebanski sauf que la
contrainte sur le champ $B$ se complique:
\beq
B^{IJ}\w B^{KL} =\f{1}{6} (B^{MN}\w B_{MN}) \eta^{[I |K|} \eta^{J]L}
+\f{\epsilon}{12}(B^{MN}\w *B_{MN}) \epsilon^{IJKL}
\label{whole}
\eeq
\beq
2a_2 B^{IJ}\w B_{IJ} -\epsilon a_1 B^{IJ}\w *B_{IJ}=0
\label{simple}
\eeq
La solution de ces contraintes, pour des champs $B$ non-d\'eg\'en\'er\'es
($B^{IJ}\w *B_{IJ}\neq 0$), sont \cite{prieto,bf1,bf2}:
\beq
B^{IJ}=\alpha *(e^I \w e^J) + \beta\, e^I \w e^J
\label{B}
\eeq
avec
\beq
\f{a_2}{a_1}=\f{\alpha^2+\epsilon\beta^2}{4\alpha\beta}
\label{a1a2}
\eeq
Ins\'erant cette solution dans l'action\Ref{actionbf}, on obtient
l'action
de Palatini g\'en\'eralis\'ee:
\beq
S=\alpha \int  *(e^I \w e^J)\w F_{IJ}  +
\beta \int e^I \w e^J \w F_{IJ}
\eeq
avec un couplage entre le secteur g\'eom\'etrique $*(e\w e)$
(relativit\'e g\'en\'erale)
et le secteur  non-g\'eometrique  $e\w e$.

Ainsi je me suis propos\'e durant ma th\`ese d'\'etudier la quantification de cette
action BF g\'en\'eralis\'ee \`a l'aide du formalisme de mousse de spin, permettant
de voir le r\^ole du param\`etre d'Immirzi: est-il un param\`etre modifiant
la th\'eorie quantique comme en \lqg ou est-il un param\`etre sans
cons\'equence physique comme dans l'analyse canonique covariante \'etudi\'ee
dans la partie III.

\medskip

Consid\'erant \Ref{a1a2}, on se rend compte qu'une fois choisi un couple
$(\alpha,\beta)$ nous avons quatre secteurs tout comme avec l'action de
Plebanski.
Dans le cas Euclidien, on peut \'echanger
$\alpha$ et $\beta$.
Sous cette transformation, le champ $B$ se change en son dual $*B$. 
Cela refl\`ete que notre action initiale se voit pas la diff\'erence entre
$B$ et $*B$.
On peut aussi \'echanger  $B\rightarrow -B$ sans affecter la physique du mod\`ele.
Cela nous fournit les quatre secteurs suivants:
\beq
(\alpha,\beta) \quad (-\alpha,-\beta) \quad (\beta,\alpha) \quad
(-\beta,-\alpha).
\eeq
Dans le cas Lorentzien, la sym\'etrie sous $*$ nous donne les quatre secteurs
suivants:
\beq
(\alpha,\beta) \quad (\beta,-\alpha) \quad
(-\alpha,-\beta) \quad (-\beta,\alpha).
\eeq
Comparant avec l'action de Palatini g\'en\'eralis\'ee, la th\'eorie
avec $a_1,a_2$ correspond \`a un param\`etre d'Immrizi $\gamma=\alpha/\beta$
donn\'e par l'\'equation:
\beq
\f{a_2}{a_1}=\f{1}{4}\left(\gamma +\f{\epsilon}{\gamma}\right).
\eeq
On remarque les deux secteurs de la th\'eorie d\'efinis par 
$\gamma$ et $\epsilon/\gamma$ correspondent \`a la sym\'etrie
\'echangeant  $\alpha$ et $\epsilon\beta$.

Ainsi, le groupe de sym\'etrie de la th\'eorie est 
$Diff(M)\times SO(4)\times Z_{2}\times Z_{2}$, avec 
 $SO(4)$ remplac\'e par $SO(3,1)$ dans le cas Lorentzien.
Le  $Z_{2}\times Z_{2}$ correspond \`a l'existence des
4 secteurs de solutions et est responsable pour
les interf\'erences au niveau quantique.

\medskip

Comme dans le cas de l'action de Plebanski, quand $B$ est non-d\'eg\'en\'er\'e,
on peut reformuler les contraintes
sur le champ $B$ \'echangeant le r\^ole des indices internes et des indices
d'espace-temps\footnotemark. Cela est possible quand $a_2\neq 0$ et 
 $\left(\f{a_1}{2a_2}\right)^2\neq\epsilon$:
\beq
\left(\epsilon_{IJMN}-\f{a_1}{a_2}\eta_{[I\mid M\mid}\eta_{J]N}
\right)
B^{MN}_{cd}B^{IJ}_{ab}=
e\,\epsilon_{abcd}
\left(1-
\epsilon\left(\f{a_1}{2a_2}\right)^2
\right)
\label{simpl2}
\eeq
avec
\beq
e=\f{1}{4!}\epsilon_{IJKL}B^{IJ}\w B^{KL}.
\eeq
Il est int\'eressant de noter que pour $(ab)=(cd)$, \Ref{simple2} donne un
\'equivalent de la contrainte \Ref{simple}:
\beq
2a_2\f{1}{2}\epsilon_{IJKL}B^{IJ}_{ab}B^{KLab}
-\,a_1B^{IJ}_{ab}B^{ab}_{IJ}=0.
\label{simple2}
\eeq

\footnotetext{J'ai donn\'e une preuve dans \cite{bf1}.}

\medskip

Il est facile de voir le lien entre l'action BF contrainte g\'en\'eralis\'ee
et l'action de Plebanski originale \cite{bf1,bf2}. En effet, effectuons le
changement de variables suivant sur \Ref{actionbf}:
\beq
\left\{
\begin{array}{ccc}
B^{IJ}& =&\alpha E^{IJ}+\epsilon\beta *E^{IJ} \\
\tl{\phi}_{IJKL}&=&
(\alpha+\epsilon\beta\f{1}{2}\epsilon_{IJ}\,^{AB})\phi_{ABCD}
(\alpha+\epsilon\beta\f{1}{2}\epsilon_{KL}\,^{CD})
\end{array}
\right.
\label{change}
\eeq
Alors l'action devient:
\beq
S\,=\,
\f{1}{|\alpha^2-\epsilon\beta^2|^3}\,
\int\,(\alpha E^{IJ}+\epsilon\beta*E^{IJ})\wedge F_{IJ}\,
-\,\f{1}{2}\tl{\phi}_{IJKL}E^{IJ}\wedge E^{KL}
\,+\,\mu\epsilon^{IJKL}\tl{\phi}_{IJKL}
\label{after}
\eeq
et on voit bien que nous avons les m\^emes contraintes que dans l'action de Plebanski
sur le champ $E$, ce qui implique que $E$ est restreint \`a \^etre {\it simple}
et s'exprime en fonction d'un champs de t\'etrade selon l'un des secteurs \Ref{secteurs}.
De plus, cela montre que l'\'etude (et la quantification) de l'action BF
g\'en\'eralis\'ee utilise les m\^emes techniques que pour l'action de Plebanski.

\subsection{Les contraintes \`a la Barrett-Crane}

Pour discr\'etiser la th\'eorie, on se place sur une triangulation de l'espace-temps
quadri-dimensionnel i.e. sur une vari\'et\'e simpliciale.
La discr\'etisation de la th\'eorie se fait comme dans le cas de la th\'eorie BF
(le lecteur trouvera une approche syst\'ematique de la discr\'etisation dans
\cite{laurent:sf,reis:sf}).
L'espace-temps est form\'e de 4-simplex, t\'etra\`edres, triangles, c\^ot\'es et points.
Du dual, on ne retient que le 2-squelette: le 2-complexe dual
form\'e des vertex duaux aux 4-simplex, des liens duaux aux t\'etra\`edres et
des faces/plaquettes duales aux triangles.
Il s'agit alors de discr\'etiser la connexion $\om$, et sa courbure $F$, et le champ
de bivecteur $B$. $F$ et $B$ \'etant des 2-formes, il est naturel de les discr\'etiser
en les associants aux triangles, ou de mani\`ere \'equivalente aux triangles.
En ce qui concerne la connexion, on la discr\'etise sur les liens duaux:
associ\'e \`a un t\'etra\`edre, elle d\'ecrit le transport parall\`ele
entre deux 4-simplex. On remplace alors la courbure $F$ sur une plaquette
par l'holonomie de la connexion (discr\`ete) autour de cette plaquette.
En ce qui concerne le champ $B$, on l'int\`egre sur les triangles pour d\'efinir
un bivecteur associ\'e aux triangles:
$$
B^{IJ}(t)=\int_t B^{IJ}.
$$
Jusque l\`a, on est dans le m\^eme cadre que pour la th\'eorie BF.
La diff\'erence est qu'il faut maintenant discr\'etiser les contraintes
sur le champ $B$ en des contraintes sur les biecteurs $B(t)$.

\medskip

Dans un premier temps, nous allons discr\'etiser la version des contraintes
donn\'ees par \Ref{simpl2}. Le plus simple est d'utiliser le changement de variables
\Ref{change} et de se servir du champ $E$ d\'efini
par $B=\alpha E+\beta *E$. Alors, les contraintes s'\'ecrivent tr\`es simplement:
\beq
\left(\alpha^{2} \,+\, \epsilon\,\beta^{2}\right) \,
\epsilon_{IJKL}\,E^{IJ}_{ab}\,E^{KL}_{cd}
\,=\,
\,e\,\epsilon_{abcd}
\label{discon}
\eeq
o\`u $e=\f{1}{4!}\epsilon_{IJKL}B^{IJ}\w B^{KL}=
\f{1}{4!}(\alpha^2+\epsilon\beta^2)\epsilon_{IJKL}E^{IJ}\w E^{KL}$
peut \^etre consid\'er\'e comme un \'el\'ement de volume (puisque le champ
$B$ est suppos\'e non-d\'eg\'en\'er\'e). En effet, 
les contraintes sur $E$ (ou sur $B$) imposent que $E$
s'\'ecrit en fonction d'une t\'etrade,
$E^{IJ}=\pm e^I\w e^J$ ou $E^{IJ}=\pm*(e^I\w e^J)$,
cette t\'etrade d\'efinissant la m\'etrique quand nous r\'esolvons les
\'equations du mouvement. Alors le scalaire $e$
est proportionnel \`a $\epsilon_{IJKL}e^{I}\wedge e^{J}\wedge e^{K}\wedge
e^{L}$. Ainsi \`a un facteur pr\`es, le 4-volume g\'en\'er\'e
par deux triangles $t$ et $t'$ d'un 4-simplex est donn\'e par:
\beq
V(t,t')\,=\,\int_{x\in t\,;\,y\in t'}\,e\,\epsilon_{abcd}\,dx^{a}\w
dx^{b}\w dy^{c}\w dy^{d}.
\eeq
Alors on peut int\'egrer la contrainte \Ref{discon} pour en avoir une
version discr\'etis\'ee:
\beq
\epsilon_{IJKL}\,E^{IJ}(t)\,E^{KL}(t')
=\f{1}{\left( \alpha^{2}\,+\,\epsilon\,\beta^{2}\right)}\,V(t,t').
\label{cons}
\eeq
Consid\'erant un unique triangle $t$, on obtient  la contrainte de simplicit\'e
du bivecteur $t$:
\beq
\epsilon_{IJKL}\,E^{IJ}(t)\,E^{KL}(t)=0.
\eeq
Et consid\'erant deux triangles $t$ et $t'$
partageant un m\^eme c\^ot\'e, on obtient
la simplicit\'e de $E(t)+E(t')$:
\beq
\epsilon_{IJKL}\,E^{IJ}(t)\,E^{KL}(t')=0.
\eeq
Ces contraintes correspondent bien aux contraintes $(2)$ et $(3)$ sur les bivecteurs
d'un 4-simplex g\'eom\'etrique. On peut les traduire sur les bivecteurs $B(t)$:
\beq
a_2\epsilon_{IJKL}\,B^{IJ}(t)\,B^{KL}(t')=a_1B^{IJ}(t)B_{IJ}(t')
\label{Bsimple}
\eeq
pour $t=t'$ ou $t$ et $t'$ adjacents (partageant un m\^eme c\^ot\'e).

\medskip

Maintenant, il s'agit de donner une version de ces contraintes au niveau quantique.
Cela se fait en deux \'etapes. Tout d'abord, on utilise l'application $\theta$
\Ref{correspondance} pour traduire les bivecteurs en des g\'en\'erateurs de $so(4)$
ou $so(3,1)$, puis on quantifie g\'eom\'etriquement\footnote{Il est \'egalement possible
d'utiliser des techniques d'int\'egrale de chemin discr\'etis\'ee comme dans
\cite{laurent:sf,alej:sf}}. Utilisant la correspondance $\theta:B\arr J$, on peut traduire
la contrainte $\Ref{Bsimple}$ ci-dessus en une contrainte sur
les repr\'esentations de  $Spin(4)$ ou $\slc$ \cite{bf1,bf2} reliant
leur deux Casimirs $C_{1,2}$:
\beq
2a_2C_2=a_1C_1
\eeq
o\`u $J^{IJ}J_{IJ}=2C_1$ et $\f{1}{2}\epsilon_{IJKL}J^{IJ}J^{KL}
=2C_2$.
On peut remplacer $C_{1,2}$ par leur valeur explicite
dans les repr\'esentations $(j_-,j_+)$ de $Spin(4)$ ou $(n,\rho)$
de $\slc$. On trouve que les contraintes n'ont pas de sens: il
n'y a presque pas de solutions. Serait-il impossible de quantifier
la th\'eorie en g\'en\'eral ($a_1,a_2$ arbitraires i.e. $\gamma$ arbitraire)?

J'ai donn\'e la  solution de ce probl\`eme dans \cite{bf1,bf2}.
Elle consiste \`a exploiter l'ambigu\"\i t\'e de la correspondance entre
$B\lrarr J$. En effet, au lieu d'utiliser l'application $\theta$,
on est libre d'utiliser une application
arbitraire $\theta\circ(a*+b)$ (o\`u $*$ est l'op\'erateur de Hodge).
On peut alors choisir $(a,b)=(-\alpha,\beta)$
(au facteur $\beta^2-\epsilon\alpha^2$ pr\`es). Cela revient \`a faire
correspondre $E$ aux g\'en\'erateurs canoniques de $so(4)$ ou $so(3,1)$
au lieu de $B\lrarr J$! Dans ce cas-l\`a, on retrouve les contraintes
de simplicit\'e classique imposant de travailler avec des repr\'esentations
simples donn\'ees par $C_2=0$.

\medskip

Une fois les contraintes sur $B$ traduites au niveau quantique, on
peut proc\'eder \`a la quantification du mod\`ele proprement dit.
On trouve bien s\^ur des 4-simplex quantiques et la question
d\'elicate est le couplage  entre ces 4-simplex i.e. le
poids attribu\'e aux t\'etra\`edres. Cette discr\'etisation a \'et\'e
effectu\'e dans \cite{dan&ruth} et d\'efinit le mod\`ele de Barrett-Crane,
que l'on peut \'egalement d\'eriver d'une th\'eorie
g\'en\'eratrice \cite{alej:e,alej:l}. Je d\'ecrirai ce mod\`ele et
sa g\'eom\'etrie en d\'etail dans la prochaine section.

\subsection{Les contraintes \`a la Reisenberger} 

Il existe une autre mani\`ere de discr\'etiser la th\'eorie BF contrainte
consistant \`a discr\'etiser l'autre version des contraintes donn\'ee
par \Ref{whole} et \Ref{simple}.
Une fois encore, on utilise le 4-volume g\'en\'er\'e par deux triangles:
$$
V(t,t')\,=\,\int_{x\in t\,;\,y\in t'}\,e\,\epsilon_{abcd}\,dx^{a}\w
dx^{b}\w dy^{c}\w dy^{d}.
$$
Cette fois-ci, on d\'ecompose la 2-forme $B$ dans le 4-simplex en tant que somme de
2-formes associ\'ees aux triangles \cite{reis:sf,bf,laurent:sf}:
\beq
B^{IJ}(x)=\sum_t B^{IJ}_t(x),
\eeq
o\`u $B^{IJ}_t(x)$ est tel que:
\beq
\int B^{IJ}_t\w D=B^{IJ}[t]\int_{t*}D,
\eeq
avec $D$  une 2-forme quelconque et $t*$ la surface duale au triangle
(plus pr\'ecis\'ement, la partie de cette surface duale \`a l'int\'erieure
du 4-simplex consid\'er\'e). Il est alors simple de remarquer que:
\beq
\int_t B^{IJ}_{t'}=\delta_{t,t'}B^{IJ}[t]
\eeq
\beq
\int B^{IJ}_t\w B^{KL}_{t'}=B^{IJ}[t]B^{KL}[t']\epsilon(t,t')
\eeq
o\`u $\epsilon(t,t')$ est le signe du volume orient\'e  $V(t,t')$.
Plus pr\'ecis\'ement $\epsilon(t,t')=\pm 1$ si $t$ et $t'$ ne partagent pas de c\^ot\'e,
sinon $\epsilon(t,t')=0$. En posant alors:
 \beq
\Omega^{IJKL}=\sum_{t,t'}B^{IJ}[t]B^{KL}[t']\epsilon(t,t'),
\label{defomega}
\eeq
on peut traduire les contraintes \Ref{whole} et \Ref{simple} sur le champ $B$
dans le contexte discret:
\beq
\tl{\Omega}^{IJKL}=\Omega^{IJKL}-
\f{1}{6}\eta^{[IK}\eta^{J]L}\Omega^{AB}_{AB}-
\f{1}{24}\epsilon^{IJKL}\epsilon_{ABCD}\Omega^{ABCD}=0,
\eeq
\beq
4a_2\Omega^{AB}_{AB}=a_1\epsilon_{ABCD}\Omega^{ABCD}.
\eeq
C'est l'extension des contraintes de Reisenberger \cite{reis:sf}
au cas avec param\`etre d'Immirzi, que j'ai explicit\'e dans \cite{bf2}.
Dans le cas de $Spin(4)$, on peut d\'ecomposer $\Omega$ en parties
self-duale et anti-self-duale, ce qui conduit aux contraintes:
\beq
\tl{\Omega}^{ij}_{++}=
\Omega^{ij}_{++}-\delta^{ij}\f{1}{3}tr(\Omega_{++}),
\eeq

\beq
\tl{\Omega}^{ij}_{--}=
\Omega^{ij}_{--}-\delta^{ij}\f{1}{3}tr(\Omega_{--}),
\eeq

\beq
\Omega^{ij}_{+-}=0,
\eeq

\beq
\Omega_0=
(\alpha-\beta)^2tr(\Omega_{++})+
(\alpha+\beta)^2tr(\Omega_{--})=0,
\eeq
o\`u on remarque que le param\`etre d'Immirzi (\`a travers les coefficients
$\alpha$ et $\beta$) n'intervient que dans la contrainte scalaire
(la derni\`ere).
On peut ensuite impl\'ementer ces contraintes au niveau quantique
soit en restreignant le domaine d'int\'egration dans la fonction de partition
\cite{reis:sf,bf2}, soit
en tant qu'\'equations diff\'erentielles sur la fonction de partition \cite{laurent:sf}.

\section{Le mod\`ele de Barrett-Crane Lorentzien}

Pour construire le mod\`ele, on peut suivre une proc\'edure de discr\'etisation de la
th\'eorie BF contrainte ou partir directement d'une th\'eorie de champs g\'en\'eratrice.
Dans les deux cas, il s'agit de d\'ecider de la mani\`ere de coller
les 4-simplex ensembles. Dans le cadre de la discr\'etisation, cela correspond au couplage
entre les 4-simplex d\'efini par la th\'eorie continue.
Dans le cadre de la th\'eorie de champs, cela correspond au choix
du propagateur ou, au niveau
des diagrammes de Feynman, \`a la mani\`ere
de coller les vertex d'interaction (repr\'esentant
les 4-simplex). Dans tous les cas, il n'y a qu'un seul test possible pouvant d\'emontrer
la validit\'e d'un couplage particulier: c'est l'\'etude des corr\'elations dans une limite
semi-classique.
N\'eanmoins, il existe un choix canonique du couplage, qui provient du recollement le plus naturel
des 4-simplex \cite{dan&ruth} ou de la th\'eorie de champs la plus simple \cite{alej:l}, que
l'on appelle le mod\`ele de Barrett-Crane. Il existe d'autres possibilit\'es de couplage
respectant les sym\'etries impos\'ees au mod\`ele -invariance des 4-simplex sous $SO(3,1)$ et invariance
des t\'etra\`edres sous $SO(3)$- qu'il serait \'egalement int\'eressant d'\'etudier.
Cependant, je me contenterai dans ce chapitre de discuter du mod\`ele usuel,
de son interpr\'etation g\'eom\'etrique
et son interpr\'etation en terme d'amplitude de transition.

\subsection{D\'efinition du mod\`ele: amplitude des mousse de spin}

On formule le mod\`ele sur des 2-complexes localement duaux \`a des triangulations de l'espace-temps:
on a des vertex 5-valents li\'es par des liens o\`u se rencontrent 4 faces, ainsi 10 faces
se rencontrent \`a chaque vertex.
Aux faces du 2-complexe (duales aux triangles)
sont attach\'ees des repr\'esentations simples de $\slc$
labell\'ees par un nombre r\'eel $\rho\ge0$. On associe aux liens
(duaux des t\'etra\`edres) les entrelaceurs simples, r\'ealis\'es en tant qu'int\'egrale sur
l'hyperbolo\"\i de ${\cal H}_+\sim\slc/\su$.
Le poids associ\'e aux faces est simplement le poids statistique de la mesure de Plancherel
de la repr\'esentation de la face, qui fournit un
\'equivalent de la dimension de la repr\'esentation
dans le cas des repr\'esentations unitaires d'un groupe non-compact (de dimension infinie).
Le poids des vertex (duaux des 4-simplex) est l'amplitude de Barrett-Crane correspondante, obtenue
en contractant (prenant la trace sur les repr\'esentations) des entrelaceurs de Barrett-Crane
associ\'es \`a ces 5 t\'etra\`edres (liens).
L'ambigu\"\i t\'e est pour le poids assign\'e aux t\'etra\`edres. Dans le mod\`ele canonique,
on prend la ``norme'' de l'entrelaceur simple i.e. on contracte l'entrelaceur simple avec lui-m\^eme en
un graphe ``oeil''.
\begin{figure}[t]
\begin{center}
\psfrag{a}{$j_1$}
\psfrag{b}{$j_2$}
\psfrag{c}{$j_3$}
\psfrag{d}{$j_4$}
\psfrag{e}{$j_5$}
\psfrag{f}{$j_6$}
\psfrag{g}{$j_7$}
\psfrag{h}{$j_8$}
\psfrag{j}{$j_9$}
\psfrag{k}{$j_{10}$}
\includegraphics[width=9cm]{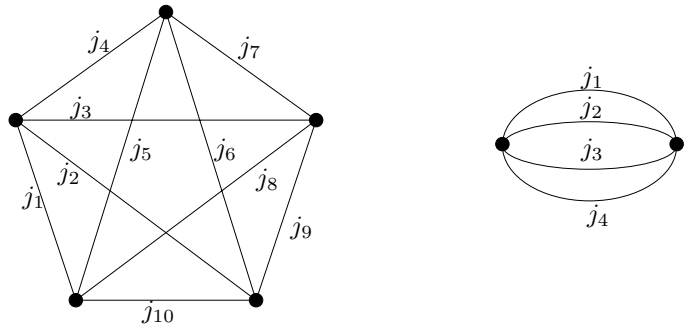}
\end{center}
\caption{Le diagramme d'un 4-simplex et le diagramme ``oeil'' d'un t\'etra\`edre,
donnants les poids associ\'es aux vertex et liens dans le mod\`ele de Barrett-Crane:
on trace les entrelaceurs simples (de Barrett-Crane) sur les repr\'esentations.}
\label{bc.eps}
\end{figure}
Ainsi la fonction de partition du mod\`ele est:
\beq
Z\,=\,(\prod_f
\int_{\rho_f}d\rho_f\,\rho_f^2)\,(\prod_{v,
e_v}\int_{H^+}dx_{e_v})\,\prod_e
\mathcal{A}_e(\rho_k)\,\prod_v\mathcal{A}_v (\rho_k, x_i),
\label{eq:Z}
\eeq
o\`u on associe une variable d'int\'egration $x$ pour chaque t\'etra\`edre dans chaque 4-simplex.
En particulier, on a 2 variables $x$ pour chaque t\'etra\`edre correspondant
aux entrelaceurs simples de chacun des 4-simplex partageant le t\'etra\`edre. Les amplitudes sont
donn\'ees en fonction du noyau $K^\rho$ ds repr\'esentations simples:
\beq
A_e(\rho_1,\rho_2,\rho_3,\rho_4)\,=\,\int_{H^+} dx_1
K^{\rho_1}(x_1,x_2)K^{\rho_2}(x_1,x_2)K^{\rho_3}(x_1,x_2)K^{\rho_4}(x_1,x_2),
\eeq
\beqs
A_v(\rho_k,
x_i)&=&K^{\rho_1}(x_1,x_2)K^{\rho_2}(x_2,x_3)K^{\rho_3}(x_3,x_4)K^{\rho_4}(x_4,x_5)K^{\rho_5}(x_1,x_5)
\nonumber \\
&&K^{\rho_6}(x_1,x_4)
K^{\rho_7}(x_1,x_3)K^{\rho_8}(x_3,x_5)K^{\rho_9}(x_2,x_4)
K^{\rho_{10}}(x_2,x_5) \label{eq:ampl}.
\eeqs
Je rappelle l'expression explicite du noyau $K$:
$$
K^\rho(x,y)=\f{2\sin\f{\rho\eta}{2}}{\rho\sh\eta},
$$
o\`u $\eta$ est la distance hyperbolique entre $x$ et $y$.
Dans la formule de la fonction de partition, il faut faire attention \`a n'int\'egrer que sur 4
(au lieu des 5) variables $x_{e_v}$ pour chaque vertex/4-simplex pour obtenir une amplitude bien d\'efinie
\cite{bc2}.

\medskip

Ceci est le couplage ``standard''. D'autres couplages possibles sont obtenus
en prenant diverses puissances
du diagramme ``oeil'' pour le poids des t\'etra\`edres. Ce changement respecte
bien toutes les sym\'etries du mod\`ele: l'invariance sous
$\slc$ des 4-simplex et l'invariance des t\'etra\`edres sous $\su$.
Cela modifie consid\'erablement
les propri\'et\'es de la fonction de partition, notamment sa convergence/divergence.
Il peut \^etre facilement pris en compte
dans la formulation en th\'eorie de champs g\'en\'eratrice \cite{dan:gluing}:
on peut \'ecrire une th\'eorie de champs avec tous les couplages possibles. Ceci
d\'efinit une famille de th\'eorie avec une infinit\'e de constantes de couplages,
qui pourrait fournir un cadre pertinant pour l'\'etude de la renormalisation de ces mod\`eles.

\medskip

La fonction de partition pr\'ec\'edente concernait une triangulation ferm\'ee.
Dans le cas d'une vari\'et\'e avec bords, il faut commencer par
d\'ecrire la fronti\`ere dans notre contexte discr\'etis\'e.
La fronti\`ere est constitu\'ee de t\'etra\`edres et est d\'ecrite
par les entrelaceurs simples qui leur sont associ\'es. Ainsi, on rajoute
\`a la fonction de partition $Z$ donn\'ee par \Ref{eq:Z} un terme
$C^{\rho_1\rho_2\rho_3\rho_4}_{(j_1k_1)(j_2k_2)(j_3k_3)(j_4k_4)}(x)$ pour
chaque t\'etra\`edre:
\beq
C^{\rho_1\rho_2\rho_3\rho_4}_{(j_1k_1)(j_2k_2)(j_3k_3)(j_4k_4)}(x)\,=
\,D^{\rho_1}_{00j_1k_1}(x)\,D^{\rho_2}_{00j_2k_2}(x)
D^{\rho_3}_{00j_3k_3}(x)\,D^{\rho_4}_{00j_4k_4}(x).
\label{xintertwiner}
\eeq
o\`u $D^{\rho}_{00jk}(g)=\la\rho 00|g|\rho j k\ra$
est  un \'el\'ement de matrice, repr\'esentant l'\'el\'ement de groupe $g$
dans la repr\'esentation $\rho$ dans la base canonique de $R^\rho$ d\'efinie par 
sa d\'ecomposition en repr\'esentations de $\su$.
Le t\'etra\`edre fronti\`ere est d\'ecrit par la variable $x\in\H_+$ et les vecteurs
$|\rho j_i k_i\ra$.

Il est facile de v\'erifier que cette convention permet de retrouver le bon poids
associ\'e \`a un t\'etra\`edre quand on colle deux vari\'et\'es le long de leurs
t\'etra\`edres communs (fronti\`ere commune).

\medskip

Ayant introduit une fronti\`ere, on peut \'egalement d\'ecrire les
\'etats de cette fronti\`ere. La fronti\`ere, form\'ee par des t\'etra\`edres,
est duale \`a un graphe (tranche du 2-complexe support de la mousse de spin)
dont les vertex $v$ sont les t\'etra\`edres et les liens $e$ les triangles les reliant.
Ce graphe est labell\'e par des repr\'esentations simples avec des entrelaceurs simples:
c'est un {\it r\'eseau de spin simple} et nous retrouvons la m\^eme
structure cin\'ematique  qu'issue de la quantification canonique covariante
effectu\'ee dans la partie III.
Dans un formalisme connexionelle, la g\'eom\'etrie de la fronti\`ere est 
d\'ecrite par une connexion discr\`ete, form\'ee par des \'el\'ements de groupe
d\'ecrivant la transport parall\`ele d'un t\'etra\`edre \`a un autre, et par
les variables $x$ associ\'ees aux t\'etra\`edres (par l'entrelaceur $C(x)$).
Donc un \'etat de la fronti\`ere est une fonction de cette connexion discr\`ete
et des variables $x$, qui a les m\^emes invariances/sym\'etries que la fonction
de partition (avec fronti\`ere). Il est facile de v\'erifier que nous demandons:
$$
\phi(g_e,x_v)=\phi(k_{s(e)}^{-1}g_ek_{t(e)},k_v.x_v)
\textrm{ pour tout } k_v\in \slc,
$$
qui est l'invariance des fonctions cylindriques (projet\'ees) demand\'ee
lors de la quantification du formalisme canonique invariant par Lorentz.
Dans ce contexte, on choisit la mesure de Haar sur $\slc$ pour d\'efinir le produit
scalaire et on trouve les r\'eseaux de spin projet\'es comme base de l'espace $L^2$.
Pour obtenir les r\'eseaux de spin simples comme base de l'espace d'Hilbert, il suffit
de se rendre compte qu'en plus de l'invariance sous $\slc$, nous imposons
\'egalement une invariance sous $\su^4$
(de l'entrelaceur $C(x)$ de \Ref{xintertwiner})
au niveau de chaque vertex/t\'etra\`edre.
Ainsi, au final, les r\'eseaux de spin simples fournissent bien une base
des \'etats-fronti\`eres de la th\'eorie.

Interpr\'etant alors la fonction de partition $Z$ comme d\'efinissant une amplitude
de transition entre \'etats fronti\`eres, les r\'eseaux de spin simples
sont nos \'etats cin\'ematiques et le mod\`ele de Barrett-Crane
d\'efinit leur dynamique. On retrouve ainsi le m\^eme cadre de travail que
celui d\'eriv\'e de la quantification canonique covariante propos\'ee en  partie III.

Notons que, d\^u au fait que la mousse de spin consid\'er\'ee dans la construction
du mod\`ele de Barrett-Crane est suppos\'ee (localement) duale \`a une vari\'et\'e
simpliciale, les r\'eseaux de spin (simples) vivants sur la fronti\`ere
sont a priori 4-valents. Une fois remarqu\'e le lien du mod\`ele avec la quantification
canonique, il est simple de g\'en\'eraliser cela et d'\'etendre le mod\`ele de Barrett-Crane
\`a un (2-)complexe cellulaire quelconque, non n\'ecessairement dual \`a une
d\'ecomposition simpliciale d'une vari\'et\'e 4d. Les \'etats fronti\`eres sont donn\'es
par des r\'eseaux de spin simples quelconques et le mod\`ele d\'efinit
des ``amplitudes de transition'' entre r\'eseaux de spin. L'amplitude est d\'efinie
par les poids associ\'es aux 4-cellules (g\'en\'eralisation des 4-simplex) et
aux 3-cellules (g\'en\'eralisation des t\'etra\`edres). Le poids assign\'e aux 4-cellules
est l'\'evaluation de leur r\'eseau de spin (simple) fronti\`ere, comme dans le cas
du 4-simplex (voir fig.\ref{bc.eps}). Similairement, le poids assign\'e aux 3-cellules
est donn\'e par la ``norme'' de l'entrelaceur simple (contraction de 2 entrelaceur simples
dont la valence est d\'etermin\'e par la 3-cellule).

\subsection{Significations G\'eom\'etriques des variables}

Les variables du mod\`ele sont les repr\'esentations $\rho$ associ\'ees
aux faces de la mousse de spin et les vecteurs $x$ associ\'es aux liens.
Tout d'abord, consid\'erant la d\'erivation du mod\`ele par les m\'ethodes
de quantification g\'eom\'etrique, $\rho$
correspondant \`a l'aire d'un triangle
dual \`a la face en question\footnote{Il est \'egalement
possible de d\'eriver ce r\'esultat
par des calculs de corr\'elations utilisant l'int\'egrale de chemin
discr\'etis\'ee de la th\'eorie BF \cite{laurent:sf}.}: l'aire du triangle
est donn\'e par $\rho\ge 0$ (ou $\sqrt{\rho^2+1}>0 $).
Ainsi les valeurs des repr\'esentations d\'efinissent les aires des triangles
de la vari\'et\'e simpliciale duale \`a la mousse de spin. Cela d\'efinit-il 
enti\`erement la triangulation? En fait, il existe une ambigu\"\i t\'e
dans la d\'efinition d'un 4-simplex \`a partir de seulement les aires de ses
triangles: bien qu'un 4-simplex est compl\`etement d\'etermin\'e par
les longueurs de ses 10 c\^ot\'es, il n'est pas exactement enti\`erement d\'etermin\'e
par les aires de ses 10 triangles (il existe certains cas pathologiques
pour lesquels la correspondance n'est pas unique).
N\'eanmoins, ces consid\'erations
concernent une potentielle interpr\'etation classique du 4-simplex quantique,
ce qui ne nuit pas \`a la l\'egitimit\'e du mod\`ele.

Maintenant, dans le contexte d'une vari\'et\'e simpliciale, quel est le r\^ole 
des variables $x$? Elles ont l'interpr\'etation naturelle de
normales aux t\'etra\`edres de la vari\'et\'e. Les t\'etra\`edres \'etant
de genre espace, ce sont des vecteurs de genre temps et
prennent bien leurs valeurs sur l'hyperbolo\"\i de $\H_+$.
Mais pourquoi y a-t-il deux normales $x$ pour chaque t\'etra\`edre -une pour chacun
des entrelaceurs simples associ\'es au t\'etra\`edres i.e. pour chacun des 4-simplex
auxquels il appartient?
En fait, le mod\`ele de Barrett-Crane reconstruit une vari\'et\'e simpliciale
et plus pr\'ecis\'ement une vari\'et\'e plate par morceaux:
on prend des morceaux d'espace-temps plat (les 4-simplex) et on
les colle ensemble. A chaque 4-simplex est attach\'e un rep\`ere local.
C'est le principe d'\'equivalence appliqu\'e aux points de l'espace-temps remplac\'e
par les 4-simplex: pour chaque 4-simplex, il existe un r\'ef\'erentiel local
pour lequel le 4-simplex est plat.
Alors les deux normales aux t\'etra\`edres sont en fait le m\^eme vecteur mais
vu des deux rep\`eres diff\'erents associ\'es aux 4-simplex. Cela permet
de prendre en compte la courbure de l'espace-temps. En effet, les rep\`eres
des 4-simplex n'\'etant pas identiques d\^u \`a la courbure, on a
besoin d'une connexion non-triviale pour passer de l'un \`a l'autre.
Cela est ainsi pris en compte dans le fait que nous travaillons avec
deux normales $x,y\in\H_+$ pour chaque t\'etra\`edre. Elles permettent
de reconstruire, \`a des \'el\'ements du sous-groupe stabilisateur $\su$ pr\`es,
la connexion (discr\`ete) $g$, alors d\'efinie par $g.x=y$.
Par cons\'equent, la donn\'ee des variables $x$ du mod\'ele de Barrett-Crane
\'equivaut \`a la donn\'ee de la connexion discr\`ete entre 4-simplex.
Ceci n\'eanmoins au groupe $\su$ pr\`es... Cependant cela semble correspondre
\`a la connexion de Lorentz $\cA$ utilis\'ee pour la quantification canonique
covariante.

Bien s\^ur, nous avons une invariance de Lorentz locale pour chaque simplex,
correspondant au niveau math\'ematique \`a l'invariance de l'amplitude d'un
4-simplex sous $\slc$. Physiquement, cela correspond au fait qu les cinq normales
$(x_A,x_B,\dots,x_E)$ des 5 t\'etra\`edres d'un 4-simplex fix\'e
sont d\'efinies \`a une transformation de Lorentz pr\`es:
$(x_A,x_B,\dots,x_E)$ est \'equivalent \`a $(g\cdot x_A,g\cdot
x_B,\dots,g\cdot x_E)$. Cela revient \`a dire que le rep\`ere local
associ\'e \`a chaque 4-simplex est d\'efini \`a une
transformation de Lorentz pr\`es.
Alors, \'etant donn\'e deux 4-simplex adjacents,
on peut les tourner afin que les deux normales d\'ecrivant
leur t\'etra\`edre commun soient align\'ees.
S'il est possible de faire de m\^eme d'une mani\`ere globale,
pour tous les 4-simplex,
et d'obtenir des normales align\'ees partout, alors cela veut dire que la
connexion (discr\`ete) est triviale et que nous avons un espace-temps plat.
Mais cette configuration n'est qu'une parmi toutes celles autoris\'ees
par le mod\`ele de Barrett-Crane, qui permet donc bien de d\'ecrire
des espaces-temps courbes.

\medskip

J'ai parl\'e de la connexion entre les 4-simplex, on peut \'egalement d\'efinir
la taille de ces 4-simplex. Le volume (d'espace-temps) d'un 4-simplex
est d\'efini par le produit vectoriel de deux bivecteurs associ\'es \`a
deux de ses triangles non-adjacents:
\beq
{\cal V}^{(4)}=
\f{1}{30}\sum_{t,t'}\f{1}{4!}
\epsilon_{IJKL}sign(t,t')B_t^{IJ}B_{t'}^{KL},
\eeq
o\`u on somme sur les couples de triangles  $(t,t')$ prenant
en compte leur orientation relative $sign(t,t')$.
Au niveau quantique, on remplace les bivecteurs par des g\'en\'erateurs de $\slc$
dans les repr\'esentations attach\'ees aux triangles:
\beq
{\cal V}^{(4)}=
\f{1}{30}\sum_{t,t'}\f{1}{4!}
\epsilon_{IJKL}sign(t,t')J_t^{IJ}J_{t'}^{KL}.
\eeq
Il existe une autre formule pour ce volume utile dans le contexte g\'eom\'etrique
du mod\`ele prenant en compte les normales aux t\'etra\`edres du 4-simplex:
\beq
({\cal V}^{(4)})^3=\f{1}{4!}\epsilon^{abcd}
N_a\wedge N_b\wedge N_c\wedge N_d,
\eeq
o\`u les normales orient\'ees ont pour norme $|N_i|=v^{(3)}_i$, le 3-volume
du t\'etra\`edre correspondant.

\medskip

Les amplitudes
des 4-simplex quantifi\'es d\'efinissent la dynamique du mod\`ele, pouvant
\^etre consid\'er\'es comme des op\'erateurs allant de l'espace d'Hilbert
des \'etats quantiques d'une hypersurface ``pass\'ee'' aux \'etats
quantiques d'une hypersurface ``future'' d\'eduite de la premi\`ere
par un mouvement de Pachner correspondant au 4-simplex \cite{fotini:sf}.

\medskip

Par contre, les diagrammes ``oeil'' attach\'es aux t\'etra\`edres peuvent \^etre
consid\'er\'es comme des {\it poids statistiques} refl\'etant le nombre d'\'etats
quantiques possibles du t\'etra\`edre d\'efini par l'aire
de ses 4 triangles.
En effet, un t\'etra\`edre classique est d\'efini par 6 nombres
(les 6 longueurs de ses c\^ot\'es par exemple) et n'est pas
enti\`erement d\'etermin\'e par les 4 aires de ses faces:
il manque les aires des 2 parallellogrammes int\'erieures \cite{bb}.
Cette ind\'etermination persiste au niveau quantique.
En fait, le t\'etra\`edre quantique est donn\'e par les 4 repr\'esentations
associ\'ees aux triangles, mais \'egalement par une variable
suppl\'ementaire correspondant \`a l'entrelaceur entre les 4
repr\'esentations. Cette variable, donn\'ee par la repr\'esentation
interne dans la d\'ecomposition du 4-vertex en vertex 3-valents,
correspond \`a l'aire d'un parallellogramme int\'erieur.
Mais, il n'est alors pas possible de sp\'ecifier l'aire d'un autre
parallellogramme \footnote{Le lecteur peut trouver l'explication
initiale dans \cite{bb} et un court r\'esum\'e de la
situation dans \cite{causal1}}.
Dans le cas Euclidien, le diagramme ``oeil'' est \'egal \`a la dimension
de l'espace des \'etats quantiques du t\'etra\`edre \cite{bb}!
Dans le cas Lorentzien, les param\`etres sont continus et cette dimension
est infinie. Malgr\'e cela, le diagramme ``oeil'' peut toujours
\^etre interpr\'et\'e comme donnant la dimension de l'espace
du t\'etra\`edre quantique, tout comme la mesure de Plancherel
peut \^etre consid\'er\'ee comme d\'efinissant la dimension
des repr\'esentations unitaires de $\slc$ qui sont de dimension infinie.
Pour cette interpr\'etation, il est plus clair de raisonner en
terme de {\it connexion}: on remplace  la donn\'ee d'aire int\'erieure
au t\'etra\`edre par la transformation de Lorentz reliant les rep\`eres
des deux 4-simplex partageant le t\'etra\`edre. Cette transformation
est d\'efinie par deux \'el\'ements de $\H_+$ (les normales).
Le diagramme ``oeil'' est alors obtenu en int\'egrant sur toutes les
connexions possibles (\'el\'ements de groupe de $\slc$)
au niveau du t\'etra\`edre.
Ainsi le diagramme ``oeil'' d\'efinit un poids pour
chaque connexion possible: il refl\`ete l'ind\'etermination
de la connexion et donc prend en compte la courbure de l'espace-temps.
On interpr\`ete ainsi le diagramme ``oeil'' comme
d\'efinissant une fluctuation quantique localis\'ee au t\'etra\`edre
que nous pouvons appeller onde r\'efractive quantique. En effet,
les {\it ondes r\'efractives} introduites dans \cite{wave} sont
une nouvelle classe d'ondes gravitationnelles, consistant
en une discontinuit\'e dans la m\'etrique alors que les aires (des
directions de genre nul) restent bien d\'efinies.
Cette condition refl\`ete parfaitement la situation dans le mod\`ele
de Barrett-Crane: les deux 4-simplex s'accordent sur la valeur
des 4 aires de leur t\'etra\`edre commun mais  ne d\'efinissent pas 
la m\^eme 3-g\'eom\'etrie (normales diff\'erentes, 3-volumes
diff\'erents). C'est cette ``ind\'etermination'' qui est
responsable du caract\`ere non-topologique du mod\`ele.
En effet, comme expliqu\'e \`a la fin du chapitre pr\'ec\'edent,
le diagramme ``oeil'' de la th\'eorie BF vaut 1: le t\'etra\`edre
est compl\`etement d\'etermin\'e.

\medskip

Une derni\`ere remarque sur la g\'eom\'etrie concerne
l'aire probable des triangles.
Il est possible de se faire une id\'ee du graphe de probabilit\'e
de l'aire $\rho$ d'un triangle donn\'e dans l'approximation d'un espace-temps
isotrope autour du triangle. Dans cette situation, on peut oublier les autres
repr\'esentations de la mousse de spin et l'amplitude de $\rho$ est donn\'ee par:
\beq {\cal A}(\rho)=\rho^2\left(
\f{\sin(\rho\theta)}{\rho\sinh(\theta)} \right)^N \qquad
\theta=\f{1}{N}(2\pi-\theta_0), \eeq
o\`u $N$ est le nombre de
4-simplex partageant ce m\^eme triangle et $\theta_0$ l'angle de d\'eficit
d\'efinissant la courbure autour du triangle,
$\theta_0=0$ correspondant \`a un espace-temps plat.
Tout d'abord, l'amplitude pour $\rho=0$ est nulle, ce qui
correspond au fait qu'un 4-simplex d\'eg\'en\'er\'e a une
probabilit\'e nulle.
Ensuite,  on obtient une s\'erie discr\`ete de maxima 
refl\'etant la possibilit\'e de g\'en\'erer dynamiquement un spectre
d'aire discret. Ce spectre change varie avec la courbure $\theta_0$.
De plus, il y a une tr\`es grande diff\'erence d'amplitude entre
le premier pic et les suivants, la diff\'erence grandissant avec $N$
(i.e. quand on raffine la triangulation). Ainsi, pour grand $N$,
l'aire la plus probable est de l'ordre de $l_P^2$, $\rho$ valant
entre 1 et 6. Par cons\'equent, on s'attend \`a ce que le r\'egime
pertinent du mod\`ele soit fourni par pleins de petits 4-simplex
(avec $\rho$ petit) et non pas dans une limite asymptotique 
$\rho\arr\infty$.

\subsection{Un formalisme discret du premier ordre}

On peut voir le mod\`ele de Barrett-Crane comme la quantification
d'un formalisme du premier ordre \`a la Regge utilisant
comme variables fondamentales les aires des triangles
et les angles dih\'edraux entre les normales entre
deux t\'etra\`edres partageant un triangle (dans un 4-simplex donn\'e).
Ces angles dih\'edraux correspondent aux termes $K^\rho(x_i,x_j)$
de l'amplitude ${\cal A}_v$ d'un 4-simplex.
Cela est en contraste avec l'approche traditionnelle d'une action
\`a la Regge pour la gravit\'e, du second ordre, o\`u les aires
et les angles sont consid\'er\'es comme des fonctions de longueurs
des c\^ot\'es de la triangulation.
Ainsi, dans le mod\`ele de Barrett-Crane, il n'y a pas de longueurs
de c\^ot\'es et il s'agit de le comprendre en tant que
 formalisme du premier ordre \footnote{Il existe une tentative d'introduire
des variables longueurs des c\^ot\'es  mais elle n'a pas encore
abouti \`a des constructions concr\`etes \cite{crane:bc}.}.

Ainsi, dans le cas Euclidien, un formalisme du premier ordre
fut propos\'e par Barrett \cite{balone} utilisant l'action
suivante:
\beq
S(l,\theta)\,=\,\sum_t\,A_t(l)\,\epsilon_t\,=\,\sum_t\,A_t(l)\,(
2\pi\,-\,\sum_{\sigma(t)}\,\theta_t(\sigma)),
\eeq
o\`u les aires $A_t$ des triangles $t$ sont suppos\'ees \^etre fonctions
des longueurs des c\^ot\'es $l$, 
$\epsilon_t$ est l'angle de d\'eficit associ\'e au 
 triangle $t$ (la mesure simpliciale de la courbure)
et
$\theta_t(\sigma)$ est l'angle dih\'edral associ\'e
au triangle $t$ dans le 4-simplex $\sigma(t)$ le contenant.
Les  angles dih\'edraux $\theta$, en tant que variables ind\'ependantes,
d\'eterminent une unique m\'etrique simpliciale pour chaque 4-simplex,
a priori diff\'erente de celle obtenue par les longueurs des c\^ot\'es.
En fait, les variables $\theta$ ne peuvent pas l\'egitimement \^etre
appel\'ees ``angles dih\'edraux'' tant que les deux g\'eom\'etries
ne co\"\i ncident pas. Cette contrainte  imposant
d'obtenir la bonne g\'eom\'etrie \`a partir des variables $\theta$
est exprim\'ee par l'identit\'e de Schl{\"a}fli:
\beq \sum_t\,A_t\,d\theta_t\,=\,0. \eeq
En effet, les variations de l'action, tenant compte de cette contrainte,
ont pour r\'esultat que les aires des triangles calcul\'ees \`a partir
des longueurs $l$ et des angles $\theta$ sont identiques:
\beq
A_t(l)\,\propto\,A_t(\theta).
\eeq

On peut voir cette contrainte comme ayant un r\^ole similaire
\`a la contrainte de simplicit\'e (de Plebanski) sur la 2-forme
$B$ imposant qu'elle d\'erive d'une t\'etrade et par cons\'equent
permettant son interpr\'etation m\'etrique.
Tenant compte de cette contrainte \`a l'aide
d'un multiplicateur de Lagrange $\lambda_\sigma$, l'action devient \cite{balone}:
\beq S\,=\,\sum_t\,A_t(l)\,\epsilon_t\,+\,
\sum_\sigma\,\lambda_\sigma\,det\Gamma_{ij}(\theta),
\eeq
la contrainte \'etant r\'e-exprim\'ee comme la nullit\'e
du d\'eterminant de la matrice
$\Gamma_{ij}=-\cos{\theta_{ij}}=-\cos{(x_i\cdot x_j)}$, o\`u on
a (r\'e)introduit les variables $x$ normales aux t\'etra\`edres
(dans le 4-simplex $\sigma$).

La diff\'erence entre ce formalisme et la situation dans le
mod\`ele de Barrett-Crane est que, dans le cadre des mousses de spin,
les aires ne sont pas consid\'er\'ees en tant que
fonctions des longueurs mais en tant que variables ind\'ependantes.
Le r\^ole des angles dih\'edraux est toutefois le m\^eme
et on peut consid\'erer cette action du premier ordre
comme ``cach\'ee'' dans le mod\`ele de mousse de spin.

Il faut cependant remarquer qu'il existe d'autres formulations du
premier ordre \`a la Regge, correspondant \`a d'autres formats des 
variables fondamentales \cite{CDM,Katsy,Katsy2}.
Egalement, l'id\'ee d'utiliser les aires comme variables
fondamentales fut d'abord pr\'econis\'ee par \cite{Rov}, puis
\'etudi\'ee dans \cite{Make1,Make2}. La possibilit\'e
de d\'ecrire la g\'eom\'etrie simpliciale en fonction de l'aire des triangles
(i.e. inverser la relation entre les aires et les longueurs) fut alors analys\'ee
dans \cite{BRW,MW,ReggeWill}.

\section{Imposer la causalit\'e pour retrouver un temps orient\'e}

La probl\'ematique de cette section est de montrer comment se restreindre
\`a un seul secteur de la th\'eorie, $II_+$ ou $II_-$, reliant le champ $B$
\`a la t\'etrade ($B=\pm*(e\w e)$), et ainsi obtenir
un mod\`ele l\'egitime d'int\'egrale de chemin pour la gravit\'e.
Changer le signe du champ $B$  revient \`a changer le signe du lapse
et donc du temps propre.  Par cons\'equent, la question de se restreindre
au secteur relativit\'e g\'en\'erale est li\'ee \`a la question d'identifier
une fl\`eche temps dans le mod\`ele de Barrett-Crane:
il s'agit d'imposer une structure causale au mod\`ele.
Dans ce cadre, on pourra l\'egitimement parler d'amplitudes de transition
(dans le temps) au lieu de simples corr\'elations, et on pourra
alors se poser la question d'une \'evolution unitaire des degr\'es de libert\'e
du mod\`ele. En effet, tel quel, le mod\`ele de Barrett-Crane, sommant sur toutes
les triangulations possibles de l'espace-temps, d\'efinit un projecteur sur les
\'etats quantiques physiques de l'hypersurface tout comme le mod\`ele
de Ponzano-Regge \footnote{Le fait de sommer sur les triangulations au lieu
de se contenter d'une triangulation fixe est d\^u au caract\`ere non-topologique
de la th\'eorie.} et ne d\'ecrit pas une \'evolution dans le temps.
Je me propose donc de construire un mod\`ele causal. Cette proposition
est issue d'une collaboration avec Daniele Oriti \cite{causal1,causal2}.

Elle repose sur une impl\'ementation non-triviale au niveau quantique
de la condition $(1)$ sur les 4-simplex imposant qu'un changement
d'orientation du mod\`ele revient \`a changer les repr\'esentations en leur
duales. Dans le mod\`ele de Barrett-Crane, cela est impl\'ement\'e
de mani\`ere triviale, le mod\`ele \'etant invariant sous changement
au dual. Il s'agit par cons\'equent de briser cette sym\'etrie.

\subsection{Les points stationnaires: des 4-simplex orient\'es}

Le noyau  $K$ utilis\'e dans la construction des r\'eseaux de spin
simples et du mod\`ele de Barrett-Crane est:
$$
K^{\rho_{ij}}(x_i,x_j)=\f{2\sin(\eta_{ij}\rho_{ij}/2)}{\rho_{ij}\sh\eta_{ij}},
$$
o\`u $\eta_{ij}$ est l'angle (ou distance hyperbolique)
entre $x_i\in\H_+$ et $x_j\in\H_+$.
Pour $x$ sur $\H_+$, on peut consid\'erer le vecteur $-x\in\H_-$ et \'etendre
le noyau $K$ \`a une fonction sur les deux hyperbolo\"\i des $\H_+$ et $\H_-$.
On remarque alors les invariances suivantes de $K$:
\beq
K^{\rho_{ij}}(x_i,x_j)\,=\,K^{-\rho_{ij}}(\eta_{ij})\,=
\,K^{\rho_{ij}}(-\eta_{ij})\,=\,K^{\rho_{ij}}(-x_i,-x_j)\,=
\,K^{\rho_{ij}}(x_j,x_i).
\eeq
Ces sym\'etries deviennent \'evidentes lorsqu'on consid\`ere la 
d\'ecomposition unique
des fonctions de repr\'esentation de $\slc$ de 1er
type en fonction des fonctions de repr\'esentations du 2nd type (voir
\cite{ruhl} pour plus de d\'etails):
\beqs
K^{\rho_{ij}}(x_i,x_j)&=&\frac{2\sin(\eta_{ij}\rho_{ij}/2)}{\rho_{ij}\sinh\eta_{ij}}
=\frac{e^{i\,\eta_{ij}\,\rho_{ij}/2}}{i\rho_{ij}\sinh\eta_{ij}}\,-
\,\frac{e^{-\,i\,\eta_{ij}\,\rho_{ij}/2}}{i\rho_{ij}\sinh\eta_{ij}}\,
\nonumber \\
&=&K^{\rho_{ij}}_{+}(x_i,x_j)+K^{\rho_{ij}}_{-}(x_i,x_j)=
K^{\rho_{ij}}_{+}(x_i,x_j)+K^{\rho_{ij}}_{+}(-x_i,-x_j)
\nonumber \\
&=&K^{\rho_{ij}}_{+}(x_i,x_j)+K^{\rho_{ij}}_{+}(x_j,x_i)=
K^{\rho_{ij}}_{+}(\eta_{ij})+K^{\rho_{ij}}_{+}(-\eta_{ij}).
\label{eq:Ksplit}
\eeqs
Ce d\'ecoupage permet de r\'e-\'ecrire la fonction de partition du mod\`ele
pour une triangulation fixe $\Delta$:
\beq
A(\Delta)=\sum_{\epsilon_t=\pm1}\int\prod_t\rho_t^2\textrm{d}\rho_t
\prod_T A_{eye}(\{\rho_t,t\in T\})
\prod_\sigma \int_{({\cal H}^+)^4} \prod_{T\in \sigma }\textrm{d}x^{(\sigma )}_T
\left(\prod_{t\in \sigma }
\f{\epsilon_t}{i\rho_t\sinh\eta_t}\right)
e^{i\sum_{t\in \sigma }\epsilon_t\rho_t\eta_t},
\eeq
o\`u on note $t$ les triangles, $T$ les t\'etra\`edres et $\sigma $ les 4-simplex.
Consid\'erant tous les termes \`a part l'exponentielle
comme la mesure du mod\`ele, on peut consid\'erer que le mod\`ele est
l'int\'egrale de chemin d'une action -action de Regge-
qui s'exprime pour un {\it 4-simplex unique
d\'ecoupl\'e}:
\beq
S_R=\sum_{t\in \sigma}\epsilon_t\rho_t\eta_t,
\eeq
o\`u les 10 angles $\eta_t$ sont des fonctions des cinq vecteurs $x_i\in\H_+$. 
Les cinq vecteurs \'etant d\'etermin\'es sur $\H_+$
\`a translation et rotation globale pr\`es, ils d\'efinissent 9 param\`etres,
il en r\'esulte une contrainte sur les angles $\eta_t$.
Elle est donn\'ee par l'identit\'e de Schl{\"a}fli, qu'il nous faut
prendre en compte pour \'etudier les variations de l'action $S$.

L'amplitude $A(\Delta)$ somme sur les deux signes de $\epsilon_t$,
ce qui rend l'amplitude invariante sous la sym\'etrie $\Z_2$ \'echangeant
les repr\'esentations $\rho_t$ avec leur
duales $\overline{\rho_t}\equiv-\rho_t$. A l'aide de l'\'etude des points stationnaires
de $S$, je vais montrer que ces $\epsilon_t$ refl\`etent
l'orientation dans le temps du 4-simplex. Il suffira alors de ne sommer que sur une seule
configuration coh\'erente des signes $\epsilon_t$ pour obtenir une amplitude
causale. Une telle configuration des $\epsilon_t$ munit alors la vari\'et\'e
simpliciale $\Delta$ d'une {\it structure causale}.

\medskip

Pour d\'eriver cela,
commen\c cons par un peu de g\'eom\'etrie simpliciale Lorentzienne
et d\'emontrons l'identit\'e de Schl{\"a}fli dans le cadre Lorentzien.
Consid\'erons un 4-simplex. Appelons $x_i\in\H_+$ les normales non-orient\'ees
aux cinq t\'etra\`edres, que l'on suppose donc de genre espace,
et $n_i=\alpha_ix_i$ les normales orient\'ees (vers l'ext\'erieur) avec
$\alpha_i=\pm$.
On d\'efinit le param\`etre de boost associ\'e au triangle partag\'e
par les t\'etra\`edres $i$ et $j$ la quantit\'e:
\beq
\eta_{ij}=\cosh^{-1}(x_i\cdot x_j)\ge 0 \label{angledef},
\eeq
et l'angle dih\'edral Lorentzien:
\beq
\theta_{ij}=\alpha_i\alpha_j\eta_{ij}=
\alpha_i\alpha_j\cosh^{-1}(\alpha_i\alpha_jn_i.n_j).
\eeq
L'identit\'e de Schl{\"a}fli est la diff\'erentielle de la contrainte
stipulant que les 10 angles dih\'edraux ont \'et\'e construits \`a
partir des vecteurs $n_i$:
\beq
\sum_{i\ne j} 
A_{ij}\textrm{d}\theta_{ij}
=\sum_{i\ne j} \alpha_i\alpha_j
A_{ij}\textrm{d}\eta_{ij}
=0
\label{schlaflieq}
\eeq
o\`u les $A_{ij}$ sont les aires des 10 triangles de l'unique 4-simplex d\'efini
par les $n_i$. Pour d\'emontrer cela, on utilise
la matrice:
\beq
\gamma_{ij}=n_in_j=\alpha_i\alpha_j\cosh\eta_{ij}
=\alpha_i\alpha_j\cosh(\alpha_i\alpha_j\theta_{ij}).
\eeq
Alors la fermeture du 4-simplex s'\'ecrit:
\beq
\sum_{i=1}^{i=5}|v^{(3)}_i|n_i=0
\eeq
o\`u $v^{(3)}_i$ est le 3-volume du t\'etra\`edre $i$.
Cela implique
\beq
\forall j,\quad
\sum_i |v^{(3)}_i|\gamma_{ij}=0,
\eeq
c'est-\`a-dire l'existence du vecteur propre $(|v^{(3)}_i|)_{i=1\dots5}$
de valeur propre nulle de la matrice $\gamma_{ij}$.
Ainsi la contrainte g\'eom\'etrique reliant les angles $\theta_{ij}$
est la nullit\'e du d\'eterminant $det(\gamma^\sigma_{ij})$. C'est sous cette forme
qu'on l'utilise dans la formulation  du premier ordre \`a la Regge.
L'identit\'e de Schl{\"a}fli s'obtient en diff\'erentiant la relation
ci-dessus puis en contractant de nouveau l'expression avec le vecteur des
3-volumes:
\beq
\sum_{i\ne j} |v^{(3)}_i||v^{(3)}_j|\textrm{d}\gamma_{ij}=
\sum_{i\ne j} |v^{(3)}_i||v^{(3)}_j|\alpha_i\alpha_j
\sinh(\eta_{ij})\textrm{d}\eta_{ij} =0. \eeq
Finalement, il est facile de montrer que
\beq
\sinh(\eta_{ij})|v^{(3)}_i||v^{(3)}_j|=
\f{4}{3}|{\cal V}^{(4)}|A_{ij}
\eeq
o\`u ${\cal V}^{(4)}$ est le 4-volume du 4-simplex, ce qui nous permet de conclure
avec  \Ref{schlaflieq}.

\medskip

Les points stationnaires de l'action pour un 4-simplex sont donc donn\'es par
l'\'equation:
\beq
\textrm{d}S_R=\sum_{t=(ij)\in s}\epsilon_t\rho_t\textrm{d}\eta_t
=\mu\times \sum_{i\ne j} \alpha_i\alpha_j
A_{ij}\textrm{d}\eta_{ij} \eeq
avec le multiplicateur de Lagrange $\mu\in\mathbf{R}$.
On en d\'eduit les points stationnaires:
\beq \left|
\begin{array}{ccc}
\epsilon_{ij}\alpha_i\alpha_j&=&sign(\mu) \\
\rho_{ij}&=&|\mu|A_{ij}
\end{array}
\right.
\label{localorient}
\eeq
Cela veut dire que les aires des triangles sont donn\'ees (\`a un
facteur d'\'echelle pr\`es) par les labels $\rho_{ij}$ des repr\'esentations
de $\slc$ et que nous avons une relation de coh\'erence
entre les orientations des t\'etra\`edres $\alpha_i$, les orientations
des triangles $\epsilon_{ij}$ et l'orientation globale
du 4-simplex $sign(\mu)$:
seules ces configurations particuli\`eres de valeurs de $\epsilon_t$,
correspondant aux points stationnaires, repr\'esentent
des g\'eom\'etries Lorentziennes (simpliciales) bien-d\'efinies.
En se restreignant \`a ces configurations coh\'erentes, on obtient
alors des 4-simplex orient\'es bien d\'efinis. Il s'agit maintenant
d'\'etendre cette orientation \`a toute la mousse de spin.

\subsection{Une amplitude de transition causale}

J'ai expliqu\'e pr\'ec\'edemment comment orienter un 4-simplex isol\'e.
Cela se traduit par des orientations coh\'erentes des triangles
\`a l'int\'erieur du 4-simplex.
Maintenant, je vais \'etendre ce choix d'orientation  \`a toute la vari\'et\'e
simpliciale. Cela consiste \`a choisir des orientations
coh\'erentes des t\'etra\`edres d'un 4-simplex \`a un autre.
En effet, l'orientation d'un 4-simplex permet de d\'efinir les
t\'etra\`edres pass\'es et futurs du 4-simplex et on impose
que si un t\'etra\`edre est pass\'e pour un 4-simplex, il
doit \^etre futur pour l'autre. Cela se traduit par
des relations entre les $\mu_v$ (orientation des 4-simplex)
et $\alpha_{T,v}$ (orientation du t\'etra\`edre $T$ dans le 4-simplex
$v$):
\beq
\forall T,\,\mu_{p(T)}\alpha_{T,p(T)}=-\mu_{f(T)}\alpha_{T,f(T)}
\label{globalorient}
\eeq
o\`u $v=p(T)$ et $v=f(T)$ sont les deux 4-simplex (pass\'e et futur)
partageant $T$.

Cela revient \`a imposer des contraintes sur les signes autour de chaque boucle
de 4-simplex \'equivalentes \`a demander que le 2-complexe (la mousse de spin)
soit orientable. Dans ce contexte, \cite{gft} propose un moyen de ne g\'en\'erer
que des configurations orientables \`a partir de la th\'eorie de
champs g\'en\'eratrice du mod\`ele de Barrett-Crane en imposant que le champ
g\'en\'erateur -champ de t\'etra\`edres quantifi\'es- est
invariant sous permutation {\it paire} de ces \'el\'ements, ce qui revient
\`a orienter les t\'etra\`edres de mani\`ere coh\'erente. 

\medskip

Maintenant, au lieu de sommer sur toutes les configurations $\epsilon_t$ possibles,
on peut se restreindre \`a ne sommer que sur les configurations coh\'erentes.
Pour une triangulation fix\'ee, cela revient \`a se choisir des orientations
compatibles $\{\mu_v,\alpha_{T,v},\epsilon_t\}$, satisfaisant
\Ref{localorient} et \Ref{globalorient}, et de d\'efinir 
une {\it amplitude causale} $A_{causal}(\Delta)$:
\beqs
A_{\{\mu_v,\alpha_{T,v},\epsilon_t\}}(\Delta)&=&\prod_s\int_{({\cal H}_+)^4}
\prod_{T\in s}\textrm{d}x^{(s)}_T
\prod_{t\in s}
\f{\epsilon_t}{i\rho_t\sinh\eta_t}\int\prod_t\rho_t^2\textrm{d}\rho_t
\prod_T A^T_{eye}(\{\rho_t\}_{t\in T})
\prod_s\,e^{i\sum_{t\in s}\epsilon_t\rho_t\eta_t} \nonumber\\
&=&\prod_s\int_{({\cal H}_+)^4} \prod_{T\in s}\textrm{d}x^{(s)}_T
\prod_{t\in s}
\f{\epsilon_t}{i\rho_t\sinh\eta_t}\int\prod_t\rho_t^2\textrm{d}\rho_t
\prod_T A^T_{eye}(\{\rho_t\}_{t\in T})\,e^{i\,\sum_t\,\rho_t\,\sum_{s|t\in
s}\theta_t(s)}\nonumber\\
&=&\prod_s\int_{({\cal H}_+)^4} \prod_{T\in s}\textrm{d}x^{(s)}_T
\prod_{t\in s}
\f{\epsilon_t}{i\rho_t\sinh\eta_t}\int\prod_t\rho_t^2\textrm{d}\rho_t
\prod_T A^T_{eye}(\{\rho_t\}_{t\in T})\,e^{i\,S_{R}}.
\label{orientampli}
\eeqs
On a ainsi bris\'e la sym\'etrie $\Z_2$ et impos\'e une
structure causale \`a la triangulation.
On peut alors d\'efinir la fonction de partition du mod\`ele en sommant
sur les vari\'et\'e orient\'ees i.e. sur les structures causales:
\beq
Z_{causal}\,=\,\sum_{\Delta,\{\mu_v,\alpha_{T,v},\epsilon_t\}}
\,\lambda(\Delta)\,A_{causal}(\Delta),
\eeq
que l'on r\'e-\'ecrire sous la forme plus explicite:
\beq
Z_{causal}\,=\,\sum_\Delta\,\lambda(\Delta)\,
\int\mathcal{D}\theta_t(\Delta)\,
\int\mathcal{D}A_t(\Delta)\,e^{i\,S_{R}^\Delta(A_t,\theta_t)}.
\eeq
On obtient ainsi une int\'egrale de chemin sur des vari\'et\'es simpliciales
et leur structures causales en fonction de variables g\'eom\'etriques
aires et angles. De plus, ce mod\`ele est construit enti\`erement alg\'ebriquement
\`a partir de la th\'eorie des repr\'esentations de $\slc$!

On peut aussi voir le changement de l'amplitude de Barrett-Crane r\'eelle
\`a l'amplitude causale en termes des diverses fonctions de corr\'elations.
En r\'esum\'e, l'amplitude r\'eelle correspond \`a la fonction de Hadamard, qui
ne prend en compte aucune orientation, et l'amplitude orient\'ee
correspond \`a la fonction de Wightman (ou au propagateur de Feynman
selon que l'on somme sur les deux directions temporelles globales ou non).
Habituellement, on peut passer de l'une \`a l'autre par un changement
de contour d'int\'egration. Cela se retrouve dans notre contexte alg\'ebrique
par le changement de contour d'int\'egration effectu\'e pour passer
des fonctions de repr\'esentations de $\slc$ de 1er type aux fonction
de 2nd type.
Le lecteur int\'eress\'e peut trouver une discussion
d\'etaill\'ee sur les propri\'et\'es
du mod\`ele causal \`a la Barrett-Crane dans \cite{causal1}.

\medskip

A l'aide de cette amplitude causale, on peut explorer la question de l'\'evolution
dans le mod\`ele de gravit\'e quantique fourni par les mousses de spin \`a la
Barrett-Crane.
Tout d'abord, il est possible de reformuler le mod\`ele
dans le langage des ensembles causaux (gr\^ace \`a la structure causale impos\'ee)
et m\^eme dans le langage des ensemble causaux quantiques
({\it quantum causal sets} ou {\it quantum causal histories}) comme
d\'efinis par Fotini Markopoulou\cite{fotini:causal}.
\begin{figure}
\begin{center}
\includegraphics[width=9cm]{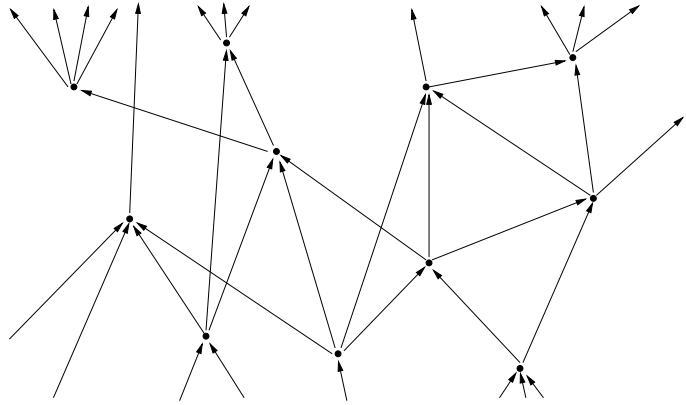}
\caption{Un exemple de structure causale dans le mod\`ele de
mousses de spin de Barrett-Crane.}
\end{center}
\end{figure}
Un ensemble causal est simplement un ensemble de points ordonn\'es
par des liens causaux (ordre partiel). Dans notre cas, les points sont les 4-simplex
et les liens les t\'etra\`edres qui permettent de passer d'un 4-simplex \`a un autre.
On peut \'egalement raisonner sur l'ensemble causal dual, qui est l'ensemble
causal des liens.

\begin{figure}
\begin{center}
\includegraphics[width=9cm]{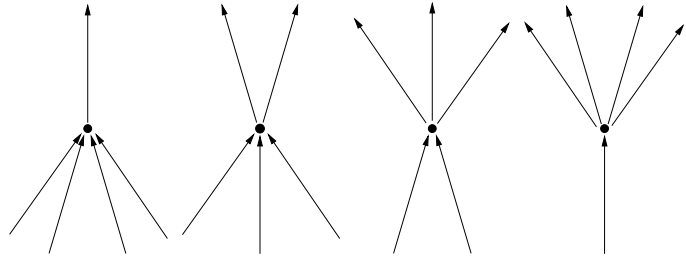}
\caption{Blocs fondamentaux des ensembles causaux du mod\`ele de Barrett-Crane:
ils correspondent aux mouvements de Pachner rempla\c cant $n$ t\'etra\`edres
par $5-n$ t\'etra\`edres.}
\end{center}
\end{figure}

\begin{figure}
\begin{center}
\includegraphics[width=9cm]{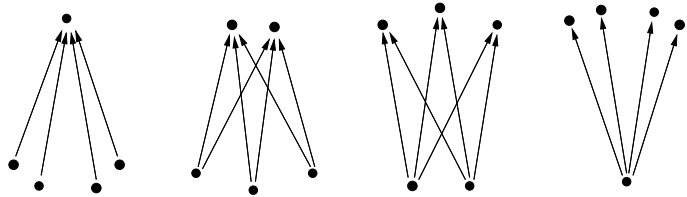}
\caption{Blocs fondamentaux des ensembles causaux duaux  du mod\`ele de Barrett-Crane:
les \'ev\`enements sont les liens/t\'etra\`edres du mod\`ele de Barrett-Crane et
un 4-simplex est repr\'esent\'e par plusieurs liens.}
\end{center}
\label{dualcausal}
\end{figure}

Un ensemble causal quantique est un ensemble causal habill\'e par des espaces d'Hilbert
sur les liens (resp. vertex) et des op\'erateurs \'evolution
au vertex (resp. sur les liens).
Une mani\`ere de faire est d'associer aux t\'etra\`edres (les liens de l'ensemble causal)
l'espace d'Hilbert des entrelaceurs simples
$C^{\rho_1,\rho_2,\rho_3,\rho_4}(x\in\H_+)$
i.e. un espace $L^2(\H_+)_{\rho_1,\rho_2,\rho_3,\rho_4}$ de fonctions $L^2$
sur l'hyperbolo\"\i de: un t\'etra\`edre est d\'efini par sa normale $x\in\H_+$.
Alors on peut consid\'erer l'amplitude (causale) d'un 4-simplex comme d\'efinissant
l'op\'erateur \'evolution. Un point essentiel est que cet
op\'erateur n'est {\bf pas unitaire} (voir \cite{causal1} pour une preuve).
Physiquement, cela est normal. En effet, c'est l'amplitude de
transition (totale) entre deux
\'etats que l'on aimerait voir unitaire, et
elle est donn\'ee non pas par un 4-simplex mais pas la somme sur toutes les triangulations
de l'espace-temps interm\'ediaire entre les deux \'etats. Ainsi, puisque la somme
de deux op\'erateurs unitaires n'est pas a priori unitaire, on n'a pas besoin
de supposer que l'amplitude d'un seul 4-simplex est unitaire. Cependant, la v\'erification
que l'amplitude globale est unitaire est pour le moment hors de port\'ee\dots

Une autre mani\`ere de voir est de travailler sur l'ensemble causal dual.
Dans ce cas, si on consid\`ere toujours les t\'etra\`edres comme \'etant d\'efini
par leur normales, alors on associe les espaces d'Hilbert $L^2(\H_+)$ aux points.
Les amplitudes des liens duaux (voir fig.\ref{dualcausal}) se lisent \`a partir
de l'amplitude causale: ce sont juste des exponentielles! Les op\'erateurs correspondants
sont de simples transformations de Lorentz boostant un vecteur $\H_+$ \`a un autre et
sont bien \'evidemment unitaires.
En fait, le mod\`ele de Barrett-Crane  s'interpr\`ete \`a ce niveau comme
une th\'eorie de particules libres sur l'hyperbolo\"\i de $\H_+$:
le mod\`ele r\'eel
associe aux liens (duaux) la fonction de Hadamard et le mod\`ele causal
effectue un changement d'int\'egrale de contour pour obtenir la fonction
de Wightman. Ceci r\'esume les propri\'et\'es d'unitarit\'e du mod\`ele de
Barrett-Crane \`a l'\'echelle microscopique.

\section*{Conclusion: une int\'egrale de chemin pour la gravit\'e}

Au final, le mod\`ele de mousse de spin
de Barrett-Crane impl\'emente une discr\'etisation de l'int\'egrale
de chemin de la gravit\'e et a un interp\'etation g\'eom\'etrique  claire
en termes de g\'eom\'etries simpliciales.
La discr\'etisation
est fond\'ee sur la reformulation de la relativit\'e g\'en\'erale
en une th\'eorie BF contrainte. On sait quantifier la th\'eorie
BF topologique en la discr\'etisant puis, en impl\'ementant
la contrainte au niveau quantique, on construit le mod\`ele de Barrett-Crane.
N\'eanmoins ces contraintes imposent que le champ $B$ d\'erive
d'une t\'etrade $B=\pm*(e \w e)$ et on obtient deux secteurs,
qui sont la relativit\'e g\'en\'erale \`a un signe pr\`es.
Au niveau classique, on peut voir ce changement de signe comme
un changement de la direction temporelle. Ainsi pour r\'eellement obtenir
la gravit\'e, il faudrait se restreindre \`a un seul secteur et imposer
au mod\`ele une structure causale. Par cons\'equent,
durant ma th\`ese, en collaboration avec Daniele Oriti, j'ai
identif\'ie la sym\'etrie $\Z_2$ responsable du changement de signe
au niveau discret des mousses de spin et d\'eduit une amplitude causale
pour le mod\`ele de Barrett-Crane, qui semble ainsi \^etre une
int\'egrale de chemin l\'egitime  de la gravit\'e, formul\'ee
comme une somme sur les vari\'et\'es causales munies d'une
structure causale. 

L'image de l'espace-temps en r\'esultant est un ensemble de 4-simplex plats
coll\'es les uns avec les autres, ou au niveau dual, un 2-complexe
fait de bulles labell\'ees par des repr\'esentations de $\slc$. Le
couplage se faisant au niveau des t\'etra\`edres (fronti\`eres entre
deux 4-simplex) sous la forme d'une onde r\'efractive ou, sur
la mousse de spin, au niveau des liens charni\`eres entre les bulles.

On peut \'egalement effectuer un splitting $3+1$ du mod\`ele afin
de d\'ecrire l'\'evolution d'une hypersurface (de genre espace).
Les \'etats quantiques de l'hypersurface sont alors donn\'es
par des r\'eseaux de spin simples, tout comme dans le formalisme canonique
covariant d\'evelopp\'e dans la partie III: les points de l'hypersurface 
sont d\'efinis par la normale \`a l'hypersurface en ces points, et les liens
entre ces points sont d\'ecrits par l'aire d'une surface duale donn\'e par
une repr\'esentation de $\slc$. Puis le mod\`ele de Barrett-Crane d\'efinit
des amplitudes de transition entre ces \'etats.

On a donc un mod\`ele complet de gravit\'e quantique. Ne reste donc
plus qu'\`a essayer d'en tirer des pr\'edictions physiques\dots


\chapter*{Conclusion: Mousses de Spin comme Gravit\'e Quantique}

Les mousses de spin ou {\it spin foams} d\'ecrivent l'espace-temps
cr\'e\'e par l'\'evolution d'\'etats de g\'eom\'etries d\'efinis par
les r\'eseaux de spin: ce sont des 2-complexes avec
des repr\'esentations de groupe attach\'ees \`a ses faces
et dont la g\'eom\'etrie est
d\'efinie par leurs entrelacements.
Elles fournissent la dynamique de mod\`eles
de \lqg et d\'eveloppent les concepts n\'ecessaires \`a son interpr\'etation
en tant que gravit\'e quantique i.e. th\'eorie quantique d'un espace-temps
dynamique.

Math\'ematiquement, les mousses de spin d\'efinissent une
int\'egrale de chemin discr\'etis\'ee et bien-d\'efinie
de la relativit\'e g\'en\'erale. La proc\'edure de construction
des mod\`eles repose actuellement sur la quantification de
la th\'eorie BF topologique, que l'on contraint afin d'obtenir
une th\'eorie \`a degr\'es de libert\'e locaux telle que la gravit\'e.

En basse dimension, j'ai \'etudi\'e les mousses de spin, qui
correspondent simplememt \`a la th\'eorie BF (sans contrainte).
Cela permet de comprendre le contexte des mousses de spin,
leur g\'eom\'etrie et comment les utiliser pour obtenir
des objets avec interpr\'etation physique tels que des
amplitudes de transition.

En quatre dimensions, la question principale consiste
en l'impl\'ementation des contraintes au niveau
quantique permettant de passer de la th\'eorie BF
\`a la relativit\'e g\'en\'erale. J'ai ainsi montr\'e
comment, en imposant une structure causale \`a la mousse de spin,
on obtient un mod\`ele l\'egitime d'int\'egrale de chemin de la
relativit\'e. Dans ce contexte, on retrouve les m\^emes structures
cin\'ematiques d\'efinies par des r\'eseaux de spin simple
que dans la quantification canonique invariante sous Lorentz
(que j'ai d\'evelopp\'e dans la partie III), ce qui
permet une traduction directe des mousses de spin
quadri-dimensionnelle dans un cadre canonique.
De plus, on retrouve un spectre continu des aires
(le mod\`ele ne contient que des surfaces de genre espace).
Et tenant compte de la structure causale du mod\`ele, il est possible
de l'\'etudier dans le contexte des ensembles causaux (quantiques), qui fournit
un cadre d'\'etude des propri\'et\'es d'unitarit\'e de l'\'evolution.

Au final, on obtient des mod\`eles compl\`etement alg\'ebriques
o\`u la g\'eom\'etrie est reconstruite \`a partir de la
th\'eorie des repr\'esentations du groupe de Lorentz.
Et il est m\^eme possible d'\'ecrire une th\'eorie de champs
sur le groupe de Lorentz g\'en\'erant les mousses de spin
par leur d\'eveloppement perturbatif en diagrammes de Feynman. Cela ouvre
la porte \`a des calculs non-perturbatifs sur les mousses
de spin utilisant la fonction de partition de cette th\'eorie
g\'en\'eratrice. J'ai donc \'etudi\'e l'effet non-perturbatif le
plus simple: les solutions classiques de la th\'eorie de champs.
Elles permettent d'introduire des g\'eom\'etries de fond -des {\it
backgrounds}- non-triviales (autre que le vide total i.e. l'absence
d'espace-temps) \`a partir desquelles j'esp\`ere comprendre la
structure non-perturbative des mod\`eles de mousses de spin. En
effet, le but serait d'\'etudier les effets des fluctuations
des mousses de spin autour d'une g\'eom\'etrie presque plate.

{\renewcommand{\thechapter}{}\renewcommand{\chaptername}{}
\addtocounter{chapter}{-1}
\chapter{Conclusion}\markboth{\sl CONCLUSION}{\sl CONCLUSION}}

La gravit\'e quantique \`a boucles, ou {\it Loop Quantum Gravity},
est une r\'ecente quantification canonique, et en cela non-perturbative,
de la relativit\'e g\'en\'erale.
Fond\'ee sur un d\'ecoupage de l'espace-temps en une hypersurface
3d \'evoluant dans le temps,
elle fournit une base d'\'etats
quantiques (cin\'ematiques),
invariants de jauge $\su$ (rotation de l'espace) et
invariants sous diff\'eomorphismes (spatiaux), repr\'esentant
la g\'eom\'etrie quantique de l'espace.
Ces \'etats sont vecteurs propres d'op\'erateurs aires et volumes
\`a spectre discret, ce qui correspond \`a l'image de
quanta de g\'eom\'etrie.
Puis la th\'eorie d\'ecrit leur \'evolution, qui g\'en\`ere
notre espace-temps.
Dans cette th\`ese, j'ai \'etudi\'e la structure (covariante)
de cet espace-temps quantique.

On appelle les \'etats cin\'ematiques r\'eseaux de spin,
ou {\it Spin Networks}, et les espace-temps r\'esultants
mousses de spin, ou {\it Spin Foams}.
Les mousses de spin permettent de d\'ecrire la dynamique
des r\'eseaux de spin. Elles fournissent ainsi un nouveau
point de vue sur les questions rencontr\'ees en
{\it Loop Quantum Gravity}, telle la pertinence du param\`etre
d'Immirzi $\gamma$ dont d\'epend la th\'eorie quantique et le spectre
des op\'erateurs g\'eom\'etriques, ou telle
l'invariance de la th\'eorie sous le groupe de Lorentz.

Mon approche a \'et\'e d'\'etudier les
possibles structures invariantes sous
le groupe de Lorentz dans un contexte de  gravit\'e
quantique \`a boucles.

\medskip

Ainsi, dans un premier temps, j'ai \'etudi\'e l'espace
des observables (partielles) invariantes sous Lorentz
de la relativit\'e g\'en\'erale dans un formalisme connexionnel.
La difficult\'e vient de la non-compacit\'e du groupe de Lorentz.
Par cons\'equent, je me suis interess\'e plus g\'en\'eralement
\`a l'espace des observables invariantes sous un groupe de jauge
non-compact. Dans ce cadre, j'ai d\'evelopp\'e,
avec Laurent Freidel, une th\'eorie des r\'eseaux de spin
pour groupes non-compacts, qui fournissent une repr\'esentation
de l'alg\`ebre des observables. Gardant \`a l'esprit l'application
de ces r\'esultats math\'ematiques \`a la th\'eorie de la gravit\'e,
j'ai appliqu\'e la th\'eorie math\'ematique aux groupes
de Lorentz en trois et quatre dimensions, $SO(2,1)$ et $SO(3,1)$.
J'ai expos\'e ces r\'esultats dans la partie II de cette th\`ese.

\medskip

Dans la partie III, j'ai explor\'e la possibilit\'e
de d\'evelopper un formalisme canonique invariant sous le groupe
de Lorentz.
Pour commencer, je me suis attaqu\'e au cas de la gravit\'e
en $2+1$ dimensions. Le formalisme canonique est relativement simple
et aboutit \`a des r\'eseaux de spin $SO(2,1)$, qui ont une interpr\'etation
g\'eom\'etrique naturelle en termes de g\'eom\'etrie simpliciale.
Dans ce contexte, j'arrive \`a la conclusion qu'en
g\'eom\'etrie quantique,
les distances spatiales sont continues et les
distances temporelles  discr\`etes.

Suivant le travail de Sergei Alexandrov, j'ai ensuite \'etudi\'e
un formalisme canonique en $3+1$ dimensions gardant explicite
l'invariance sous le groupe de Lorentz. Ce formalisme
prend en compte explicitement
le plongement de l'hypersurface dans l'espace-temps.
Il y a alors deux mani\`eres naturelles de quantifier la th\'eorie.

La premi\`ere permet de retrouver la gravit\'e quantique \`a boucles
usuelle, avec des spectres discrets des op\'erateurs g\'eom\'etriques.
En fait, la th\'eorie quantique est une simple extension des structures
$\su$ au groupe de Lorentz. Malheureusement, il s'av\`ere que les structures
ne se transforment pas normalement (habituellement)
sous diff\'eomorphismes dans la direction
temporelle: cela pourrait impliquer des anomalies dans la dynamique
quantique de la th\'eorie.

La seconde quantification utilise, par contre, des structures covariantes
et pr\'evoit des spectres continus des op\'erateurs g\'eom\'etriques.
De plus, la th\'eorie se r\'ev\`ele ind\'ependante du param\`etre d'Immirzi!
On obtient ainsi une {\it Loop Quantum Gravity} covariante.
N\'eanmoins, l'\'etude rigoureuse de la th\'eorie est compliqu\'ee
par le fait de variables de configuration non-commutatives.

\medskip

Dans la partie IV, je me suis finalement interess\'e aux mousses
de spin elles-m\^emes.
Dans un premier temps, j'ai d\'ecrit la logique des mod\`eles de mousses
de spin, leur g\'eom\'etrie et leur structure non-perturbative
\`a l'aide de th\'eories en basses dimensions (deux et trois).

Puis, j'ai \'etudi\'e le mod\`ele de Barrett-Crane.
J'ai rappel\'e comment il est  d\'eriv\'e \`a partir d'une discr\'etisation
et d'une quantification g\'eom\'etrique de l'espace-temps.
Il se r\'ev\`ele \^etre une th\'eorie quantique d'une th\'eorie
BF (topologique) contrainte, qui est \'equivalent classiquement
\`a la relativit\'e g\'en\'erale dans un certain secteur.
Dans ce cadre, en partant d'une action de type BF g\'en\'eralis\'ee,
j'ai montr\'e que le param\`etre d'Immirzi ne modifiait
en rien la th\'eorie quantique.
Puis, dans un travail r\'ealis\'e avec Daniele Oriti, j'ai montr\'e
qu'en imposant une structure causale \`a l'espace-temps (quantique)
du mod\`ele, on obtenait une int\'egrale de chemin (discr\'etis\'ee)
bien-d\'efinie de la relativit\'e g\'en\'erale.
Finalement, point important, il s'av\`ere que les \'etats cin\'ematiques
du mod\`ele sont exactement les m\^emes que ceux de la
{\it Loop Quantum Gravity} covariante d\'evelopp\'ee en partie III. Ainsi
les structures canoniques d'espace et celles d'espace-temps
co\"\i ncident et nous fournissent une vision coh\'erente d'une g\'eom\'etrie
quantique.

\medskip

A travers mon travail, j'esp\`ere convaincre que les mod\`eles de mousses
de spin (\`a la Barrett-Crane) forment des mod\`eles de gravit\'e quantique
viables, ces structures covariantes s'interpr\'etant \`a la fois
dans un cadre canonique
-une gravit\'e quantique \`a boucles covariante- et
en tant qu'int\'egrale de chemin de la relativit\'e g\'en\'erale.
Ces mod\`eles d\'ecrivant l'espace-temps \`a l'\'echelle de Planck,
il s'agit maintenant de voir \`a quelles structures macroscopiques
elles m\`enent afin d'obtenir des pr\'edictions physiques r\'eelles,
ce qui permettrait \'eventuellement de les valider en tant que
(v\'eritable) gravit\'e quantique.
Une fa\c con  de proc\'eder est d'\'etudier la renormalisation de ces
mousses de spin, c'est-\`a-dire de regarder leur comportement sous
transformation d'\'echelle. Il s'agit dans ce cadre de comprendre
tout d'abord les caract\'eristiques g\'eom\'etriques des mousses de spin,
les informations qu'il faut raffiner quand on regarde \`a plus petite \'echelle
et qu'il faut oublier quand on regarde \`a plus grande \'echelle. Cela
permettra de d\'eterminer les choses que nous devons calculer, les
questions que nous pouvons nous poser. La formulation sous forme
de th\'eorie de champ g\'en\'eratrice ouvre la porte \'egalement \`a une
\'etude de la renormalisation comme en th\'eorie quantique des champs:
on \'ecrit tous les couplages possibles du mod\`ele de Barrett-Crane
(au niveau des t\'etra\`edres) et on regarde par exemple leur renormalisation
en fonction du nombre de 4-simplexes triangulant une r\'egion de 
l'espace-temps.

\medskip

Il reste n\'eanmoins de nombreuses questions \'evoqu\'ees ici mais auxquelles
je n'ai pas pu r\'epondre enti\`erement, concernant le temps (le dynamique en 
gravit\'e \`a boucles) et le spectre des op\'erateurs g\'eom\'etriques. En 
effet, la \lqg a un spectre discret d\'ependant du param\`etre d'Immirzi.
Le point de vue covariant ignore le param\`etre d'Immirzi et fournit un
spectre continu des distances spatiales (tout en identiquant une
quantification des intervalles temporels). On peut donc s'interroger
sur la pertinence de $\gamma$ et sur l'incompatibilit\'e a priori de l'approche 
\lqg et des mousses de spin. Tout d'abord, les op\'erateurs g\'eom\'etriques
sont impl\'ement\'es au niveau cin\'ematique et ne refl\'ete pas les 
v\'eritables observables cens\'ees prendre en compte la dynamique et r\'esoudre 
\'egalement la contrainte Hamiltonienne. De plus, il y a la question de la 
fixation de jauge, car on a vu \`a travers l'exemple des variables \`a la 't 
Hooft que changer de fixation de jauge modifie les relations canoniques et les
propri\'et\'es des op\'erateurs g\'eom\'etriques. Ainsi, en \lqg, la jauge est 
d\'ej\`a fix\'ee, ce qui peut poser un probl\`eme pour exprimer la dynamique 
(covariante) et les changements d'observateurs. Par contre, dans le cadre
de la \lqg covariante et des mousses de spin, la jauge n'est pas fix\'ee, ce 
qui peut poser des probl\`emes d'interpr\'etration car un observateur donn\'e 
d\'efinit un feuilletage de l'espace-temps et donc une fixation de jauge, une 
dynamique particuli\`ere. Peut-\^etre retrouvera-t-on un spectre discret dans 
ce contexte-l\`a.

Li\'e \`a cela est le probl\`eme des conditions aux bords, qui d\'eterminent 
le cadre physique et les questions que nous pouvons poser. Ainsi, les mousses
de spin ont des \'etats fronti\`eres donn\'es pas les r\'eseaux de spin 
simples, qui d\'efinissent l'hypersurface (canonique) d'une mani\`ere
intrins\`equement diff\'erente des r\'eseaux de spin $SU(2)$. Alors, comment
un r\'eseau de spin simple d\'efinit-il la g\'eom\'etrie de la tranche 
spatiale 3d? Le probl\`eme est li\'e au fait que, dans le mod\`ele de 
Barrett-Crane, un t\'etra\`edre n'est pas enti\`erement d\'etermin\'e par 
lui-m\^eme mais par le 4-simplex auquel il appartient: il s'agit alors 
d'\'etudier les g\'eom\'etries 3d les plus probables et de calculer des 
corr\'elations entre g\'eom\'etriqes 3d aux fronti\`eres en fonction de 
la structure de l'espace-temps.

Finalement, je ne vois pas d'oppositions r\'eelles entre la \lqg $SU(2)$ et
l'approche covariante  par mousses de spin: il me semble que ce soit
une diff\'erence de point de 
vue. Maintenant que le cadre math\'ematique est fix\'e, il s'agit de comprendre
le contexte physique: la question de l'observateur et la prise en compte de la 
mati\'ere. Et il est certain que cette approche par quantification directe
de la relativit\'e g\'en\'erale fournira des points de vue int\'eressants sur
ces questions tout comme elle fournit d\'ej\`a une description pertinente
de l'espace-temps quantique.

\appendix
\chapter{Le groupe de Lorentz $\slc$}

\section*{Les repr\'esentations de $\slc$}

Notons $T_X$ les g\'en\'erateurs de l'alg\`ebre de Lie $sl(2,\C)$ et
$f_{XY}^Z$ les constantes de structure:
\beq
[T_X, T_Y]=f_{XY}^Z T_Z.
\eeq
Usuellement on note
 $T_X=(iK_a,iJ_a)$ avec $\vec{K}$ les g\'en\'erateurs des boosts
et $\vec{J}$ les g\'en\'erateurs des rotations (d'espace).
On introduit \'egalement
\beqs
&H_+=J_1+iJ_2, \qquad H_-=J_1-iJ_2, \qquad H_3=J_3,& \\
&F_+=K_1+iK_2, \qquad F_-=K_1-iK_2, \qquad F_3=K_3.&
\eeqs
Les relations de commutation s'\'ecrivent alors:
\beqs
& [H_+,H_3]=-H_+, \qquad [H_-,H_3]=H_-, \qquad [H_+,H_-]=2H_3, &
\nonumber \\
& [H_+,F_+]=[H_-,F_-]=[H_3,F_3]=0, & \nonumber \\
& [H_+,F_3]=-F_+, \qquad  [H_-,F_3]=F_-, & \\
& [H_+,F_-]=-[H_-,F_+]=2F_3, & \nonumber \\
& [F_+,H_3]=-F_+, \qquad [F_-,H_3]=F_-, & \nonumber \\
& [F_+,F_3]=H_+, \qquad [F_-,F_3]=-H_-, \qquad [F_+,F_-]=-2H_3. &
\nonumber
\eeqs
$\slc$ a deux op\'erateurs Casimir $C_1=\vec{J}^2-\vec{K}^2$ et
$C_2=\vec{J}.\vec{K}$.

Une repr\'esentation irr\'eductible du groupe de Lorentz
est caract\'eris\'ee par deux nombres $(n\in \N/2,\mu\in \C)$.
Une base orthonormale utile
de l'espace d'Hilbert correspondant $\H_{n,\mu}$
est donn\'e par sa d\'ecomposition en repr\'esentations
$V^j$ de $su(2)$ (g\'en\'er\'e  par les $\vec{J}$):
\beq
\{ \xi_{j,m}\},\qquad m=-j,-j+1,\dots,j-1,j, \quad
j=n,n+1,\dots
\eeq
L'action des g\'en\'erateurs sur cette base est:
\beqs
H_3\xi_{j,m}&=& m\xi_{j,m}, \nonumber \\
H_+\xi_{j,m}&=& \sqrt{(j+m+1)(j-m)}\xi_{j,m+1}, \label{gauss-rep} \\
H_-\xi_{j,m}&=& \sqrt{(j+m)(j-m+1)}\xi_{j,m-1}, \nonumber \\
F_3\xi_{j,m}&=& \gamma_{(j)}\sqrt{j^2-m^2}\xi_{j-1,m}+\beta_{(j)}m\xi_{j,m}
-\gamma_{(j+1)}\sqrt{(j+1)^2-m^2}\xi_{j+1,m}, \nonumber \\
F_+\xi_{j,m}&=&
\gamma_{(j)}\sqrt{(j-m)(j-m-1)}\xi_{j-1,m+1}+\beta_{(j)}\sqrt{(j-m)(j+m+1)}
\xi_{j,m+1} \nonumber \\
&+& \gamma_{(j+1)}\sqrt{(j+m+1)(j+m+2)}\xi_{j+1,m+1}, \label{boosts-rep} \\
F_-\xi_{j,m}&=&
-\gamma_{(j)}\sqrt{(j+m)(j+m-1)}\xi_{j-1,m-1}+\beta_{(j)}\sqrt{(j+m)(j-m+1)}
\xi_{j,m-1} \nonumber \\
&-& \gamma_{(j+1)}\sqrt{(j-m+1)(j-m+2)}\xi_{j+1,m-1}, \nonumber
\eeqs
avec
\beq
\beta_{(j)}=-\frac{in\mu}{j(j+1)}, \qquad
\gamma_{(j)}=\frac{i}{2j}
\sqrt{\frac{(j^2-n^2)(j^2-\mu^2)}{
\left(j-\f{1}{2}\right)
\left(j+\f{1}{2}\right)
}}.
\eeq
Les Casimirs valent $C_1=n^2+\mu^2-1$ et $C_2=in\mu$.
\makeatletter
\renewcommand{\theenumi}{\arabic{enumi}}
\renewcommand{\labelenumi}{\theenumi.}
Les repr\'esentations unitaires correspondent aux deux cas:
\begin{enumerate}
\item la s\'erie principale:
$(n,\mu)=(n,i\rho)$ avec $n\in \N/2$ et $\rho\in \R$
\item la s\'erie compl\'ementaire:
$(n,\mu)=(0,\rho)$ avec $|\rho|<1$ et $\rho\in \R$
\end{enumerate}
Les repr\'esentations de la s\'erie principale sont
celles intervenant dans la formule de Plancherel d\'ecomposant
des fonctions $L^2$ sur $\slc$, et donc celles que l'on utilise pour
les r\'eseaux de spin $\slc$ (base de l'espace $L^2$ des fonctions invariantes)
tels qu'introduits dans la partie II.
Plus pr\'ecis\'ement, pour une fonction $f\in L^2(\slc)$, on a
\beq
f(g)=\f{1}{8 \pi^4}\sum_n \int {\rm
Tr}[F(n,\rho)D^{n,\rho}(g^{-1})](n^2+\rho^2) d\rho
\eeq
o\`u $(n^2+\rho^2) d\rho$ est la mesure de Plancherel et
la transform\'ee de Fourier $F$ est d\'efinie par:
\beq
F(n,\rho)=\int f(g) D^{n,\rho}(g) dg.
\eeq

Les repr\'esentations simples sont les repr\'esentations de la
s\'erie principale avec un Casimir nul $C_2=n\times\rho=0$.
Elles sont de deux types: $(n,0)$ and $(0,i\rho)$.
Dans les deux cas, on a $\beta_{(j)}=0$ pour tout $j$.
Toutefois, les repr\'esentations $(0,i\rho)$ ont la particularit\'e
qu'elles poss\`edent un vecteur $\xi_{j=m=0}$ invariant sous $\su$.

\section*{Le quotient $SO(3,1)/SO(3)$}

Pour $x$ sur l'hyperbolo\"\i de futur ${\cal H}^+$ i.e
un vecteur unit\'e de genre temps orient\'e vers le futur,
on d\'efinit son stabilisateur sous l'action
$x\arr g.x$ de $SO(3,1)$:
\beq
SO(3)_x=\{h\in SO(3,1) / h.x = x\}
\eeq
La loi de transformation de $SO(3)_x$ sous l'action de
 $SO(3,1)$ est
\beq
\forall g\in SO(3,1), \, SO(3)_{g.x}=gSO(3)_xg^{-1}
\label{so3tf}
\eeq
Alors ${\cal H}^+$  peut \^etre vu comme le quotient $SO(3,1)/SO(3)$
de $SO(3,1)$ par l'action \`a gauche du sous-groupe $SO(3)$
-le sous-groupe canonique  $SO(3)$  i.e le stabilisateur de
$x_0=(1,0,0,0)$.
Alors pour $[g_0]\in  SO(3,1)/SO(3)$ et $g\in SO(3,1)$, on a:
\beq
SO(3)_{[g_0]}=\{k\in SO(3,1) / k.[g_0]=[g_0]\}
=\{k\in SO(3,1) / SO(3)g_0k= SO(3)g_0\}
\eeq
\beq
SO(3)_{g.[g_0]}=SO(3)_{[g_0g^{-1}]}=gSO(3)_{[g_0]}g^{-1}
\eeq
Maintenant, si on consid\`ere une repr\'esentation unitaire  $R^\I$
de $SO(3,1)$, on peut la d\'ecomposer sur les repr\'esentations
irr\'eductibles unitaires $SO(3)_x$
pour un vecteur fix\'e $x$:
\beq
R^\I=\bigoplus_{j}V^j_{(x)}
\eeq
Ainsi on peut choisir une base $|\I x j m\rangle$ de $R^\I$ compatible avec
la d\'ecomposition en somme directe des espaces $V^j_{(x)}$.
Alors, \`a la vue de \Ref{so3tf}, les bases pour diff\'erents choix de $x$
sont reli\'ees par
\beq
|\I (g.x) j m\rangle =D^\I(g)|\I x j m\rangle.
\eeq

\medskip
Maintenant, si on regarde les fonctions sur $\H_+$,
une base de l'espace $L^2(\H_+)$ est fournie par
 les repr\'esentations simples $(0,\rho)$. Ces repr\'esentations
contiennent un vecteur invariant sous $\su$, not\'e $|\rho \,j=0\rangle$,
et se d\'ecompose sur
{\it toutes} les repr\'esentations de $\su$:
\beq
\forall x\in\H_+\,
R^{(0,\rho)}=\bigoplus_{j\ge0}V^j_{(x)}.
\eeq
On peut alors d\'efinir un {\it noyau} bi-invariant sous $\su$:
\beq
K_\rho(g)=\langle \rho \,0|
D^{(0,\rho)}(g)|\rho \,0\rangle
\eeq
D\^u \`a son invariance, c'est une fonction de seul l'angle de boost
de $g$. Il est tr\`es utile quand on \'etudie les fonctions $L^2$ sur
${\cal H}_+$ (c'est une base des fonctions $L^2$ bi-invariantes)
et correspond (plus ou moins) au propagateur
d'une particle libre sur l'hyperbolo\"\i de.
Une expression explicite du noyau est
\beq
K_\rho(x,y)=
K_\rho(x^{-1}y)
=\f{\sin\rho \eta}{\rho\sinh\eta}
\eeq
o\`u $\eta$ est la distance hyperbolique (param\`etre de boost)
entre $x$ et $y$.
On peut aussi v\'erifier que
\beq
\int_0^\infty \rho^2 d\rho \, K_\rho(x,y)=
2\pi^2\delta_{{\cal H}_+}(x,y)
\eeq
o\`u $d\mu(\rho)=\rho^2d\rho$ est la mesure de Plancherel
restreinte aux repr\'esentations simples $(n=0,\rho)$.
De plus
\beq
\int_{{\cal H}_+}dy K_\rho(x,y)
K_{\rho'}(y,z)=\f{\delta(\rho-\rho')}{\rho^2}
K_\rho(x,z)
\eeq
ce qui signifie que le noyau $K$ est le projecteur sur la composante
de Fourier $\rho$ sur l'hyperbolo\"\i de ${\cal H}_+$.

\chapter{Quantification G\'eom\'etrique d'un Bivecteur}

Je d\'ecris la quantification (g\'eom\'etrique) d'un bivecteur et d'un t\'etr\`edre
en trois dimensions et en quatre dimensions utilisant la structure de Poisson de
Kirillov-Konstant sur les alg\`egres $so(3)$ et $so(4)$ suivant le travail de Baez et
Barrett \cite{bb}.

Tout d'abord, \'etant donn\'e une alg\`ebre de Lie ${\bf g}$, le commutateur sur ${\bf g}$
induit une structure de Poisson sur son dual ${\bf g}^*$, dite de Kirillov-Konstant.
Plus pr\'ecis\'ement, des \'el\'ements $a,b\in{\bf g}$ sont des fonctions
lin\'eaires sur le dual ${\bf g}^*$. Cela permet de d\'efinir un crochet de Poisson sur
les fonctions lin\'eaires, qui s'\'etend de fa\c con unique \`a l'alg\`ebre des
fonctions ${\cal C}^\infty$. En effet, il existe un champ de bivecteurs $\Omega$ sur
${\bf g}^*$ tel que
$$
[a,b]=\Omega(da,db),
$$
puis le crochet de Poisson est
$$
\{f,g\}=\Omega(df,dg).
$$
Pour $f$ donn\'ee, on peut alors d\'efinir un champs de vecteurs $f^\#$ v\'erifiant
$$
\forall g,\,\{f,g\}=dg(f^\#)=f^\#g.
$$
Une 2-forme $\omega$ est dite compatible avec la structure de Poisson ssi:
$$
\om(a^\#,b^\#)=\{a,b\}.
$$
$\om$ est bien d\'efinie et est symplectique sur chaque orbite co-adjointe,
\'egalement appell\'ee feuille symplectique\footnote{En effet, $\Omega$ est tangent \`a
chaque feuille et non-d\'eg\'en\'er\'ee en tant que forme bilin\'eaire sur le
fibr\'e cotangent \`a chaque feuille. Alors $\om$ est son inverse.}.
Puis, il s'agit de quantifier ce crochet de Poisson. On suit la proc\'edure de quantification
g\'eom\'etrique.
Sur chaque feuille, on choisit une structure complexe $J$ compatible avec $\omega$,
qui munit la feuille d'une structure de K\"ahler. On choisit alors un fibr\'e en droite complexe
et holomorphe munie d'une connexion de courbure $\omega$.
Cela n'est possible que si la classe de cohomologie de $\omega$ est un entier.
Puis on obtient un espace d'Hilbert $H$ en consid\'erant l'espace des sections
de carr\'e int\'egrable.

\section{Bivecteur Quantique 3d}
En utilisant l'isomorphisme entre $\Lambda^2\R^n$ et $so(n)^*$, \'etudier l'espace
des bivecteurs revient \`a \'etudier l'alg\`ebre $so(n)^*$. En trois dimensions, on regarde
donc $so(3)^*$, qui est de dimension 3.
Les feuilles symplectiques de classe de cohomlogie enti\`ere
sont les sph\`eres ${\cal S}^2$ (centr\'ees sur l'origine) de rayon
$2\pi j$, $j\in \N$. Chaque sph\`ere induit alors la repr\'esentation de spin $j$ de $SO(3)$.
Et sommant sur toutes les orbites, on obtient
{\it l'espace d'Hilbert d'un bivecteur quantique en 3 dimensions}:
\beq
{\cal H}=\bigoplus_j V^j.
\eeq
Choisissant la base canonique de l'alg\`ebre de Lie $so(3)$, le crochet de
Poisson est:
\beq
\{E^1,E^2\}=E^3 \qquad \{E^2,E^3\}=E^1 \qquad \{E^3,E^1\}=E^2.
\eeq
Quantifiant, on obtient les relations de commutation:
\beq
[\what{E}^1,\what{E}^2]=i\what{E}^3 \qquad
[\what{E}^2,\what{E}^3]=i\what{E}^1 \qquad
[\what{E}^3,\what{E}^1]=i\what{E}^2.
\eeq

\medskip

A partir du cas du bivecteur, on peut former un t\'etra\`edre en tant que 4 copies
de $so(3)^*$.
En effet, un t\'etra\`edre peut \^etre d\'efini \`a partir de 4 bivecteurs $E_{1,..4}$
satisfaisant une relation de {\it fermeture}:
$$
E_1+E_2+E_3+E_4=0.
$$
En fait, il faut aussi une condition de positivit\'e \'etant
donn\'e que:
$$
U=E_1\cdot(E_2\w E_3)=V^2\ge 0
$$
o\`u $V$ est (6 fois) le volume (orient\'e) du t\'etra\`edre.
On munit $(so(3)^*)^4$ de la structure de Poisson de Kirillov-Konstant et on
consid\`ere  la sous-vari\'et\'e contrainte
$C=\{E_1+E_2+E_3+E_4=0\}$. La contrainte g\'en\`ere l'action diagonale de $SO(3)$ sur
les $E_i$. On  obtient donc l'espace des phases r\'eduit $T$ (comme t\'etra\`edre) en
quotientant $C$ par cette action.

Pour quantifier le t\'etra\`edre, on commence par quantifier les 4 copies de $so(3)^*$.
On obtient donc des op\'erateurs $\what{E}^i_1,..,\what{E}^i_4$ sur ${\cal H}^{\otimes 4}$:
$$
\begin{array}{c}
\what{E}^i_1=\what{E}^i\otimes Id \otimes Id \otimes Id \\
\what{E}^i_2=Id \otimes\what{E}^i\otimes Id \otimes Id \\
\what{E}^i_3=Id \otimes Id \otimes\what{E}^i\otimes Id\\
\what{E}^i_4=Id \otimes Id \otimes Id \otimes\what{E}^i.
\end{array}
$$
Puis on impose la contrainte de fermeture en s\'electionnant les \'etats de
${\cal H}^{\otimes 4}$ annulant l'action de
$\what{E}^i_1+\what{E}^i_2+\what{E}^i_3+\what{E}^i_4$. Ces \'etats sont pr\'ecis\'ement ceux
incvariant sous l'action (diagonale) de $SO(3)$. On obtient ainsi l'espace d'Hilbert:
\beq
{\cal T}={\rm Inv}({\cal H}^{\otimes 4})
=\bigoplus_{j_1,..,j_4}{\rm Inv}(V^{j_1}\otimes V^{j_2}\otimes V^{j_3}\otimes V^{j_4}),
\eeq
qui est en fait l'espace des entrelaceurs 4-valents.
Pour la condition de positivit\'e, il s'agit de quantifier $U$. L'op\'erateur correspondant est
$$
\what{U}=\sum_{i,j,k}\epsilon_{ijk}\what{E}^i_1\what{E}^j_2\what{E}^k_3.
$$
Il faudrait alors se restreindre aux vecteurs propres correspondant aux valeurs
propres positives.

\section{Bivecteur Quantique 4d}

Maintenant, il s'agit de quantifier $so(4)^*$. Pour cela, on utilise que
$so(4)=so(3)\oplus so(3)$. Il suffit donc de copier les r\'esultats pr\'ec\'edents.
N\'eanmoins, il y a une ambigu\"it\'e dans l'isomorphisme entre $\Lambda^2\R^4$ et $so(4)^*$.
Ainsi, choisir $\theta$ ou $\theta\circ \star$ ($\star$ est l'op\'erateur de Hodge sur
$\Lambda^2\R^4$) modifie la structure complexe. Au final, cela revient \`a consid\'erer
l'espace d'Hilbert
$$
\H\otimes\H=\bigoplus_{j,k}V^j\otimes V^k
$$
ou l'espace d'Hilbert
$$
\H\otimes\overline{\H}=\bigoplus_{j,k}V^j\otimes \overline{V^k}.
$$
Nous allons voir que c'est le second choix qui est le plus utile pour l'\'etude du
t\'etra\`edre. Concentrons-nous donc sur ce cas. Choisissant la base canonique
$E^{\pm i}$ de $so(4)$, le crochet de Poisson est:
\beq
\{E^{+i},E^{+j}\}=\epsilon^{ijk}E^{+k} \qquad
\{E^{-i},E^{-j}\}=\epsilon^{ijk}E^{-k} \qquad
\{E^{+i},E^{-j}\}=0,
\eeq
qui se quantifie de mani\`ere \'evidente.

Pour l'application au t\'etra\`edre et au 4-simplex, nous nous int\'eressons
particuli\`erement \`a l'espace des bivecteurs simples. Rappelons qu'un bivecteur quelconque
se d\'ecompose en parties selfdual et antiselfdual $b=b^++b^-$ (avec $\star b^\pm=\pm b^\pm$)
et qu'un bivecteur simple est tel que $|b^+|^2=|b^-|^2$.
On peut donc introduire les op\'erateurs Casimirs
$$
K^\pm=\sum_{i=1,2,3}\left(\what{E}^{\pm i}\right)^2,
$$
et l'{\it espace d'Hilbert d'un bivecteur simple} est l'espace des vecteurs $\psi$ tel que
$K^+\psi=K^-\psi$, ce qui s\'electionne le sous-espace
\beq
\bigoplus_j V^j\otimes \overline{V^j} \subset \H\otimes\overline{\H}.
\eeq

\medskip

Un t\'etra\`edre en quatre dimensions est maintenant form\'e de 4 bivecteurs satisfaisant
\makeatletter
\renewcommand{\theenumi}{\arabic{enumi}}
\renewcommand{\labelenumi}{(\theenumi)}
\begin{enumerate}
\item une contrainte de fermeture $E_1+E_2+E_3+E_4=0$.
\item une contrainte de simplicit\'e sur chaque bivecteur $|E_i^+|^2=|E_i^-|^2$.
\item une contrainte de simplicit\'e sur chaque somme de bivecteurs
$|E_i^++E_j^+|^2=|E_i^-+E_j^-|^2$.
\item une contrainte de chiralit\'e $U^++U^-=0$, avec
$U^\pm=\pm\la E^\pm_1,[E^\pm_2,E^\pm_3]\ra$.
\end{enumerate}
La contrainte de chiralit\'e permet de distinguer si le t\'etra\`edre est form\'e
\`a partir des bivecteurs $E_i$ (ou $-E^i$) ou des bivecteurs $\star E_i$ (ou $-\star E_i$).
Imposant la contrainte de fermeture sur l'espace des phases, on obtient un espace des phases
$T^+\times T^-$ simple double copie de l'espace des phases $T$ du t\'etra\`edre 3d.
Puis il est simple de v\'erifier que les contraintes de simplicit\'e $C_i=|E_i^+|^2-|E_i^-|^2$
ont un crochet de Poisson nul avec toutes les fonctions sur $T^+\times T^-$.
On peut donc se placer imm\'ediatement sur l'espace $\{C_i=0\}\subset T^+\times T^-$
Par contre, pour les contraintes $C_{ij}=|E_i^++E_j^+|^2-|E_i^-+E_j^-|^2$, on obtient
\beq
\{C_{ij},C_{ik}\}=4(U^++U^-).
\eeq
Ainsi, imposer les contraintes de simplicit\'e, on obtiendra
gratuitement la contrainte de chiralit\'e.

Quantifiant, imposant la contrainte de fermeture et les contraintes de simplicit\'e $C_i$,
on se r\'eduit \`a l'espace d'Hilbert:
\beq
\bigoplus_{j_1,..j_4}{\rm Inv}
\left((j_1,j_1)\otimes(j_2,j_2)\otimes(j_3,j_3)\otimes(j_4,j_4)\right)
\subset {\cal T}\otimes \overline{{\cal T}}
\subset (\H\otimes\overline{\H})^{\otimes 4},
\eeq
o\`u on note $(j,j)=V^j\otimes\overline{V^j}$.
Imposant les contraintes $\what{C}_{ij}$, on r\'eduit cet espace d'entrelaceurs \`a
un espace uni-dimensionnel, qui est l'{\it espace d'Hilbert du
t\'etra\`edre quantique en 4 dimensions}. Cet espace prend en compte gratuitement
la contrainte de chiralit\'e puisque:
\beq
4i[\what{C}_{ij},\what{C}_{ik}]=\what{U}^++\what{U}^-,
\eeq
o\`u $\what{U}^\pm=\sum_{i,j,k}\epsilon_{ijk}\what{E}^{\pm i}_1
\what{E}^{\pm j}_2\what{E}^{\pm k}_3$.
Ainsi, tout \'etat satisfaisant les contraintes de simplicit\'e r\'esoud \'egalement
la contrainte de chiralit\'e.


\bibliographystyle{amsordx}

\begin{thebibliography}{10}
\bibitem{carlo:hist1}
C Rovelli, {\it Strings, loops and others: a critical survey of the present approaches to quantum gravity}, gr-qc/9803024

\bibitem{carlo:hist2}
C Rovelli, {\it Notes for a brief history of quantum gravity},
gr-qc/0006061

\bibitem{lee:book}
L Smolin, {\it Three Roads to Quantum Gravity},
Weidenfeld \& Nicolson General, Orion Publishing Co, 2000

\bibitem{carlo&lee}
C Rovelli, L Smolin,
{\it Discreteness of area and volume in quantum gravity},
gr-qc/9411005,
Nucl.Phys. {\bf B442} (1995) 593-622; Erratum-ibid. {\bf B456} (1995) 753

\bibitem{carlo:bh1}
C Rovelli,
{\it Black Hole Entropy from Loop Quantum Gravity},
 gr-qc/9603063,Phys.Rev.Lett. {\bf 77} (1996) 3288-3291

\bibitem{carlo:bh2}
M Barreira, M Carfora, C Rovelli,
{\it Physics with nonperturbative quantum gravity: radiation from a quantum black hole},
gr-qc/9603064, Gen.Rel.Grav. {\bf 28} (1996) 1293-1299

\bibitem{ash:qg}
A Ashtekar, {\it Quantum Mechanics of Geometry},
gr-qc/9901023

\bibitem{carlo:lqg}
C Rovelli, {\it Loop Quantum Gravity},
gr-qc/9710008, Living Rev.Rel. {\bf 1} (1998) 1

\bibitem{carlo:cours}
M Gaul, C Rovelli, {\it Loop Quantum Gravity and the Meaning of Diffeomorphism Invariance}, gr-qc/9910079, Lect.Notes Phys. {\bf 541} (2000) 277-324

\bibitem{thiemann:long}
T Thiemann, {\it Introduction to Modern Canonical Quantum General Relativity},
gr-qc/0110034

\bibitem{thiemann:short}
T Thiemann, {\it Lectures on Loop Quantum Gravity}, gr-qc/0210094

\bibitem{lee:lqg}
L Smolin, {\it Quantum gravity with a positive cosmological constant},
hep-th/0209079

\bibitem{dirac}
HJ Matschull, {\it Dirac's Canonical Quantization Programme},
quant-ph/9606031

\bibitem{holst}
S Holst, {\it Barbero's Hamiltonian derived from a generalized Hilbert-Palatini action},
gr-qc/9511026,
Phys.Rev. {\bf D53} (1996) 5966-5969

\bibitem{reis:constraints}
M Reisenberger, {\it New constraints for canonical general relativity},
gr-qc/9505044,
Nucl.Phys. {\bf B457} (1995) 643-687

\bibitem{barros}
N Barros e Sa, {\it Hamiltonian analysis of General relativity with the Immirzi parameter},
gr-qc/0006013, Int.J.Mod.Phys. {\bf D10} (2001) 261-272

\bibitem{3+1}
S Alexandrov, ER Livine,
{\it $\su$ Loop Quantum Gravity seen from Covariant Theory},
gr-qc/0209105,
Phys.Rev. {\bf D67} (2003) 044009

\bibitem{carlo:report}
C Rovelli, {\it Ashtekar formulation of general relativity and
  loop-space non-perturbative quantum gravity: a report}, with
corrections and modifications by M Reisenberger, 1998

\bibitem{sergei1}
S Alexandrov,
{\it $SO(4,C)$-covariant Ashtekar-Barbero gravity and the Immirzi
  parameter},
gr-qc/0005085,
 Class.Quant.Grav. {\bf 17} (2000) 4255-4268

\bibitem{barbero}
JF Barbero,
{\it Reality Conditions and Ashtekar Variables: a Different Perspective},
gr-qc/9410013 ,
Phys.Rev. {\bf D51} (1995) 5498-5506;
{\it Real Ashtekar Variables for Lorentzian Signature Space-times},
gr-qc/9410014,
Phys.Rev. {\bf D51} (1995) 5507-5510

\bibitem{immirzi}
G Immirzi,
{\it Real and complex connections for canonical gravity},
gr-qc/9612030,
Class.Quant.Grav. {\bf 14} (1997) L177-L181

\bibitem{carlo&tom}
 C Rovelli, T Thiemann,
{\it  The Immirzi parameter in quantum general relativity},
gr-qc/9705059,
Phys.Rev. {\bf D57} (1998) 1009-1014

\bibitem{bh}
A Ashtekar, J Baez, K Krasnov,
{\it Quantum Geometry of Isolated Horizons and Black Hole Entropy},
gr-qc/0005126,
Adv.Theor.Math.Phys. {\bf 4} (2000) 1-94

\bibitem{olaf}
O Dreyer,
{\it Quasinormal Modes, the Area Spectrum, and Black Hole Entropy},
gr-qc/0211076

\bibitem{samuel}
J Samuel,
{\it Is Barbero's Hamiltonian formulation a Gauge Theory of Lorentzian
  Gravity?},
gr-qc/0005095,
Class.Quant.Grav. {\bf 17} (2000) L141-L148

\bibitem{sergei2}
S Alexandrov, D Vassilevich,
{\it Area spectrum in Lorentz covariant loop gravity},
gr-qc/0107071,
Phys.Rev. {\bf D64} (2001) 044023

\bibitem{sergei3}
S Alexandrov,
{\it On choice of connection in loop quantum gravity},
gr-qc/0107071,
Phys.Rev. {\bf D65} (2002) 024011

\bibitem{holo-flux}
H Sahlmann, {\it When Do Measures on the Space of Connections Support
  the Triad Operators of Loop Quantum Gravity?}, gr-qc/0207112; \\
A Okolow, J Lewandowski,
{\it Diffeomorphism covariant representations of the holonomy-flux star-algebra},
gr-qc/0302059

\bibitem{carlo:area}
C Rovelli, P Upadhya,
{\it Loop quantum gravity and quanta of space: a primer},
 gr-qc/9806079

\bibitem{ash-lew}
A Ashtekar, J Lewandowski,
{\it Projective Techniques and Functional Integration},
gr-qc/9411046,
J.Math.Phys. {\bf 36} (1995) 2170-2191

\bibitem{ash-lew2}
A Ashtekar, J Lewandowski,
{\it Representation Theory of Analytic Holonomy $C*$ Algebras},gr-qc/9311010,
publi{\'e} dans ``Knots and Quantum Gravity'' (ed. J.Baez, Oxford U.Press)

\bibitem{gns}
M Arnsdorf, S Gupta,
{\it Loop Quantum Gravity on Non-Compact Spaces},
gr-qc/9909053,
Nucl. Phys. {\bf B577} (2000) 529-546

\bibitem{grot}
N Grot, C Rovelli,
{\it Moduli-space structure of knots with intersections},
gr-qc/9604010,
J.Math.Phys. {\bf 37} (1996) 3014-3021

\bibitem{thiemann:qsd}
T Thiemann,
{\it Quantum Spin Dynamics (QSD)},
gr-qc/9606089,
Class.Quant.Grav. {\bf 15} (1998) 839-873;
{\it Quantum Spin Dynamics (QSD) II},
 gr-qc/9606090,
Class.Quant.Grav. {\bf 15} (1998) 875-905

\bibitem{borissov}
R Borissov,
{\it Graphical Evolution of Spin Network States},
gr-qc/9606013,
Phys.Rev. {\bf D55} (1997) 6099-6111

\bibitem{lee:hamil}
L Smolin,
{\it The classical limit and the form of the hamiltonian constraint in nonperturbative quantum gravity},
gr-qc/9609034 

\bibitem{gaul}
M Gaul, C Rovelli,
{\it A generalized Hamiltonian Constraint Operator in Loop Quantum Gravity and its simplest Euclidean Matrix Elements},
gr-qc/0011106,
Class.Quant.Grav. {\bf 18} (2001) 1593-1624

\bibitem{borissov&gupta}
R Borissov, S Gupta,
{\it Propagating spin modes in canonical quantum gravity},
gr-qc/9810024,
Phys.Rev. {\bf D60} (1999) 024002

\bibitem{holovariation}
J Lewandowski, ET Newman, C Rovelli, {\it Variation of the parallel
propagator and holonomy operator and the Gauss law constraint},
J Math Phys 34 (1993) 4646.

\bibitem{alekseev}
 A Alekseev, AP Polychronakos, M Smedb\"ack,
{\it On area and entropy of a black hole},
hep-th/0004036

\bibitem{laurent:sf}
 L Freidel, K Krasnov,
{\it Spin Foam Models and the Classical Action Principle},
hep-th/9807092, 
Adv.Theor.Math.Phys. {\bf 2} (1999) 1183-1247

\bibitem{carlo:vol}
R De Pietri, C Rovelli,
{\it Geometry Eigenvalues and Scalar Product from Recoupling Theory in Loop Quantum Gravity},
gr-qc/9602023,
Phys.Rev. {\bf D54} (1996) 2664-2690

\bibitem{vol:thesis}
M Seifert,
{\it Angle and Volume Studies in Quantized Space},
gr-qc/0108047

\bibitem{kodama}
H Kodama,
{\it Quantum Gravity by the Complex Canonical Formulation},
gr-qc/9211022,
Int.J.Mod.Phys. {\bf D1} (1992) 439-524

\bibitem{soo}
C Soo,
{\it Wave function of the Universe and Chern-Simons Perturbation Theory},
gr-qc/0109046,
Class.Quant.Grav. {\bf 19}
 (2002) 1051-1064

\bibitem{carlo&reis}
M Reisenberger, C Rovelli,
{\it ``Sum over Surfaces'' form of Loop Quantum Gravity},
gr-qc/9612035,
 Phys.Rev. {\bf D56} (1997) 3490-3508

\bibitem{baez:sf}
J Baez,
{\it Spin Foam Models},
gr-qc/9709052,
Class.Quant.Grav. {\bf 15} (1998) 1827-1858

\bibitem{psn}
ER Livine,
{\it Projected Spin Networks for Lorentz connection: Linking Spin Foams and Loop Gravity},
gr-qc/0207084,
Class.Quant.Grav. {\bf 19} (2002) 5525-5542



\bibitem{spinnet}
L Freidel, ER Livine, {\it Spin Networks for Non-Compact Groups},
hep-th/0205268,
J.Math.Phys. {\bf 44} (2003) 1322-1356

\bibitem{brion}
M Brion, {\it Invariants et covariants des groupes alg{\'e}briques
r{\'e}ductifs}, Notes de cours de l'{\'e}cole d'{\'e}t{\'e} de Monastir (Juillet-Aout
1996)

\bibitem{spring}
TA Springer {\it Invariant theory}
Lecture Notes in Mathematics, N 585

\bibitem{baez:spinnet}
JC Baez, gr-qc/9504036,
{\it Spin networks in nonperturbative quantum gravity},
The Interface of Knots and Physics, ed. Louis Kauffman, A.M.S., Providence,
1996, pp. 167-203

\bibitem{frolich}
J Frohlich, K Gawedzki, hep-th/9310187,
{\it Conformal Field Theory and Geometry of Strings}, Vancouver
1993, Proceedings, Mathematical quantum theory, vol. 1, p.57-97

\bibitem{karim}
E Buffenoir, K Noui, P Roche,
{\it Hamiltonian Quantization of Chern-Simons theory with $SL(2,\C)$ Group},
hep-th/0202121,
Class.Quant.Grav. {\bf 19} (2002) 4953


\bibitem{gomb}
A Gomberoff, D Marolf,
{\it On Group Averaging for SO(n,1) }, gr-qc/9902069,
 Int.J.Mod.Phys. {\bf D8} (1999) 519-535

\bibitem{vara}
VS Varadarajan,
{\it An introduction to harmonic analysis on semisimple Lie g
roups} Cambridge studies in Adv.Math. {\bf 16}, Cambridge University Press




\bibitem{carlip}
S Carlip,
{\it Quantum Gravity in 2 +1 Dimensions},
Cambridge University Press (1998)

\bibitem{witten:2+1}
E Witten,
{\it $(2+1)$-Dimensional Gravity as an Exactly Soluble System},
Nucl.Phys. {\bf B311} (1988) 46 

\bibitem{2+1bis}
ER Livine, C Rovelli,
{\it Length versus time-interval discreteness in 2+1 Lorentzian canonical quantum gravity}, rapport interne CPT-2001/P.4176

\bibitem{2+1}
L Freidel, ER Livine, C Rovelli,
{\it Spectra of Length and Area in 2+1 Lorentzian Loop Quantum Gravity},
gr-qc/0212077,
Class.Quant.Grav. {\bf 20} (2003) 1463

\bibitem{2+1toymodel}
A Ashtekar, V Husain, C Rovelli, J Samuel, L Smolin,
{\it $2+1$-Quantum Gravity as a Toy Model for the $3+1$ Theory},
Class.Quant.Grav. {\bf 6} (1989) 185 

\bibitem{jacek}
J Wisniewski,
{\em 2+1 General Relativity: Classical and Quantum},
PhD Thesis (PennState University, 2002)

\bibitem{davids}
S Davids,
{\it Semiclassical Limits of Extended Racah Coefficients},
gr-qc/9807061, J.Math.Phys. {\bf 41} (2000) 924-943;
{\it A State Sum Model for (2+1) Lorentzian Quantum Gravity},
gr-qc/0110114

\bibitem{laurent:2+1}
L Freidel,
{\it A Ponzano-Regge model of Lorentzian 3-Dimensional gravity},
gr-qc/0102098,
Nucl.Phys.Proc.Suppl. {\bf 88} (2000) 237-240

\bibitem{thiemann:2+1}
T Thiemann,
{\it QSD IV: $2+1$ Euclidean Quantum Gravity as a model
to test $3+1$ Lorentzian Quantum Gravity},
gr-qc/9705018,
Class.Quant.Grav. {\bf 15} (1998) 1249-1280

\bibitem{ooguri}
H Ooguri,
{\it Discrete and Continuum Approaches to Three-Dimensional Quantum Gravity},
hep-th/9108006,
 Mod. Phys. Lett. {\bf A6} (1991) 3591-3600;
{\it Partition Functions and Topology-Changing Amplitudes in the 3D Lattice Gravity of Ponzano and Regge},
hep-th/9112072,
Nucl.Phys. {\bf B382} (1992) 276-304

\bibitem{laurent:asymp}
L Freidel, D Louapre,
{\it  Asymptotics of 6j and 10j symbols},
hep-th/0209134

\bibitem{hooft}
G 't Hooft,
{\it Canonical Quantization of Gravitating Point Particles
in $2+1$ Dimensions}, gr-qc/9305008,
Class.Quant.Grav. {\bf 10} (1993) 1653-1664;
{\it The Evolution of Gravitating Point Particles in $(2+1)$-dimensions},
Class.Quant.Grav. {\bf 10} (1993) 1023-1038;
{\it Causality in $(2+1)$-dimensional Gravity},
Class.Quant.Grav. {\bf 9} (1992) 1335-1348 

\bibitem{knapp}
AW Knapp
{\it Representation of semi-simple groups},
Princeton University Press

\bibitem{sergei:pathintegral}
S Alexandrov, D Vassilevich,
{\it Path integral for the Hilbert-Palatini and Ashtekar gravity},
gr-qc/9806001,
Phys.Rev. {\bf D58} (1998) 124029



\bibitem{dan:review}
D Oriti,
{\it Spacetime geometry from algebra: spin foam models for non-perturbative quantum gravity},
gr-qc/0106091, Rept.Prog.Phys. {\bf 64} (2001) 1489-1544

\bibitem{alej:review}
A Perez,
{\it Spin Foam Models for Quantum Gravity}
gr-qc/0301113,
Class.Quant.Grav. {\bf 20} (2003) R43

\bibitem{bc1}
JW Barrett, L Crane,
{\it Relativistic spin networks and quantum gravity},
gr-qc/9709028 ,
J.Math.Phys. {\bf 39} (1998) 3296-3302

\bibitem{bc2}
JW Barrett, L Crane,
{\it A Lorentzian Signature Model for Quantum General Relativity},
gr-qc/9904025,
Class.Quant.Grav. {\bf 17} (2000) 3101-3118

\bibitem{bf}
R De Pietri, L Freidel,
{\it $so(4)$ Plebanski Action and Relativistic Spin Foam Model},
gr-qc/9804071,
Class.Quant.Grav. {\bf 16} (1999) 2187-2196

\bibitem{bb}
JW Barrett, JC Baez,
{\it The Quantum Tetrahedron in 3 and 4 Dimensions},
gr-qc/9903060,
Adv.Theor.Math.Phys. {\bf 3} (1999) 815-850

\bibitem{causal1}
ER Livine, D Oriti,
{\it Implementing causality in the spin foam quantum geometry},
gr-qc/0210064

\bibitem{2d}
ER Livine, A Perez, C Rovelli, {\it  2d manifold-independent spinfoam theory},
gr-qc/0102051, rapport interne CPT-2001/P.4138 

\bibitem{diffeo}
L Freidel, D Louapre,
{\it Diffeomorphisms and spin foam models},
gr-qc/0212001

\bibitem{fotini:sf}
F Markopoulou,
{\it Dual formulation of spin network evolution},
 gr-qc/9704013

\bibitem{asymp:ruth&barrett}
JW Barrett, RM Williams,
{\it The asymptotics of an amplitude for the 4-simplex},
gr-qc/9809032,
Adv.Theor.Math.Phys. {\bf 3} (1999) 209-215

\bibitem{causal2}
ER Livine, D Oriti,
{\it Causality in spin foam models for quantum gravity}, gr-qc/0302018,
Proceedings of the 15th SIGRAV Conference on General Relativity
and Gravitational Physics (2003)

\bibitem{asymp:baez}
JC Baez, JD Christensen, G Egan,
{\it Asymptotics of 10j symbols},
gr-qc/0208010,
Class.Quant.Grav. {\bf 19} (2002) 6489

\bibitem{asymp:barrett}
JW Barrett, CM Steele,
{\it Asymptotics of Relativistic Spin Networks},
 gr-qc/0209023


\bibitem{ooguri:gft}
H Ooguri,
{\it Schwinger-Dyson equation in three-dimensional simplicial quantum gravity},
hep-th/9210028,
Prog.Theor.Phys. {\bf 89} (1993) 1-22

\bibitem{summation}
L Freidel, D Louapre,
{\it Non-perturbative summation over 3D discrete topologies},
hep-th/0211026

\bibitem{gft}
R De Pietri, L Freidel, K Krasnov, C Rovelli,
{\it Barrett-Crane model from a Boulatov-Ooguri field theory over a homogeneous space},
hep-th/9907154,
Nucl.Phys. {\bf B574} (2000) 785-806

\bibitem{alej:e}
A Perez, C Rovelli,
{\it A spin foam model without bubble divergences},
gr-qc/0006107,
Nucl.Phys. {\bf B599} (2001) 255-282

\bibitem{alej:l}
A Perez, C Rovelli,
{\it Spin foam model for Lorentzian General Relativity},
gr-qc/0009021,
Phys.Rev. {\bf D63} (2001) 041501

\bibitem{carlo:gft}
MP Reisenberger, C Rovelli,
{\it Spin foams as Feynman diagrams},
gr-qc/0002083;
{\it Spacetime as a Feynman diagram: the connection formulation},
gr-qc/0002095,
Class.Quant.Grav. {\bf 18} (2001) 121-140

\bibitem{finite:alej}
A Perez,
{\it Finiteness of a spinfoam model for euclidean quantum general relativity},
gr-qc/0011058,
Nucl.Phys. {\bf B599} (2001) 427-434

\bibitem{finite:bb}
JW Barrett, JC Baez,
{\it Integrability for Relativistic Spin Networks},
gr-qc/0101107,
Class.Quant.Grav. {\bf 18} (2001) 4683-4700

\bibitem{finite}
L Crane, A Perez, C Rovelli,
{\it A finiteness proof for the Lorentzian state sum spinfoam model for quantum general relativity},
gr-qc/0104057

\bibitem{observables}
A Perez, C Rovelli,
{\it Observables in quantum gravity},
gr-qc/0104034

\bibitem{bf1}
ER Livine,
{\it Immirzi parameter in the Barrett-Crane model?},
gr-qc/0103081, rapport interne CPT-2001/P.4220 

\bibitem{bf2}
ER Livine, D Oriti,
{\it Barrett-Crane spin foam model from generalized BF-type action for gravity},gr-qc/0104043,
Phys.Rev. {\bf D65} (2002) 044025

\bibitem{2dsurface}
P Kramer, M Lorente,
{\it Surface Embedding, topology and dualization for spin networks},
J.Phys.A:Math.Gen. {\bf 35} (2002) 8563-8574

\bibitem{laurent:vol3d}
L Freidel, K Krasnov,
{\it Discrete Space-Time Volume for 3-Dimensional BF Theory and 
Quantum Gravity},
hep-th/9804185,
Class.Quant.Grav. {\bf 16} (1999) 351-362

\bibitem{boulatov}
DV Boulatov,
{\it A model of three-dimensional Lattice Gravity},
hep-th/9202074,
Int.J.Mod.Phys. {\bf A6} (1991) 79-87

\bibitem{pullin}
R Gambini, J Pullin,
{\it A finite spin-foam-based theory of three and four dimensional quantum gravity},
gr-qc/0111089,
Phys.Rev. {\bf D66} (2002) 024020

\bibitem{reis:bc}
M Reisenberger,
{\it On relativistic spin network vertices},
gr-qc/9809067,
J.Math.Phys. {\bf 40} (1999) 2046-2054

\bibitem{puzio}
L Freidel, K Krasnov, R Puzio,
{\it BF Description of Higher-Dimensional Gravity Theories},
hep-th/9901069,
Adv.Theor.Math.Phys. {\bf 3} (1999) 1289-1324

\bibitem{pleb}
JF Plebanski, {\it On the separation of einstenian
substructures}, J. Math. Phys. {\bf 12}, 2511 (1977)

\bibitem{prieto}
VA Prieto,
{\it Formulaciones reales de la gravedad},
M.S. Thesis, CINVESTAV-IPN (M\'exico D.F., 1997) \\
R Capovilla, M Montesinos,
VA Prieto, E Rojas,
{\it BF gravity and the Immirzi parameter},
gr-qc/0102073,
Class.Quant.Grav. {\bf 18} (2001) L49; Erratum-ibid. {\bf 18} (2001)
1157

\bibitem{reis:sf}
MP Reisenberger,
{\it A lattice worldsheet sum for 4-d Euclidean general relativity},
gr-qc/9711052

\bibitem{alej:sf}
A Perez,
{\it Spin foam quantization of $SO(4)$ Plebanski's action},
gr-qc/0203058,
Adv.Theor.Math.Phys. {\bf 5} (2002) 947-968

\bibitem{dan&ruth}
D Oriti, R Williams,
{\it Gluing 4-simplices: a derivation of the Barrett-Crane spin
foam model for Euclidean quantum gravity},
gr-qc/0010031,
Phys.Rev. {\bf D63} (2001) 024022

\bibitem{dan:gluing}
D Oriti,
{\it Boundary terms in the Barrett-Crane spin foam model and consistent gluing},gr-qc/0201077,
Phys.Lett. {\bf B532} (2002) 363-372

\bibitem{wave}
JW Barrett,
{\it Refractive gravitational waves and quantum fluctuations},
gr-qc/0011051

\bibitem{crane:bc}
L Crane, D Yetter,
{\it A More Sensitive Lorentzian State Sum},
gr-qc/0301017

\bibitem{balone}
JW Barrett,
{\it First order Regge calculus},
hep-th/9404124,
Class. Quant. Grav. {\bf 11}  (1994) 2723


\bibitem{CDM}
M Caselle, A D'Adda, L Magnea,
{\it Regge calculus
as a local theory of the Poincare group},
Phys. Lett. {\bf B232} (1999) 457 

\bibitem{Katsy} VM Khatsymovsky,
{\it Continuous time Regge
gravity in the tetrad connection variables},
Class. Quant. Grav. {\bf 8}  (1991) 1205

\bibitem{Katsy2}
VM Khatsymovsky, {\it Regge calculus in the
canonical form},
gr-qc/9310004,
Gen. Rel. Grav. {\bf 27} (1995) 583

\bibitem{Rov}
C Rovelli, {\it The basis of the
Ponzano-Regge-Turaev-Viro-Ooguri quantum gravity model is the loop
representation basis},
hep-th/9304164,
Phys. Rev. {\bf D48} (1993) 2702

\bibitem{Make1}
J Makela, {\it On the phase space coordinates and
the Hamiltonian constraint of Regge calculus},
Phys. Rev. {\bf D49} (1994) 2882

\bibitem{Make2} J Makela,
{\it Variation of area variables in
Regge calculus},
gr-qc/9801022,
Class. Quant. Grav. {\bf 17} (2000) 4991)

\bibitem{BRW}
JW Barrett, M Rocek, RM Williams,
{\it A
note on area variables in Regge calculus},
gr-qc/9710056,
Class. Quant. Grav. {\bf 16} (1999) 1373 

\bibitem{MW}
J Makela, RM Williams,
{\it Constraints on area
variables in Regge calculus},
gr-qc/0011006,
Class. Quant. Grav. {\bf 18} (2001) L43 

\bibitem{ReggeWill} T Regge, RM Williams,
{\it Discrete
structures in gravity},
gr-qc/0012035,
J. Math. Phys. {\bf 41} (2000) 3964 

\bibitem{ruhl}
W Ruhl, {\it The Lorentz group and harmonic
analysis}, W. A. Benjamin Inc., New York (1970) 

\bibitem{fotini:causal}
F Markopoulou,
{\it An insider's guide to quantum causal histories},
hep-th/9912137,
Nucl. Phys. Proc. Suppl. {\bf 88} (2000) 308;
{\it Quantum causal histories},
hep-th/9904009,
Class. Quant. Grav. {\bf 17} (2000) 2059;
{\it The internal description of a causal set: What the universe
looks like from the inside},
gr-qc/9811053,
Commun. Math. Phys. {\bf 211} (2000) 559


\end{thebibliography}
\end{document}